\RequirePackage{fix-cm} % Allow Computer Modern fonts at nonstandard sizes.

\newif\ifsubmit     % hide comments and move the whole paper to the appendix
\newif\ifllncs      % blarf
\newif\ifexabs      % extended abstract
\newif\ifblind  

\newif\ifProofAudit

\newif\ifhidecomments
\hidecommentstrue

\ifhidecomments
  \ProofAuditfalse
\else
  \ProofAudittrue
\fi

\documentclass[11pt]{article}
\usepackage{libertine}
\usepackage{fullpage}
\usepackage[letterpaper,top=2cm,bottom=2cm,left=3cm,right=3cm,marginparwidth=1.75cm]{geometry}

\usepackage{amsthm}
\usepackage{threeparttable}
\usepackage[T1]{fontenc}

\usepackage{enumitem}
\usepackage{graphicx} % Required for inserting images
\usepackage{physics}
\usepackage{breakcites}
\usepackage{amsmath,bm}
\usepackage{amssymb}

\ifllncs
\else
  \usepackage[libertine]{newtxmath}
\fi

\usepackage{bm}
\usepackage[colorlinks=true, allcolors=blue]{hyperref}
\usepackage[dvipsnames]{xcolor}
\usepackage{bbm}
\usepackage{cleveref}
\usepackage{mathtools}
\usepackage{algorithm, algpseudocode, float, algorithmicx}
\usepackage{mdframed}

\ifllncs
\else
  \usepackage{caption}
\fi

\usepackage{multirow}
\usepackage[normalem]{ulem}
\usepackage{tikz}
\usepackage{quantikz}
\usetikzlibrary{tikzmark}

\ifllncs
\else
\newtheorem{prop}{Proposition}
\newtheorem{remark}{Remark}
\fi
\usepackage[english]{babel}

\ifhidecomments
\newcommand{\justin}[1]{}
\newcommand{\bhaskar}[1]{}
\newcommand{\omri}[1]{}
\newcommand{\jiahui}[1]{}
\else
\newcommand{\justin}[1]{{\color{blue}\noindent[\textbf{Justin:} #1]}}
\newcommand{\authnote}[3]{\textcolor{#3}{[{\footnotesize {\bf #1:} { {#2}}}]}}
\newcommand{\bhaskar}[1]{\authnote{Bhaskar}{#1}{ForestGreen}}
\newcommand{\omri}[1]{\authnote{Omri}{#1}{Red}}
\newcommand{\jiahui}[1]{\authnote{Jiahui}{#1}{magenta}}
\fi

\ifllncs

  \spnewtheorem{claim}{Claim}{\bfseries}{\rmfamily}
  \crefname{claim}{claim}{claims}
  \Crefname{claim}{Claim}{Claims}
\else

\theoremstyle{definition}
\newtheorem{definition}{Definition}[section]
\newtheorem{theorem}{Theorem}[section]
\newtheorem{lemma}{Lemma}[section]
\newtheorem{corollary}{Corollary}[section]
\newtheorem{claim}{Claim}[section]
\crefname{claim}{claim}{claims}
\Crefname{claim}{Claim}{Claims}
\theoremstyle{definition}
\fi

\newcommand{\secp}{\lambda}
\newcommand{\negl}{\mathsf{negl}}
\newcommand{\getsr}{\overset{\$}{\leftarrow}}
\newcommand{\bit}{\{0,1\}}
\newcommand{\pk}{\mathsf{pk}}
\newcommand{\sk}{\mathsf{sk}}

\newcommand{\mode}{\mathsf{mode}}
\newcommand{\Transfer}{\mathsf{Transfer}}
\newcommand{\KeyGen}{\mathsf{KeyGen}}
\newcommand{\TokenGen}{\mathsf{TokenGen}}
\newcommand{\Sign}{\mathsf{Sign}}
\newcommand{\Ver}{\mathsf{Ver}}
\newcommand{\Eval}{\mathsf{Eval}}
\newcommand{\OTP}{\mathsf{OTP}}
\newcommand{\OSP}{\mathsf{OSP}}
\newcommand{\OSS}{\mathsf{OSS}}
\newcommand{\SigToken}{\mathsf{SigToken}}
\newcommand{\BUS}{\mathsf{BUS}}
\newcommand{\sfP}{\mathsf{P}}
\newcommand{\Prove}{\mathsf{Prove}}
\newcommand{\Sim}{\mathsf{Sim}}
\newcommand{\Hyb}{\mathsf{Hyb}}

\newcommand{\PGen}{\mathsf{PGen}}
\newcommand{\Decomp}{\mathsf{Decomp}}
\newcommand{\CO}{\mathsf{CO}}
\newcommand{\SEQ}{\mathsf{SEQ}}

\newcommand{\aux}{\mathsf{aux}}
\newcommand{\crs}{\mathsf{crs}}
\newcommand{\adv}{\mathsf{Adv}}
\newcommand{\PRF}{\mathsf{PRF}}

\newcommand{\bbN}{\mathbb{N}}
\newcommand{\bbR}{\mathbb{R}}

\newcommand{\cA}{\mathcal{A}}
\newcommand{\cB}{\mathcal{B}}
\newcommand{\cC}{\mathcal{C}}
\newcommand{\calC}{\mathcal{C}}
\newcommand{\cD}{\mathcal{D}}

\newcommand{\cF}{\mathcal{F}}
\newcommand{\cG}{\mathcal{G}}

\newcommand{\cM}{\mathcal{M}}

\newcommand{\cO}{\mathcal{O}}
\newcommand{\cP}{\mathcal{P}}
\newcommand{\cQ}{\mathcal{Q}}
\newcommand{\cR}{\mathcal{R}}
\newcommand{\cS}{\mathcal{S}}

\newcommand{\cW}{\mathcal{W}}
\newcommand{\cX}{\mathcal{X}}
\newcommand{\cY}{\mathcal{Y}}
\newcommand{\cZ}{\mathcal{Z}}

\newcommand{\StateD}{\mathsf{State}}
\newcommand{\MaxEntry}{\mathsf{MaxEntry}}
\newcommand{\st}{\mathsf{st}}

\newcommand{\Sig}{\mathsf{Sig}}
\newcommand{\QLP}{\mathsf{QLP}}
\newcommand{\concat}{\|}
\newcommand{\LEQ}{\mathsf{LEQ}}
\newcommand{\REQ}{\LEQ}

\newcommand{\SKE}{\mathsf{SKE}}

\newcommand{\Enc}{\mathsf{Enc}}
\newcommand{\Dec}{\mathsf{Dec}}

\newcommand{\inp}{\mathsf{inp}}
\newcommand{\outp}{\mathsf{outp}}

\newcommand{\Reflect}{\mathsf{Refl}}

\newcommand{\inputcache}{\mathsf{InpC}}
\newcommand{\outputcache}{\mathsf{OutC}}
\newcommand{\serialindex}{\mathsf{SI}}
\newcommand{\early}{\mathsf{early}}

\newcommand{\proofaudit}[3]{%
  \ifProofAudit
    \par\addvspace{0.6\baselineskip}%
    \begingroup\normalfont\small\color{red}%
    \hypersetup{linkcolor=orange,citecolor=orange,urlcolor=orange}%
    \noindent\textbf{[Proof audit #1 --- #2]}\quad\ignorespaces#3\par
    \endgroup\addvspace{0.6\baselineskip}%
  \fi
}
\DeclareRobustCommand{\proofauditloc}[2]{%
  \ifProofAudit
    \ifhmode\unskip\space\else\noindent\fi
    \begingroup\normalfont\footnotesize\color{red}%
    \hypersetup{linkcolor=red,citecolor=red,urlcolor=red}%
    \mbox{\textbf{[Audit #1:}}\space\ignorespaces#2\textbf{]}%
    \endgroup
  \fi\space
}
\title{Query-Limited RAM Programs and their Applications}
\author{Jiahui Liu\\Fujitsu Research \and Justin Raizes\\NTT Research \and Bhaskar Roberts\\NTT Research \and Omri Shmueli\thanks{This work was done in part while the author was a visiting scientist at the Simons Institute for the Theory of Computing at UC Berkeley.}\\NTT Research}
\date{}

\begin{document}
\pagenumbering{roman} 

\maketitle
\begin{abstract}

Quantum one-time programs (Broadbent, Gutoski and Stebila, CRYPTO 2013) or OTPs for short, enable a functionality to be encoded into a quantum token that can be evaluated on a single chosen input and then becomes unusable. While powerful, this primitive is inherently stateless and tied to a setting in which a quantum token needs to be issued and distributed for every single evaluation of a circuit. This raises a natural question: can the one-time computation paradigm be extended to richer, stateful forms of controlled access, and would such an extension offer inherent advantages beyond standard OTPs?

%Quantum one-time programs (Broadbent, Gutoski and Stebila, CRYPTO 2013) or OTPs for short, enable a functionality to be encoded into a quantum token that can be evaluated on a single chosen input and then becomes unusable. While powerful, this primitive is inherently stateless and tied to a setting in which a single generator produces and distributes a quantum state. This raises a natural question: can the one-time computation paradigm be extended to richer, stateful forms of controlled access, and would such an extension offer inherent advantages beyond standard OTPs?

%\ifllncs\vspace{-1mm}\fi
We introduce \emph{query-limited RAM programs} (QLPs), a RAM-generalization of QOTPs that supports structured, stateful computation under bounded or policy-driven access. QLPs allow controlled sequences of evaluations while preventing adversarial forking or rollback of computational state. This enables new applications beyond stateless one-time programs, including quantum tokens for Turing Machines whose size depends only on \emph{code length} (and not runtime), transferable $k$-time or budget-limited programs, and low-communication mechanisms for delegating computation in settings such as Software-as-a-Service.

%In this work, we introduce \emph{query-limited programs} (QLPs), a RAM-generalization of quantum one-time programs that supports structured, stateful computation under bounded or policy-driven access. QLPs allow controlled sequences of evaluations while preventing adversarial forking or rollback of computational state. We show that QLPs enable new applications beyond stateless one-time programs, including quantum $k$-time or budget programs, which can be transferred from user to user with classical communication, minimizing (classical) communication for use cases of Software as a Service (SaaS), and more.

%\ifllncs\vspace{-5mm}\fi
To construct QLPs, we introduce \emph{one-shot programs}, unifying one-shot signatures (Amos, Georgiou, Kiayias and Zhandry, STOC 2020) with the single effective query paradigm (Gupte, Liu, Raizes, Roberts and Vaikuntanathan, STOC 2025). We prove that one-shot programs generically imply query-limited programs, demonstrating that the strengthened unclonability guarantees of one-shot signatures translate into enhanced functionality. Along the way, we clarify the relationship between signature-token primitives and quantum one-time programs via generic constructions, essentially showing that one-time signing programs imply one-time general computation.

\end{abstract}
\normalsize

\ifllncs \else \newpage \tableofcontents \pagebreak \fi
%\pagebreak
\pagenumbering{arabic}

\section{Introduction} \label{section:intro}
One-time programs, introduced by Goldwasser, Kalai, and Rothblum \cite{goldwasser2008one}, aim to distribute an encoded program for a functionality $f$ that can be evaluated on a single user-chosen input and then becomes useless. In the classical world, this goal is unattainable without trusted hardware (e.g., one-time memories) to enforce single use. Quantum information suggests a radically different possibility. Broadbent, Gutoski, and Stebila \cite{BGS13} proposed \emph{quantum} one-time programs (which we refer to simply as OTPs), in which the program is embodied in an unclonable quantum state. Yet formalizing security against quantum adversaries proved subtle: a quantum adversary may evaluate a program coherently, partially extract information, and then uncompute, producing behaviors that do not correspond to “one query’’ in any physical sense. Recent work of Gupte et al.~\cite{GLRRV25} introduced the simulation paradigm of \emph{single effective query} (SEQ), which measures not how many times an interface is invoked, but how many times an adversary \emph{effectively learns} an input/output pair.

In the SEQ ideal world, the oracle may be called many times coherently; nevertheless, it maintains a database of effective queries and refuses to answer once it detects knowledge of two distinct inputs.  This shift enabled meaningful simulation-based security for OTPs relative to a classical oracle, together with a construction achieving such guarantees \cite{GLRRV25}. 

\paragraph{Stateless vs. Stateful Computation.}
The above developments still concern, at their core, the one-time evaluation of a \emph{stateless} functionality. On the other hand, computation most generally may be stateful. As a basic motivating example, consider the standard scenario of Software-as-a-Service (SaaS): an entity wants to allow restricted access to some algorithm (e.g., a powerful AI model), in the form of oracle query access. The state of the computation, in that case, records not only how many queries have been made thus far but also the inner memory state of the conversation, etc. The classical solution is that the server holds the code of the AI model and for each query, the client sends its input\footnote{If privacy is wanted, at least theoretically, we can use tools like fully-homomorphic encryption.} and the server computes the output. Quantum OTPs open the possibility of the server issuing a quantum token for each query\footnote{Note that even using OTPs, it isn't immediately clear how to prevent a malicious user from abusing this service by forking the inner state, which is created between different queries. In the classical solution from before this is not possible, since the server is keeping the state of itself, and gives the user only the answer.}.

Both the classical and quantum solutions above suffer from two fundamental limitations. First, computational complexity: the server just wants to issue access, but needs to execute proportionally to the computation of the client (and at the very least, proportionally to the number of queries). Second, communication complexity: a $k$-query computation will take $\Omega(k)$ communication, while in principle, there was no real need for two entities to communicate at all.

Putting the above example aside, more basically, assume we want to allow one-time access to a piece of \emph{code}, e.g., in the form of a Turing Machine (TM), instead of a circuit. OTPs allow to issue a one-time token for running the TM, but where the quantum token's size is proportional to the \emph{running time} of that TM\footnote{More precisely, the worst-case running time, taken over all inputs of some fixed size $n$.}. This leaves open a basic question that arises throughout cryptography: what happens when the hidden functionality is not naturally a circuit that is meant to be evaluated once, but rather a stateful program whose future behavior depends on how it has already been used? In many cryptographic settings, such as indistinguishability obfuscation and fully homomorphic encryption, lifting results from circuits to RAM-style computation is notoriously nontrivial and often incurs significant overhead \cite{garg2016candidate,CCS:WanChaShi15,C:HHWW19,STOC:LinMooWic23,STOC:CHJV15}. These considerations naturally raise a question:
\begin{quote}
    \centering
    \textit{Can the quantum one-time paradigm be extended beyond a single stateless evaluation to richer, stateful forms of controlled access?}
\end{quote}

\vspace{1mm}
Ideally, we would like to issue a quantum token where its size (and optimistically, even the computational work for generating it) is proportional to the actual hidden information, namely, the size of the program's code, rather than depending on the amount of possible future computation. This is relevant for both; the above example of restricted query-access to an AI model, or generally issuing tokens for Turing Machines. To this end, we introduce the notion of \emph{query-limited programs} (QLPs). A local query-limited program offers a qualitatively different guarantee: the user can evaluate the program \emph{locally}, while the access policy is enforced by the program itself rather than by an always-online provider.

\paragraph{Leveraging One-Shot Signatures for One-Time Stateful Computation.}
A different line of work in quantum cryptography studies a particularly strong form of unclonability, embodied by \emph{one-shot signatures} (OSS) \cite{AGKZ20, SZ25}. On one hand, OSS can be viewed as OTPs for a restricted class of functionalities -- namely, signing algorithms, closely related to tokenized (one-time) signatures \cite{BDS23}. On the other hand, OSS strengthen the no-cloning requirement in a striking way: given only a (classical) common reference string, the signing capability is a quantum state that is unclonable \emph{even to the quantum algorithm that generates it}. In particular, any quantum polynomial-time generator cannot produce two usable signing states for the same public serial number. While the canonical OSS application is one-shot \emph{signing}, the generator-resistant flavor of unclonability they capture is exactly the kind of ``decentralized'' control that seems necessary to push beyond stateless OTPs.

Along different axes, OTPs and OSS provide complementary guarantees: OTPs emphasize one-time obfuscated functionality, while OSS emphasize generator-resistant unclonability. In this work we explore whether the stronger unclonability inherent in OSS can be leveraged to fundamentally strengthen the \emph{functionality} of quantum one-time programs. To this end we develop a unified framework that bridges these notions and show new functionalities enabled by their interaction. Along the way, we investigate their formal relationship and ask whether unclonable \emph{signing} can serve as a generic mechanism for unclonable \emph{general computation}. In particularly, we investigate whether decentralized unclonable signing capability suggests a quantum capability can gate not just a single output, but a \emph{sequence} of state transitions, without relying on trusted hardware or an online authority.

\subsection{Results and Future Work}
\label{sec:our_results}
The results of this work consist of a chain of new notions, their construction and applications, which answer the above question positively. Our work has three main parts. The first two parts introduce two new notions; (1) one-shot programs and (2) query-limited programs, where the latter is built on the former. The third part of our work discusses examples of applications. We further elaborate below on our contributions. 

\paragraph{First Part: Defining and Constructing One-Shot Programs.}
We begin by putting forth and defining the notion of \emph{one-shot programs} ($\OSP$), which upgrades quantum one-time programs by extending their no-cloning guarantee to that of one-shot signatures. At a high level, a one-shot program $\OSP$ scheme consists of two efficient algorithms; a classical algorithm $\PGen$ for compiling programs and a quantum algorithm $\TokenGen$ for generating tokens. Given a classical functionality $f$ we can classically compile it $P_f \gets \PGen(f)$. The program $P_f$ can be thought of as an ideal obfuscation of a function which depends on $f$ (and executes it, conditioned on some initial verification procedure on the input). Then, $\TokenGen$ allows to sample one-time tokens $\left( \ket{T}, s \right) \gets \TokenGen$, which in turn consist of an unclonable state $\ket{T}$ and a classical serial number $s$. Correctness says that one can use $P_f$ and the token to execute $f$ on any chosen input $x$ and sampled serial number $s$. Informally, security guarantees that once the instance $P_f$ is sampled honestly, it is computationally hard to execute $f$ twice using the same serial number $s$, on any two inputs $x, x'$.

\textit{Some intuition behind one-shot programs.}
At a first glance, the move from OTPs to OSPs seems rather strange: the whole point of OTPs is to give one time access to a functionality. However, if an adversary can generate tokens $\left( \ket{T}, s \right) \gets \TokenGen\left( \pk \right)$ locally and on demand, it can effectively execute $f$ many times on different values of $x$. The resolution to this, is in the fact that the function $f$ acts on \emph{both} parts; $x$ and the serial number $s$. While $x$ is to the choosing of the evaluator, $s$ is not. To elaborate, the previous interface of functions used for OTPs was $f\left( x ; r_{x} \right)$, where $f$ is a randomized functionality, the evaluator chooses $x$, and $r_{x}$ is an $x$-dependent randomness. Here, the interface is $f\left( x, s ; r_{x} \right)$ where $x, r_{x}$ are as before, but $s$ is independently derived from the token generation. Intuitively, this exact combination seem to create our divergence from OTPs. On the one hand, we let the (possibly malicious) evaluator sample its own tokens. On the other hand, each token is necessarily tied to a serial number $s$, which in turn plays two roles: (1) it acts as a second input to $f$ and thus influences its execution, and furthermore (2) enforces a "collision-resistance" property against the first part of its input, $x$, which the evaluator \emph{does} have control over.

%These two changes together seem to create a divergence from OTPs -- on the one hand we let the (possibly malicious) evaluator sample its own tokens, but on the other hand we give it no control over a second part of its input, the serial number $s$.

\vspace{2mm}
We formally define OSPs in Definition \ref{def:osp-compiler} (which is based on Definition \ref{def:generalized-otp-compiler}), and show how to construct these generically from any one-shot signature \cite{AGKZ20, SZ25}. Furthermore, our proofs are modular in a way that creates a correspondence between one-time signing capabilities and one-time general computation. By essentially the same construction and security proof, we show:
\begin{itemize}
    \item
    \textbf{One-Shot Programs from One-Shot Signatures:} In a classical oracle model, a one-shot program compiler can be constructed from any one-shot signature scheme (Theorem \ref{theorem:OSP_from_OSS}).

    \item
    \textbf{One-Time Programs from Signature Tokens:} In a classical oracle model, a one-time program compiler can be constructed from any quantum signature token scheme (Theorem \ref{thm:generalized-otp}).
\end{itemize}
Purely in the context of OTPs, the above improves our understanding: the previous work of \cite{GLRRV25} construct OTPs directly and by using techniques based on coset states. Coset states have a specific algebraic structure which the previous proofs use, and additionally, one can construct quantum signature tokens from them \cite{BDS23, coladangelo2021hidden}. Here, we show that \emph{any} quantum signature token scheme can be used to construct a OTP compiler, in particular, not relying on the algebraic structure of coset states. Finally, as an additional sanity check for our definition, we show that one-shot programs imply one-shot signatures in a classical oracle model (Section \ref{sec:OSS-from-BUS}). The oracles can all be heuristically instantiated efficiently using indistinguishability obfuscation and one-way functions.

\paragraph{Second Part: Defining and Constructing Query-Limited RAM Programs.}
We next define a new notion called Query-Limited RAM Programs (QLPs), and its security. As briefly explained in the introduction, a QLP compiler is a generalization of a one-time program compiler, in the sense that  it allows the generation of a quantum one-time token for a stateful functionality. A QLP compiler consists of two efficient quantum algorithms; a program compiler $\widetilde{P_0} \gets \PGen\left( P \right)$ for a classical program $P$, and an evaluation algorithm $\left( y, \widetilde{P_{\st'}} \right) \gets \Eval\left( x, \widetilde{P_{\st}} \right)$. The evaluation algorithm is given a classical input $x$ and a quantum token $\widetilde{P_{\st}}$ of a stateful program (here, the program is in "state $\st$"). It outputs a classical $y$ and advances the program's state. Similarly to OSPs, also here we give the (possibly malicious) evaluator delicate control: while it can choose $x$ arbitrarily, the security guarantee of QLPs prevent it from changing the inner state in any illegal way. We formally define QLPs in Definition \ref{def:RAM-program-compiler} and construct them in the classical oracle model. Specifically, we do this in the following steps.
\begin{enumerate}
    \item
    \textbf{Introducing the Limited-Effective Query (LEQ) Oracle.}
    To define the security of $\QLP$, we introduce the \emph{limited}-effective query (LEQ) oracle, as a RAM-generalization of the previous \emph{single}-effective query (SEQ) oracle. The SEQ oracle was introduced in \cite{GLRRV25} as part of their resolution addressing barriers for constructing OTPs \cite{BGS13}. Our new LEQ oracle plays a central role for realizing QLP compilers. In the technical overview we elaborate on its definition and the main ideas behind it.

    \item 
    \textbf{QLPs from OSPs.}
    We construct a Query-Limited Program compiler from a One-Shot Program compiler. We provide our construction and a formal analysis and security proof with respect to a classical oracle in Section \ref{sec:query-limited-RAM-programs} (Theorem \ref{thm:QLP_main_them}), and explain the main ideas in our technical overview. 
\end{enumerate}

\paragraph{Third Part: Applications.}
Finally, we give preliminary results demonstrating the power of QLPs as an abstract primitive as well as our construction's framework, through several new applications. First, as mentioned in the introduction, QLP compilers directly yield one-time programs for code/RAM programs and, more generally, $k$-time programs, which allow a possibly stateful program to be executed for up to $k$ steps. Importantly, both the token size and, in our construction, its generation time, depend only polylogarithmically on $k$.

\medskip
\noindent
\textit{Pay-per-use (PPU) programs.}
An additional application we show is the ability to \emph{retroactively} determine the bound $k$, which is captured by our pay-per-use (PPU) programs. We consider a setting where a user repeatedly evaluates a program and later proves (to a verifier) an upper bound on the \emph{total cost} incurred.
Each step of a PPU evaluation advances the internal state, which is kept hidden, and produces an output $y$. Separately, the user can run an explicit proof interface that yields a (classically verifiable) proof that the program's accumulated cost equals (or is bounded by) $c$.

\medskip
\noindent
\textit{New forms of issuing code access.}
Both our $k$-time programs and pay-per-use programs enable new capabilities for delegating restricted access to computation while using minimal communication and computation. For example, both suggest, for the first time, different solutions for enabling Software-as-a-Service while keeping resources small, as was explained in our Introduction (Section \ref{section:intro}).

\medskip
\noindent
\textbf{Semi-quantum query-limited programs.}
Finally, following a sequence of works on semi-quantum public-key cryptographic primitives \cite{gavinsky2014classicalqm, radian2019semi, AGKZ20, shmueli2022public, shmueli2022semi}, we show how to construct a \emph{semi-quantum} $\QLP$ compiler (Definition \ref{def:semi-quantum-query-limited-program-compiler}). A semi-quantum QLP compiler can have a purely classical issuer that transfers the $\QLP$ restricted capability over a classical channel. Our construction does \emph{not} transform any $\QLP$ generically into a semi-quantum version. Instead, we focus on an intermediate structure which is generically implied by one-shot programs in order to construct it.

\smallskip
\noindent\emph{Definition (specifics).} A \emph{semi-quantum} $\QLP$ compiler (cf.\ \Cref{def:semi-quantum-query-limited-program-compiler}) augments $(\PGen,\Eval)$ with a classical-message interactive protocol $\Transfer$ between a sender holding $\tilde P$ and a receiver, after which the receiver obtains a new program $\tilde P'$ that continues from the same logical RAM state. Correctness is required to hold for arbitrary polynomial-length \emph{interleavings} of $\Eval$ operations and $\Transfer$ operations (where $\Transfer$ is modeled as a no-op on the ideal RAM state).

\begin{theorem}[Informal: Semi-quantum $\QLP$]\label{thm:informal-semi-quantum}
From any $\QLP$ compiler based on $\OSP$ tokens, there exists a semi-quantum $\QLP$ compiler supporting a classical-message $\Transfer$ protocol (cf.\ \Cref{fig:semi-quantum-QLP-compiler-construction}).
\end{theorem}

The protocol is conceptually simple: the receiver generates a fresh next-token serial number $s'$, the sender consumes its current token to execute a special ``transfer'' step that authenticates $s'$, and the receiver thereby obtains the updated classical state plus a fresh token enabling the next evaluation.

\paragraph{Future work.} For future work, we can consider two directions.
\begin{enumerate}
    \item \label{future_directions_bullet_1}
    Formulating the security definition for query-limited access for stateful/RAM programs where constructions exist in the plain model (without oracles), without running into impossibilities.\footnote{Since OTP is a special case of our QLP, the impossibilities of realizing the SEQ simulation security in the plain model, as discussed in \cite{GLRRV25}, apply to our QLP as well.}

    \item \label{future_directions_bullet_2}
    Building on item \ref{future_directions_bullet_1}, constructions in the plain model from indistinguishability obfuscation and more desirably, weaker standard assumptions, for generic or specialized functionalities .
\end{enumerate}

\begin{table}[pt] % better as table, since this is tabular data
\centering
\small
\setlength{\tabcolsep}{3pt}
\begin{threeparttable}
\begin{tabular}{|p{0.10\linewidth}|p{0.23\linewidth}|p{0.23\linewidth}|p{0.23\linewidth}|p{0.17\linewidth}|}
\hline
\textbf{Work} &
\textbf{Security Notion} & \textbf{Assumptions and Tools} & \textbf{Applications} & \textbf{Classical Vendor?} \\
\hline
\cite{GLRRV25} & OTP, SEQ model, sim-based security
&
Classical oracle (or iO for weaker security), coset states
&
One-time PRFs, NIZKs, signatures & No, fully quantum
\\
\hline
\cite{gunn2024quantum} & OTP, game-based (implied by \cite{GLRRV25}'s security)
&
Classical oracle, signature tokens
&
Same \cite{GLRRV25} (with corresponding security)
& No, fully quantum
\\
\hline
\cite{stambler2025cryptography} & OTP, single-physical query\tnote{a}, sim-based
&
\emph{Classically-accessible} classical oracle\tnote{b}, semi-quantum signature tokens
&
\cite{stambler2025cryptography} RAM programs and applications (below)
& Yes, classical vendor and communication
\\
\hline
\cite{stambler2025cryptography} & RAM programs, single-physical query, sim-based
&
\emph{Classically-accessible} classical oracle, \cite{stambler2025cryptography} OTP
&
Long-lived OTP and copy protection
& Yes, classical vendor and communication
\\
\hline
\textcolor{red}{This work} & OSP (OTP resp.), SEQ model, sim-based
&
Classical oracle, OSS (signature tokens resp.)
&
Same as \cite{GLRRV25}; QLP and its applications (see below)
& Yes, classical vendor and communication
\\
\hline
\textcolor{red}{This work} & QLP, limited-effective query (LEQ) model, sim-based security
&
OSP, classical oracle, signatures
& OTP for RAM/code,
$k$-time programs, budget programs, pay-per-use programs, semi-quantum QLP, OSS
& Yes, classical vendor and communication
\\
\hline
\end{tabular}

\begin{tablenotes}[flushleft]
\footnotesize
\item[a] As discussed in the introduction, a single physical query oracle allows literally one query. Un-computing and re-querying on the same input are not allowed. A single-effecive query oracle allows multiple queries as long as its database records no more than one distinct query.
\item[b] A classically-accessible oracle is an oracular interface that allows no superposition queries and measures any incoming query before answering. A classical oracle is a a black-box classical function that allows superposition queries.
\end{tablenotes}
\medskip
\caption{Comparison of our results with most relevent previous works. See Section~\ref{sec:related_works} for detailed discussion.}
\label{fig:definitions_impossibilities_figure}
\end{threeparttable}
\end{table}

\subsection{Related Works}
\label{sec:related_works}
%\jiahui{TODO}

%works in security notions, assumptions/tools, applications and whether the preparation and communication can be dequantized.
\ifllncs
Apart from \cite{BGS13,GLRRV25}'s results mentioned in the introduction,
\else
As mentioned in the introduction,
\cite{BGS13} put forward the definition for quantum one time programs for quantum functionalities and show a strong impossibility result. \cite{GLRRV25} proposes the SEQ oracle model and gets around \cite{BGS13}'s impossibility result, giving a construction for all classical functions (meaningful only for high-entropy functions) in the classical oracle/ideal obfuscation model. 
\fi
\cite{GLRRV25} additionally proposes some weaker operational security definitions where they obtain one-time PRFs, NIZKs, signatures in the plain model from iO with respect to this operational security. \cite{gunn2024quantum} proposes a game-based security where a somewhat natural class of attackers cannot produce a second output as long as the functionality samples a high min-entropy output on every input. They construct one-time programs from any signature tokens under this definition. However, it does not rule out a more malicious attacker that can still produce multiple outputs; one example is the impossibility of \cite{GLRRV25}. The follow-up works \cite{cananth2026noninteractive,ananth2026quantum} improve the assumptions in the plain model for OTP of PRFs and explores going from OTP to non-interactive MPC with preprocessing.

The work of \cite{stambler2025cryptography} also propose RAM oracles similar to our $\QLP$ notion and semi-quantum OTPs from semi-quantum signature tokens \cite{shmueli2022semi}, satisfying strong simulation-based security which allows only single-physical queries in the ideal world. The program vendor and communication can also be made classical. However, their constructions use classical measured oracles. A measured oracle is an oracle where adversaries' query registers have to be measured (in the computational basis) by the program interface upon querying, which gives a different model and relies on additional hardware assumptions. Their applications include OTP and semi-quantum copy protection with no need of long-term quantum memories, also in the classically-accessible measured oracle model. Besides, compared to the semi-quantum token based scheme in \cite{stambler2025cryptography} where the program vendor has to communicate with every incoming user/receiver, our OSS based $\OSP$ and $\QLP$ has an additional ``decentralized'' feature that allows anyone to generate their own tokens and transfer programs to others via classical communication.
%The work of \cite{stambler2025cryptography} also propose RAM oracles similar to our $\QLP$ notion and semi-quantum OTPs from semi-quantum signature tokens \cite{shmueli2022semi}, satisfying strong simulation-based security which allows only single-physical queries in the ideal world. The program vendor and communication can also be made classical. However, their constructions use classical-only access to oracles, where adversaries' query registers have to be measured by the program interface upon querying. This is a hardware assumption/heuristics. Their applications include OTP and semi-quantum copy protection with no need of long-term quantum memories, also in the classically-accessible oracle model. Besides, compared to the semi-quantum token based scheme in \cite{stambler2025cryptography} where the program vendor has to communicate with every incoming user/receiver, our OSS based $\OSP$ and $\QLP$ has an additional ``decentralized'' feature that allows anyone to generate their own tokens and transfer programs to others via classical communication.

Apart from the above, \cite{stambler2025information} studies information-theoretic security for a specialized adversarial model called geometric local $\mathsf{QNC_0}$ adversaries. \cite{liu2020quantum} studies the security of classical one-time memory against quantum superposition attacks. \cite{liu2023depth} constructs quantum one-time memory in the quantum random-oracle model with depth-bounded adversaries. %the honest party only needs short-term quantum memory, while an adversary that tries to keep quantum memory for longer cannot successfully attack because its quantum depth is bounded.

\section{Technical Overview}
\label{sec:tech_overview_osp}
In this section we explain the main ideas behind our new notions, their definitions and constructions. Before going into these details, we first recall the idea of one-time program (OTP) compilers, their security definition and construction.

\subsection{Previous Work on Quantum One-Time Programs} \label{section:overview_previous_work}
We begin by recalling quantum one-time programs for classical randomized functionalities. Let $f : \cX \times \cR \to \cY$. An OTP compiler consists of two efficient quantum algorithms,
$$
\widetilde f \leftarrow \PGen\left( 1^\lambda, f \right)
\;
,
\enspace 
y \leftarrow \Eval\left( \widetilde f, x \right) \enspace.
$$
$\PGen\left( 1^\lambda, f \right)$ samples a quantum state $\widetilde f$, and then $\Eval\left( \widetilde f, x \right)$ takes the quantum token and a later-chosen input $x$ and computes a classical output $y$. Correctness requires that honest evaluation on any chosen input $x$ produce an output statistically close to $f(x;r)$ over a uniformly random $r$. Importantly, the evaluator chooses $x$, but not $r$.

On the side of security, we would ideally like to enforce that "the quantum token $\widetilde f$ gives exactly one-query access to $f$". The work of Broadbent, Gutoski, and Stebila~\cite{BGS13} defined this as
$$
\OTP(f) \approx_c \Sim^{f_1} \enspace ,
$$
where $\Sim$ is a simulator and $f_1$ denotes a single-query access to $f$, which may be a superposition (coherent) query. In other words, the quantum token can be indistinguishably simulated by a simulator that has access to one physical query. This definition was formalized and subsequently proved impossible by \cite{BGS13}. An issue observed by \cite{BGS13} as part of their impossibility is that, for example, if $f$ is a deterministic function, such definition cannot hold. This is due to the evaluator's ability to measure only the output, which, due to the determinism of $f$, will incur a non-disruptive measurement and keep the quantum token as good as new, allowing a second query. On the other hand, a one-time physical access to $f$ by the simulator will not be able to answer a second query.

\paragraph{The Single-Effective Query Approach.}
A solution to this tension proposed in \cite{GLRRV25} is to give a definition referring to how much "effective information" overall was extracted from one token. Formally, the authors define the Single-Effective Query (SEQ) oracle (which gives more information than $f_1$), and then prove that access to the quantum token is no better than access to the oracle. Concretely, the SEQ oracle $O_f^{\mathsf{SEQ}}$ (defined in Figure \ref{fig:seq-oracle}) implements $f(x;H(x))$, where $H:\cX \to \cR$ is an internal random oracle represented using Zhandry’s compressed-oracle technique \cite{Zha18}. The use of compressed oracles allows us to not only simulate $H$, but to also monitor the queries made thus far, which we keep in a database register $D$. The register $D$ is initially empty and remains supported on databases containing at most one entry $(x,r)$. On a query $x$, the SEQ oracle does nothing in branches where the database already records a different input $x' \neq x$. Otherwise, it coherently computes $r=H(x)$, XORs $f(x;r)$ into the output register, and uncomputes its temporary workspace.

To observe one clear point of difference between the SEQ setting and the single physical query setting of $f_1$, consider what happens when the simulator queries on the same input $x$ as the one recorded in the database. More generally, for an arbitrary $f$, consider a simulator which makes a superposition query in the form
$$
\sum_x \alpha_{x} \cdot \ket{x}_{\mathsf{input}}\ket{u}_{\mathsf{output}}
\rightarrow
\sum_x \alpha_{x} \cdot \ket{x}_{\mathsf{input}}\ket{u+f(x)}_{\mathsf{output}}
\enspace ,
$$
without measuring the output. In the setting of $f_1$, oracle access to the simulator now shuts down, while in the SEQ setting the simulator can query the SEQ oracle again (using the same two registers) to un-compute the output register and goes back to the state $\sum_x \alpha_{x} \ket{x}_{\mathsf{input}} \ket{u}_{\mathsf{output}}$.
%Using ideal obfuscation, \cite{GLRRV25} constructs quantum one-time program protocol such that a simulator interacting with the above SEQ interface will mimic the real-world interaction between a quantum adversary and the program.

SEQ security requires an efficient simulator that reproduces the encoded program using only this restricted oracle access. Informally, modulo security parameters and public auxiliary information, we ask for the computational indistinguishability
$$
\left( f,\mathsf{OTP}(f) \right)
\approx_{c}
\left( f,\mathsf{Sim}^{O_f^{\mathsf{SEQ}}} \right) \enspace .
$$

\ifllncs
\else

\noindent
\textit{Connection to the zero-knowledge simulation paradigm.}
Intuitively, the above somewhat follows the zero-knowledge simulation paradigm, where one separates the information which the interaction thinks of a private, from the actual information sent during the interaction. Given some minimal information (which we think of as public) only the latter needs to be simulated and be indistinguishable even given arbitrary side information. For example, in zero knowledge, the simulator has access to the verifier's code (which we think of as minimal information needed for simulation) and should simulate the real interaction with the prover. The indistinguishability should hold even given arbitrary independent side information, for example, even given a witness for the instance. Analogously, in the setting of OTPs, the simulator gets access to some minimal information (here, it is access to the SEQ oracle) and simulates a quantum state that's indistinguishable from the real quantum OTP token. Indistinguishability needs to stay intact even given the side information of the circuit of $f$.

\fi

\paragraph{Previous construction of One-Time Program Compilers.}
Previous work by \cite{GLRRV25} realizes SEQ security in the classical-oracle model using hidden subspace states. For simplicity, first consider a one-bit input $x\in\{0,1\}$. The compiler samples a uniformly random subspace $A \subseteq \mathbb F_2^n$ of dimension $n/2$ and provides one copy of
$$
\ket{A} = \frac{1}{\sqrt{|A|}} \sum_{ u \in A } \ket{u} \enspace .
$$

Writing $A^0=A$ and $A^1=A^\perp$, the program additionally provides access to a quantumly accessible classical oracle that acts as
$$
\ket{x,v,u}
\longmapsto
\begin{cases}
\ket{x,v,u \oplus f(x;G(v))}\; , \;
   & v\in A^x \setminus\{0\}, \\
\ket{x,v,u} \;,
   &\text{otherwise},
\end{cases}
$$
where $G$ is an internal random oracle. To evaluate on $x=0$, the evaluator uses $\ket{A}$ directly; for $x=1$, it first applies a Hadamard transform, obtaining $\ket{A^\perp}$. Querying the oracle and measuring the output produces the desired sample, except with negligible correctness error. For an $m$-bit input, the construction uses one independent subspace state per bit, checks the resulting tuple $\mathbf v=(v_1,\ldots,v_m)$ coordinate-wise, and derives the randomness as $G(\mathbf v)$. So, the information of $\mathbf{v}$ contains the information of the input $x$, and this is used to derive the randomness $r$.

In a nutshell, the previous security proof shows that recording queries inside the SEQ oracle implies disturbance to these subspace states. In the real construction randomness is indexed by the vectors $\mathbf v$, while in the ideal functionality it is indexed by $x$. A coherent caching procedure bridges these representations. Finally, direct product hardness of subspace states implies the impossibility of finding any two vectors $\mathbf v := (v_1, \ldots, v_m)$, $\mathbf v' := (v'_1, \ldots, v'_m) \in \mathbb{F}_{2}^{n \cdot m}$ where there is some $i \in [m]$ and $b \in \{ 0,1 \}$ such that $v_i \in A^b$ and $v'_i \in A^{\lnot b}$. This eventually rules out the recording of any two incompatible outputs.

\subsection{One-Shot Programs}  \label{section:overview_one-shot_programs}
We next explain the definition and construction of our new primitive of a One-Shot Program (OSP) compiler, formally given in Definition \ref{def:osp-compiler}. For the functions which are encoded by OSP compilers we consider $f$ having a slightly different input interface. Before, $f$ had a main input $x$ and an auxiliary randomness input $r$. Here, the main input itself is divided into two parts, $x$ and $s$:
\[
    f : \mathcal S_\lambda \times \mathcal X \times \mathcal R
    \rightarrow \mathcal Y \enspace .
\]

An OSP compiler consists of four algorithms:
\[
\begin{aligned}
    \KeyGen(1^\lambda)
        &\longrightarrow \mathsf{crs} \; ,\\
    \PGen(\mathsf{crs},f)
        &\longrightarrow P_f \; ,\\
    \TokenGen(\mathsf{crs})
        &\longrightarrow (\ket{T},s) \; ,\\
    \Eval( P_f,s,x,\ket{T} )
        &\longrightarrow y \enspace .
\end{aligned}
\]
Here, $P_f$ is classical and may include oracle access. The issuer compiles $f$, whereas the evaluator can generate tokens locally using only the public $\crs$. These procedures are separate: program compilation does not require a token or its serial number. Correctness requires that evaluating with an honestly generated token produce a distribution statistically close to $f(s,x;r)$ for uniform $r$.

\paragraph{Security: single-effective query per serial number.}
We generalize the preceding SEQ oracle by giving it inputs $(s,x)$ and using an internal random oracle $H(s,x)$ to generate the evaluation randomness. Its compressed database now records tuples $(s,x,r)$. On a query $(s,x)$, it acts as the identity in branches containing a record $(s,x',r')$ with $x'\neq x$. Otherwise, it coherently computes $f(s,x;H(s,x))$ and uncomputes its workspace, exactly as before. The database may therefore contain many entries, but \emph{at most one entry for each serial number}.

Security (Definition \ref{def:generalized-otp-compiler}) requires that $P_f$, including its oracle interface, be simulatable using only this generalized SEQ access:
\[
    P_f
    \approx_c
    \Sim^{O_f^{\mathsf{SEQ}}}
    ( \crs, \mathsf{aux}_f ) \enspace .
\]
The above indistinguishability should hold given public parameters, the description of $f$, and a common token and serial. The fact that indistinguishability holds even with this extra information, makes the security definition be at least as strong as arbitrarily simulating the quantum token (because the simulator can hold the key that allows to generate the quantum token).

As we briefly explained in Section \ref{sec:our_results}, allowing public token generation may initially seem contrary to the purpose of one-time programs: an evaluator can generate many tokens and perform many evaluations. The distinction is that \emph{the serial number is an input to $f$} and influences its execution. A new token enables evaluation at its associated serial, while the security restriction applies separately to each serial. Thus, we seek one effective choice of $x$ per $s$, rather than one effective evaluation overall. As a side note, this combination, of publicly generated tokens and serial-dependent functionality, is what we will later exploit for one-time stateful computation.

\paragraph{Construction from one-shot signatures.}
To get the right intuition for constructing OSP compilers, we recall the previous construction of OTP compilers given above, which uses subspace states. The subspace states in the preceding construction really play the role of one-time signature tokens: the input $x$ specifies the message to sign, and the vectors supplied to the classical oracle constitute its signature. Here we will use signature tokens directly. Specifically, since we are dealing with one-shot program compilers rather than one-time program compilers, we will use one-shot signatures (Definition \ref{def:one-shot-signature}).
%Specifically, we will use one-shot signatures (Definition \ref{def:one-shot-signature}) for the case of one-shot program compilers and standard signature tokens (Definition \ref{def:signature-token}) for standard one-time program compilers.

The construction uses the same "sign, then evaluate" structure of the previous OTP construction. Setup and token generation are inherited from a strongly unforgeable one-shot signature scheme $\OSS$. To compile $f$, we sample an internal random oracle
\[
    G: \mathcal S_\lambda \times \cX \times \Sigma_\lambda
    \longrightarrow
    \cR
    \enspace ,
\]
and provide quantum access to the classical oracle
\[
\begin{aligned}
    &\ket{s,x,\sigma,u} \longmapsto \\
    &\quad
    \begin{cases}
        \ket{ s,x,\sigma,u\oplus f(s,x;G(s,x,\sigma)) }
             \enspace &\mathsf{OSS.Ver}(\mathsf{crs},s,x,\sigma)=1\\
        \ket{s,x,\sigma,u}
            &\text{otherwise}
    \end{cases}
\end{aligned}
\]
An honest evaluator uses $\ket{T}$ to sign $x$, obtaining $\sigma$, and queries this oracle. Signature correctness and the independent randomness of $G$ give the required output distribution. Importantly, the randomness is indexed by the signature as well as the serial and input.

\paragraph{One-shot programs security proof overview.}
The intuition for security is that the oracle evaluates $f$ only when the query contains a valid signature $\sigma$ on $x$ under serial number $s$. However, ruling out the production of two signatures is not enough: simulation must reproduce arbitrary coherent queries, partial measurements, and un-computation. The main challenge is that the real-world adversary has access to the stateless oracle from our construction, whereas the simulator has access only to the stateful SEQ oracle (defined in Figure \ref{fig:seq-oracle-with-serial-numbers}).

First, let us better understand the mismatch between the real and ideal worlds. The real oracle derives its randomness as $r=G(s,x,\sigma)$, whereas the ideal oracle uses $r=H(s,x)$. Recall that the SEQ oracle maintains a coherent database of records $(s,x,r)$, with at most one record per serial. On a query $(s,x)$, it does nothing in branches containing $(s,x',r')$ with $x'\neq x$; otherwise, it coherently evaluates $f(s,x;H(s,x))$ and uncomputes its workspace. Records may be created or erased by these operations. We therefore want to reach a hybrid where the randomness indexing and the SEQ-style check can \emph{ignore $\sigma$}, all at the same time without losing the correlations that the real oracle creates with the signature register.

Our proof builds on the coherent caching approach of~\cite{GLRRV25}, but must extend it from a single record to many records indexed by serial numbers. In their proof, the cache synchronizes a single tuple of subspace vectors with the one effective query recorded by the SEQ oracle. In our setting, many effective queries with distinct serial numbers may remain recorded simultaneously, and the adversary may query and un-compute across these serials in superposition. We therefore need a more general reversible cache that maintains the correspondence between each record $(s,x,r)$ and its signature information $(s,x,\sigma)$ as records are created or erased. Together with a reversible database-splitting operation, this will allow us to remove $\sigma$ from the ideal oracle's database without losing the information needed for un-computing. We do this in three main steps.

\vspace{2mm}
\textbf{Step 1: Introducing an SEQ-style check.} We first implement $G$ as a compressed oracle (see Section~\ref{sec:compressed-oracle}), whose database records tuples $(s,x,\sigma,r)$. Our next step is to add a check that refuses a valid query $(s,x,\sigma)$ whenever the database already records $(s,x',\sigma',r')$ with $(x',\sigma')\neq(x,\sigma)$. If this changed the adversary's view noticeably, we could extract the current query and a conflicting database entry to obtain two distinct valid message-signature pairs under the same serial, contradicting strong unforgeability of the OSS. Thus, we may restrict the database to at most one recorded pair $(x,\sigma)$ per serial.

\textbf{Step 2: Separating the signature information reversibly.} We now want the compressed database to record only $(s,x,r)$, as in the ideal oracle. Simply deleting $\sigma$ would not preserve the joint quantum state. Instead, we introduce a quantum cache register $C$, initially empty, and a reversible change of representation $\mathsf{Split}$. On databases whose serial numbers are all distinct, this operation separates each record into two matching parts:
\[
\begin{aligned}
\{( s_i, x_i, \sigma_i, r_i )\}_i
\quad
\longleftrightarrow
\quad
\Bigl( &\{ (s_i, x_i, r_i) \}_i, \; \{ (s_i, x_i, \sigma_i) \}_i \Bigr)
\enspace .
\end{aligned}
\]

The first component is the smaller database and the second is the cache. The transformation is reversible because the distinct serial numbers identify which entries belong together, allowing the original database to be reconstructed. Thus, signature information is \emph{moved, not discarded}. We show that the compressed-oracle operations on the original database can be implemented using the smaller database together with reversible cache updates before and after evaluation. The cache records the signature associated with an outstanding evaluation, but allows that record to be erased when the evaluation is un-computed.

We can then make the SEQ-style check ignore $\sigma$ as well. The old check rejects a different pair $(x',\sigma')$ under the same serial; the new check rejects only a different input $x'\neq x$. The checks differ only when $x'=x$ but $\sigma'\neq\sigma$, with both signatures valid. A noticeable difference would again violate \emph{strong} unforgeability, this time by producing two signatures on the same message. After this change, the smaller database and its check have exactly the form required by the ideal SEQ oracle.

\textbf{Step 3: Implementing the cache using only SEQ access.} It remains to implement the cache updates without reading the ideal oracle's private database. For each valid query $(s,x,\sigma)$, the simulator makes a coherent probe on $\ket{s,\bar x,0,0}$, where $\bar x=x\oplus1\neq x$ denotes $x$ with a fixed bit flipped. The last register is the probe's answer bit $b$. If $b=0$, the simulator toggles membership of $(s,x,\sigma)$ in $C$, inserting it if absent and removing it if present; it then uncomputes the probe. This implements a coherent test for a record $(s,x',r')$ with $x'\neq\bar x$. On branches where the main query is allowed, such a record can only be at $x$, so the probe supplies precisely the control needed for the cache. Neither the answer bit nor the database is measured.

The simulator performs this cache routine, makes the main SEQ query on $\ket{s,x,u}\otimes\ket{+}_B$, and performs the cache routine again. Here $u$ is the adversary's output accumulator; using $\ket{+}_B$ for the unused answer target prevents it from retaining acceptance information. The cache routine is its own inverse, but the two applications need not cancel when the main query changes the database: their purpose is to track that change reversibly. On rejected branches, the database is unchanged and the two cache operations do cancel. The simulator therefore needs only public signature verification, its own cache, and SEQ access to $f$.

\paragraph{One-time Programs from Any Signature Tokens}
We also construct one-time programs from any signature tokens via a similar manner. The construction satisfies the simulation based SEQ security in \cite{GLRRV25} and meanwhile uses any signature token instead of the subspace coset states used in \cite{GLRRV25}. Note that \cite{gunn2024quantum} also constructs one-time programs from any signature tokens but they prove it under a weaker security notion where the challenger will enforce measurements on the adversary's registers. We therefore achieve the best of both worlds proposed by these past works. The security proof resembles the one-shot programs security proof and we omit the discussions here.

\subsection{Query-Limited RAM Programs}
\label{sec:tech_overview_qlp}
We are now ready to define and construct our quey-limited programs for code/RAM functionalities, using one-shot programs.

\paragraph{RAM Programs.}
A randomized RAM program consists of a function $P:\mathcal M\times\mathcal X\times\mathcal R\to\mathcal M\times\mathcal Y$ and an initial memory state $\mathsf{st}_1=\bot$. Each evaluation takes the current state, a user-chosen input, and fresh randomness, and computes
\[
P(\mathsf{st}_i,x_i;r_i)=(\mathsf{st}_{i+1},y_i).
\]
A query-limited program (QLP) compiler (Definition \ref{def:RAM-program-compiler}) consists of two efficient quantum algorithms, and encodes $P$ into an object that supports these successive evaluations:
\[
\widetilde P_1\leftarrow\mathsf{PGen}(1^\lambda,P),
\qquad
(y_i,\widetilde P_{i+1})\leftarrow\mathsf{Eval}(\widetilde P_i,x_i). \enspace ,
\]
where the encodings $\widetilde P_i$ are quantum states. For every polynomial-length input sequence, correctness requires that the joint distribution of outputs from honest evaluation be statistically close to that of the underlying RAM program. The evaluator chooses the inputs, but does not directly choose the internal memory states.

The first thing to notice is that this interface captures simple access policies. For a $k$-time program, the state includes a counter recording how many evaluations have been completed, and the program rejects once the counter reaches $k$. For a more general budget-limited program, it instead tracks the accumulated cost and rejects a query that would exceed the budget. Even more generally, the state can contain the program's working memory, not merely an access counter.

\paragraph{Operational Security Definition.}
Before introducing our general security definition, consider the intuitive goal of preventing an evaluator from obtaining two incompatible executions. An execution has the form
\[
\mathsf{exe}=\left( \left(  \left( \mathsf{st}_1,x_1,r_1 \right), y_1 \right), \ldots, \left( \left( \mathsf{st}_n,x_n,r_n \right), y_n \right) \right) \enspace ,
\]
where $P(\mathsf{st}_i,x_i;r_i)=(\mathsf{st}_{i+1},y_i)$ at every step. We would like an adversary given $\mathsf{QLP}(P)$ to be unable to produce two valid executions from the same initial state that differ at a common step. Of course, outputting one execution and a shorter prefix of it should not constitute an attack.

However, such an operational guarantee cannot hold for every family of programs. In particular, one can  sometimes successfully obtain deterministic outputs while the corresponding computation is uncomputed. We therefore seek a simulation-based definition that captures the access provided by the encoded program, while leaving functionality-specific operational guarantees to a separate analysis, as in the SEQ framework of~\cite{GLRRV25}.

\paragraph{The Limited-Effective-Query (LEQ) Model for RAM Programs.}
Our ideal functionality is the LEQ oracle $O_P^{\mathsf{LEQ}}$. Informally, a QLP compiler is secure if its entire encoded program can be simulated using only LEQ access:
\[
\mathsf{QLP}(P) \approx_c \mathsf{Sim}^{O_P^{\mathsf{LEQ}}}.
\]
The comparison includes the program's oracle interfaces and any auxiliary information allowed by Definition \ref{def:RAM-program-compiler}. $\Sim$ can interact with the LEQ oracle by submitting queries $(i, x_i)$ and receiving an output $y_i$, which $O_P^{\LEQ}$ computes using the $i$'th memory state $\st_i$. We emphasize that the memory state $\st_i$ is \emph{NOT} exposed to $\Sim$ at any point\footnote{This notion does not conflict with some applications where the users may want to read part of the memory state information (e.g. remaining queries, budget) where we can put this information in the output.}.

The LEQ oracle meaningfully limits access to $P$ while still faithfully modeling the fact that a real quantum adversary can run the program forward and then uncompute it (as long as it does not measure).
In contrast to the SEQ oracle, which limits the evaluator to one ``effective query'', the LEQ oracle allows the evaluator to repeatedly evaluate the program as the state updates. Instead, it limits the evaluator to one ``effective execution path'' by maintaining a superposition over totally ordered databases of queries and program states
\[
    D = 
    \left\{
    \begin{array}{c}
         1:\ (\st_1,\ x_1,\ r_1,\ y_1),  \\
         2: (\st_2,\ x_2,\ r_2,\ y_2), \\
         \dots \\
         i_{{\max}}:\ (\st_{i_{\max}},\ x_{i_{\max}},\ r_{i_{\max}},\ y_{i_{\max}})
    \end{array}
    \right\}
\]
% \begin{gather*}
%     (1, \st_1, x_1, r_1)
%     \\
%     (2, \st_2, x_2, r_2)
%     \\
%     \dots
% \end{gather*}
which are internally consistent:
\[
    P(\st_i, x_i; r_i) = (\st_{i+1}, y_i) \quad \text{for all } i
\]
Because the database is totally ordered, the adversary cannot simultaneously explore different branches of the program. If $D$ contains $(i, \st_i, x_i, r_i)$, then the adversary cannot also evaluate on a different $(i, x'_i)$, or evaluate the same $(i, x_i)$ using a different program state $\st_i'$.
% know an evaluation corresponding to a different input $(i, \st_i, {\color{red}x_i'}, r_i)$ or a different memory state $(i, {\color{red}\st_i'}, x_i, r_i)$.

\paragraph{Relation to the Operational Definition.}
We briefly pause to describe the relation of the general and operational definitions before continuing with the mechanics of maintaining the database. 
Suppose we could show for a particular class of programs that any execution the adversary finds must be recorded in the LEQ database. Then if the adversary found two executions, the LEQ database would contain two executions -- against its definition. 

Although we have not yet described \emph{how} the LEQ database records queries, there are two intuitive criteria for when knowledge of an execution implies its presence in the database.
First, the adversary should not be able to ``forge'' query responses by computing them without actually querying the LEQ oracle. Second, each query's response should be sufficiently randomized, or else the program state will not change significantly by measuring the query's response and uncomputing the query. The precise conditions are quite similar to those for the SEQ oracle and we refer the reader to \cite{GLRRV25} for an in-depth discussion.

\justin{This could also reasonably go at the end of the definition, but I think it's useful to compare the high-order bits of each definition. Could use some editing potentially.}

\paragraph{Definition: Recording Queries in $O_P^{\LEQ}$.}
.\justin{I expanded this paragraph and broke it into 3 as part of explaining the stack structure more fundamentally.}
The main challenge of defining $O_P^{\LEQ}$ is the mechanism for recording an ``effective execution path'' corresponding to the evaluations that the evaluator knows.
$O_P^{\LEQ}$ will maintain a register $\cS$ containing the current state of the program and a sequence of ancilla registers $\cA_1, \dots, \cA_k$ corresponding to the individual steps of the computation. When it receives a basis state query $\ket{i, x, u}$, it will use $\cA_{i}$ as a workspace to answer the query. If $\ket{i, x, u}$ is an ``effective query'' then $(i, x_i)$ should end up recorded in $\cA_{i}$. As an overly simplified view, we can imagine the database as a sequence of memory states
\[
    \st_1 \rightarrow \st_2 \rightarrow \st_3 \rightarrow \dots
\]
corresponding to the evaluations that the evaluator knows.

\cite{GLRRV25} already gives a solution to recording a single effective query by using Zhandry's compressed oracle~\cite{zhandry19compressed}. Indeed, a natural approach to the LEQ oracle is to imagine internally running a sequence of SEQ oracles for each recorded $i$; when the $i$'th SEQ oracle records a new query, create a new SEQ oracle for $i+1$.
% \cite{GLRRV25}'s solution to opportunistically recording a single query is to use a compressed oracle~\cite{zhandry19compressed}. 
However, maintaining an ordered database of queries requires more care. There are two issues we need to address: 1) maintaining an unbroken chain of memory states and 2) preventing the oracle from ``getting stuck''.
% We could try using a compressed oracle~\cite{zhandry19compressed} to opportunistically record queries like in the SEQ oracle, but this turns out to be insufficent. There are two subtleties in maintaining an ordered database which require more care.

\paragraph{Definition: Maintaining order.}
First, we need to ensure that the database contains a complete, ordered history of the execution with no gaps.
The subtlety is that a quantum adversary can always erase its own memory of an evaluation by making another evaluation, as observed by \cite{BGS13,gunn2024quantum,GLRRV25}. 
If we permit an evaluator to make a second query to $\st_1$, for example, it might \emph{erase its own memory of evaluating on $\st_1$}, which would cause the 1st SEQ oracle to in turn \emph{erase the record of $\st_2$ in $D$}. Now there is a gap in the database:
\[
    \st_1 \rightarrow {\color{red} ???} \rightarrow \st_3 \rightarrow \dots
\]

Preventing gaps is a simple as restricting the adversary from accessing middle memory states. In other words, $O_P^{\LEQ}$ provides access to $D$ as a stack.
The evaluator has two options for its queries: either repeat the most recent query $(i, x_{i_{\max}})$, or make a new query $(i_{\max}+1, x_{i_{\max}+1})$ to be evaluated using the next memory state $\st_{i_{\max}+1}$ (obtained from $P(\st_{i_{\max}}, x_{i_{\max}}; r_{i_{\max}})$). 
They are not allowed to access the middle of the stack, ensuring that the database maintains a complete history of the computation so far.

Of course, an evaluator who has access to the real $\QLP(P)$ can always re-evaluate middle queries. Ensuring indistinguishability from the LEQ oracle which forbids middle queries becomes an important step in proving any $\QLP$ construction secure.
% \justin{Expand this to make the stack more clearly fundamental. Maybe explain the definition first as a "stack of SEQ oracles", then go into the details. Key point to make: need to allow uncomputation queries. If the adversary does this to a middle query, we might end up with a broken history $(\st_1, ???, \st_3)$.}

\paragraph{Definition: Preventing Sticking.}
Second, there is a subtle issue with implementing $D$ naively using a compressed oracle: it can get stuck after a deterministic query.  
Suppose that an evaluator honestly makes subsequent queries to $x_1$, then $x_2$. They expect the second query to use memory state $\st_2$, which may depend on $x_1$.
However, a key feature of \cite{GLRRV25}'s SEQ oracle is that it does \emph{not} record ``gentle'' queries. 
If $P$ were deterministic, then $x_1$ is never recorded and $O_P^\LEQ$ would answer the second query $x_2$ using the \emph{first} memory state $\st_1$.\footnote{For example, $P$ might be a deterministic counting program where the program outputs $y_i = i$, maintaining state $\st_i = i$ and ignoring $x_i$. Here, the evaluator would expect outputs $1$, then $2$. If we implement $D$ naively for $O_P^\LEQ$, then any simulator would only see $1$ forever and not know how to properly answer the second query.} 
Since $\Sim$ can never learn the correct response to the second query, a naive implementation of $D$ results in an unrealizable security notion.

To ensure that LEQ security is possible to achieve, we implement the database $D$ differently. 
\footnote{A particularly simple solution is to append a truly random value $r_2 \in \{0,1\}^\secp$ to $P$'s output. Now if an honest evaluator (or the simulator) wants to avoid getting stuck, it just has to remember $r_2$. 
However, appending $r_2$ is unsatisfactory because then the ideal functionality depends on the security parameter $\secp$.
Despite this, this idea provides a useful conceptual stepping-stone for our proofs.}
Instead, we implement $D$ using ideas from \cite{eprint:GLQRRZ26}'s recent generalization of one-time programs to quantum functionalities. 
To answer a query to $\ket{i, x_i, u}$, the LEQ oracle swaps $\cA_i$ between a fully un-used state $\ket{0^n,0}$ and a fresh run of $P_{\OSP}$:
\[
    \ket{i, x_i, u}_{\cQ} \otimes \left(\ket{\st_i}_{\cS} \otimes \ket{0^{n+1}}_{\cA_i} \right)
    \leftrightarrow
    \sum_{r_i} \ket{i, x_i, u\oplus y_{i, r_i}}_{\cQ} \otimes \left(\ket{\st_{i+1}}_{\cS}\otimes \ket{(\st_i, x_i, r_i), 1}_{\cA_i} \right)
\]
where $(\st_{i+1, r_i}, y_{i, r_i}) = P(\st, x_i; r)$ and $n$ is a placeholder length. Essentially, this approach strips out the compressed oracle's final attempt to re-compress the database after each query, allowing $x_i$ to be recorded and the computation to continue even if $y_i$ ignored $r_i$.

\paragraph{Construction.} Our construction is conceptually simple: give the evaluator a one-shot program for the functionality $P_{\OSP}$ that
\begin{enumerate}
    \item allows them to use a token $\ket{T_i}$ to evaluate $(\st_{i+1}, y_i) = P(\st_i, x_i; r_i)$ if they have a signature $\sigma$ on the memory state $\st_i$ together with the matching serial number $s_i$,
    \item then signs the new state $\st_{i+1}$ together with the next serial number $s_{i+1}$.\footnote{In the construction, we hide the memory from the adversary states $\st_{i+1}$ by encrypting them, so that we achieve a stronger security. We omit this step here for simplicity of exposition.}
\end{enumerate}
Explicitly, the evaluator chooses the next serial number $s_{i+1}$, then evaluates the OSP on input $(x_i, \st_i, \sigma_i, s_{i+1})$ using serial number $s_i$.
Additionally providing them with a signature on a starting serial number and initial memory state $s_1\concat \st_1$ allows the evaluator to kick-start its computation.

At any time the evaluator only holds a \emph{single} valid chain element
$(\st_{i},\sigma_i)$ tied to the current serial number $s_i$; advancing to the next step requires
consuming the current one-shot token $\ket{T_i}$ and obtaining from $P_{\OSP}$ a new signature
$\sigma_{i+1}$ on a fresh serial $s_{i+1}$.
Thus, any attempt to fork the computation (e.g., reuse the same state under two different next inputs)
would force either reusing the same one-shot serial $s$ on two distinct effective inputs, or forging a
fresh signature on some new $(s_{i+1},\st_{i+1})$-message.

To prevent forging a signature for a forked computation, we use \cite{GLRRV25}'s one-time signature construction: to sign $m$ using randomness $r$, sign $m\concat r \concat G_{\Sign}(m\concat r)$ where $G_{\Sign}$ is a random oracle. We show that their construction also gives useful guarantees for stateful programs where the OSP signs different messages depending on $\st_i$.

\paragraph{Security: Overview.} 
To prove security, we need to lift the one-time guarantees provided by the OSP to a stateful program.
The SEQ oracle for the OSP from the construction maintains a database containing many queries $(i, x_i, \st_i, \sigma_i, s_{i+1})$, along with the associated serial number $s_i$, and the answer $y_i$ and a signature $\sigma_{i+1}$ associated the next serial number $s_{i+1}$ with the next memory state $\st_{i+1}$.
Ultimately, we need to enforce that the database is \emph{ordered} and \emph{consistent}:\footnote{This database has been simplified for the purposes of this overview.}
\[
    D = 
    \left\{
    \begin{array}{rccl}
    [\text{serial} = {{\color{red}s_1}},
    &\text{input} = (1, x_1, {\color{red}\st_1}, \sigma'_1, \tikzmarknode{s2entry1}{\color{blue}s_2}), 
    &\text{output} = (y_1, \sigma_{2}),
    &\text{state} = \tikzmarknode{st2entry1}{\color{blue}\st_{2}}],
    \\
    &&&
    \\
    &&&
    \\{}
    [\text{serial} = \tikzmarknode{s2entry2}{\color{blue}s_2},
    &\text{input} = (2, x_2, \tikzmarknode{st2entry2}{\color{blue}\st_2}, \sigma'_2, {\color{teal}s_3}), 
    &\text{output} = (y_2, \sigma_{3}),
    &\text{state} = {\color{teal}\st_{3}}],
    \\
    &\dots&&
    % \\
    % \big[\text{serial} = {\color{blue}s_1}, 
    % &\text{input} = (2, x_2, {\color{blue}\sigma_2}, {\color{teal}s_3})), 
    % &\text{output} = (y_2, r_{1,x}\concat {\color{teal}r'_{2}})
    % \\
    % &\dots
    % \\
    % (s_{i}, 
    % &((x_{i}, \tilde{\st_i}, *, r'_{i}\concat *, s_{i+1})), 
    % &r_{i,x}\concat r'_{i+1})
    \end{array}
    \right\}
\] 
\begin{tikzpicture}[remember picture,overlay]
  \draw[->, thick, shorten >=3pt, shorten <=3pt] (s2entry1.south west) -- (s2entry2.north east);
  \draw[->, thick] (st2entry1.south west) -- (st2entry2.north east);
\end{tikzpicture}
Note that the above ordering asks that the $s_2$ and $\st_2$ recorded in the first row exactly match those in the second row, establishing continuity of the computation.

To prove security, we need to 1) establish this consistent ordering and 2) translate between the evaluator's queries (which are recorded directly in the SEQ oracle) and the LEQ oracle's interface. Although prior works provide a useful starting point, the greater structure of the LEQ oracle requires new ideas.

\paragraph{Ordering the Database: Dynamically Constrained Signatures.}
The evaluator can compute on a program state $\st$ if and only if it has a signature $\sigma$ on $\st$ together with a serial number $s$. 
Intuitively, the evaluator's behavior can be controlled by constraining the signatures to a certain set. If the program were stateless, then this is sufficient.
However, for RAM programs, the kinds of messages the program is willing to sign \emph{change} as the program is queried. 
% Hence, we need to ``dynamically constrain'' the signatures.

The key to dynamically constraining signatures lies in the structure of entry 2's input signature $\sigma'_2$: it contains some randomness $r'_{2,\Sign} = r'_{2}\concat G_\Sign(r'_2\concat m)$, along with a baseline signature on $m\concat r'_{2,\Sign}$. We use this structure to show that entry 1's output signature $\sigma_2$ also contains the \emph{same} randomness $r'_{2,\Sign}$, which is only possible if the same $s_2$ and $\st_2$ were also recorded in the first entry.

We show this in two steps. First, we argue that knowledge of $\sigma_{i+1}$ implies that $G_{\Sign}$ records $s_{i+1}$, $\st_{i+1}$, and the randomness used by the OSP for the first query.
Second, we strengthen the compressed oracle chaining lemma from \cite{GLRRV25} to establish a strong functional connection from the signed message to the first entry. 
As compared to their original chaining lemma, the strengthened version has much greater flexibility in the relation between the SEQ oracle's database entry and the $G_\Sign$ entry. 
We refer the reader to \Cref{sec:LEQ-proof-ordering} for more details.

% \justin{The first part is important, but maybe we don't need so much detail about the proof}

\paragraph{Translating Between SEQ and LEQ.}
The final step is to translate between the SEQ oracle, which handles all the cryptographic tools used to enforce consistent computation, and the LEQ oracle, which purely deals with the program functionality. The translation is handled via a carefully designed cache system. For SEQ security, it was sufficient to just cache the evaluator's inputs, but more is necessary for proving LEQ security.

The primary difference between the SEQ oracle and the LEQ oracle is the way that the LEQ enforces a consistent history by providing access only to the most recent query $i$ or a new query $i+1$, like a stack. 
It doesn't answer old queries $i-1$, even though a real evaluator would certainly be able to re-evaluate old queries. To match the evaluator's view, the simulator needs to answer old queries, which necessitates \emph{also} caching the output of each query. However, the input cache \emph{depends} on whether the output is known outside of the LEQ oracle. Ensuring that the output cache does not interfere with the input cache requires some additional care. We refer the reader to \Cref{sec:LEQ-security} for more details.
\subsection{Applications}
As we have discussed, the notion of query-limited RAM programs already captures applications such as $k$-time program and programs with a prefixed budget. Here, we demonstrate more applications.

\paragraph{Pay-Per-Use Programs}
We propose the following paradigm.
A pay-per-use program can be queried any number of times, and afterward, the user pays for
the cost of the queries that they made. At the time of the payment, the user has to provide a proof that its previous \emph{effective} queries (queries whose information is successfully obtained; uncomputed queries should not count) actually cost the value $c$ as they claim. Soundness is defined via effective queries, formalized through the REQ oracle: after using the program, the user gives a challenger its claimed cost $c$ and a proof; the challenger first interacts with the user to verify the validity of the the proof and if the proof is valid, the challenger then reads from the database of the LEQ oracle to check if it is equal to $c$.

%\jiahui{added construction}
In our construction, the RAM program has two modes: the regular $\Eval$ mode that evaluates functionality $P$; and a $\Prove$ mode that outputs the cost $c'$, a random string $r_1$, and a signature on $(c', r_1)$. During the cost verification protocol, the verifier will check if the alleged cost matches the signed message and the validity of the signature. After one uses the RAM program in $\Prove$ mode, it will enter a $\mathsf{Halt}$ mode that rejects to answer anything. This prevents the user from a "prove-then-query-more" attack.

%\jiahui{Add more details and proof summary later}

\paragraph{Semi-Quantum Query-Limited Programs} 
A semi-quantum query-limited programs allows one user to transfer the capability of evaluating a program to another user via only \emph{classical communication}. In addition to the original RAM program syntax, the semi-quantum query-limited programs have a $\mathsf{Transfer}$ algorithm that lets a sender send a program to a receiver via classical interactions. Our construction also consists of all components in the RAM program construction, with an additional $\Transfer$ protocol.

The $\Transfer$ protocol roughly works as follows: the receiver generates a token $\ket{T'}$ and corresponding serial number $s'$ after receiving the $\crs$ from the sender; the sender, after getting the receiver's serial number $s'$, evaluates the program in $\Transfer$ mode with a dummy input (e.g. a 0-string) using its own token $\ket{T}$ and the new serial $s'$. The sender then sends the output of the $\Transfer$ evaluation, the classical part of the program (i.e. the classical oracle) to the 
receiver. 
After the transfer, the sender's token will be consumed and the program the receiver obtains will work correctly under its new token $\ket{T'}$ . The receiver will acquire a program that inherits the sender's program's RAM state. In particular, the receiver will be able to evaluate the "remaining queries" in the program. Suppose the original program is a $k$-time program and the sender has used $k'$ queries, then the receiver will obtain a $(k-k')$-time program. Similar scenario applies to programs with budget/cost policies.
%\jiahui{Check this? Is the sender allowed to use part of the program and then send the program witj "remaining queries" to receiver? }
%\fi

%\iffalse
\paragraph{One-Shot Signatures from Blind-Unforgeable Signature Scheme} 
Finally, we show that $\OSP$ can be used not only as an input primitive (via $\OSS$) but also as an \emph{amplifier} that upgrades standard signature assumptions into one-shot signatures with more flexible signature formats.

Given a blind-unforgeable signature scheme $\BUS$, which one can obtain from post-quantum subexponentially secure signature schemes and a one-shot program compiler $\OSP$, there exists a one-shot signature scheme $\OSS$ whose verification algorithm is essentially $\BUS$'s verification. Here the one-shot program implements a signing functionality for $\BUS$ keyed by a serial number; $\OSP$ security enforces one-shotness per serial number, while blind-unforgeability supplies the underlying post-quantum unforgeability needed for the resulting signature format.
%\fi
\ifllncs \else \section{Preliminaries}
\label{sec:preliminaries}
A distribution over oracles (a.k.a. an oracle distribution) samples parameters $\mathsf{pp}$ from a given distribution and then implements the oracle $O_\mathsf{pp}$ defined by those parameters. Next, let us define the indistinguishability of two oracle distributions. Two oracle distributions are \textbf{perfectly indistinguishable} if any \textit{computationally unbounded} distinguisher, given quantum query access to an oracle drawn from one of the distributions, has $0$ advantage at distinguishing the two distributions. Next, two oracle distributions are \textbf{computationally indistinguishable} if any \textit{QPT} distinguisher, given quantum query access to an oracle drawn from one of the distributions, has $\negl(\secp)$ advantage at distinguishing the two distributions.

We use $*$ to denote wildcards in strings. For example, $x_1\concat *$ denotes any state from the set $\{x_1\concat x_2\}_{x_2}$.

\subsection{Compressed Oracles}\label{sec:compressed-oracle}
Compressed oracles are a technique for implementing a random oracle that allow the oracle to record queries \cite{Zha18}. Essentially, the oracle purifies its randomness, maintaining a superposition over functions in the distribution and responding to queries by operating coherently on this superposition.

\paragraph{Database Register $D$:} Let $H$ be a random oracle mapping $\cX \to \cY$. The compressed oracle has a database register $D$ that stores a partial truth table $d$ for the function. $d$ is a set of tuples of the form $(x, y) \in \cX \times \cY$. Every tuple in $d$ has a unique $x$-value. If $(x, y) \in d$, we say that $x \in d$ (slightly abusing notation) and $d(x) = y$. Alternatively, for a given $x$, if there exists no $y$ such that $(x, y) \in d$, we say that $x \notin d$ and $d(x) = \bot$. 

Next, let us define a useful eigenbasis for register $D$. For any $x \notin d$, any $\tilde{y} \in \cY$,
    \begin{equation}\label{eq:psi}
        \text{let } \ket{\psi(d, x, \tilde{y})}_D = \frac{1}{\sqrt{\abs{\cY}}} \cdot \sum_{y \in \cY} (-1)^{\langle y, \tilde{y}\rangle} \ket{d \cup \{(x,y)\}}_D
    \end{equation}
For any fixed $x \in \cX$, we will define an eigenbasis as follows. Our eigenbasis will include $\ket{d}$ and $\ket{\psi(d, x, \tilde{y})}$ for every database $d$ such that $x \notin d$ and every $\tilde{y} \in \cY$.

\paragraph{Subroutines:} The compressed oracle uses the following subroutines: $\{\Decomp_x\}_{x \in \cX}$, $\Decomp$, $\CO'$. They are unitaries that operate on the database register $D$ and sometimes also on a query register $Q$, which stores values of the form $(x,y) \in \cX \times \cY$.
\begin{itemize}
    \item $\{\Decomp_x\}_{x \in \cX}$: 
    For a given $x \in \cX$, let $d$ be a database such that $x \notin d$, and let $\tilde{y} \in \cY \backslash \{0\}$.
    \begin{align*}
        \ket{d} &\overset{\Decomp_x}{\longrightarrow} \ket{\psi(d, x, 0)}\\
        \ket{\psi(d, x, 0)} &\overset{\Decomp_x}{\longrightarrow} \ket{d}\\
        \ket{\psi(d, x, \tilde{y})} &\overset{\Decomp_x}{\longrightarrow} \ket{\psi(d, x, \tilde{y})}
    \end{align*}
    \item $\Decomp$: Let $(x, y) \in \cX \times \cY$, and let $d$ be a database.
    \begin{align*}
        \ket{x,y}_\cQ \otimes \ket{d}_\cD \overset{\Decomp}{\longrightarrow} \ket{x,y}_\cQ \otimes \left(\Decomp_x\ket{d}_\cD\right)
    \end{align*}
    \item $\CO'$: Let $(x, y) \in \cX \times \cY$, and let $d$ be a database such that $x \in d$. Then $\CO'$ acts as follows.
    \begin{align*}
        \ket{x,y}_\cQ \otimes \ket{d}_\cD \overset{\CO'}{\longrightarrow} \ket{x,y \oplus d(x)}_\cQ \otimes \ket{d}_\cD
    \end{align*}
    $\CO'$ acts as the identity on all states orthogonal to the ones above.
\end{itemize}

\paragraph{Compressed Oracle:} Now we can describe how the compressed oracle operates. The $D$ register is initialized to $\ket{\emptyset}$. On each query, the user submits a query register $Q = Q_x \times Q_y$. Then the oracle applies the following sequence of operations to $Q \times D$:
\[\CO := \Decomp \circ \CO' \circ \Decomp\]
Finally, the $Q$ register is returned to the user.

\paragraph{A Useful Eigenbasis:} Let us define a useful basis for $Q_x \times Q_y \times D$. The basis is easy to work with because queries to the compressed oracle simply permute these basis vectors. First, $Q_x$ (which stores the $x$-value of $Q$) will be in the computational basis. Second, $Q_y$ (which stores the $y$-value of $Q$) will be in the Fourier basis. The Fourier basis comprises the following basis vectors. For any $\tilde{y} \in \cY$,
\begin{equation}\label{eq:phi}
    \text{let } \ket{\phi(\tilde{y})}_{Q_y} = \frac{1}{\sqrt{\abs{\cY}}} \cdot \sum_{y \in \cY} (-1)^{\langle y, \tilde{y}\rangle} \ket{y}_{Q_y}
\end{equation}

Third, the basis for $D$ will be the one defined above (\cref{eq:psi}).

In total, the basis for $Q_x \times Q_y \times D$ comprises all vectors of the form
\begin{equation}\label{eq:basis}
\begin{split}
    &\ket{x}_{Q_x} \ket{\phi(\tilde{y})}_{Q_y} \ket{d}_D, \text{ or }\\
    &\ket{x}_{Q_x} \ket{\phi(\tilde{y})}_{Q_y} \ket{\psi(d,x,\tilde{y}')}_D\\
    &\text{where } x \in \cX, \quad \tilde{y}, \tilde{y}' \in \cY, \quad \text{$d$ is a database s.t. } x \notin d
\end{split}
\end{equation}

\begin{lemma}\label{thm:orthonormal-basis}
    The basis vectors in \cref{eq:basis} are an orthonormal basis for $Q_x \times Q_y \times D$.
\end{lemma}
\justin{Perhaps we should move the proof of this and the next lemma to the appendix and save the preliminaries just for theorem/lemma statements.}
\begin{proof}
    First, the vectors $\{\ket{\phi(\tilde{y})}\}_{\tilde{y} \in \cY}$ are orthonormal. For any $\tilde{y}, \tilde{y}' \in \cY$,
    \begin{align*}
        \bra{\phi(\tilde{y})}\ket{\phi(\tilde{y}')} &= \frac{1}{\abs{\cY}} \cdot \sum_{y, y' \in \cY} (-1)^{\langle y, \tilde{y} \rangle} \cdot (-1)^{\langle y', \tilde{y}'\rangle} \bra{y}\ket{y'}\\
        &= \frac{1}{\abs{\cY}} \cdot \sum_{y \in \cY} (-1)^{\langle y, \tilde{y} \rangle + \langle y, \tilde{y}' \rangle} = \frac{1}{\abs{\cY}} \cdot \sum_{y \in \cY} (-1)^{\langle y, \tilde{y} + \tilde{y}' \rangle}\\
        &= \frac{1}{\abs{\cY}} \cdot \abs{\cY} \cdot \mathbbm{1}_{\tilde{y} + \tilde{y}' = 0}\\
        &= \mathbbm{1}_{\tilde{y} = \tilde{y}'}
    \end{align*}
    We used the fact that when $\tilde{y} + \tilde{y}' \neq 0$, then $\langle y, \tilde{y} + \tilde{y}' \rangle = 0$ for exactly half of the $y$-values in $\cY$, and $\langle y, \tilde{y} + \tilde{y}' \rangle = 1$ for the other half. We also used the fact that $\tilde{y} + \tilde{y}' \neq 0$ if and only if $\tilde{y} = \tilde{y}'$. This is because we are working over the additive group of binary strings, so every vector is its own inverse.

    Second, $\ket{d}$ is orthogonal to $\ket{\psi(d,x,\tilde{y}')}$ for any $x \in \cX$, $\tilde{y}' \in \cY$, and any database $d$ for which $x \notin d$. This is because $d$ does not contain $x$, but $\ket{\psi(d,x,\tilde{y}')}$ is a superposition only over databases that contain $x$. $\ket{\psi(d,x,\tilde{y}')}$ is a superposition over databases of the form $\ket{d \cup \{(x,y)\}}$ for some $y \in \cY$. Such databases do contain $x$.

    Third, the vectors $\{\ket{\psi(d, x, \tilde{y})}\}_{d, x, \tilde{y}}$ (where $x \notin d$) are orthonormal. For any $\tilde{y}, \tilde{y}' \in \cY$,
    \begin{align*}
        \bra{\psi(d,x,\tilde{y})}\ket{\psi(d,x,\tilde{y}')} &= \frac{1}{\abs{\cY}} \cdot \sum_{y, y' \in \cY} (-1)^{\langle y, \tilde{y} \rangle} \cdot (-1)^{\langle y', \tilde{y}'\rangle} \bra{d \cup \{(x,y)\}}\ket{d \cup \{(x,y')\}}\\
        &= \frac{1}{\abs{\cY}} \cdot \sum_{y \in \cY} (-1)^{\langle y, \tilde{y} \rangle + \langle y, \tilde{y}' \rangle}\\
        &= \mathbbm{1}_{\tilde{y} = \tilde{y}'}
    \end{align*}

    Fourth, for any $y \in \cY$, $\ket{y}$ can be written as a linear combination of $\ket{\phi(\tilde{y})}$ vectors as follows:
    \[\ket{y} = \frac{1}{\sqrt{\abs{\cY}}} \cdot \sum_{\tilde{y} \in \cY} (-1)^{\langle y, \tilde{y}\rangle} \ket{\phi(\tilde{y})}\]
    This is because
    \begin{align*}
        \frac{1}{\sqrt{\abs{\cY}}} \cdot \sum_{\tilde{y} \in \cY} (-1)^{\langle y, \tilde{y}\rangle} \ket{\phi(\tilde{y})} &= \frac{1}{\sqrt{\abs{\cY}}} \cdot \sum_{\tilde{y} \in \cY} (-1)^{\langle y, \tilde{y}\rangle} \cdot \left(\frac{1}{\sqrt{\abs{\cY}}} \cdot \sum_{y' \in \cY} (-1)^{\langle y', \tilde{y}\rangle} \ket{y'}\right)\\
        &= \sum_{y' \in \cY} \ket{y'} \cdot \left(\frac{1}{\abs{\cY}} \cdot \sum_{\tilde{y} \in \cY} (-1)^{\langle y+y', \tilde{y} \rangle}\right)\\
        &= \sum_{y' \in \cY} \ket{y'} \cdot \left(\frac{1}{\abs{\cY}} \cdot \abs{\cY} \cdot \mathbbm{1}_{y+y' = 0}\right) = \sum_{y' \in \cY} \ket{y'} \cdot \mathbbm{1}_{y = y'}\\
        &= \ket{y}
    \end{align*}

    Fifth, for any $(x,y) \in \cX \times \cY$ and any database $d$ that does not contain $x$, the combined database $\ket{d \cup \{(x,y)\}}$ can be written as a linear combination of $\ket{\phi(d, x, \tilde{y})}$ vectors as follows:
    \[\ket{d \cup \{(x,y)\}} = \frac{1}{\sqrt{\abs{\cY}}} \cdot \sum_{\tilde{y} \in \cY} (-1)^{\langle y, \tilde{y}\rangle} \ket{\phi(d, x, \tilde{y})}\]

    This is because
    \begin{align*}
        \frac{1}{\sqrt{\abs{\cY}}} \cdot \sum_{\tilde{y} \in \cY} (-1)^{\langle y, \tilde{y}\rangle} \ket{\phi(d, x, \tilde{y})} &= \frac{1}{\sqrt{\abs{\cY}}} \cdot \sum_{\tilde{y} \in \cY} (-1)^{\langle y, \tilde{y}\rangle} \cdot \left(\frac{1}{\sqrt{\abs{\cY}}} \cdot \sum_{y' \in \cY} (-1)^{\langle y', \tilde{y}\rangle} \ket{d \cup \{(x,y')\}}\right)\\
        &= \sum_{y' \in \cY} \ket{d \cup \{(x,y')\}} \cdot \left(\frac{1}{\abs{\cY}} \cdot \sum_{\tilde{y} \in \cY} (-1)^{\langle y+y', \tilde{y} \rangle}\right)\\
        &= \sum_{y' \in \cY} \ket{d \cup \{(x,y')\}} \cdot \mathbbm{1}_{y=y'} = \ket{d \cup \{(x,y)\}}
    \end{align*}

    Sixth, the previous parts imply that the vectors in \cref{eq:basis} each have unit norm, and the vectors are orthogonal to each other. Furthermore, any computational basis eigenstate $\ket{x}_{Q_x} \ket{y}_{Q_y} \ket{d}_D$ can be written as a linear combination of the basis vectors.
\end{proof}

The next lemma says that a query to the compressed oracle just permutes the basis vectors of \cref{eq:basis}, as long as $\tilde{y}' \neq 0$, and furthermore, the query only modifies the value of the $D$ register, when viewed in this basis.

\begin{lemma}\label{thm:CO-query-behavior}
    For any $x \in \cX$, any $\tilde{y}, \tilde{y}' \in \cY$ such that $\tilde{y}' \neq 0$, and any database $d$ such that $x \notin d$,
    \begin{align*}
        \ket{x}_{Q_x} \ket{\phi(\textcolor{red}{0})}_{Q_y} \ket{\textcolor{red}{d}}_D &\overset{\CO}{\longrightarrow} \ket{x}_{Q_x} \ket{\phi(\textcolor{red}{0})}_{Q_y} \ket{\textcolor{red}{d}}_D \quad \text{($\tilde{y} = 0$)}\\
        \ket{x}_{Q_x} \ket{\phi(\textcolor{blue}{\tilde{y}})}_{Q_y} \ket{\textcolor{red}{d}}_D &\overset{\CO}{\longrightarrow} \ket{x}_{Q_x} \ket{\phi(\textcolor{blue}{\tilde{y}})}_{Q_y} \ket{\psi(d,x,\textcolor{blue}{\tilde{y}})}_D \quad \text{(if $\tilde{y} \neq 0$)}\\
        \ket{x}_{Q_x} \ket{\phi(\textcolor{ForestGreen}{\tilde{y}'})}_{Q_y} \ket{\psi(d,x,\textcolor{ForestGreen}{\tilde{y}'})}_D &\overset{\CO}{\longrightarrow} \ket{x}_{Q_x} \ket{\phi(\textcolor{ForestGreen}{\tilde{y}'})}_{Q_y} \ket{\textcolor{red}{d}}_D \quad (\tilde{y} = \tilde{y}')\\
        \ket{x}_{Q_x} \ket{\phi(\textcolor{blue}{\tilde{y}})}_{Q_y} \ket{\psi(d,x,\textcolor{ForestGreen}{\tilde{y}'})}_D &\overset{\CO}{\longrightarrow} \ket{x}_{Q_x} \ket{\phi(\textcolor{blue}{\tilde{y}})}_{Q_y} \ket{\psi(d,x,\textcolor{blue}{\tilde{y}} + \textcolor{ForestGreen}{\tilde{y}'})}_D \quad (\text{if }\tilde{y} \neq \tilde{y}')
    \end{align*}
\end{lemma}
\begin{proof}
    First, let us examine how $\CO'$ acts. For any $x \in \cX$, any $\tilde{y}, \tilde{y}' \in \cY$, and any database $d$ that does not contain $x$, $\CO'$ maps 
    \[\ket{x}\ket{\phi(\tilde{y})}\ket{\psi(d,x,\tilde{y}')} \overset{\CO'}{\longrightarrow} \ket{x}\ket{\phi(\tilde{y})}\ket{\psi(d,x,\tilde{y}'+\tilde{y})}\]
    This is because:
    \begin{align*}              
        \ket{x}\ket{\phi(\tilde{y})}\ket{\psi(d,x,\tilde{y}')} &= \ket{x} \left(\frac{1}{\sqrt{\abs{\cY}}} \cdot \sum_{y \in \cY} (-1)^{\langle y, \tilde{y}\rangle} \ket{y}\right)\left(\frac{1}{\sqrt{\abs{\cY}}} \cdot \sum_{y' \in \cY} (-1)^{\langle y', \tilde{y}'\rangle} \ket{d \cup \{(x,y')\}}\right)\\
        &= \frac{1}{\abs{\cY}} \cdot \sum_{y, y' \in \cY} (-1)^{\langle y, \tilde{y}\rangle + \langle y', \tilde{y}'\rangle} \cdot \ket{x}\ket{y}\ket{d \cup \{(x,y')\}}\\
        &\overset{\CO'}{\longrightarrow} \frac{1}{\abs{\cY}} \cdot \sum_{y, y' \in \cY} (-1)^{\langle y, \tilde{y}\rangle + \langle y', \tilde{y}'\rangle} \cdot \ket{x}\ket{y+y'}\ket{d \cup \{(x,y')\}}\\
        &= \frac{1}{\abs{\cY}} \cdot \sum_{y', y'' \in \cY} (-1)^{\langle y''-y', \tilde{y}\rangle + \langle y', \tilde{y}'\rangle} \cdot \ket{x}\ket{y''}\ket{d \cup \{(x,y')\}} \quad \text{where $y'' = y+y'$}\\
        &= \frac{1}{\abs{\cY}} \cdot \sum_{y', y'' \in \cY} (-1)^{\langle y'', \tilde{y}\rangle + \langle y', \tilde{y}'-\tilde{y}\rangle} \cdot \ket{x}\ket{y''}\ket{d \cup \{(x,y')\}}\\
        &= \frac{1}{\abs{\cY}} \cdot \sum_{y', y'' \in \cY} (-1)^{\langle y'', \tilde{y}\rangle + \langle y', \tilde{y}'+\tilde{y}\rangle} \cdot \ket{x}\ket{y''}\ket{d \cup \{(x,y')\}}\\
        &= \ket{x} \left(\frac{1}{\sqrt{\abs{\cY}}} \cdot \sum_{y'' \in \cY} (-1)^{\langle y'', \tilde{y}\rangle} \ket{y''}\right)\left(\frac{1}{\sqrt{\abs{\cY}}} \cdot \sum_{y' \in \cY} (-1)^{\langle y', \tilde{y}'+\tilde{y}\rangle} \ket{d \cup \{(x,y')\}}\right)\\
        &= \ket{x} \ket{\phi(\tilde{y})}\ket{\psi(d,x,\tilde{y}'+\tilde{y})}
    \end{align*}
    We used the fact that $\tilde{y} = -\tilde{y}$ because we are working over the additive group of binary strings.

    Second, let us examine how $\CO$ acts on $\ket{x} \ket{\phi(0)} \ket{d}$ (where $\tilde{y} = 0)$.
    \begin{align*}
        \ket{x} \ket{\phi(0)} \ket{d} &\overset{\Decomp}{\longrightarrow} \ket{x} \ket{\phi(0)} \ket{\psi(d,x,0)}\\
        &\overset{\CO'}{\longrightarrow} \ket{x} \ket{\phi(0)} \ket{\psi(d,x,0)}\\
        &\overset{\Decomp}{\longrightarrow} \ket{x} \ket{\phi(0)} \ket{d}
    \end{align*}

    Third, let us examine how $\CO$ acts on $\ket{x} \ket{\phi(\tilde{y})} \ket{d}$ when $\tilde{y} \neq 0$.
    \begin{align*}
        \ket{x} \ket{\phi(\tilde{y})} \ket{d} &\overset{\Decomp}{\longrightarrow} \ket{x} \ket{\phi(\tilde{y})} \ket{\psi(d,x,0)}\\
        &\overset{\CO'}{\longrightarrow} \ket{x} \ket{\phi(\tilde{y})} \ket{\psi(d,x,\tilde{y})}\\
        &\overset{\Decomp}{\longrightarrow} \ket{x} \ket{\phi(\tilde{y})} \ket{\psi(d,x,\tilde{y})}
    \end{align*}

    Fourth, let us examine how $\CO$ acts on $\ket{x} \ket{\phi(\tilde{y}')} \ket{\psi(d,x,\tilde{y}')}$ (where $\tilde{y} = \tilde{y}'$ and $\tilde{y}' \neq 0$).
    \begin{align*}
        \ket{x} \ket{\phi(\tilde{y}')} \ket{\psi(d,x,\tilde{y}')} &\overset{\Decomp}{\longrightarrow} \ket{x} \ket{\phi(\tilde{y}')} \ket{\psi(d,x,\tilde{y}')}\\
        &\overset{\CO'}{\longrightarrow} \ket{x} \ket{\phi(\tilde{y}')} \ket{\psi(d,x,0)}\\
        &\overset{\Decomp}{\longrightarrow} \ket{x} \ket{\phi(\tilde{y}')} \ket{d}
    \end{align*}

    Fifth, let us examine how $\CO$ acts on $\ket{x} \ket{\phi(\tilde{y})} \ket{\psi(d,x,\tilde{y}')}$ when $\tilde{y} \neq \tilde{y}'$ (and $\tilde{y}' \neq 0$).
    \begin{align*}
        \ket{x} \ket{\phi(\tilde{y})} \ket{\psi(d,x,\tilde{y}')} &\overset{\Decomp}{\longrightarrow} \ket{x} \ket{\phi(\tilde{y})} \ket{\psi(d,x,\tilde{y}')}\\
        &\overset{\CO'}{\longrightarrow} \ket{x} \ket{\phi(\tilde{y})} \ket{\psi(d,x,\tilde{y}+\tilde{y}')}\\
        &\overset{\Decomp}{\longrightarrow} \ket{x} \ket{\phi(\tilde{y})} \ket{\psi(d,x,\tilde{y}+\tilde{y}')}
    \end{align*}
\end{proof}

% \begin{lemma}
%     For any $x \in \cX$, let 
%     $\Pi_x$ project onto all databases that record nothing $d = \emptyset$ or a single input $x$. 
%     \[\Pi_x = \ketbra{\emptyset} + \sum_{u \in \cY} \ketbra{\psi(\emptyset,x,u)}\] 
%     Let the query register contain $x$ along with some superposition over $y$-values: 
%     \[\ket{q} = \sum_{y \in \cY} \alpha_y \ket{x, y}_Q\]
%     If before
    
% \end{lemma}

Next, we show a particular situation in which the compressed oracle acts almost as if it were just the $\CO'$ operation, without $\Decomp$. Intuitively, this occurs whenever an input-output pair $(x, H(x))$ is known external to the oracle. Furthermore, it also holds when only part of $H(x)$ is known, proportional to how much is known.

\jiahui{Original proof is a little problematic with a wrong use of triangle inequality. Fixed}

\iffalse
\proofaudit{P01}{Incorrect inequality; direct repair}{%
The split of diagonal and cross terms uses an inequality of the form $|A+B|^2\geq |A|^2-|B|^2$, which is false (take $B=-0.5, A=1$).  The lemma itself has a short repair. For a normalized $\ket{\psi}$, let $P_+$ project the queried database cell onto its uniform non-$\bot$ state, controlled on the query input. Orthogonality of the external $(x,r_2)$ records and Cauchy--Schwarz give $p:=\|P_+\psi\|^2\leq1/|\cR_2|$. There is initially no $\bot$ component in that cell. Since $\Decomp$ swaps $\ket{\bot}$ and the uniform state and fixes their orthogonal complement,
\[
 \langle\psi|\Decomp|\psi\rangle=1-p,\qquad
 \mathsf{TD}(\Decomp\psi,\psi)=\sqrt{2p-p^2}
 \leq\sqrt{2/|\cR_2|}.
\]
This proves a slightly stronger bound and also gives the useful vector estimate $\|(\Decomp-I)\psi\|\leq\sqrt{2/|\cR_2|}$. Replace the present calculation, including its final negative radicand, by this argument; state normalization explicitly:

}
\fi

% Replacement for Lemma \ref{lem:decomp-approx-identity} and its proof
% in 03-preliminaries.tex. Uses the existing project macros and environments.

\begin{lemma}\label{lem:decomp-approx-identity}
Let $H:\cX\to\cR_1\times\cR_2$ be a random oracle implemented as a
compressed oracle, where $\cR_1,\cR_2$ are finite and nonempty.
Let $\ket{\psi}$ be a normalized state of the form
\[
    \ket{\psi}
    =\sum_{x,r_2,D'_x,u}
      \alpha_{x,r_2,D'_x,u}\ket{\phi_{x,r_2}}\ket{x,u}
      \otimes\sum_{r_1\in\cR_1}
      \beta_{u,D'_x,x,r_1,r_2}
      \ket{D'_x\cup\{(x,r_1\concat r_2)\}},
\]
where the sum ranges over $x\in\cX$, $r_2\in\cR_2$, databases $D'_x$
with $D'_x(x)=\bot$, and computational-basis labels $u$.
Suppose that
$\braket{\phi_{x,r_2}}{\phi_{x',r'_2}}=0$
whenever $(x,r_2)\ne(x',r'_2)$, and that $\Decomp$ is controlled by
the displayed query-input register containing $x$.
Then, with trace distance defined using the factor $1/2$,
\[
    \mathsf{TD}(\Decomp\ket{\psi},\ket{\psi})
    \leq \sqrt{\frac{2}{|\cR_2|}-\frac{1}{|\cR_2|^2}}
    \leq \sqrt{\frac{2}{|\cR_2|}}.
\]
Moreover,
\[
    \bigl\|(\Decomp-I)\ket{\psi}\bigr\|
    \leq \sqrt{\frac{2}{|\cR_2|}}.
\]
\end{lemma}

\begin{proof}
Write $N_1=|\cR_1|$, $N_2=|\cR_2|$, and $N=N_1N_2$.
For $d(x)=\bot$, define
\[
    \ket{+_{x,d}}
    :=\frac{1}{\sqrt N}
      \sum_{a\in\cR_1,\,b\in\cR_2}
      \ket{d\cup\{(x,a\concat b)\}}.
\]
For fixed $x$, these states are orthonormal as $d$ varies. Each is
orthogonal to every database basis state whose cell at $x$ is $\bot$.
Let
\[
    P_{+,x}:=\sum_{d:\,d(x)=\bot}\ket{+_{x,d}}\bra{+_{x,d}},
    \qquad
    P_+:=\sum_x\ket{x}\bra{x}_{Q_x}\otimes P_{+,x},
\]
with identity operators on all other registers understood.
Thus $P_+$ projects the queried database cell onto its uniform
non-$\bot$ state.

For brevity, let $j=(x,r_2,D'_x,u)$,
$\alpha_j=\alpha_{x,r_2,D'_x,u}$,
$\beta_{j,a}=\beta_{u,D'_x,x,a,r_2}$, and
\[
    w_j:=|\alpha_j|^2\bigl\|\ket{\phi_{x,r_2}}\bigr\|^2.
\]
The orthogonality assumption, the computational-basis registers $x,u$,
and the database basis give the normalization identity
\begin{equation}\label{eq:decomp-repair-normalization}
    1=\|\ket{\psi}\|^2
     =\sum_j w_j\sum_{a\in\cR_1}|\beta_{j,a}|^2.
\end{equation}
In particular, no separate normalization of $\alpha$ or $\beta$ is
needed. The factors $\|\ket{\phi_{x,r_2}}\|^2$ can be omitted when
these external vectors are normalized.

For each $a\in\cR_1$,
\[
    P_{+,x}\ket{D'_x\cup\{(x,a\concat r_2)\}}
    =\frac{1}{\sqrt N}\ket{+_{x,D'_x}}.
\]
Consequently,
\[
    P_+\ket{\psi}
    =\frac{1}{\sqrt N}\sum_j
      \alpha_j\left(\sum_{a\in\cR_1}\beta_{j,a}\right)
      \ket{\phi_{x,r_2}}\ket{x,u}\ket{+_{x,D'_x}}.
\]
The vectors indexed by distinct $j$ in this expression are mutually
orthogonal: different $x$ or $u$ are distinguished by the query
registers; different $r_2$ for fixed $x$ are distinguished by the
external vectors; and different $D'_x$ for fixed $x$ give orthogonal
uniform database states. Therefore, by Cauchy--Schwarz and
\eqref{eq:decomp-repair-normalization},
\begin{align*}
    p:=\|P_+\ket{\psi}\|^2
    &=\frac1N\sum_j w_j
      \left|\sum_{a\in\cR_1}\beta_{j,a}\right|^2\\
    &\leq\frac{N_1}{N}\sum_jw_j
      \sum_{a\in\cR_1}|\beta_{j,a}|^2
     =\frac1{N_2}.
\end{align*}

Let $P_\bot$ project onto databases whose queried cell equals $\bot$,
controlled on $Q_x$. 
\[
    P_{\bot,x}
    :=\sum_{d:\,d(x)=\bot}\ket d\bra d,
    \qquad
    P_\bot
    :=\sum_{x\in\cX}
      \ket{x}\bra{x}_{Q_x}\otimes P_{\bot,x},
\]
with identity operators on all other registers understood.
%Thus $P_\bot$ projects onto the subspace in which the database cell selected by the query-input register $Q_x$ equals $\bot$.
Every term in the prescribed state $\ket{\psi}$ has a non-$\bot$
entry at its queried input, so $P_\bot\ket{\psi}=0$.

The prescribed form of $\ket{\psi}$ implies
$P_\bot\ket{\psi}=0$.
By definition, $\Decomp_x$ swaps $\ket{+_{x,d}}$ with $\ket d$ and
fixes their orthogonal complement, for every $d(x)=\bot$.
Set
\[
    \ket v=P_+\ket{\psi},\qquad
    \ket z=(I-P_+)\ket{\psi},\qquad
    \ket b=\Decomp\ket v.
\]

Since $P_+$ is the orthogonal projector onto the uniform
non-$\bot$ subspace of the queried cell, the definition of $\ket z$
removes precisely this component:
\[
    P_+\ket z
    =P_+(I-P_+)\ket{\psi}
    =(P_+-P_+^2)\ket{\psi}
    =0.
\]
Moreover, $P_\bot P_+=0$ because the uniform non-$\bot$ subspace
is orthogonal to the $\bot$ subspace; together with
$P_\bot\ket{\psi}=0$, this gives
\[
    P_\bot\ket z
    =P_\bot(I-P_+)\ket{\psi}
    =P_\bot\ket{\psi}-P_\bot P_+\ket{\psi}
    =0.
\]
Thus $\ket z$ is orthogonal to both subspaces exchanged by
$\Decomp$, so $\Decomp\ket z=\ket z$.

%The vector $\ket z$ has neither a uniform component nor a $\bot$ component in the queried cell, so $\Decomp\ket z=\ket z$.
Moreover, $\ket b$ lies in the range of $P_\bot$, whereas $\ket v$
and $\ket z$ lie in its kernel. Thus $\ket v,\ket z,\ket b$ are
pairwise orthogonal and
\[
    \ket{\psi}=\ket z+\ket v,\qquad
    \Decomp\ket{\psi}=\ket z+\ket b,
    \qquad
    \|\ket v\|^2=\|\ket b\|^2=p,
    \qquad \|\ket z\|^2=1-p.
\]
It follows exactly that
\[
    \bra{\psi}\Decomp\ket{\psi}=1-p,
    \qquad
    \bigl\|(\Decomp-I)\ket{\psi}\bigr\|^2
      =\|\ket b-\ket v\|^2=2p.
\]
Using the trace-distance identity for normalized pure states,
\[
    \mathsf{TD}(\Decomp\ket{\psi},\ket{\psi})
    =\sqrt{1-|\bra{\psi}\Decomp\ket{\psi}|^2}
    =\sqrt{2p-p^2}.
\]
Since $p\leq 1/N_2\leq1$ and $2p-p^2$ is increasing on $[0,1]$,
\[
    \mathsf{TD}(\Decomp\ket{\psi},\ket{\psi})
    \leq\sqrt{\frac2{N_2}-\frac1{N_2^2}} = \sqrt{\frac2{|\cR_2|}-\frac1{|\cR_2|^2}} \leq \sqrt{\frac2{|\cR_2|}},
    \qquad
    \bigl\|(\Decomp-I)\ket{\psi}\bigr\|
    \leq\sqrt{\frac2{N_2}}.
\]
Therefore we also have
 \[
        |\bra{\psi}\Decomp\ket{\psi}|^2 \geq 1 - \frac{1}{|\cR_2|} 
    \]
\end{proof}

We mention a few other useful lemmas which were introduced by prior works. Zhandry showed that if a quantum algorithm would find $(x, H(x))$ pairs after a polynomial number of queries, then $H$ records a corresponding $(x, r)$ pair with overwhelming probability when it is implemented as a compressed oracle.

\begin{lemma}[\cite{zhandry19compressed}, Lemma 5]
   \label{lem:zhandry_lemma5} 
   Consider a quantum algorithm $\cA$ making queries to a random oracle
$H$ and outputting tuples $(x_1, \cdots , x_k, y_1, \cdots , y_k)$
. Let $R$ be a collection of such
tuples. Suppose with probability $p$, $\cA$ outputs a tuple such that (1) the tuple is in
$R$ and (2) $H(x_i) = y_i$ for all $i$. Now consider running $\cA$ with the compressed oracle,
and suppose the database $D$ is measured after $\cA$ produces its output. Let $p'$
be the probability that (1) the tuple is in $R$, and (2) $D(x_i) = y_i$ for all $i$ (and in
particular $D(x_i) \neq \bot$). Then:
   $$ \sqrt{p} \leq \sqrt{p'} + \sqrt{k/\vert \cY \vert}$$
\end{lemma}

\cite{GLRRV25} observed that the compressed oracle treats all inputs identically, and so it equivalent up to renaming.

\begin{claim}[\cite{GLRRV25}]\label{claim:compressed-oracle-renaming}
    Let $\mathsf{SWITCH}_{x_1,x_2}$ be the unitary which maps any database $D$ to the unique $D'$ defined by $D'(x_1) = D(x_2)$, $D'(x_2) = D(x_1)$, and $D'(x_3) = D(x_3)$ for all $x_3\notin \{x_1, x_2\}$ and acts as the identity on all orthogonal states. Let $U_{x_1,x_2}$ be the unitary which maps $\ket{x_1}\mapsto \ket{x_2}$, $\ket{x_2}\mapsto \ket{x_1}$ and acts as the idenity on all orthogonal states. Then
    \[
        \CO  
        = 
        (U_{x_1,x_2} \otimes I \otimes \mathsf{SWITCH}_{x_1,x_2}) \CO (U_{x_1,x_2} \otimes I \otimes \mathsf{SWITCH}_{x_1,x_2})
    \]
    % operates on $D$ by switching any $(x_1, y_1)\in D$ to  $(x_2, y_1)$ and switching $(x_2, y_2)\in D$ to $(x_1, y_2)$ and otherwise operating as the identity. 
\end{claim}
\begin{proof}
    This follows from the observation that $(U_{x_1,x_2} \otimes I \otimes \mathsf{SWITCH}_{x_1,x_2})$ commutes with both $\Decomp$ and $\CO'$ and it is self-inverse.
\end{proof}

\subsection{Pseudorandom Functions}
Here we define pseudorandom functions (PRFs), following definition 3.24 of \cite{KatLin}.

A \textbf{keyed function} $F: \bit^* \times \bit^* \to \bit^*$ takes as input a key $k$ and a main input $x$ and outputs $y$. The function is \textbf{length preserving} if for any key length $\secp \in \bbN$, and any $k \in \bit^\secp$, $F(k, \cdot)$ is only defined for $x$-values in $\bit^\secp$, and for any such $x$, $F(k, x) \in \bit^\secp$.

Next, let $\mathsf{Func}_\secp$ be the set of all functions mapping $\bit^\secp \to \bit^\secp$.

\begin{definition}[Pseudorandom Function]\label{def:PRF}
    An efficient, length preserving, keyed function $\PRF: \bit^* \times \bit^* \to \bit^*$ is a \textbf{pseudorandom function} if for all QPT distinguishers $D$, there is a negligible function $\negl$ such that for any $\secp \in \bbN$:
    \[\abs{\Pr\left[b = 1 : \begin{array}{ll}
         k &\getsr \bit^\secp  \\
         b &\gets D^{\PRF(k, \cdot)}(1^\secp)
    \end{array}\right] - \Pr\left[b = 1 : \begin{array}{ll}
         f &\getsr \mathsf{Func}_\secp  \\
         b &\gets D^{f(\cdot)}(1^\secp)
    \end{array}\right]} \leq \negl(\secp)\]
\end{definition}

\subsection{Blind-Unforgeable Signatures}

We recall here the definition of a (classical) signature scheme and the definition of blind unforgeability, a notion of post-quantum security. 

The following definition of a signature scheme comes from \cite{KatLin}\bhaskar{Cite the specific definition in \cite{KatLin}}.

\begin{definition}[Signature Scheme]\label{def:classical-signature-scheme}
    A (classical) signature scheme for message space $\cM$ comprises three PPT algorithms $(\KeyGen, \Sign, \Ver)$ with the following \textbf{syntax}:
    \begin{itemize}
        \item $\KeyGen(1^\secp) \to (\sk, \pk)$: Takes a security parameter $\secp \in \bbN$ and generates a secret signing key $\sk$ and a public verification key $\pk$. We assume that $\sk, \pk$ are each at least $\secp$ bits long and that $\secp$ can be determined from $\sk$ or $\pk$.
        \item $\Sign(\sk, m) \to \sigma$: Takes a message $m \in \cM$ and outputs a signature $\sigma$.
        \item $\Ver(\pk, m, \sigma) \to \bit$: Verifies the signature $\sigma$ on message $m$. $\Ver$ outputs $1$ for valid signatures and $0$ for invalid signatures.
    \end{itemize}
    All variables $(\secp, \sk, \pk, m, \sigma)$ are classical strings.
    
    Additionally, the signature scheme satisfies \textbf{perfect correctness}, which says that for any $\secp \in \bbN$ and any message $m \in \cM$,
    \[\Pr\left[b = 1 : \substack{(\sk, \pk) \gets \KeyGen(1^\secp)\\
    \sigma \gets \Sign(\sk, m)\\
    b \gets \Ver(\pk, m, \sigma)}\right] = 1\]
\end{definition}

Next, here is \cite{GLRRV25}'s variant on the notion of blind unforgeability for signatures~\cite{EC:AMRS20}, verbatim.

\begin{definition}[Quantum Blind Unforgeability]\label{def:blind-unforge}
    A signature scheme $(\KeyGen, \Sign, \Ver)$ for message space $\cM$ (\cref{def:classical-signature-scheme}) is \emph{blind-unforgeable} if for every QPT adversary $\cA$ and blinding set $B\subset \cM$,
    \[
        \Pr\left[
                m\in B
                \land 
                \Ver(\pk, m, \sigma) = \mathsf{Accept} 
            :
            \begin{array}{c}
                (\sk, \pk) \gets \KeyGen(1^\secp)
                \\
                (m, \sigma) \gets \cA^{\Sign_B(\sk, \cdot)}(\pk)
            \end{array}
        \right] = \negl(\lambda),
    \]
    where $\Sign_B(\sk, \cdot)$ denotes a (quantumly-accessible) signature oracle that signs messages $m$ using $\sk$ if $m\notin B$, and otherwise outputs $\bot$.
\end{definition}

This definition differs from the original in that the adversary may choose its blinding set $B$. \cite{EC:AMRS20} show that the hardness of this task is polynomially related to their original definition, which samples $B$ uniformly at random.

We note that any sub-exponentially secure signature scheme is blind-unforgeable\footnote{Here we need the scheme to be secure against subexponential time and queries, and we also need that the message space is smaller than this subexponential bound.}, since the adversary could simply query for all signatures in $\cM \backslash B$, then simulate the blind-unforgeability experiment. \cite{EC:AMRS20} also gives several other signature schemes which are blind-unforgeable (under the original definition).

\subsection{Signature Tokens}
A signature token is a quantum state that can be used to sign one and only one message \cite{BDS23}. Here, we expand the definition of signature token schemes to encompass one-shot signatures as well.

In our definition (\cref{def:signature-token}), a signature token also comes with a unique serial number $s$, and the security guarantee is that an adversary cannot produce two valid message-signature pairs \textit{for the same serial number}. This guarantee resembles the guarantee of one-shot signatures. 

Indeed the one-shot signature primitive is a special case of  \cref{def:signature-token} where $\sk = \pk$ so that signature tokens can be generated publicly. Additionally, our definition is general enough to capture regular signature tokens as well. A regular signature token scheme simply enforces that there is only one valid serial number. Along with the strong unforgeability property, this ensures that an adversary cannot generate two valid message-signature pairs.

The following definition of a generalized signature token scheme is adapted from \cite{BDS23} and \cite{SZ25} definition 9.
\begin{definition}[Generalized Signature Token Scheme]\label{def:signature-token}
$ $
\begin{itemize}
    \item \textbf{Syntax:} A generalized signature token scheme comprises the following QPT algorithms:
    \begin{itemize}
        \item $\KeyGen(1^\secp) \to (\sk, \pk)$: Takes a security parameter $\secp \in \bbN$ and generates a secret signing key $\sk$ and a public verification key $\pk$.
        \item $\TokenGen(\sk) \to (\ket{T}, s)$: Generates a quantum signing token $\ket{T}$ and a classical serial number $s \in \cS_\secp \cup \{\bot\}$ associated with $\ket{T}$.
        $\TokenGen$ may output $(\ket{T},s) = (\bot, \bot)$ with negligible probability.
        \item $\Sign(\ket{T}, m) \to \sigma$: Takes a signing token $\ket{T}$ and a message $m$ from the message space $\cM_\secp$ and outputs a signature $\sigma \in \Sigma_\secp$.
        \item $\Ver(\pk, s, m, \sigma) \to \bit$: Verifies the signature $\sigma$ on message $m$ with serial number $s$. $\Ver$ outputs $1$ for valid signatures and $0$ for invalid signatures.
    \end{itemize}
    The variables $(\secp, \sk, \pk, s, m, \sigma)$ are classical, and $\ket{T}$ is a quantum state.

    % \item \textbf{Correctness:} The scheme is \textit{correct} if there is a negligible function $\varepsilon(\secp)$ such that for any $\secp \in \bbN$ and any message $m \in \cM_\secp$,
    % \[\Pr\Bigg[1 \gets \Ver(\pk, s, m, \sigma): 
    % \substack{
    %     (\sk, \pk) \gets \KeyGen(1^\secp)\\
    %     (\ket{T}, s) \gets \TokenGen(\sk)\\
    %     \sigma \gets \Sign(\ket{T}, m) 
    % }
    % \Bigg] \geq 1 - \varepsilon(\secp)\]
    \item \textbf{Correctness:} The scheme is \textit{correct} if there is a negligible function $\negl(\secp)$ such that for all $\secp \in \bbN$,
    \[\Pr\left[(\ket{T},s) = (\bot, \bot) : \substack{
        (\sk, \pk) \gets \KeyGen(1^\secp)\\
        (\ket{T}, s) \gets \TokenGen(\sk)}\right] \leq \negl(\secp)\]
    and there exists a negligible function $\negl'(\secp)$ such that for all $\secp \in \bbN$, all $m \in \cM_\secp$, all $(\sk, \pk)$ in the support of $\KeyGen(1^\secp)$, and all $(\ket{T}, s)$ in the support of $\TokenGen(\sk)$ \textit{for which $(\ket{T}, s) \neq (\bot, \bot)$},
    \begin{align*}
        \Pr\Bigg[b = 1 : 
        \substack{
        \sigma \gets \Sign(\ket{T}, m)\\
        b \gets \Ver(\pk, s, m, \sigma)
        }
    \Bigg] \geq 1 - \negl'(\secp)
    \end{align*}
    \bhaskar{Check that this notion of correctness is actually achieved by \cite{BDS23,SZ25}.}
    \item \textbf{Strong Unforgeability:} The scheme is \textit{strongly unforgeable} if for any QPT adversary $\cA$, there exists a negligible function $\negl(\secp)$ such that for any $\secp \in \bbN$,
    \[\Pr\Bigg[
    \substack{
        (m_0, \sigma_0) \neq (m_1, \sigma_1) \land\\
        1 \gets \Ver(\pk, s', m_0, \sigma_0) \land\\
        1 \gets \Ver(\pk, s', m_1, \sigma_1)
    }:
    \substack{
        (\sk, \pk) \gets \KeyGen(1^\secp)\\
        (\ket{T}, s) \gets \TokenGen(\sk)\\
        (m_0, m_1, \sigma_0, \sigma_1, s') \gets \cA(1^\secp, \pk, \ket{T}, s) 
    }
    \Bigg] \leq \negl(\secp)\]
\end{itemize}
% \bhaskar{I suspect that we only need a MAC rather than a signature.}
\end{definition}

\bhaskar{The following discussion should be elaborated further. In particular, we should argue that our precise notion of correctness is achieved and that \cite{SZ25}'s scheme satisfies strong unforgeability, not just regular unforgeability.} 
The signature tokens defined in \cite{BDS23} (definitions 4-6, 25) are a special case of the tokens in \cref{def:signature-token} where $s$ has the same value $s^*$ for every token and $\Ver$ requires that $s = s^*$. This ensures that no adversary can generate two different valid message-signature pairs. We call such a scheme a \textit{one-time} signature token scheme. In particular, we would use a one-time signature token scheme with strong unforgeability.

\begin{definition}[One-Time Signature Token Scheme]\label{def:one-time-signature-token}
    A \textbf{one-time signature token} scheme ($\KeyGen$, $\TokenGen$, $\Sign$, $\Ver$) is a type of generalized signature token scheme (\cref{def:signature-token}) where for any $\secp \in \bbN$, the set of valid serial numbers $\cS_\secp$ contains just one element $s^*$. Then for any $\secp \in \bbN$ and any $(\sk, \pk)$ in the support of $\KeyGen(1^\secp)$,
    \begin{itemize}
        \item $\TokenGen(\sk)$ always outputs $s \in \{s^*,\bot\}$.
        \item For any $(s, m, \sigma)$ such that $s \neq s^*$, $\Ver(\pk, s, m, \sigma)$ always outputs $0$.
    \end{itemize}
\end{definition}

One-shot signatures are a special case of \cref{def:signature-token} in which $\TokenGen$ is a public procedure ($\sk = \pk$). 
\begin{definition}[One-Shot Signature Scheme]\label{def:one-shot-signature}
    A \textbf{one-shot signature} scheme ($\KeyGen$, $\TokenGen$, $\Sign$, $\Ver$) is a type of generalized signature token scheme (\cref{def:signature-token}) where for any $\secp \in \bbN$ and any $(\sk, \pk)$ in the support of $\KeyGen(1^\secp)$,
    \[\sk = \pk.\]
\end{definition}
\cite{SZ25} construct one-shot signatures, which satisfy all the properties of \cref{def:signature-token} and satisfy $\sk = \pk$.\footnote{We can map the notation of \cite{SZ25}, definition 9, to the notation of our \cref{def:signature-token} as follows: $\mathsf{Setup} \to \KeyGen$, $\mathsf{Gen} \to \TokenGen$, $\mathsf{CRS} \to \pk$, $\mathsf{CRS} \to \sk$, $\pk \to s$, $(\pk, \ket{\sk}, \mathsf{CRS}) \to \ket{T}$.

We also note that \cite{SZ25} defines a weaker notion of security (regular unforgeability), but their construction satisfies strong unforgeability as well. %Their notion of security says that it is hard to find a valid signature on two different messages. This is weaker than strong unforgeability, which says it is also hard to find two different signatures on the same message. Fortunately, their construction satisfies strong unforgeability as well. \bhaskar{To do: explain why their construction is strongly secure}}%, $\sigma \to (\pk, \sigma)$
}

% \bhaskar{\cite{SZ25,BDS23}'s signature tokens are strongly secure, although for \cite{SZ25}, the strong security doesn't follow immediately from the definition of one-shot signatures.}

% \subsection{One-Shot Signatures}
% \label{sec:one_shot_prelim}

\subsection{One-Time Programs}
The following definitions come from \cite{GLRRV25} section 4.

\paragraph{Syntax:} Let $\cF$ be a family of classical randomized functions $f: \cX \times \cR \to \cY$. $\cX$, $\cR$, $\cY$ are some sets whose descriptions are included in the description of $f$. $f$ may be associated with some public, classical auxiliary information $\mathsf{aux}_f$. Next, each $f \in \cF$ takes a user-defined input $x \in \cX$, samples $r \getsr \cR$, and outputs $y = f(x;r)$. 

A one-time program compiler for a function family $\cF$ takes a description of an $f \in \cF$ and outputs a quantum program $\sfP_f$, which allows the user to evaluate $f$.

\begin{definition}[One-Time Program Compiler for $\cF$]\label{def:otp-syntax}
A one-time program compiler for $\cF$ is a pair of QPT algorithms $(\PGen, \Eval)$, which have the following syntax:
\begin{itemize}
    \item $\sfP_f \gets \PGen(1^\secp, f)$: Takes as input the security parameter $1^\secp$ and a description of the function $f \in \cF$, and outputs a quantum program $\sfP_f$, which may comprise a quantum state and an oracle.
    \item $y \gets \Eval(\sfP_f, x)$: Takes a quantum program $\sfP_f$ and a classical input $x \in \cX$ and outputs a classical value $y \in \cY$.
\end{itemize}
\end{definition}

\paragraph{Correctness:} The one-time program is correct if querying the program on any input $x$ results in a distribution over $y$-values that is statistically close to the output of the classical randomized functionality.

\begin{definition}[Correctness, \cite{GLRRV25} definition 4.2]\label{def:otp-correctness}
    A one-time program compiler for $\cF$ is \textbf{correct} if for every $f \in \cF$ and $x \in \cX$, the following distributions over $y$-values have negligible statistical distance:
    \begin{align*}
        \left\{\Eval\left[\PGen(1^\secp, f), x\right]\right\} \quad \text{and} \quad \left\{f(x;r)\right\}
    \end{align*}
    The randomness in the second distribution is over the sampling of $r \getsr \cR$.
\end{definition}

\paragraph{Security:} We will use the notion of single-effective query (SEQ) security, which is a simulation-based notion of security introduced in \cite{GLRRV25} that is designed to model uncomputation attacks. A real-world adversary can always uncompute their query and return to the original program state if they did not make any destructive measurements, so our ideal-world functionality, called the SEQ oracle, allows the adversary to uncompute their queries as well. The SEQ oracle uses the compressed oracle technique to record the adversary's queries and stops answering new queries once one query has been recorded. Security says that the real one-time program for $f$ can be simulated given only the SEQ oracle for $f$.

The SEQ oracle $O^\SEQ_f$ implements the function mapping $x \to f(x; H(x))$, where $H$ is a random oracle mapping $\cX \to \cR$. The SEQ oracle implements $H$ as a compressed oracle, which allows it to record queries, and it refuses to answer any query that would increase the number of recorded queries to $2$ or more. The following definition of the SEQ oracle comes from \cite{GLRRV25}, section 4.2.

$O^\SEQ_f$ has an internal database register $D$, which can store classical values of the form $d = (x,r) \in \cX \times \cR$ or $d = \emptyset$. $D$ is initialized to $\ket{\emptyset}$. $O^\SEQ_f$ also acts on a query register $Q = Q_x \times Q_u \times Q_B$, which stores classical values of the form $(x, u, b) \in \cX \times \cY \times \bit$. On each query, $O^\SEQ_f$ acts as a unitary on $Q \times D$ and an ancilla register $Q'$ according to the procedure given in \cref{fig:seq-oracle}.

\begin{figure}[H]
\begin{mdframed}
    To respond to a query, $O^\SEQ_f$ acts as a unitary on $Q \times D$ and an ancilla register $Q'$. Given some basis state $\ket{x,u,b}_Q \otimes \ket{d}_D$, $O^\SEQ_f$ does the following:
    \begin{enumerate}
        \item If $d = (x', r)$ for some $x' \neq x$, then skip the following steps. Otherwise, continue.
        \item Prepare the state $\ket{x,0}$ in an ancillary register $Q'$. Query $H$ by applying the compressed oracle unitary $\CO$ to $Q' \times D$.\label{step:seq-oracle-queries-CO}
        \item Apply the following isometry to registers $Q \times Q'$:
        \begin{align*}
            \ket{x, u, b}_Q \otimes \ket{x,r}_{Q'} \to \ket{x, u \oplus f(x;r), b \oplus 1}_Q \otimes \ket{x,r}_{Q'}
        \end{align*}
        \item Uncompute \cref{step:seq-oracle-queries-CO}.
    \end{enumerate}
\end{mdframed}
\caption{Querying the SEQ Oracle $O^\SEQ_f$}\label{fig:seq-oracle}
\end{figure}

Now we can define our notion of security, which says that the one-time program for $f$ can be simulated given only the SEQ oracle $O^\SEQ_f$.
\begin{definition}[SEQ Security, \cite{GLRRV25} definition 4.8]\label{def:seq-security}
    Let $\secp$ be the security parameter. Let $f$ be a function in a function family $\cF$ that possibly comes with a public, classical auxiliary input $\mathsf{aux}_f$. Let $D$ be a QPT distinguisher. 
    
    A one-time program compiler for $\cF$ satisfies \textbf{SEQ security} if there exists a QPT simulator $\Sim$\footnote{We allow $\Sim$ to take $1^\secp$ as input. We believe \cite{GLRRV25} should have included this feature in their definition of SEQ security because the simulator that they construct in figure 4 does take $1^\secp$ as input.} such that for every $f \in \cF$ and all distinguishers $D$, there exists a negligible function $\negl(\cdot)$ such that for all $\secp \in \mathbb{N}$,
    \begin{align*}
        \abs{\Pr\left[1 \gets D\left(\PGen(1^\secp, f), f, \mathsf{aux}_f\right)\right] - \Pr\left[1 \gets D\left(\Sim^{O^\SEQ_f}(1^\secp, \mathsf{aux}_f), f, \mathsf{aux}_f\right)\right]} \leq \negl(\secp)
    \end{align*}
    \bhaskar{I changed the definition to allow $\Sim$ to know $1^\secp$. I believe that this should have been included in \cite{GLRRV25}'s definition because the simulator that they construct takes $\secp$ as an input.}
\end{definition}

\subsection{qIND-qCPA Secure  Secret Key Encryption}
In this section, we introduce a security notion for SKE where the quantum adversary an make superposition queries in both the query phase and challenge phase.
The definition is formalized in \cite{gagliardoni2016semantic}.

\begin{definition}[Quantum encryption oracle]
Let $\mathcal{E} = (\KeyGen,\Enc,\Dec)$ be a secret-key encryption scheme, and let
$k \leftarrow \KeyGen(1^n)$.
The quantum encryption oracle associated with $\Enc_k$ is the unitary
\[
U_{\Enc_k} :
\sum_{x,y} \alpha_{x,y}\ket{x}\ket{y}
\;\mapsto\;
\sum_{x,y} \alpha_{x,y}\ket{x}\ket{y \oplus \Enc_k(x)}.
\]
Within a single oracle query, the same encryption randomness is used coherently
across all basis states in superposition; for each new query, fresh independent
randomness is used.
\end{definition}

\begin{definition}[qCPA learning phase]
In the qCPA learning phase, the adversary $\mathcal{A}$ is given oracle access
to $U_{\Enc_k}$.
\end{definition}

\begin{definition}[Type-(2) encryption operator]
For length-preserving encryption, define the type-(2) operator
\[
U^{(2)}_{\Enc_k} :
\sum_x \alpha_x \ket{x}
\;\mapsto\;
\sum_x \alpha_x \ket{\Enc_k(x)}.
\]
More generally, if encryption expands the message length, one may view
$U^{(2)}_{\Enc_k}$ as a unitary acting on an input register together with an
ancilla register initialized to $\ket{0}$, such that applying it to
$\rho \otimes \ket{0}\!\bra{0}$ yields the honest encryption of $\rho$.
\end{definition}

\begin{definition}[qIND challenge phase]
The qIND challenge phase proceeds as follows.
\begin{enumerate}
    \item The adversary $\mathcal{A}$ chooses two $m$-qubit states
    $\rho_0,\rho_1$ having efficient classical descriptions, and sends these
    classical descriptions to the challenger $\mathcal{C}$.
    
    \item The challenger samples a uniformly random bit
    \[
    b \xleftarrow{\$} \{0,1\}.
    \]
    
    \item The challenger replies with the encryption of $\rho_b$ obtained by
    applying the type-(2) operator $U^{(2)}_{\Enc_k}$ to $\rho_b$.
    
    \item The adversary outputs a bit $b'$ and wins iff $b' = b$.
\end{enumerate}
\end{definition}

\paragraph{Security Game  $\mathsf{qIND\mbox{-}qCPA}^{\mathcal{A}}_{\mathcal{E}}(n)$}
\begin{enumerate}
    \item Generate a key:
    \[
    k \xleftarrow{\$} \KeyGen(1^n).
    \]
    
    \item Give $\mathcal{A}$ oracle access to $U_{\Enc_k}$ (the qCPA learning phase).
    
    \item $\mathcal{A}$ outputs %classical descriptions of 
    two $m$-qubit states
    $\rho_0,\rho_1$.
    
    \item Sample
    \[
    b \xleftarrow{\$} \{0,1\}.
    \]
    
    \item Return to $\mathcal{A}$ the challenge ciphertext state
    \[
    \sigma \;:=\; U^{(2)}_{\Enc_k}\,\rho_b\,\bigl(U^{(2)}_{\Enc_k}\bigr)^\dagger.
    \]
    
    \item $\mathcal{A}$ outputs a bit $b'$.
    
    \item Output $1$ iff $b' = b$, and output $0$ otherwise.
\end{enumerate}

\begin{definition}[qIND-qCPA security]\label{def:qIND-qCPA}
A secret-key encryption scheme $\mathcal{E}$ is \emph{qIND-qCPA secure} if, for
every quantum probabilistic polynomial-time adversary $\mathcal{A}$,
\[
\Pr\!\left[
\mathsf{qIND\mbox{-}qCPA}^{\mathcal{A}}_{\mathcal{E}}(n)=1
\right]
\;\le\;
\frac{1}{2} + \negl(n).
\]
Equivalently, the advantage
\[
\adv^{\mathsf{qIND\mbox{-}qCPA}}_{\mathcal{E}}(\mathcal{A},n)
:=
\left|
\Pr\!\left[
\mathsf{qIND\mbox{-}qCPA}^{\mathcal{A}}_{\mathcal{E}}(n)=1
\right]
-\frac{1}{2}
\right|
\]
is negligible in $n$.
\end{definition}

\begin{theorem}[\cite{gagliardoni2016semantic}]
    Assuming post-quantum secure pseudorandom permutations (PRP), there exists qIND-qCPA secure SKE.
\end{theorem} \fi
\section{Generalized One-Time Programs}
The first construction of an OTP compiler that satisfied SEQ security \cite{GLRRV25} used the specific tokenized signature scheme of \cite{BDS23}. The intuition for the construction is that the user is required to sign their input in order to obtain a response, and the signature token scheme limits them to signing just one input.

Here, we generalize the construction and security definition in two ways. First, we show that an SEQ-secure OTP compiler can be constructed from \textit{any} one-time signature token scheme (\cref{def:one-time-signature-token}), not just \cite{BDS23}'s scheme. Second, we show that if the signature token is replaced with a one-shot signature scheme (\cref{def:one-shot-signature}), the result is a new primitive, called a one-shot program (\cref{def:osp-compiler}), that allows the user to query the program on many inputs as long as they have distinct serial numbers. One-shot programs will be used to construct query-limited RAM programs in \cref{sec:query-limited-RAM-programs}.

Both of the results above have similar constructions and security proofs, so we construct a generalized one-time program primitive (\cref{def:generalized-otp-compiler}) to capture both results.

\ifllncs
\subsection{Organization}
Due to limitation of space, we first give some preliminaries such as lemmas on compressed oracle and useful security notions to section~\ref{sec:preliminaries}. 

We leave the full definition and construction for generalized OTP compiler to sections~\ref{sec:generalized_otp_definition} and ~\ref{sec:generalized_otp_construction}, their security proof to section~\ref{sec:generalized-otp-security-proof}, one-time programs from any signature tokens and one-shot programs to section~\ref{sec:otp_from_sig_token}. 
\else
\ifllncs
\section{Generalized One-Time Programs: Definition and Construction}
\else
\fi

\subsection{Definition}
\label{sec:generalized_otp_definition}
Here we define a generalized one-time program compiler. Informally, the compiler generates a program $P_f$ that implements a randomized functionality $f$. The program also limits the user to making one effective query for each possible serial number.

\paragraph{The Function Family.} Let $\cF$ be a family of randomized functions mapping $\cS_\secp \times \cX \times \cR \to \cY$, for some given sets $\cS_\secp, \cX, \cR, \cY$. Each $f \in \cF$ takes as input a serial number $s \in \cS_\secp$ and a main input $x \in \cX$. $f$ also takes a uniformly random string $r \getsr \cR$ and then computes $y = f(s,x;r) \in \cY$. Additionally, each $f$ may be associated with some public, classical auxiliary information $\mathsf{aux}_f$.

% in \cref{sec:OSS-from-BUS}
For example, $f$ may generate a signature on $(s,x,r)$. In this case, $f$ is parametrized by a signing key $\sk$ for a blind-unforgeable signature scheme, and $f(s, x)$ computes $y = [r,\Sign(\sk, (s, x, r))]$. Additionally, $\aux_f$ is the $\pk$ of the signature scheme. Alternatively, $f$ may require $s$ to be some fixed value $s^*$ if we are constructing regular one-time programs.

The guarantee of the generalized OTP is that the adversary cannot evaluate any given $f$ on the same $s$ with two different $x$-values.

\paragraph{The SEQ Oracle.} 

\Cref{fig:seq-oracle-with-serial-numbers} describes an oracle $O^{SEQ}_{f}$ that is a variant of the SEQ oracle for $f \in \cF$. $O^{SEQ}_{f}$ takes $(s, x)$ as input and computes $f(s, x; H(s, x))$ as long as its database does not already record $(s, x', r)$ for an $x' \neq x$. This means the database may record many queries, but they will have distinct $s$-values.
\begin{figure}
\begin{mdframed}
    To respond to a query, the oracle acts as a unitary on $\cQ \times \cD$ and an ancilla register $\cQ'$. Given some basis state $\ket{s,x,u,b}_\cQ \otimes \ket{D}_\cD$, the oracle does the following:
    \begin{enumerate}
        \item If the database $D$ contains an entry of the form $(s', x', r')$ such that $s' = s$ but $x' \neq x$, then skip the following steps. Otherwise, continue.\label{step:seq-oracle-SEQ-check}
        \item Prepare the state $\ket{s, x, 0}$ in an ancillary register $\cQ'$. Query $H(s, x)$ by applying the compressed oracle unitary $\CO$ to $\cQ' \times \cD$.\label{step:seq-oracle-with-serial-number-queries-CO}
        \item Apply the following isometry to registers $\cQ \times \cQ'$:
        \begin{align*}
            \ket{s, x, u, b}_\cQ \otimes \ket{s, x,r}_{\cQ'} \to \ket{s, x, u \oplus f(s, x; r), b \oplus 1}_\cQ \otimes \ket{s,x,r}_{\cQ'}
        \end{align*}
        \item Uncompute \cref{step:seq-oracle-with-serial-number-queries-CO}.
    \end{enumerate}
\end{mdframed}
\caption{The SEQ Oracle $O^{SEQ}_{f}$}\label{fig:seq-oracle-with-serial-numbers}
\end{figure}

\paragraph{Generalized One-Time Program Compiler.} This primitive generates quantum program tokens using 
\[(\ket{T}, s) \gets \TokenGen(\sk)\] 
and each token allows one evaluation of the program. The primitive also generates a \textit{classical} program 
\[P_f \gets \PGen(\pk, f)\]
which enables the evaluation of $f$ by any user who possesses a program token. We have chosen to separate the program's classical component $P_f$ from its quantum component $\ket{T}$, as it will simplify the construction of semi-quantum query-limited programs, which can be sent over classical channels (\cref{sec:semi-quantum-QLP}).

\begin{definition}[Generalized One-Time Program Compiler for $\cF$]\label{def:generalized-otp-compiler}
$ $
\begin{itemize}
    \item \textbf{Syntax}: A generalized one-time program compiler for $\cF$ comprises the following QPT algorithms:
    \begin{itemize}
        \item $\KeyGen(1^\secp) \to (\sk, \pk)$: Takes a security parameter $\secp \in \bbN$ and outputs a classical secret key $\sk$ and classical public key $\pk$.
        \item $\TokenGen(\sk) \to (\ket{T}, s)$: Generates a program token $\ket{T}$ and the corresponding serial number $s \in \cS_\secp \cup \{\bot\}$. $\TokenGen$ may output $(\ket{T}, s) = (\bot, \bot)$ with negligible probability.
        \item $\PGen(\pk, f) \to P_f$: Takes the key $\pk$ and a function $f \in \cF$ and outputs a \emph{classical} program $P_f$, which may include an oracle.
        \item $\Eval(P_f, s, x, \ket{T}) \to y$: Takes a program $P_f$, a serial number $s \in \cS_\secp$, a main input $x \in \cX$, and a signature token $\ket{T}$ that is associated with $s$. Then $\Eval$ outputs $y \in \cY$. For a correct evaluation, $y = f(s,x;r)$ for a uniformly random $r \getsr \cR$. 
    \end{itemize}

    \item \textbf{Correctness:}
    A generalized one-time program compiler for $\cF$ is \textbf{correct} if there is a negligible function $\negl(\cdot)$ such that for all $\secp \in \bbN$,
    \[\Pr\left[(\ket{T},s) = (\bot, \bot) : \substack{
        (\sk, \pk) \gets \KeyGen(1^\secp)\\
        (\ket{T}, s) \gets \TokenGen(\sk)}\right] \leq \negl(\secp)\]
    and there exists a negligible function $\negl'(\cdot)$ such that for all $(\secp, f, x) \in \bbN \times \cF \times \cX$, all $(\sk, \pk)$ in the support of $\KeyGen(1^\secp)$, and all $(\ket{T}, s)$ in the support of $\TokenGen(\sk)$ \textit{for which $(\ket{T}, s) \neq (\bot, \bot)$}, the following distributions over $y$-values are $\negl'(\secp)$-close in statistical distance:
    \begin{align*}
        \left\{\substack{
            P_f \gets \PGen(\pk, f)\\
            y \gets \Eval\left(P_f, s, x, \ket{T}\right)
        }\right\} \quad \text{and} \quad \left\{\substack{
            r \getsr \cR\\
            y = f(s, x;r)
        }\right\}
    \end{align*}
    \bhaskar{To do: check that the definition above is achievable from one-shot signatures, and update the construction and correctness proof accordingly.}

    \item \textbf{Security:} A generalized one-time program compiler for $\cF$ is \textbf{SEQ-secure} if there exists a QPT simulator $\Sim$ such that for all $f \in \cF$ and all QPT distinguishers $D$, there exists a negligible function $\negl(\cdot)$ such that for all $\secp \in \mathbb{N}$,
    \begin{align*}
        \Bigg|\Pr\Bigg[b = 1 &: \substack{
            (\sk,\pk) \gets \KeyGen(1^\secp)\\
            (\ket{T}, s) \gets \TokenGen(\sk)\\
            P_f \gets \PGen(\pk, f)\\
            b \gets D\left(P_f, \ket{T}, s, \pk, f, \aux_f\right)
        }\Bigg]
        - \Pr\Bigg[b = 1 &: \substack{
            (\sk, \pk) \gets \KeyGen(1^\secp)\\
            (\ket{T}, s) \gets \TokenGen(\sk)\\
            P_f^\Sim \gets \Sim^{O^\SEQ_{f}}(1^\secp, \pk, \aux_f)\\
            b \gets D\left(P_f^\Sim, \ket{T}, s, \pk, f, \aux_f\right)
        }\Bigg]\Bigg| \leq \negl(\secp)
    \end{align*}
    where $O^\SEQ_f$ is the oracle in \cref{fig:seq-oracle-with-serial-numbers}.
    \jiahui{should we specify where did $\aux_f$ come from and give an example for $\aux_f$ }
    \justin{Do we need to have an $\aux_f$? We quantify over all $f$ and all $D$, so $D$ can just have $\aux_f$ hardcoded.}\bhaskar{I agree that it seems like we can omit $\aux_f$ from the inputs to $D$.}
\end{itemize}    
\end{definition}

\subsection{Construction}
\label{sec:generalized_otp_construction}
Here we construct a generalized one-time program compiler for any randomized function family $\cF$ (\cref{def:generalized-otp-compiler}) using a generalized signature token scheme (\cref{def:signature-token}). At a high level, the construction works as follows. The program includes an oracle that computes $f(s, x; r)$ on any inputs $(s, x)$ as long as the user can provide a valid signature $\sigma$ on message $x$ with serial number $s$. 
The user cannot evaluate the program on two inputs with the same $s$ value, by the security of the generalized signature token scheme. Furthermore, if $f$ is highly random, any evaluation of the oracle will consume the signature token because the random string $r$ is chosen by hashing $G(s, x, \sigma)$.\\

\Cref{fig:generalized-otp-compiler-construction} describes the construction in detail. Let $\SigToken.(\KeyGen, \TokenGen, \Sign, \Ver)$ be a generalized signature token scheme (\cref{def:signature-token}). For any $\secp \in \bbN$, let $\Sigma_\secp$ be the domain of signatures, and let $\cS_\secp$ be the domain of the serial numbers.

\begin{figure}
\begin{mdframed}
\begin{itemize}    
    \item $\KeyGen(1^\secp)$: Compute and output
    \[(\sk, \pk) \gets \SigToken.\KeyGen(1^\secp)\]
        
    \item $\TokenGen(\sk)$: Compute and output
    \[(\ket{T}, s) \gets \SigToken.\TokenGen(\sk)\]
    
    \item $\PGen(\pk, f)$:
    \begin{enumerate}
        \item Sample a random oracle $G: \cS_\secp \times \cX \times \Sigma_\secp \to \cR$.
        \item Generate an oracle $O_{f,G,\pk}$ that acts as follows:
        \begin{mdframed}
        \begin{enumerate}
            % \itemindent=-10pt
            \item[] \textbf{Hardcoded:} $f,G,\pk$
            \item[] \textbf{Inputs:} Takes a query register $Q = Q_s \times Q_x \times Q_\sigma \times Q_u$ with eigenstates of the form $\ket{s, x, \sigma, u}$ where $(s, x, \sigma, u) \in \cS_\secp \times \cX \times \Sigma_\secp \times \cY$. Here, $s$ is a serial number, $x$ is the evaluator's input, $\sigma$ is a signature on $x$, and $u$ is in the register that stores the output of the function.
            \item Check that $\mathsf{SigToken}.\Ver(\pk, s, x, \sigma) = 1$. If so, then continue. If not, then abort and output register $Q$.
            \item Query $G$ to compute $r = G(s,x,\sigma)$:\label{OSP-construction-step:query-RO}
            \[\ket{s,x,\sigma}_{Q_s \times Q_x \times Q_\sigma} \otimes \ket{0}_R \to \ket{s,x,\sigma} \otimes \ket{r}\]
            \item Compute $y = f(s,x;r)$, and CNOT $y$ onto $Q_u$:
            \[\ket{s, x, \sigma, u}_Q \otimes \ket{r}_R \to \ket{s, x, \sigma, u \oplus f(s,x;r)}_Q \otimes \ket{r}_R\]
            \item Uncompute item \ref{OSP-construction-step:query-RO}.
            \item Output register $Q$.
        \end{enumerate}
        \end{mdframed}
        \item Output $P_f = O_{f,G,\pk}$.
    \end{enumerate}
    \item $\Eval(P_f, s, x, \ket{T})$:
    \begin{enumerate}
        \item Compute $\sigma = \mathsf{SigToken}.\Sign(\ket{T}, x)$.
        \item Query $O_{f,G,\pk}$ on input $(s, x, \sigma, 0)$ to obtain output $(s, x, \sigma, y)$. Then output $y$.
    \end{enumerate}
\end{itemize}
\end{mdframed}
\caption{Generalized One-Time Program Compiler Construction}\label{fig:generalized-otp-compiler-construction}
\end{figure}

Now we prove that the construction in \cref{fig:generalized-otp-compiler-construction} is a generalized one-time program compiler for $\cF$.

\begin{theorem}\label{thm:generalized-otp}
    The construction in \cref{fig:generalized-otp-compiler-construction} is a generalized one-time program compiler for $\cF$ (\cref{def:generalized-otp-compiler}).
\end{theorem}
\begin{proof}
    First, the construction in \cref{fig:generalized-otp-compiler-construction} clearly satisfies the syntax of a generalized one-time program compiler for $\cF$. 
    
    Second, to begin the proof of correctness, we must show that
    \[\Pr\left[(\ket{T},s) = (\bot, \bot) : \substack{
        (\sk, \pk) \gets \KeyGen(1^\secp)\\
        (\ket{T}, s) \gets \TokenGen(\sk)}\right] \leq \negl(\secp)\]
    This follows directly from the correctness property of $\SigToken$ (\cref{def:signature-token}).

    Third, to finish the proof of correctness, we must show that for any $(\secp, f, x, \sk, \pk, \ket{T}, s)$ for which $(\ket{T}, s) \neq (\bot, \bot)$, the following distributions over $y$-values are statistically close.
    \begin{equation}\label{eq:distributions-in-correctness-definition}
        \left\{\substack{
            P_f \gets \PGen(\pk, f)\\
            y \gets \Eval\left(P_f, s, x, \ket{T}\right)
        }\right\} \quad \text{and} \quad \left\{\substack{
            r \getsr \cR\\
            y = f(s, x;r)
        }\right\}
    \end{equation}
    The left distribution of \cref{eq:distributions-in-correctness-definition} is equivalent to the following distribution $\cD$:
    \begin{align*}
            \sigma &\gets \SigToken.\Sign(\ket{T}, x)\\
            b &\gets \SigToken.\Ver(\pk, s, x, \sigma)\\
            r &\getsr \cR\\
            y &= \begin{cases}
                f(s, x; r), & b = 1\\
                0, & b = 0
            \end{cases}
    \end{align*}
    This follows by unpacking $\PGen$ and $\Eval$.

    Next, $\cD$ is statistically close to the right distribution of \cref{eq:distributions-in-correctness-definition}. This is because $\SigToken.\Ver(\pk, s, x, \sigma)$ equals $1$ with overwhelming probability, by the correctness of $\SigToken$ (\cref{def:signature-token}). This shows that the left and right distributions of \cref{eq:distributions-in-correctness-definition} are statistically close, and it completes the proof of correctness.
    
    Fourth, the proof of security (\cref{thm:generalized-otp-security}) is given in \cref{sec:generalized-otp-security-proof}.
\end{proof}
\ifllncs
\section{One-Time Programs From Any Signature Token}
\label{sec:otp_from_sig_token}
\else
\subsection{One-Time Programs From Any Signature Token}
\label{sec:otp_from_sig_token}
\fi
We will use the generalized one-time program compiler constructed in \cref{fig:generalized-otp-compiler-construction} to construct a (regular) one-time program compiler (\cref{def:otp-syntax}).

Let us instantiate $\SigToken$ in \cref{fig:generalized-otp-compiler-construction} with a one-time signature token scheme (\cref{def:one-time-signature-token}). Additionally, the $\KeyGen$ and $\TokenGen$ functions of the generalized OTP compiler are merged into $\PGen$.\\

\Cref{fig:OTP-construction} constructs the OTP compiler for $\cF$ formally. Let $\mathsf{GeneralOTP}.$$(\KeyGen$, $\TokenGen$, $\PGen$, $\Eval)$ be the generalized OTP compiler constructed in \cref{fig:generalized-otp-compiler-construction}, where $\SigToken$ is instantiated with a one-time signature token scheme (\cref{def:one-time-signature-token}). Additionally, for every $f \in \cF$, let $f : \cX \times \cR \to \cY$.
\begin{figure}
\begin{mdframed}
\begin{itemize}
\item $\PGen(1^\secp, f)$:
\begin{enumerate}
    \item Compute $(\sk, \pk) \gets \mathsf{GeneralOTP}.\KeyGen(1^\secp)$
    \item Compute $(\ket{T}, s) \gets \mathsf{GeneralOTP}.\TokenGen(\sk)$
    \item Let $f' : \cS_\secp \times \cX \times \cR \to \cY$ be the following function that ignores its first input:
    \[f'(s, x; r) = f(x;r)\]
    \item Compute $P_{f'} \gets \mathsf{GeneralOTP}.\PGen(\pk, f')$
    \item Output $P_f = (\ket{T}, s, \pk, P_{f'})$.
\end{enumerate}
\item $\Eval(P_f, x)$:
\begin{enumerate}
    \item Parse $P_f = (\ket{T}, s, \pk, P_{f'})$.
    \item Compute and output $y \gets \mathsf{GeneralOTP}.\Eval(P_{f'}, s, x, \ket{T})$.
\end{enumerate}
\end{itemize}
\end{mdframed}
\caption{Construction of a One-Time Program Compiler for $\cF$}\label{fig:OTP-construction}
\end{figure}

\begin{theorem}
    The construction in \cref{fig:OTP-construction} is a one-time program compiler for $\cF$, which satisfies the syntax (\cref{def:otp-syntax}), correctness (\cref{def:otp-correctness}), and SEQ security (\cref{def:seq-security}) properties.
\end{theorem}
\begin{proof}
    First, it is clear by inspection that the construction satisfies the syntax of a one-time program compiler.

    Second, correctness follows from the correctness of $\mathsf{GeneralOTP}$. The correctness of $\mathsf{GeneralOTP}$ implies that for any $f, x$, the following two distributions over $y$-values are negligibly close in statistical distance:
    \begin{align*}
        \left\{\substack{
            (\sk, \pk) \gets \mathsf{GeneralOTP}.\KeyGen(1^\secp)\\
            (\ket{T}, s) \gets \mathsf{GeneralOTP}.\TokenGen(\sk)\\
            P_{f'} \gets \mathsf{GeneralOTP}.\PGen(\pk, f')\\
            y \gets \mathsf{GeneralOTP}.\Eval\left(P_{f'}, s, x, \ket{T}\right)
        }\right\} \quad \text{and} \quad \left\{\substack{
            r \getsr \cR\\
            y = f(x;r)
        }\right\}
    \end{align*}
    The left distribution is the same as
    \begin{align*}
        \left\{\substack{
            P_f \gets \PGen(1^\secp, f)\\
            y \gets \Eval\left(P_f, x\right)
        }\right\}
    \end{align*}
    This proves the correctness of the OTP compiler.

    Third, to prove SEQ security, let us start with the SEQ security property of $\mathsf{GeneralOTP}$, which says the following: there exists a simulator $\Sim'$ such that for all $f \in \cF$ and all QPT distinguishers $D$, there exists a negligible function $\negl(\cdot)$ such that for all $\secp \in \bbN$,
    \begin{equation}\label{eq:generalized-seq-security-of-otp-construction}
      \Bigg|\Pr\Bigg[b = 1: \substack{
            (\sk,\pk) \gets \mathsf{GeneralOTP}.\KeyGen(1^\secp)\\
            (\ket{T}, s) \gets \mathsf{GeneralOTP}.\TokenGen(\sk)\\
            P_{f'} \gets \mathsf{GeneralOTP}.\PGen(\pk, f')\\
            b \gets D\left(P_{f'}, \ket{T}, s, \pk, f', \aux_{f'}\right)
        }\Bigg]
        - \Pr\Bigg[b = 1: \substack{
            (\sk, \pk) \gets \mathsf{GeneralOTP}.\KeyGen(1^\secp)\\
            (\ket{T}, s) \gets \mathsf{GeneralOTP}.\TokenGen(\sk)\\
            P_{f'}^\Sim \gets \Sim'^{O^\SEQ_{f'}}(1^\secp, \pk, \aux_{f'})\\
            b \gets D\left(P_{f'}^\Sim, \ket{T}, s, \pk, f', \aux_{f'}\right)
        }\Bigg]\Bigg| \leq \negl(\secp)
    \end{equation}
    where $O^\SEQ_{f'}$ is the oracle in \cref{fig:seq-oracle-with-serial-numbers}.

    We can simplify \cref{eq:generalized-seq-security-of-otp-construction} to be:
    \begin{align*}
        \Bigg|\Pr\Bigg[b = 1 &: \substack{
            P_f \gets \PGen(1^\secp, f)\\
            b \gets D\left(P_f, f', \aux_{f'}\right)
        }\Bigg]
        - \Pr\Bigg[b = 1 &: \substack{
            (\sk, \pk) \gets \mathsf{GeneralOTP}.\KeyGen(1^\secp)\\
            (\ket{T}, s) \gets \mathsf{GeneralOTP}.\TokenGen(\sk)\\
            P_{f'}^\Sim \gets \Sim'^{O^\SEQ_{f'}}(1^\secp, \pk, \aux_{f'})\\
            b \gets D\left(P_{f'}^\Sim, \ket{T}, s, \pk, f', \aux_{f'}\right)
        }\Bigg]\Bigg| \leq \negl(\secp)
    \end{align*}
    
    Fourth, let us reconcile the difference in SEQ oracles. The OTP compiler uses $O^\SEQ_f$ from \cref{fig:seq-oracle} to define security, whereas the generalized OTP compiler uses $O^\SEQ_{f'}$ from \cref{fig:seq-oracle-with-serial-numbers}. $O^\SEQ_f$ is similar to $O^\SEQ_{f'}$, except that $O^\SEQ_f$ does not take $s$ as input, does not record $s$ in the database, and does not consider $s$ when deciding to answer the query. However, note that because $\SigToken$ is a one-time signature token, $\cS_\secp$ contains a single element $s^*$. Therefore, every query to $O^\SEQ_{f'}$ has $s = s^*$ and every entry of the oracle database records $s = s^*$. Furthermore, $O^\SEQ_{f'}$ answers a query $x$ if and only if the database does not record $(s^*, x', r')$ for some $x' \neq x$. Then querying $O^\SEQ_{f}$ on input $(x, u, b)$ is equivalent to querying $O^\SEQ_{f'}$ on input $(s^*, x, u, b)$ and ignoring the $s^*$ value after the query.
    
    Fifth, let us define the simulator $\Sim$ for the OTP compiler. 
    \begin{enumerate}
        \item Given query access to $O^\SEQ_{f}$, $\Sim$ simulates queries to $O^\SEQ_{f'}$ as follows. Take as input a query $(s, x, u, b)$. If $s \notin \cS_\secp$, then do not answer the query. Otherwise, query $O^\SEQ_{f}$ on input $(x, u, b)$ to obtain $(x, u', b')$, and output $(s^*, x, u', b')$.
        \item $\Sim$ computes
        \begin{align*}
            (\sk, \pk) &\gets \mathsf{GeneralOTP}.\KeyGen(1^\secp)\\
            (\ket{T}, s) &\gets \mathsf{GeneralOTP}.\TokenGen(\sk)\\
            P_{f'}^\Sim &\gets \Sim'^{O^\SEQ_{f'}}(1^\secp, \pk, \aux_{f'})
        \end{align*}
        where $O^\SEQ_{f'}$ is simulated using the procedure above.

        \item $\Sim$ outputs $P_f^\Sim = (\ket{T}, s, \pk, P_{f'}^\Sim)$.
    \end{enumerate} 

    With this choice of $\Sim$, \cref{eq:generalized-seq-security-of-otp-construction} simplifies further to:
    \begin{align*}
        \Bigg|\Pr\Bigg[b = 1 : \substack{
            P_f \gets \PGen(1^\secp, f)\\
            b \gets D\left(P_f, f', \aux_{f'}\right)
        }\Bigg]
        - \Pr\Bigg[b = 1 : \substack{
            P_{f}^\Sim \gets \Sim^{O^\SEQ_{f}}(1^\secp, \aux_{f'})\\
            b \gets D\left(P_{f}^\Sim, f', \aux_{f'}\right)
        }\Bigg]\Bigg| \leq \negl(\secp)
    \end{align*}
    Furthermore, we can set $\aux_{f'} = \aux_f$, and we can give $D$ access to $f$ instead of $f'$ because they provide equivalent information. Then we have the following: there exists a simulator $\Sim$ such that for all $f \in \cF$ and all QPT distinguishers $D$, there exists a negligible function $\negl(\cdot)$ such that for all $\secp \in \bbN$,
    \begin{align*}
        \Bigg|\Pr\Bigg[b = 1 : \substack{
            P_f \gets \PGen(1^\secp, f)\\
            b \gets D\left(P_f, f, \aux_f\right)
        }\Bigg]
        - \Pr\Bigg[b = 1 : \substack{
            P_{f}^\Sim \gets \Sim^{O^\SEQ_{f}}(1^\secp, \aux_f)\\
            b \gets D\left(P_{f}^\Sim, f, \aux_f\right)
        }\Bigg]\Bigg| \leq \negl(\secp)
    \end{align*}
    This proves the SEQ security of the OTP compiler.
\end{proof}

\subsection{One-Shot Programs}\label{sec:OSP}
A one-shot program compiler is a type of generalized OTP compiler where $\sk = \pk$. This property gives the evaluator significant power. Now, if they possess the public key $\pk$, then they can generate many program tokens $(\ket{T},s)$, albeit with unique serial numbers, and evaluate the program once per token.

\begin{definition}[One-Shot Program Compiler for $\cF$]\label{def:osp-compiler}
    A \textbf{one-shot program compiler} for $\cF$ ($\KeyGen$, $\TokenGen$, $\PGen$, $\Eval$) is a type of generalized one-time program compiler for $\cF$ (\cref{def:generalized-otp-compiler}) where for any $\secp \in \bbN$ and any $(\sk, \pk)$ in the support of $\KeyGen(1^\secp)$,
    \[\sk = \pk.\]
\end{definition}

\begin{remark}
    For one-shot programs, we sometimes introduce the name $\crs = \sk = \pk$, and we may notate the output of $\KeyGen$ as 
    \[\crs \gets \KeyGen(1^\secp)\]
\end{remark}

Our construction of one-shot programs is the same as the construction in \cref{fig:generalized-otp-compiler-construction}, except we instantiate $\SigToken$ with a one-shot signature scheme (\cref{def:one-shot-signature}).

\begin{theorem} \label{theorem:OSP_from_OSS}
    Given the construction in \cref{fig:generalized-otp-compiler-construction}, let $\SigToken$ be instantiated with a one-shot signature scheme (\cref{def:one-shot-signature}). Then the construction is a one-shot program compiler for $\cF$ (\cref{def:osp-compiler}).
\end{theorem}
\begin{proof}
    The one-shot signature token scheme used in the construction is a type of generalized signature token scheme. Then by \cref{thm:generalized-otp}, the construction is a generalized one-time program compiler.

    Next, note that when $\SigToken$ is instantiated with a one-shot signature scheme, the $\KeyGen(1^\secp)$ function in the construction always outputs $(\sk, \pk)$ such that $\sk = \pk$. Therefore, the construction is a one-shot program compiler.
\end{proof}

\fi
\ifllncs \else  \ifsubmit
    \section{Proof of Security for Generalized OTPs}\label{sec:generalized-otp-security-proof}
\else
    \subsection{Proof of Security for Generalized OTPs}\label{sec:generalized-otp-security-proof}
\fi

\begin{theorem}[Security]\label{thm:generalized-otp-security}
    The construction in \cref{fig:generalized-otp-compiler-construction} satisfies the security property of a generalized one-time program compiler for $\cF$ (\cref{def:generalized-otp-compiler}).
\end{theorem}

\noindent The rest of \cref{sec:generalized-otp-security-proof} proves \cref{thm:generalized-otp-security}.

\paragraph{Simulator:} We will construct the SEQ simulator $\Sim$ in \cref{fig:SEQ-simulator}, which simulates $P_f$.
\begin{figure}[H]
\begin{mdframed}
$ $
\begin{enumerate}
    \item \textbf{Inputs}: $\Sim$ gets quantum query access to the SEQ oracle $O^\SEQ_f$ and also receives $(1^\secp, \pk, \aux_f)$.
    \item Prepare a \emph{cache register} $C$, which holds a set of any number of tuples of the form $(s,x,\sigma) \in \cS_\secp \times \cX \times \Sigma_\secp$ and is initialized to $\ket{\emptyset}$ (empty).
    % \bhaskar{Could we make $C$ record $(s, x, \sigma)$, not just $(s, \sigma)$?}
    % \justin{That should be fine. Our prior work does this, and the OSP SEQ oracle is essentially a compressed version of exponentially many OTPs.}
    \item Prepare an oracle $O^{O^\SEQ_f}_{\Sim}$, which acts as a unitary on classical eigenstates as follows:%\bhaskar{Is it a problem that $O^{O^\SEQ_f}_{\Sim}$ initializes a fresh $B$ register on each query and never uncomputes the register?}
    %\justin{Can $b$ ever get entangled with the adversary's query? For example if the adversary does a superposition between a valid query and an invalid one.}
    \begin{enumerate}
        \item $O^{O^\SEQ_f}_{\Sim}$ takes a query register $Q = Q_s \times Q_x \times Q_\sigma \times Q_u$ with eigenstates of the form $\ket{s, x, \sigma, u}$ where $(s, x, \sigma, u) \in \cS_\secp \times \cX \times \Sigma_\secp \times \cY$.
        \item Check that $\SigToken.\Ver(\pk, s, x, \sigma) = 1$. %and $S$ contains the value $0$ or $\sigma$. 
        If so, then continue. If not, then abort and output register $Q$.
        \item \textbf{Cache 1}: \label{sim-step:cache-1}
        \begin{enumerate}
            \item Prepare $\ket{s, x \oplus 1, 0, 0}$ on an internal register $Q'$, and query $O^\SEQ_f$ on $Q'$. Let $\ket{s, x \oplus 1, y, b}$ be the final state of $Q'$. \label{step:query-SEQ-oracle}
            \item If $b = 0$, then do the following: \label{step:CNOT}
            \begin{enumerate}
                \item If $C$ contains entry $(s, x, \sigma)$, then remove this entry from $C$.
                % \item Else if $C$ contains an entry of the form $(s, \sigma')$ for some $\sigma' \in \Sigma_\secp \backslash \{\sigma\}$, then overwrite this entry with $(s, \sigma' \oplus \sigma)$.\bhaskar{We can probably omit this step.}
                \item Else, add $(s, x, \sigma)$ to $C$.
            \end{enumerate}
            \item Uncompute \cref{step:query-SEQ-oracle}.\label{step:Uncompute} %\jiahui{Do you mean 3(c)i? }
        \end{enumerate}
        \item Initialize $\ket{+}_B$. Query $O^\SEQ_f$ on $\ket{s, x, u}_{Q_s \times Q_x \times Q_u} \otimes \ket{+}_B$.
        \item \textbf{Cache 2}: Repeat \cref{sim-step:cache-1}.
        \item Output register $Q$.
    \end{enumerate}
    \item Output the program $P_f^\Sim = O^{O^\SEQ_f}_{\Sim}$.
\end{enumerate}
\end{mdframed}
\caption{SEQ Simulator $\Sim^{O^\SEQ_f}$}\label{fig:SEQ-simulator}
\end{figure}

\bhaskar{I made some changes, and I believe that P02 is now fixed.}

Next, the following sequence of hybrids transforms the oracle $O_{f,G,\pk}$ from the real-world program $P_f$ into the oracle $O^{O^\SEQ_f}_\Sim$ from the ideal-world simulator $\Sim^{O^\SEQ_f}$. The hybrids are computationally indistinguishable.

Let us be more precise about the distinguisher's inputs. At the start of each hybrid, a challenger takes $\secp, f$ and samples:
\begin{align*}
    (\sk, \pk) &\gets \KeyGen(1^\secp)\\ (\ket{T}, s) &\gets \TokenGen(\sk)
\end{align*}
Then the challenger samples any additional parameters of the hybrid's oracle, such as $G$, and implements the oracle $\mathcal{O}_{f, G, \pk}$ that is defined by the parameters above. Next, the distinguisher's inputs are  
\[\mathcal{O}_{f, G, \pk}, \ket{T}, s, \pk, f, \aux_f\]
Note that these are the inputs given to the distinguisher in the SEQ security definition.

We say that two hybrids are \textit{computationally indistinguishable} if any QPT distinguisher, given access to the input distribution above, has negligible advantage at distinguishing the hybrids. We say that two hybrids are \textit{perfectly indistinguishable} if any computationally unbounded distinguisher, given access to the input distribution above, has $0$ advantage at distinguishing the hybrids. 

Finally, for each hybrid description below, any differences from the previous hybrid will be written in \textcolor{red}{red}.

\paragraph{Hybrid 1.} The real-world oracle $O_{f,G,\pk}$ from program $P_f$ (\cref{fig:generalized-otp-compiler-construction}).
\begin{itemize}
    \item \textbf{Additional Parameters:} The oracle is parametrized by a random oracle $G$ sampled uniformly at random. 
\end{itemize}
\begin{enumerate}
    \item \textbf{Inputs:} The oracle takes a query register $Q = Q_s \times Q_x \times Q_\sigma \times Q_u$ with eigenstates of the form $\ket{s, x, \sigma, u}$ where $(s, x, \sigma, u) \in \cS_\secp \times \cX \times \Sigma_\secp \times \cY$.
    \item Check that $\SigToken.\Ver(\pk, s, x, \sigma) = 1$. If so, then continue. If not, then abort and output register $Q$.
    \item Query $G$ to compute $r = G(s,x,\sigma)$:\label{hyb-1-step:query-RO}
        \[\ket{s,x,\sigma}_{Q_s \times Q_x \times Q_\sigma} \otimes \ket{0}_R \to \ket{s,x,\sigma} \otimes \ket{r}\]
    \item Compute $y = f(s,x;r)$, and CNOT $y$ onto $Q_u$:
    \[\ket{s, x, \sigma, u}_Q \otimes \ket{r}_R \to \ket{s, x, \sigma, u \oplus f(s,x;r)}_Q \otimes \ket{r}_R\]
    \item Uncompute item \ref{hyb-1-step:query-RO}.
    \item Output register $Q$.
\end{enumerate}

\paragraph{Hybrid 2.} Implement the random oracle $G$ as a compressed oracle.
\begin{itemize}
    \item {\color{red}The oracle has a database register $D$ that stores a (possibly empty) set of tuples of the form $(s', x', \sigma', r') \in \cS_\secp \times \cX \times \Sigma_\secp \times \cR$. $D$ is initialized to the empty list $\ket{\emptyset}$.}
\end{itemize}
\begin{enumerate}
    \item \textbf{Inputs:} The oracle takes a query register $Q = Q_s \times Q_x \times Q_\sigma \times Q_u$ with eigenstates of the form $\ket{s, x, \sigma, u}$ where $(s, x, \sigma, u) \in \cS_\secp \times \cX \times \Sigma_\secp \times \cY$.
    \item Check that $\SigToken.\Ver(\pk, s, x, \sigma) = 1$. If so, then continue. If not, then abort and output register $Q$.
    \item Query $G$ \textcolor{red}{on $\ket{s,x,\sigma}_{Q_s \times Q_x \times Q_\sigma} \otimes \ket{0}_R \otimes \ket{D}$ by applying the following sequence of operations (from right to left):}\label{hyb-2-step:query-RO}
        \textcolor{red}{\[\Decomp_{s, x, \sigma} \circ \CO' \circ \Decomp_{s, x, \sigma}\]}
    \item Compute $y = f(s,x;r)$, and CNOT $y$ onto $Q_u$:
    \[\ket{s, x, \sigma, u}_Q \otimes \ket{r}_R \to \ket{s, x, \sigma, u \oplus f(s,x;r)}_Q \otimes \ket{r}_R\]
    \item Uncompute item \ref{hyb-2-step:query-RO}.
    \item Output register $Q$.
\end{enumerate}

\begin{lemma}
    Hybrids 1 and 2 are perfectly indistinguishable.
\end{lemma}
\begin{proof}
    The only difference between the hybrids is that the random oracle $G$ is implemented as a compressed oracle in hybrid 2. The two hybrids are perfectly indistinguishable by \cite{Zha18}, lemma 4.
\end{proof}

\paragraph{Hybrid 3.} Add an SEQ check to ensure that the compressed oracle never records two distinct entries with the same $s$-value.
\begin{itemize}
    \item The oracle has a database register $D$ that stores a (possibly empty) set of tuples of the form $(s', x', \sigma', r') \in \cS_\secp \times \cX \times \Sigma_\secp \times \cR$. $D$ is initialized to the empty list $\ket{\emptyset}$. \textcolor{red}{The oracle initializes a control qubit $B$ to $\ket{0}$. At the start of every query, $B = \ket{0}$.}
\end{itemize}
\begin{enumerate}
    \item \textbf{Inputs:} The oracle takes a query register $Q = Q_s \times Q_x \times Q_\sigma \times Q_u$ with eigenstates of the form $\ket{s, x, \sigma, u}$ where $(s, x, \sigma, u) \in \cS_\secp \times \cX \times \Sigma_\secp \times \cY$.
    \item Check that $\SigToken.\Ver(\pk, s, x, \sigma) = 1$. If so, then continue. If not, then abort and output register $Q$.\label{hyb-3-step:verify-signature}
    \item {\color{red} \textbf{SEQ Check:} Check that $D$ contains no entry $(s', x', \sigma', r')$ such that $s' = s$ but $(x', \sigma') \neq (x, \sigma)$. If the check passes, then flip $B$ to $\ket{1}$.}\label{hyb-3-step:check-equality-of-signatures}
    \item \textcolor{red}{If $B = \ket{1}$:}
    \begin{enumerate}
        \item Query $G$ on $\ket{s,x,\sigma}_{Q_s \times Q_x \times Q_\sigma} \otimes \ket{0}_R \otimes \ket{D}$ by applying the following sequence of operations (from right to left):\label{hyb-3-step:query-RO}
        \[\Decomp_{s, x, \sigma} \circ \CO' \circ \Decomp_{s, x, \sigma}\]
    \item Compute $y = f(s,x;r)$, and CNOT $y$ onto $Q_u$:
    \[\ket{s, x, \sigma, u}_Q \otimes \ket{r}_R \to \ket{s, x, \sigma, u \oplus f(s,x;r)}_Q \otimes \ket{r}_R\]
    \item Uncompute item \ref{hyb-3-step:query-RO} by applying the following sequence of operations (from right to left):\label{hyb-3-step:uncompute-RO-query}
        \[\Decomp_{s, x, \sigma} \circ \CO' \circ \Decomp_{s, x, \sigma}\]
    \end{enumerate}
    \item {\color{red} Uncompute step \ref{hyb-3-step:check-equality-of-signatures}.}\label{hyb-3-step:check-equality-of-signatures-2}
    \item Output register $Q$.
\end{enumerate}

\begin{lemma}\label{thm:hyb-2-3-indistinguishable}
    Hybrids 2 and 3 are computationally indistinguishable.
\end{lemma}
\begin{proof}
    The difference between the hybrids is that hybrid 3 adds an SEQ check (step \ref{hyb-3-step:check-equality-of-signatures}), which aborts if $D$ already records a value $(s, x', \sigma')$ such that $(x', \sigma') \neq (x, \sigma)$. If hybrids 2 and 3 were not computationally indistinguishable, then we could break the strong security of the signature scheme (\cref{def:signature-token}). If the hybrids are distinguishable, then with non-negligible probability, the SEQ check in hybrid 3, step \ref{hyb-3-step:check-equality-of-signatures} fails to flip $B$ to $\ket{1}$. If this occurs, we can measure the $Q$ and $D$ registers to obtain values $(x, \sigma) \neq (x', \sigma')$. 
    
    Furthermore, both values are valid message-signature pairs for serial number $s$. We know that $\Ver(\pk, s, x, \sigma) = 1$ because otherwise, the oracle would have aborted in step \ref{hyb-3-step:verify-signature} before reaching step \ref{hyb-3-step:check-equality-of-signatures}. We also know that $\Ver(\pk, s, x', \sigma') = 1$ because $(s, x', \sigma')$ are recorded on the database $D$. The only values of $(s, x', \sigma')$ recorded on $D$ are values that were previously queried to the compressed oracle in steps \ref{hyb-3-step:query-RO} or \ref{hyb-3-step:uncompute-RO-query}. The only values that were queried in those steps must satisfy $\Ver(\pk, s, x', \sigma') = 1$ because otherwise, the query would have aborted in step \ref{hyb-3-step:verify-signature}.

    Let us explain the attack above in more detail. Assume toward contradiction that there is a QPT adversary that queries the hybrid 3 oracle and with non-negligible probability, at least one of these queries distinguishes hybrids 2 and 3 with non-negligible advantage. Next, in the strong unforgeability game (\cref{def:signature-token}), the challenger samples 
    \begin{align*}
        (\sk, \pk) &\gets \SigToken.\KeyGen(1^\secp)\\
        (\ket{T}, s) &\gets \SigToken.\TokenGen(\sk)
    \end{align*}
    and gives the signature adversary $(1^\secp, \pk, \ket{T}, s)$. Then our signature adversary does the following:
    \begin{enumerate}
        \item Implement the oracle described in hybrid 3 using $(\pk, f)$.
        \item Run the distinguisher, giving it query access to the hybrid 3 oracle.
        \item If step \ref{hyb-3-step:check-equality-of-signatures} of the oracle ever ends with $B = \ket{0}$, then measure the $Q$ and $D$ registers to obtain $(s,x,\sigma)$ and $(s, x', \sigma')$.
        \item Output $(x, x',\sigma, \sigma', s)$.
    \end{enumerate}
    First, $\pk$ has the same distribution as it does in hybrid 3. Second, our signature adversary runs in polynomial time because so does the distinguisher. Third, with non-negligible probability, step \ref{hyb-3-step:check-equality-of-signatures} ends with $B = \ket{0}$ because we've correctly simulated hybrid 3, and the distinguisher distinguishes the two hybrids with non-negligible advantage. Fourth, when step \ref{hyb-3-step:check-equality-of-signatures} ends with $B = \ket{0}$, the signature adversary outputs a correct answer, which satisfies:
    \begin{align*}
        (x, \sigma) &\neq (x', \sigma')\\
        1 &= \Ver(\pk, s, x, \sigma)\\
        1 &= \Ver(\pk, s, x', \sigma')
    \end{align*}
    Therefore, the signature adversary breaks the strong security of the OSS scheme.
    
    In summary, we've shown that hybrids 2 and 3 are computationally indistinguishable, assuming that the OSS scheme is strongly secure.
\end{proof}

\paragraph{Hybrid 4.} We will split our database $D$ into two registers. A smaller database register $D$ stores tuples of the form $(s',x',r')$, and a cache register $C$ stores tuples of the form $(s',x',\sigma')$. Previously, the database stored all these values together: $(s',x',\sigma',r')$.

First, the combined database $d'$ should have unique $s'$-values in order to be splittable. Let $d$ be a set of tuples of the form $(s',x',\sigma', r')$. $d$ is \textbf{splittable} if no two entries of $d$ have the same $s'$ value.

Second, here is an algorithm $\mathsf{Split}$ that splits $d$ into a smaller database $d'$ and cache $c'$.
\begin{enumerate}
    \item $\mathsf{Split}$ takes as input an (empty) cache $c$ and a combined database $d$.
    \item If $c \neq \emptyset$ or $d$ is not splittable, then output $(c,d)$ and halt. Otherwise continue.
    \item Initialize $c' = d' = \emptyset$.
    \item For every entry $(s',x',\sigma',r') \in d$, add $(s',x',\sigma')$ to $c'$, and add $(s',x',r')$ to $d'$.
    \item Output $(c',d')$.
\end{enumerate}

\begin{lemma}
    $\mathsf{Split}$ is invertible.
\end{lemma}
\begin{proof}
    Let us consider how $\mathsf{Split}(c,d)$ acts on the following exhaustive cases:
    \begin{enumerate}
        \item Case 1: $(c,d) = (\emptyset, \emptyset)$. Then $\mathsf{Split}(c,d)$ outputs $(c',d') = (\emptyset, \emptyset) = (c,d)$.
        \item Case 2: $c \neq \emptyset$. Then $\mathsf{Split}(c,d)$ outputs $(c',d') = (c,d)$.
        \item Case 3: $c = \emptyset$ and $d \neq \emptyset$ and $d$ is not splittable. Then $\mathsf{Split}(c,d)$ outputs $(c',d') = (c,d)$.
        \item Case 4: $c = \emptyset$ and $d \neq \emptyset$ and $d$ is splittable. Then $\mathsf{Split}(c,d)$ follows through and actually splits $(c,d)$. $d'$ is a non-empty database containing tuples of the form $(s',x',r')$, where every $s'$ is unique. $c'$ is a non-empty database containing tuples of the form $(s',x',\sigma')$. Furthermore, there is a perfect matching between entries of $c'$ and entries of $d'$, where the matching pairs up entries with the same $s'$ value.
    \end{enumerate}

    Next, here is how to invert $\mathsf{Split}(c,d)$ given its output $(c',d')$. First, check whether $d'$ is a non-empty database containing tuples with three entries each: $(s',x',r')$. If so, then we must be in case 4. For cases 1 - 3, $d' = d$, so $d'$ is either empty or it contains tuples with four entries each: $(s',x',\sigma',r')$.

    Second, if we are in case 4, then we set $c = \emptyset$ and construct $d$ as follows. For each pair of entries $(s',x',\sigma') \in c'$ and $(s',x',r') \in d'$ with the same $s'$-value, add $(s',x',\sigma',r')$ to $d$. This constructs $d$ correctly because every entry of $d$ has a unique $s'$-value.

    Third, if we are not in case 4, then set $(c,d) = (c',d')$. This correctly inverts $\mathsf{Split}$ on cases 1-3.
\end{proof}

Next, let $U_{\mathsf{Split}}$ be a unitary that computes $\mathsf{Split}(c,d)$ in-place:
\[\ket{c, d}_{C \times D} \to \ket{\mathsf{Split}(c,d)}_{C \times D}\]
$U_\mathsf{Split}$ is a unitary because $\mathsf{Split}$ is invertible.

$ $\\

Finally, let us define the oracle for hybrid 4.

\begin{itemize}
    \item \textcolor{red}{The oracle has a database register $D$, a cache register $C$, and control qubit $B$. $B \times C \times D$ are initialized to $\ket{0, \emptyset, \emptyset}$.}
\end{itemize}
\begin{enumerate} 
    \item \textbf{Inputs:} The oracle takes a query register $Q = Q_s \times Q_x \times Q_\sigma \times Q_u$ with eigenstates of the form $\ket{s, x, \sigma, u}$ where $(s, x, \sigma, u) \in \cS_\secp \times \cX \times \Sigma_\secp \times \cY$.
    \item Check that $\SigToken.\Ver(\pk, s, x, \sigma) = 1$. If so, then continue. If not, then abort and output register $Q$.\label{hyb-4-step:verify-signature}
    \item \textbf{SEQ Check:} \label{hyb-4-step:check-equality-of-signatures}
    \begin{enumerate}
        \item \textcolor{red}{Apply $U_\mathsf{Split}^\dag$ to $C \times D$.}
        \item Check that $D$ contains no entry $(s', x', \sigma', r')$ such that $s' = s$ but $(x', \sigma') \neq (x, \sigma)$. If the check passes, then flip $B$ to $\ket{1}$.
        \item \textcolor{red}{Apply $U_\mathsf{Split}$ to $C \times D$.}
    \end{enumerate}
    \item If $B = \ket{1}$:
    \begin{enumerate}
        \item Query $G$ on $\ket{s,x,\sigma}_{Q_s \times Q_x \times Q_\sigma} \otimes \ket{0}_R \otimes \textcolor{red}{\ket{c,d}_{C \times D}}$ by applying the following sequence of operations (from right to left):\label{hyb-4-step:query-RO}
            \[\textcolor{red}{U_\mathsf{Split}} \circ \Decomp_{s, x, \sigma} \circ \CO' \circ \Decomp_{s, x, \sigma} \circ \textcolor{red}{U_\mathsf{Split}^\dag}\]
        \item Compute $y = f(s,x;r)$, and CNOT $y$ onto $Q_u$:\label{hyb-4-step:compute-f}
            \[\ket{s, x, \sigma, u}_Q \otimes \ket{r}_R \to \ket{s, x, \sigma, u \oplus f(s,x;r)}_Q \otimes \ket{r}_R\]
        \item Uncompute item \ref{hyb-4-step:query-RO} by applying the following sequence of operations (from right to left):\label{hyb-4-step:uncompute-RO-query}
            \[\textcolor{red}{U_\mathsf{Split}} \circ \Decomp_{s, x, \sigma} \circ \CO' \circ \Decomp_{s, x, \sigma} \circ \textcolor{red}{U_\mathsf{Split}^\dag}\]
    \end{enumerate}
    \item Uncompute step \ref{hyb-4-step:check-equality-of-signatures}.\label{hyb-4-step:check-equality-of-signatures-2}
    \item Output register $Q$.
\end{enumerate}

\begin{lemma}
    Hybrids 3 and 4 are perfectly indistinguishable.
\end{lemma}
\begin{proof}
    Hybrid 4 differs from hybrid 3 in the following ways. Hybrid 4 includes a cache register $C$, and the oracle applies $U_\mathsf{Split}^\dag$ and $U_\mathsf{Split}$ many times. We will show that most of these applications of $U_\mathsf{Split}^\dag$ and $U_\mathsf{Split}$ cancel out with each other.

    First, the first time that the oracle applies $U_\mathsf{Split}^\dag$ or $U_\mathsf{Split}$ in hybrid 4 is during step \ref{hyb-4-step:check-equality-of-signatures} on the first query that reaches that step without aborting. Initially, $C \times D$ are in the state $\ket{\emptyset, \emptyset}$, and the application of $U_\mathsf{Split}^\dag$ maps this state to $\ket{\emptyset, \emptyset}$. Note that the state of $D$ after this application ($D = \ket{\emptyset}$) is the same as the initial state of $D$ in hybrid 3.

    Second, the final time that the oracle applies $U_\mathsf{Split}^\dag$ or $U_\mathsf{Split}$ is during step \ref{hyb-4-step:check-equality-of-signatures-2}. This application of $U_\mathsf{Split}$ does not affect the distinguisher's view because the distinguisher makes no further queries, and $U_\mathsf{Split}$ acts on internal registers of the oracle.

    Third, we claim that all other applications of $U_\mathsf{Split}^\dag$ or $U_\mathsf{Split}$ (between the first and last applications) cancel out with each other. It suffices to consider branches of the superposition on which $B$ is flipped to $\ket{1}$ because otherwise, steps \ref{hyb-4-step:check-equality-of-signatures} and \ref{hyb-4-step:check-equality-of-signatures-2} cancel out. The application of $U_\mathsf{Split}$ in step \ref{hyb-4-step:check-equality-of-signatures} cancels with the application of $U_\mathsf{Split}^\dag$ in step \ref{hyb-4-step:query-RO}. Next, the application of $U_\mathsf{Split}$ in \ref{hyb-4-step:query-RO} cancels with the application of $U_\mathsf{Split}^\dag$ in \ref{hyb-4-step:uncompute-RO-query}. $U_\mathsf{Split}^\dag$ commutes with step \ref{hyb-4-step:compute-f} because $U_\mathsf{Split}^\dag$ acts on registers $C \times D$, and step \ref{hyb-4-step:compute-f} acts on a disjoint set of registers $Q \times R$. The application of $U_\mathsf{Split}$ in \ref{hyb-4-step:uncompute-RO-query} cancels out with the application of $U_\mathsf{Split}^\dag$ in step \ref{hyb-4-step:check-equality-of-signatures-2}. Finally, the application of $U_\mathsf{Split}$ in step \ref{hyb-4-step:check-equality-of-signatures-2} cancels out with the application of $U_\mathsf{Split}^\dag$ in step \ref{hyb-4-step:check-equality-of-signatures} of the next query. $U_\mathsf{Split}^\dag$ (which acts on $C \times D$) commutes with step \ref{hyb-4-step:verify-signature} (which acts on $Q$) because they act on disjoint registers.
\end{proof}

\paragraph{Hybrid 5.} In this hybrid, the compressed oracle operations $\Decomp$ and $\CO'$ act on a smaller database that only records tuples of the form $(s', x', r')$ (and which do not include $\sigma'$).

Since the compressed oracle no longer records $\sigma'$, we must change our $\Decomp$ and $\CO'$ procedures. Instead of using $\Decomp_{s,x,\sigma}$, we use $\Decomp_{s,x}$, which is defined analogously (\cref{sec:compressed-oracle}). $\Decomp_{s,x}$ acts on databases $d$ comprising tuples of the form $(s', x', r')$. Given any database $d$ that does not contain a tuple of the form $(s, x, *)$, $\Decomp_{s,x}$ swaps the states
\[\ket{d} \text{ and } \ket{\psi(d, (s, x), 0)}\]
and acts as the identity on all other states.

Additionally, let $\CO'$ be replaced with $\CO''$ that ignores $\sigma$. $\CO''$ takes a query $(s,x,r)$ and a database $d$ that contains a tuple of the form $(s, x, r')$, and does the following:
    \[\ket{s,x,r}_{Q \times R} \otimes \ket{d}_D \to \ket{s,x,r \oplus r'}_{Q \times R} \otimes \ket{d}_D\]

Next, let us define a unitary $\mathsf{Cache}_{s, x, \sigma}$ that updates register $C$ to record $(s, x, \sigma)$ whenever $D$ records $(s, x, r')$. $\mathsf{Cache}_{s, x, \sigma}$ resembles the chacheing step of $\Sim$. $\mathsf{Cache}_{s, x, \sigma}$ acts on $C \times D$ as follows:
\begin{enumerate}
    \item Check if $D$ contains an entry $(s', x', r')$ such that $s' = s$ and $x' \neq x \oplus 1$. If so, then do the following:
    \begin{enumerate}
        \item If $C$ contains the entry $(s, x, \sigma)$, then remove this entry from $C$.
        \item Else, add $(s, x, \sigma)$ to $C$.
    \end{enumerate}
\end{enumerate}
$ $

Finally, let us define the oracle distribution for hybrid 5.

\begin{itemize}
    \item The oracle has a database register $D$, a cache register $C$, and control qubit $B$. $B \times C \times D$ are initialized to $\ket{0, \emptyset, \emptyset}$. 
\end{itemize}
\begin{enumerate}
    \item \textbf{Inputs:} The oracle takes a query register $Q = Q_s \times Q_x \times Q_\sigma \times Q_u$ with eigenstates of the form $\ket{s, x, \sigma, u}$ where $(s, x, \sigma, u) \in \cS_\secp \times \cX \times \Sigma_\secp \times \cY$.
    \item Check that $\SigToken.\Ver(\pk, s, x, \sigma) = 1$. If so, then continue. If not, then abort and output register $Q$.\label{hyb-5-step:verify-signature}
    \item \textbf{SEQ Check:} \label{hyb-5-step:check-equality-of-signatures}
    \begin{enumerate}
        \item Apply $U_\mathsf{Split}^\dag$ to $C \times D$.
        \item Check that $D$ contains no entry $(s', x', \sigma', r')$ such that $s' = s$ but $(x', \sigma') \neq (x, \sigma)$. If the check passes, then flip $B$ to $\ket{1}$.\label{hyb-5-step:bitflip}
        \item Apply $U_\mathsf{Split}$ to $C \times D$.
    \end{enumerate}
    % \item \textbf{SEQ Check:} Initialize a qubit $\ket{0}_B$. Check that $C \times D$ contains no entry \textcolor{red}{$(s', x', r')$} such that $s' = s$ but $(x', \sigma') \neq (x, \sigma)$. If the check passes, then flip $B$ to $\ket{1}$.\label{hyb-5-step:check-equality-of-signatures}
    % \item \textbf{SEQ Check:} \label{hyb-5-step:check-equality-of-signatures}
    % \begin{enumerate}
    %     \item $U_\mathsf{Combine}$
    %     \item Check that $D$ contains no entry $(s', x', \sigma', r')$ such that $s' = s$ but $(x', \sigma') \neq (x, \sigma)$. If the check passes, then continue. If it fails, then abort and output register $Q$.
    %     \item $U_\mathsf{Combine}^\dag$
    % \end{enumerate}
    \item If $B = \ket{1}$:
    \begin{enumerate}
        \item Query $G$ on $\ket{s,x,\sigma}_{Q_s \times Q_x \times Q_\sigma} \otimes \ket{0}_R \otimes \ket{c,d}_{C \times D}$ by applying the following sequence of operations (from right to left):\label{hyb-5-step:query-RO}
            \[\textcolor{red}{\mathsf{Cache}_{s,x,\sigma}^\dag \circ \Decomp_{s, x} \circ \CO'' \circ \Decomp_{s, x} \circ \mathsf{Cache}_{s,x,\sigma}}\]
        \item Compute $y = f(s,x;r)$, and CNOT $y$ onto $Q_u$:\label{hyb-5-step:answer-query}
            \[\ket{s, x, \sigma, u}_Q \otimes \ket{r}_R \to \ket{s, x, \sigma, u \oplus f(s,x;r)}_Q \otimes \ket{r}_R\]
        \item Uncompute item \ref{hyb-5-step:query-RO} by applying the following sequence of operations (from right to left):\label{hyb-5-step:uncompute-RO-query}
            \[\textcolor{red}{\mathsf{Cache}_{s,x,\sigma}^\dag \circ \Decomp_{s, x} \circ \CO'' \circ \Decomp_{s, x} \circ \mathsf{Cache}_{s,x,\sigma}}\]
    \end{enumerate}
    \item Uncompute step \ref{hyb-5-step:check-equality-of-signatures}.\label{hyb-5-step:check-equality-of-signatures-2}
    \item Output register $Q$.
\end{enumerate}

\begin{lemma}\label{thm:cache-is-involution}
    $\mathsf{Cache}_{s,x,\sigma}$ is its own inverse: $\mathsf{Cache}_{s,x,\sigma}^\dag = \mathsf{Cache}_{s,x,\sigma}$.
\end{lemma}
\begin{proof}
    $\mathsf{Cache}_{s,x,\sigma}$ changes the state on register $C$ (by adding or removing $(s, \sigma)$) controlled on the state of register $D$. Applying this operation a second time will undo the action of the first. If the first $\mathsf{Cache}_{s,x,\sigma}$ added $(s, \sigma)$ to $C$, then the second $\mathsf{Cache}_{s,x,\sigma}$ removes $(s, \sigma)$, and vice versa.
\end{proof}

\begin{lemma}
    Hybrids 4 and 5 are perfectly indistinguishable.
\end{lemma}
\begin{proof}
    The only difference between the hybrids is how they implement queries to $G$. See steps \ref{hyb-4-step:query-RO} and \ref{hyb-4-step:uncompute-RO-query} of hybrid 4 and steps \ref{hyb-5-step:query-RO} and \ref{hyb-5-step:uncompute-RO-query} of hybrid 5. We will show that each of these steps computes the same function as their analog in hybrid 4.\\

    \textit{First}, let us define $\Pi_{s, x, \sigma}$ to project the $C \times D$ registers onto \textit{valid} states, where $C \times D$ are the result of splitting some combined database $d$, and $(s,x,\sigma)$ is an allowed query for this database. Formally, $\Pi_{s, x, \sigma}$ projects onto all eigenstates of the following form.
	\begin{equation}\label{eq:pi}
    \begin{split}
        &\ket{c'}_C \otimes \ket{d'}_{D}, \quad \text{or} \\
        &\ket{c' \cup \{(s,x,\sigma)\}}_C \otimes \ket{\psi(d', (s,x), \tilde{r}')}_{D}\\\\
        &\text{where $\tilde{r}' \in \cR \backslash \{0\}$, $(c',d') = \mathsf{Split}(\emptyset, d)$,}\\ 
        &\text{and $d$ is a splittable database that does not contain}\\ 
        &\text{any tuple $(s',x',\sigma',r')$ for which $s' = s$}
    \end{split}
    \end{equation}
    
    Recall from \cref{eq:psi} that 
    \begin{align*}
        \ket{\psi(d', (s,x), \tilde{r}')} &= \frac{1}{\sqrt{\abs{\cR}}} \cdot \sum_{r' \in \cR} (-1)^{\langle r', \tilde{r}'\rangle} \ket{d' \cup \{(s,x,r')\}}
    \end{align*}
    % Then any state of the form $\ket{c'}_C \otimes \ket{d'}_D$, where $(c',d') = \mathsf{Split}(\emptyset, d)$ and $d$ is a splittable database that does not contain $(s,*,*,*)$, is in the span of $\Pi_{s,x,\sigma}$.\\

    \textit{Second}, let us assume that we are given a state in the span of $\Pi_{s, x, \sigma}$ and let us apply to the state either the operation in hybrid 4, step \ref{hyb-4-step:query-RO}
    \begin{equation}\label{eq:hyb-4-query-RO}
        U_\mathsf{Split} \circ \Decomp_{s, x, \sigma} \circ \CO' \circ \Decomp_{s, x, \sigma} \circ U_\mathsf{Split}^\dag
    \end{equation}
    or the operation in hybrid 5, step \ref{hyb-5-step:query-RO}
    \begin{equation}\label{eq:hyb-5-query-RO}
        \mathsf{Cache}_{s,x,\sigma}^\dag \circ \Decomp_{s, x} \circ \CO'' \circ \Decomp_{s, x} \circ \mathsf{Cache}_{s,x,\sigma}
    \end{equation}
    We will show the two operations act equivalently on the state and that the final state is also in the span of $\Pi_{s, x, \sigma}$.

    Note that $\Decomp_{s, x} \circ \CO'' \circ \Decomp_{s, x}$ is simply the query operation $\CO$ to a compressed oracle that stores tuples of the form $(s,x,r)$. Likewise, $\Decomp_{s, x, \sigma} \circ \CO' \circ \Decomp_{s, x, \sigma}$ is simply the query operation $\CO$ to a compressed oracle that stores tuples of the form $(s,x,\sigma,r)$. We can use \cref{thm:CO-query-behavior} to describe how these operations act on eigenstates of $\Pi_{s,x,\sigma}$.

    Furthermore, let us express the $R$ register in the $\{\ket{\phi(\tilde{r})}\}_{\tilde{r} \in \cR}$ basis. Recall from \cref{eq:phi} that
    \begin{align*}
        \ket{\phi(\tilde{r})} &= \frac{1}{\sqrt{\abs{\cR}}} \cdot \sum_{r \in \cR} (-1)^{\langle r, \tilde{r}\rangle} \ket{r}
    \end{align*}
    Additionally, \cref{thm:orthonormal-basis} shows that $\{\ket{\phi(\tilde{r})}\}_{\tilde{r} \in \cR}$ is an orthonormal basis for register $R$. 

    Now, let us consider the following cases:

    \begin{enumerate}
        \item Case 1: $R \times C \times D$ contains $\ket{\phi(\tilde{r})}_R \otimes \ket{c'}_C \otimes \ket{d'}_{D}$ where $\tilde{r} = 0$ and $(c', d')$ satisfy the conditions of \cref{eq:pi}.

        Let us apply the operation from hybrid 5 (\cref{eq:hyb-5-query-RO}):
        \begin{align*}
            \ket{\phi(0)}_R \otimes \ket{c'}_C \otimes \ket{d'}_D &\overset{\mathsf{Cache}_{s,x,\sigma}}{\longrightarrow} \ket{\phi(0)}_R \otimes \ket{c'}_C \otimes \ket{d'}_D\\
            &\overset{\Decomp_{s,x} \circ \CO'' \circ \Decomp_{s,x}}{\longrightarrow} \ket{\phi(0)}_R \otimes \ket{c'}_C \otimes \ket{d'}_D\\
            &\overset{\mathsf{Cache}_{s,x,\sigma}^\dag}{\longrightarrow} \ket{\phi(0)}_R \otimes \ket{c'}_C \otimes \ket{d'}_D
        \end{align*}
        
        Likewise, let us apply the operation from hybrid 4 (\cref{eq:hyb-4-query-RO}):
        \begin{align*}
            \ket{\phi(0)}_R \otimes \ket{c'}_C \otimes \ket{d'}_{D} &\overset{U_\mathsf{Split}^\dag}{\longrightarrow} \ket{\phi(0)}_R \otimes \textcolor{red}{\ket{\emptyset}_C \otimes \ket{d}_{D}}\\
            &\overset{\Decomp_{s,x,\sigma} \circ \CO' \circ \Decomp_{s,x,\sigma}}{\longrightarrow} \ket{\phi(0)}_R \otimes \ket{\emptyset}_C \otimes \ket{d}_{D}\\
            &\overset{U_\mathsf{Split}}{\longrightarrow} \ket{\phi(0)}_R \otimes \textcolor{red}{\ket{c'}_C \otimes \ket{d'}_{D}}
        \end{align*}
        This shows that the operations in \cref{eq:hyb-4-query-RO,eq:hyb-5-query-RO} act equivalently on this state. Furthermore, the final state $\ket{c'}_C \otimes \ket{d'}_{D}$ is still in the span of $\Pi_{s,x,\sigma}$.
        
        \item Case 2: $R \times C \times D$ contains $\ket{\phi(\tilde{r})}_R \otimes \ket{c'}_C \otimes \ket{d'}_{D}$ where $\tilde{r} \neq 0$ and $(c', d')$ satisfy the conditions of \cref{eq:pi}. 
        
        Let us apply the operation from hybrid 5 (\cref{eq:hyb-5-query-RO}):
        % LNCS layout fix: Case 2, hybrid 5; original lines 558--562.
        \ifllncs
        \begin{align*}
            &\ket{\phi(\tilde{r})}_R \otimes \ket{c'}_C \otimes \ket{d'}_D\\
            &\overset{\mathsf{Cache}_{s,x,\sigma}}{\longrightarrow} \ket{\phi(\tilde{r})}_R \otimes \ket{c'}_C \otimes \ket{d'}_D\\
            &\overset{\Decomp_{s,x} \circ \CO'' \circ \Decomp_{s,x}}{\longrightarrow} \ket{\phi(\tilde{r})}_R \otimes \ket{c'}_C \otimes \textcolor{red}{\ket{\psi(d', (s,x), \tilde{r})}_D}\\
            &\overset{\mathsf{Cache}_{s,x,\sigma}^\dag}{\longrightarrow} \ket{\phi(\tilde{r})}_R \otimes \textcolor{red}{\ket{c' \cup \{(s, x, \sigma)\}}_C} \otimes \ket{\psi(d', (s,x), \tilde{r})}_D
        \end{align*}
        \else
        \begin{align*}
            \ket{\phi(\tilde{r})}_R \otimes \ket{c'}_C \otimes \ket{d'}_D &\overset{\mathsf{Cache}_{s,x,\sigma}}{\longrightarrow} \ket{\phi(\tilde{r})}_R \otimes \ket{c'}_C \otimes \ket{d'}_D\\
            &\overset{\Decomp_{s,x} \circ \CO'' \circ \Decomp_{s,x}}{\longrightarrow} \ket{\phi(\tilde{r})}_R \otimes \ket{c'}_C \otimes \textcolor{red}{\ket{\psi(d', (s,x), \tilde{r})}_D}\\
            &\overset{\mathsf{Cache}_{s,x,\sigma}^\dag}{\longrightarrow} \ket{\phi(\tilde{r})}_R \otimes \textcolor{red}{\ket{c' \cup \{(s, x, \sigma)\}}_C} \otimes \ket{\psi(d', (s,x), \tilde{r})}_D
        \end{align*}
        \fi
        
        Likewise, let us apply the operation from hybrid 4 (\cref{eq:hyb-4-query-RO}):
        % LNCS layout fix: Case 2, hybrid 4; original lines 565--569.
        \ifllncs
        \begin{align*}
            &\ket{\phi(\tilde{r})}_R \otimes \ket{c'}_C \otimes \ket{d'}_{D}\\
            &\overset{U_\mathsf{Split}^\dag}{\longrightarrow} \ket{\phi(\tilde{r})}_R \otimes \textcolor{red}{\ket{\emptyset}_C \otimes \ket{d}_{D}}\\
            &\overset{\Decomp_{s,x,\sigma} \circ \CO' \circ \Decomp_{s,x,\sigma}}{\longrightarrow} \ket{\phi(\tilde{r})}_R \otimes \ket{\emptyset}_C \otimes \textcolor{red}{\ket{\psi(d, (s,x,\sigma), \tilde{r})}_D}\\
            &\overset{U_\mathsf{Split}}{\longrightarrow} \ket{\phi(\tilde{r})}_R \otimes \textcolor{red}{\ket{c' \cup \{(s, x, \sigma)\}}_C \otimes \ket{\psi(d', (s,x), \tilde{r})}_{D}}
        \end{align*}
        \else
        \begin{align*}
            \ket{\phi(\tilde{r})}_R \otimes \ket{c'}_C \otimes \ket{d'}_{D} &\overset{U_\mathsf{Split}^\dag}{\longrightarrow} \ket{\phi(\tilde{r})}_R \otimes \textcolor{red}{\ket{\emptyset}_C \otimes \ket{d}_{D}}\\
            &\overset{\Decomp_{s,x,\sigma} \circ \CO' \circ \Decomp_{s,x,\sigma}}{\longrightarrow} \ket{\phi(\tilde{r})}_R \otimes \ket{\emptyset}_C \otimes \textcolor{red}{\ket{\psi(d, (s,x,\sigma), \tilde{r})}_D}\\
            &\overset{U_\mathsf{Split}}{\longrightarrow} \ket{\phi(\tilde{r})}_R \otimes \textcolor{red}{\ket{c' \cup \{(s, x, \sigma)\}}_C \otimes \ket{\psi(d', (s,x), \tilde{r})}_{D}}
        \end{align*}
        \fi
        This shows that the operations in \cref{eq:hyb-4-query-RO,eq:hyb-5-query-RO} act equivalently on this state. Furthermore, the final state $\ket{c' \cup \{(s, x, \sigma)\}}_C \otimes \ket{\psi(d', (s,x), \tilde{r})}_{D}$ is still in the span of $\Pi_{s,x,\sigma}$.
        
        \item Case 3: $R \times C \times D$ contains $\ket{\phi(\tilde{r})}_R \otimes \ket{c' \cup \{(s,x,\sigma)\}}_C \otimes \ket{\psi(d', (s,x), \tilde{r}')}_{D}$, where $\tilde{r} = \tilde{r}'$, $\tilde{r}' \neq 0$, and $(c',d')$ satisfy the conditions of \cref{eq:pi}.

        Let us apply the operation from hybrid 5 (\cref{eq:hyb-5-query-RO}):
        % LNCS layout fix: Case 3, hybrid 5; original lines 575--579.
        \ifllncs
        \begin{align*}
            &\ket{\phi(\tilde{r}')}_R \otimes \ket{c' \cup \{(s,x,\sigma)\}}_C \otimes \ket{\psi(d', (s,x), \tilde{r}')}_{D}\\
            &\overset{\mathsf{Cache}_{s,x,\sigma}}{\longrightarrow} \ket{\phi(\tilde{r}')}_R \otimes \ket{\textcolor{red}{c'}}_C \otimes \ket{\psi(d', (s,x), \tilde{r}')}_{D}\\
            &\overset{\Decomp_{s,x} \circ \CO'' \circ \Decomp_{s,x}}{\longrightarrow} \ket{\phi(\tilde{r}')}_R \otimes \ket{c'}_C \otimes \ket{\textcolor{red}{d'}}_{D}\\
            &\overset{\mathsf{Cache}_{s,x,\sigma}^\dag}{\longrightarrow} \ket{\phi(\tilde{r}')}_R \otimes \ket{c'}_C \otimes \ket{d'}_{D}
        \end{align*}
        \else
        \begin{align*}
            \ket{\phi(\tilde{r}')}_R \otimes \ket{c' \cup \{(s,x,\sigma)\}}_C \otimes \ket{\psi(d', (s,x), \tilde{r}')}_{D} &\overset{\mathsf{Cache}_{s,x,\sigma}}{\longrightarrow} \ket{\phi(\tilde{r}')}_R \otimes \ket{\textcolor{red}{c'}}_C \otimes \ket{\psi(d', (s,x), \tilde{r}')}_{D}\\
            &\overset{\Decomp_{s,x} \circ \CO'' \circ \Decomp_{s,x}}{\longrightarrow} \ket{\phi(\tilde{r}')}_R \otimes \ket{c'}_C \otimes \ket{\textcolor{red}{d'}}_{D}\\
            &\overset{\mathsf{Cache}_{s,x,\sigma}^\dag}{\longrightarrow} \ket{\phi(\tilde{r}')}_R \otimes \ket{c'}_C \otimes \ket{d'}_{D}
        \end{align*}
        \fi
        
        Likewise, let us apply the operation from hybrid 4 (\cref{eq:hyb-4-query-RO}):
        % LNCS layout fix: Case 3, hybrid 4; original lines 582--588.
        \ifllncs
        \begin{align*}
            &\ket{\phi(\tilde{r}')} \ket{c' \cup \{(s,x,\sigma)\}} \ket{\psi(d', (s,x), \tilde{r}')}\\
            &= \ket{\phi(\tilde{r}')} \left(\frac{1}{\sqrt{\abs{\cR}}} \cdot \sum_{r' \in \cR} (-1)^{\langle r', \tilde{r}'\rangle} \cdot \ket{c' \cup \{(s,x,\sigma)\}} \ket{d' \cup \{(s,x,r')\}}\right)\\
            &\overset{U_\mathsf{Split}^\dag}{\longrightarrow} \ket{\phi(\tilde{r}')} \left(\frac{1}{\sqrt{\abs{\cR}}} \cdot \sum_{r' \in \cR} (-1)^{\langle r', \tilde{r}'\rangle} \cdot \ket{\textcolor{red}{\emptyset}} \ket{\textcolor{red}{d \cup \{(s,x,\sigma, r')\}}}\right)\\
            &= \ket{\phi(\tilde{r}')} \ket{\emptyset} \ket{\psi(d, (s,x,\sigma), \tilde{r}')}\\
            &\overset{\Decomp_{s,x,\sigma} \circ \CO' \circ \Decomp_{s,x,\sigma}}{\longrightarrow} \ket{\phi(\tilde{r}')} \ket{\emptyset} \ket{\textcolor{red}{d}}\\
            &\overset{U_\mathsf{Split}}{\longrightarrow} \ket{\phi(\tilde{r}')} \textcolor{red}{\ket{c'} \ket{d'}}
        \end{align*}
        \else
        \begin{align*}
            \ket{\phi(\tilde{r}')} \ket{c' \cup \{(s,x,\sigma)\}} \ket{\psi(d', (s,x), \tilde{r}')} &= \ket{\phi(\tilde{r}')} \left(\frac{1}{\sqrt{\abs{\cR}}} \cdot \sum_{r' \in \cR} (-1)^{\langle r', \tilde{r}'\rangle} \cdot \ket{c' \cup \{(s,x,\sigma)\}} \ket{d' \cup \{(s,x,r')\}}\right)\\
            &\overset{U_\mathsf{Split}^\dag}{\longrightarrow} \ket{\phi(\tilde{r}')} \left(\frac{1}{\sqrt{\abs{\cR}}} \cdot \sum_{r' \in \cR} (-1)^{\langle r', \tilde{r}'\rangle} \cdot \ket{\textcolor{red}{\emptyset}} \ket{\textcolor{red}{d \cup \{(s,x,\sigma, r')\}}}\right)\\
            &= \ket{\phi(\tilde{r}')} \ket{\emptyset} \ket{\psi(d, (s,x,\sigma), \tilde{r}')}\\
            &\overset{\Decomp_{s,x,\sigma} \circ \CO' \circ \Decomp_{s,x,\sigma}}{\longrightarrow} \ket{\phi(\tilde{r}')} \ket{\emptyset} \ket{\textcolor{red}{d}}\\
            &\overset{U_\mathsf{Split}}{\longrightarrow} \ket{\phi(\tilde{r}')} \textcolor{red}{\ket{c'} \ket{d'}}
        \end{align*}
        \fi
        This shows that the operations in \cref{eq:hyb-4-query-RO,eq:hyb-5-query-RO} act equivalently on this state. Furthermore, the final state $\ket{c'} \ket{d'}$ is still in the span of $\Pi_{s,x,\sigma}$.
        
        \item Case 4: $R \times C \times D$ contains $\ket{\phi(\tilde{r})}_R \otimes \ket{c' \cup \{(s,x,\sigma)\}}_C \otimes \ket{\psi(d', (s,x), \tilde{r}')}_{D}$, where $\tilde{r} \neq \tilde{r}'$, $\tilde{r'} \neq 0$, and $(c',d')$ satisfy the conditions of \cref{eq:pi}.

        Let us apply the operation from hybrid 5 (\cref{eq:hyb-5-query-RO}):
        % LNCS layout fix: Case 4, hybrid 5; original lines 594--598.
        \ifllncs
        \begin{align*}
            &\ket{\phi(\tilde{r})}_R \otimes \ket{c' \cup \{(s,x,\sigma)\}}_C \otimes \ket{\psi(d', (s,x), \tilde{r}')}_{D}\\
            &\overset{\mathsf{Cache}_{s,x,\sigma}}{\longrightarrow} \ket{\phi(\tilde{r})}_R \otimes \ket{\textcolor{red}{c'}}_C \otimes \ket{\psi(d', (s,x), \tilde{r}')}_{D}\\
            &\overset{\Decomp_{s,x} \circ \CO'' \circ \Decomp_{s,x}}{\longrightarrow} \ket{\phi(\tilde{r})}_R \otimes \ket{c'}_C \otimes \ket{\textcolor{red}{\psi(d',(s,x),\tilde{r} + \tilde{r}')}}_{D}\\
            &\overset{\mathsf{Cache}_{s,x,\sigma}^\dag}{\longrightarrow} \ket{\phi(\tilde{r})}_R \otimes \ket{\textcolor{red}{c' \cup \{(s,x,\sigma)\}}}_C \otimes \ket{\psi(d',(s,x),\tilde{r} + \tilde{r}')}_{D}
        \end{align*}
        \else
        \begin{align*}
            \ket{\phi(\tilde{r})}_R \otimes \ket{c' \cup \{(s,x,\sigma)\}}_C \otimes \ket{\psi(d', (s,x), \tilde{r}')}_{D} &\overset{\mathsf{Cache}_{s,x,\sigma}}{\longrightarrow} \ket{\phi(\tilde{r})}_R \otimes \ket{\textcolor{red}{c'}}_C \otimes \ket{\psi(d', (s,x), \tilde{r}')}_{D}\\
            &\overset{\Decomp_{s,x} \circ \CO'' \circ \Decomp_{s,x}}{\longrightarrow} \ket{\phi(\tilde{r})}_R \otimes \ket{c'}_C \otimes \ket{\textcolor{red}{\psi(d',(s,x),\tilde{r} + \tilde{r}')}}_{D}\\
            &\overset{\mathsf{Cache}_{s,x,\sigma}^\dag}{\longrightarrow} \ket{\phi(\tilde{r})}_R \otimes \ket{\textcolor{red}{c' \cup \{(s,x,\sigma)\}}}_C \otimes \ket{\psi(d',(s,x),\tilde{r} + \tilde{r}')}_{D}
        \end{align*}
        \fi
        
        Likewise, let us apply the operation from hybrid 4 (\cref{eq:hyb-4-query-RO}):
        % LNCS layout fix: Case 4, hybrid 4; original lines 601--609.
        \ifllncs
        \begin{align*}
            &\ket{\phi(\tilde{r})} \ket{c' \cup \{(s,x,\sigma)\}} \ket{\psi(d', (s,x), \tilde{r}')}\\
            &= \ket{\phi(\tilde{r})} \left(\frac{1}{\sqrt{\abs{\cR}}} \cdot \sum_{r' \in \cR} (-1)^{\langle r', \tilde{r}'\rangle} \cdot \ket{c' \cup \{(s,x,\sigma)\}} \ket{d' \cup \{(s,x,r')\}}\right)\\
            &\overset{U_\mathsf{Split}^\dag}{\longrightarrow} \ket{\phi(\tilde{r})} \left(\frac{1}{\sqrt{\abs{\cR}}} \cdot \sum_{r' \in \cR} (-1)^{\langle r', \tilde{r}'\rangle} \cdot \ket{\textcolor{red}{\emptyset}} \ket{\textcolor{red}{d \cup \{(s,x,\sigma, r')\}}}\right)\\
            &= \ket{\phi(\tilde{r})} \ket{\emptyset} \ket{\psi(d, (s,x,\sigma), \tilde{r}')}\\
            &\overset{\Decomp_{s,x,\sigma} \circ \CO' \circ \Decomp_{s,x,\sigma}}{\longrightarrow} \ket{\phi(\tilde{r})} \ket{\emptyset} \ket{\textcolor{red}{\psi(d,(s,x,\sigma),\tilde{r}+\tilde{r}')}}\\
            &= \ket{\phi(\tilde{r})} \left(\frac{1}{\sqrt{\abs{\cR}}} \cdot \sum_{r' \in \cR} (-1)^{\langle r', \tilde{r}+\tilde{r}'\rangle} \cdot \ket{\emptyset} \ket{d \cup \{(s,x,\sigma,r')\}}\right)\\
            &\overset{U_\mathsf{Split}}{\longrightarrow} \ket{\phi(\tilde{r})} \left(\frac{1}{\sqrt{\abs{\cR}}} \cdot \sum_{r' \in \cR} (-1)^{\langle r', \tilde{r}+\tilde{r}'\rangle} \cdot \ket{\textcolor{red}{c' \cup \{(s,x,\sigma)\}}} \ket{\textcolor{red}{d' \cup \{(s,x,r')\}}}\right)\\
            &= \ket{\phi(\tilde{r})} \ket{c' \cup \{(s,x,\sigma)\}}\ket{\psi(d',(s,x),\tilde{r}+\tilde{r}')}
        \end{align*}
        \else
        \begin{align*}
            \ket{\phi(\tilde{r})} \ket{c' \cup \{(s,x,\sigma)\}} \ket{\psi(d', (s,x), \tilde{r}')} &= \ket{\phi(\tilde{r})} \left(\frac{1}{\sqrt{\abs{\cR}}} \cdot \sum_{r' \in \cR} (-1)^{\langle r', \tilde{r}'\rangle} \cdot \ket{c' \cup \{(s,x,\sigma)\}} \ket{d' \cup \{(s,x,r')\}}\right)\\
            &\overset{U_\mathsf{Split}^\dag}{\longrightarrow} \ket{\phi(\tilde{r})} \left(\frac{1}{\sqrt{\abs{\cR}}} \cdot \sum_{r' \in \cR} (-1)^{\langle r', \tilde{r}'\rangle} \cdot \ket{\textcolor{red}{\emptyset}} \ket{\textcolor{red}{d \cup \{(s,x,\sigma, r')\}}}\right)\\
            &= \ket{\phi(\tilde{r})} \ket{\emptyset} \ket{\psi(d, (s,x,\sigma), \tilde{r}')}\\
            &\overset{\Decomp_{s,x,\sigma} \circ \CO' \circ \Decomp_{s,x,\sigma}}{\longrightarrow} \ket{\phi(\tilde{r})} \ket{\emptyset} \ket{\textcolor{red}{\psi(d,(s,x,\sigma),\tilde{r}+\tilde{r}')}}\\
            &= \ket{\phi(\tilde{r})} \left(\frac{1}{\sqrt{\abs{\cR}}} \cdot \sum_{r' \in \cR} (-1)^{\langle r', \tilde{r}+\tilde{r}'\rangle} \cdot \ket{\emptyset} \ket{d \cup \{(s,x,\sigma,r')\}}\right)\\
            &\overset{U_\mathsf{Split}}{\longrightarrow} \ket{\phi(\tilde{r})} \left(\frac{1}{\sqrt{\abs{\cR}}} \cdot \sum_{r' \in \cR} (-1)^{\langle r', \tilde{r}+\tilde{r}'\rangle} \cdot \ket{\textcolor{red}{c' \cup \{(s,x,\sigma)\}}} \ket{\textcolor{red}{d' \cup \{(s,x,r')\}}}\right)\\
            &= \ket{\phi(\tilde{r})} \ket{c' \cup \{(s,x,\sigma)\}}\ket{\psi(d',(s,x),\tilde{r}+\tilde{r}')}
        \end{align*}
        \fi
        This shows that the operations in \cref{eq:hyb-4-query-RO,eq:hyb-5-query-RO} act equivalently on this state. Furthermore, the final state $\ket{c' \cup \{(s,x,\sigma)\}}\ket{\psi(d',(s,x),\tilde{r}+\tilde{r}')}$ is still in the span of $\Pi_{s,x,\sigma}$.
    \end{enumerate}

    \textit{Third}, the SEQ check (step \ref{hyb-4-step:check-equality-of-signatures} in hybrid 4 or step \ref{hyb-5-step:check-equality-of-signatures} in hybrid 5) is the same in the two hybrids. Likewise, step \ref{hyb-4-step:check-equality-of-signatures-2} in hybrid 4 or step \ref{hyb-5-step:check-equality-of-signatures-2} in hybrid 5 are the same. Furthermore, if the state of $C \times D$ is in the span of $\Pi_{s,x,\sigma}$ at the beginning of the SEQ check, then it will still be in the span of $\Pi_{s,x,\sigma}$ at the end of the SEQ check. Intuitively, this is because the SEQ check always flips $B$ in step \ref{hyb-5-step:bitflip} if the initial state is in the span of $\Pi_{s,x,\sigma}$, and the applications of $U_\mathsf{Split}$ and $U_\mathsf{Split}^\dag$ cancel out.
    
    If a state of $C \times D$ is in the span of $\Pi_{s,x,\sigma}$, then it is in the span of states of the form:
    \begin{align*}
        &\ket{c'}_C \otimes \ket{d'}_{D}\\
        &\ket{c' \cup \{(s,x,\sigma)\}}_C \otimes \ket{d' \cup \{(s,x,r')\}}_D
    \end{align*}
    where $(c',d') = \mathsf{Split}(\emptyset, d)$ and $d$ is a splittable database that does not contain $(s,*,*,*)$.
    
    Let us consider how the SEQ check acts on each of the states listed above.
    \begin{enumerate}
        \item Case 1: $B \times C \times D$ contains $\ket{0}_B \otimes \ket{c'}_C \otimes \ket{d'}_{D}$. Then the SEQ check acts on this state as follows:
        \begin{align*}
            \ket{0}_B \otimes \ket{c'}_C \otimes \ket{d'}_D &\overset{U_\mathsf{Split}^\dag}{\longrightarrow} \ket{0}_B  \otimes \ket{\emptyset}_C \otimes \ket{d}_D\\
            &\overset{\text{Step \ref{hyb-5-step:bitflip}}}{\longrightarrow} \ket{1}_B \otimes \ket{\emptyset}_C \otimes \ket{d}_D\\
            &\overset{U_\mathsf{Split}}{\longrightarrow} \ket{1}_B \otimes \ket{c'}_C \otimes \ket{d'}_D
        \end{align*}
        The reason that step \ref{hyb-5-step:bitflip} flips $B$ is that $d$ does not contain any entry of the form $(s, *,*,*)$.

        \item Case 2: $B \times C \times D$ contains $\ket{0}_B \otimes \ket{c' \cup \{(s,x,\sigma)\}}_C \otimes \ket{d' \cup \{(s,x, r')\}}_{D}$. Then the SEQ check acts on this state as follows:
        % LNCS layout fix: SEQ check, Case 2; original lines 633--637.
        \ifllncs
        \begin{align*}
            &\ket{0}_B \otimes \ket{c' \cup \{(s,x,\sigma)\}}_C \otimes \ket{d' \cup \{(s,x, r')\}}_{D}\\
            &\overset{U_\mathsf{Split}^\dag}{\longrightarrow} \ket{0}_B \otimes \ket{\emptyset}_C \otimes \ket{d \cup \{(s,x,\sigma, r')\}}_{D}\\
            &\overset{\text{Step \ref{hyb-5-step:bitflip}}}{\longrightarrow} \ket{1}_B \otimes \ket{\emptyset}_C \otimes \ket{d \cup \{(s,x,\sigma, r')\}}_{D}\\
            &\overset{U_\mathsf{Split}}{\longrightarrow} \ket{1}_B \otimes \ket{c' \cup \{(s,x,\sigma)\}}_C \otimes \ket{d' \cup \{(s,x, r')\}}_{D}
        \end{align*}
        \else
        \begin{align*}
            \ket{0}_B \otimes \ket{c' \cup \{(s,x,\sigma)\}}_C \otimes \ket{d' \cup \{(s,x, r')\}}_{D} &\overset{U_\mathsf{Split}^\dag}{\longrightarrow} \ket{0}_B \otimes \ket{\emptyset}_C \otimes \ket{d \cup \{(s,x,\sigma, r')\}}_{D}\\
            &\overset{\text{Step \ref{hyb-5-step:bitflip}}}{\longrightarrow} \ket{1}_B \otimes \ket{\emptyset}_C \otimes \ket{d \cup \{(s,x,\sigma, r')\}}_{D}\\
            &\overset{U_\mathsf{Split}}{\longrightarrow} \ket{1}_B \otimes \ket{c' \cup \{(s,x,\sigma)\}}_C \otimes \ket{d' \cup \{(s,x, r')\}}_{D}
        \end{align*}
        \fi
        The reason that step \ref{hyb-5-step:bitflip} flips $B$ is that $(s,x,\sigma, r')$ is the only entry of $d \cup \{(s,x,\sigma, r')\}$ with that particular $s$-value. Then there is no entry $(s',x',\sigma',r')$ of $d \cup \{(s,x,\sigma, r')\}$ such that $s = s'$ but $(x',\sigma')\neq(x,\sigma)$. 
    \end{enumerate}
    This shows that if the state of $C \times D$ is in the span of $\Pi_{s,x,\sigma}$ at the start of the SEQ check, then the check will flip $B$ to $\ket{1}$ with certainty, and the final state of $C \times D$ will be the same as its initial state.\\
    
    % The SEQ check does not modify the computational-basis value of $C \times D$. Let's consider how a computational-basis eigenstate is affected by the SEQ check. First, the SEQ check applies $U_\mathsf{Split}^\dag$, which permutes the computational basis state. Second, it flips $B$ depending on the computational-basis value of $D$. Third, it applies the reverse permutation $U_\mathsf{Split}$, which restores the original computational-basis state of $C \times D$.

    \textit{Fourth}, at the start of any application of \cref{eq:hyb-4-query-RO} in hybrid 4 or \cref{eq:hyb-5-query-RO} in hybrid 5, the state of the $C \times D$ registers is in the span of $\Pi_{s,x,\sigma}$. We will prove this inductively. For the base case: initially, $C \times D$ is in the state $\ket{\emptyset}\ket{\emptyset}$.

    For the inductive case: the only steps that act on the $C \times D$ registers are the SEQ check (steps \ref{hyb-5-step:check-equality-of-signatures} and \ref{hyb-5-step:check-equality-of-signatures-2}) or the queries to the compressed oracle (steps \ref{hyb-5-step:query-RO} and \ref{hyb-5-step:uncompute-RO-query}). We've already shown that these steps map states in the span of $\Pi_{s,x,\sigma}$ to states in the span of $\Pi_{s,x,\sigma}$. Therefore, at the start of any query to the compressed oracle, the state of $C \times D$ will be in the span of $\Pi_{s,x,\sigma}$.\\

    \textit{Fifth}, we know that the compressed oracle queries in hybrid 4 (\cref{eq:hyb-4-query-RO}) and hybrid 5 (\cref{eq:hyb-5-query-RO}) act equivalently on any state in the span of $\Pi_{s,x,\sigma}$. The hybrids are identical except for their compressed oracle queries. Therefore, hybrids 4 and 5 perform equivalent operations.
\end{proof}

\paragraph{Hybrid 6.} We cancel out adjacent applications of $\mathsf{Cache}_{s,x,\sigma}$ and $\mathsf{Cache}_{s,x,\sigma}^\dag$, and we explicitly switch to a compressed oracle $H$ that takes queries of the form $(s,x)$ instead of $(s,x,\sigma)$.

Previously, we implemented a compressed oracle $G: \cS_\secp \times \cX \times \Sigma_\secp \to \cR$. Now we implement a compressed oracle $H: \cS_\secp \times \cX \to \cR$.

\begin{itemize}
    \item \textcolor{red}{The oracle has a database register $D$ implementing a compressed random oracle $H: \cS_\secp \times \cX \to \cR$.} The oracle also has a cache register $C$, and control qubit $B$. $B \times C \times D$ are initialized to $\ket{0, \emptyset, \emptyset}$.
\end{itemize}
\begin{enumerate}
    \item \textbf{Inputs:} The oracle takes a query register $Q = Q_s \times Q_x \times Q_\sigma \times Q_u$ with eigenstates of the form $\ket{s, x, \sigma, u}$ where $(s, x, \sigma, u) \in \cS_\secp \times \cX \times \Sigma_\secp \times \cY$.
    \item Check that $\SigToken.\Ver(\pk, s, x, \sigma) = 1$. If so, then continue. If not, then abort and output register $Q$.\label{hyb-6-step:verify-signature}
    \item \textbf{SEQ Check:} \label{hyb-6-step:check-equality-of-signatures}
    \begin{enumerate}
        \item Apply $U_\mathsf{Split}^\dag$ to $C \times D$.
        \item Check that $D$ contains no entry $(s', x', \sigma', r')$ such that $s' = s$ but $(x', \sigma') \neq (x, \sigma)$. If the check passes, then flip $B$ to $\ket{1}$.\label{hyb-6-step:bitflip}
        \item Apply $U_\mathsf{Split}$ to $C \times D$.
    \end{enumerate}
    \item If $B = \ket{1}$:
    \begin{enumerate}
        \item \label{hyb-6-step:cache-1}\textcolor{red}{\textbf{Cache 1}: Check if $D$ contains an entry $(s', x', r')$ such that $s' = s$ and $x' \neq x \oplus 1$. If so, then do the following:
        \begin{enumerate}
            \item If $C$ contains the entry $(s, x, \sigma)$, then remove this entry from $C$.
            \item Else, add $(s, x, \sigma)$ to $C$.
        \end{enumerate}}
        \item \textcolor{red}{Query $H$ on $\ket{s,x}_{Q_s \times Q_x} \otimes \ket{0}_R \otimes \ket{d}_{D}$.}\label{hyb-6-step:query-RO}
        \item Compute $y = f(s,x;r)$, and CNOT $y$ onto $Q_u$:
            \[\ket{s, x, \sigma, u}_Q \otimes \ket{r}_R \to \ket{s, x, \sigma, u \oplus f(s,x;r)}_Q \otimes \ket{r}_R\]
        \item Uncompute item \ref{hyb-6-step:query-RO}.\label{hyb-6-step:uncompute-RO-query}
        \item \textcolor{red}{\textbf{Cache 2}: Repeat step \ref{hyb-6-step:cache-1}.}\label{hyb-6-step:cache-2}
    \end{enumerate}
    \item Uncompute step \ref{hyb-6-step:check-equality-of-signatures}.\label{hyb-6-step:check-equality-of-signatures-2}
    \item Output register $Q$.
\end{enumerate}

\begin{lemma}
    Hybrids 5 and 6 are perfectly indistinguishable.
\end{lemma}
\begin{proof}
    We can change hybrid 5 into hybrid 6 without changing the function that the hybrid computes.

    First, the $\mathsf{Cache}_{s,x,\sigma}$ operation at the start of hybrid 5 step \ref{hyb-5-step:query-RO} is now written as its own step in hybrid 6 step \ref{hyb-6-step:cache-1}. Additionally, the $\mathsf{Cache}_{s,x,\sigma}^\dag$ operation at the end of hybrid 5 step \ref{hyb-5-step:uncompute-RO-query} is written as its own step in hybrid 6 step \ref{hyb-6-step:cache-2}. Note that $\mathsf{Cache}_{s,x,\sigma} = \mathsf{Cache}_{s,x,\sigma}^\dag$ (\cref{thm:cache-is-involution}), so the two chache steps of hybrid 6 (steps \ref{hyb-6-step:cache-1} and \ref{hyb-6-step:cache-2}) compute the same function. 

    Second, in hybrid 5, the $\mathsf{Cache}_{s,x,\sigma}^\dag$ operation at the end of step \ref{hyb-5-step:query-RO} cancels out with the $\mathsf{Cache}_{s,x,\sigma}$ operation at the start of step \ref{hyb-5-step:uncompute-RO-query}. Note that $\mathsf{Cache}_{s,x,\sigma}$ commutes with the intervening step \ref{hyb-5-step:answer-query} because they act on disjoint registers. $\mathsf{Cache}_{s,x,\sigma}$ acts on registers $C \times D$ whereas step \ref{hyb-5-step:answer-query} acts on registers $Q \times R$.
    
    Third, the operation 
    \[\Decomp_{s, x} \circ \CO'' \circ \Decomp_{s, x}\]
    in hybrid 5 steps \ref{hyb-5-step:query-RO} and \ref{hyb-5-step:uncompute-RO-query} is replaced by a query to the compressed oracle $H$ in hybrid 6, steps \ref{hyb-6-step:query-RO} and \ref{hyb-6-step:uncompute-RO-query}. This is just a notational change because $\Decomp_{s, x} \circ \CO'' \circ \Decomp_{s, x}$ are anyways the operation that the compressed oracle query performs.
    
    These changes transform the hybrid 5 oracle into the hybrid 6 oracle without changing the function that the oracle computes.
\end{proof}

% The next lemma says that whenever the SEQ check is executed, $C \times D$ can be merged into a splittable database.
% \begin{lemma}\label{thm:database-is-splittable}
%     On any oracle query in hybrid 6, at the start of the SEQ check (step \ref{hyb-6-step:check-equality-of-signatures}), $C \times D$ are in the span of states of the form $\ket{c'}\ket{d'}$, where $(c',d') = \mathsf{Split}(\emptyset, d)$, and $d$ is a splittable database.
% \end{lemma}
% \begin{proof}
    
% \end{proof}

\paragraph{Hybrid 7.} Change the SEQ check to only act on $D$ and not $C$.

\begin{itemize}
    \item The oracle has a database register $D$ implementing a compressed random oracle $H: \cS_\secp \times \cX \to \cR$. The oracle also has a cache register $C$, and control qubit $B$. $B \times C \times D$ are initialized to $\ket{0, \emptyset, \emptyset}$.
\end{itemize}
\begin{enumerate}
    \item \textbf{Inputs:} The oracle takes a query register $Q = Q_s \times Q_x \times Q_\sigma \times Q_u$ with eigenstates of the form $\ket{s, x, \sigma, u}$ where $(s, x, \sigma, u) \in \cS_\secp \times \cX \times \Sigma_\secp \times \cY$.
    \item Check that $\SigToken.\Ver(\pk, s, x, \sigma) = 1$. If so, then continue. If not, then abort and output register $Q$.\label{hyb-7-step:verify-signature}
    \item \textbf{SEQ Check:} \textcolor{red}{Check that $D$ contains no entry $(s', x', r')$ such that $s' = s$ but $x' \neq x$. If the check passes, then flip $B$ to $\ket{1}$.} \label{hyb-7-step:check-equality-of-signatures}
    \item If $B = \ket{1}$:
    \begin{enumerate}
        \item \label{hyb-7-step:cache-1}\textbf{Cache 1}: Check if $D$ contains an entry $(s', x', r')$ such that $s' = s$ and $x' \neq x \oplus 1$. If so, then do the following:
        \begin{enumerate}
            \item If $C$ contains the entry $(s, x, \sigma)$, then remove this entry from $C$.
            \item Else, add $(s, x, \sigma)$ to $C$.
        \end{enumerate}
        \item Query $H$ on $\ket{s,x}_{Q_s \times Q_x} \otimes \ket{0}_R \otimes \ket{d}_{D}$.\label{hyb-7-step:query-RO}
        \item Compute $y = f(s,x;r)$, and CNOT $y$ onto $Q_u$:
            \[\ket{s, x, \sigma, u}_Q \otimes \ket{r}_R \to \ket{s, x, \sigma, u \oplus f(s,x;r)}_Q \otimes \ket{r}_R\]
        \item Uncompute item \ref{hyb-7-step:query-RO}.\label{hyb-7-step:uncompute-RO-query}
        \item \textbf{Cache 2}: Repeat step \ref{hyb-7-step:cache-1}.\label{hyb-7-step:cache-2}
    \end{enumerate}
    \item Uncompute step \ref{hyb-7-step:check-equality-of-signatures}.\label{hyb-7-step:check-equality-of-signatures-2}
    \item Output register $Q$.
\end{enumerate}

\begin{lemma}
    Hybrids 6 and 7 are computationally indistinguishable.
\end{lemma}
\begin{proof}
    The main difference between the hybrids is that the SEQ check in hybrid 6 depends on $(s',x', \sigma')$, whereas the SEQ check in hybrid 7 only depends on $(s',x')$. We can transform the SEQ check in hybrid 6 into the SEQ check in hybrid 7 in a way that gives a distinguisher negligible distinguishing advantage. 
    
    First, change the check in hybrid 6 step \ref{hyb-6-step:bitflip} to the following:
    \begin{equation}\label{eq:new-bitflip}
    \begin{split}
        \text{Check that $D$ contains no entry $(s', x', \sigma', r')$ such that}\\ 
        \text{$s' = s$ but \textcolor{red}{$x' \neq x$}. If the check passes, then flip $B$ to $\ket{1}$.}
    \end{split}
    \end{equation}
    Now we check that $x' \neq x$ instead of checking that $(x',\sigma') \neq (x,\sigma)$. 
    
    Second, any PPT distinguisher has negligible advantage at distinguishing an oracle with the old check from an oracle with a new check. The only time that the two checks give different results is when $Q$ contains $(s,x,\sigma, u)$ and $D$ contains $(s',x',\sigma', r')$ such that $s' = s$, $x' = x$, but $\sigma' \neq \sigma$. In this case, we could measure the $Q$ and $D$ registers to obtain an answer $(x,x',\sigma,\sigma',s)$ that breaks the strong unforgeability of the OSS scheme. The formal proof of this follows the same argument as the proof of \cref{thm:hyb-2-3-indistinguishable}.

    Third, the SEQ check with the change in \cref{eq:new-bitflip} is the following:
    \begin{itemize}
        \item \textbf{SEQ Check:} 
    \begin{enumerate}
        \item Apply $U_\mathsf{Split}^\dag$ to $C \times D$.
        \item Check that $D$ contains no entry $(s', x', \sigma', r')$ such that $s' = s$ but $x' \neq x$. If the check passes, then flip $B$ to $\ket{1}$.
        \item Apply $U_\mathsf{Split}$ to $C \times D$.
    \end{enumerate}
    \end{itemize}

    We claim that this SEQ check is equivalent to the SEQ check from hybrid 7:
    \begin{itemize}
        \item \textbf{SEQ Check:} Check that $D$ contains no entry $(s', x', r')$ such that $s' = s$ but $x' \neq x$. If the check passes, then flip $B$ to $\ket{1}$.
    \end{itemize}

    This is because the check in \cref{eq:new-bitflip} only depends on the $(s',x')$ values in $D$, and $\mathsf{Split}$ does not change the $(s',x')$ values in $D$. If the database is $d$ is splittable, then $\mathsf{Split}(\emptyset, d)$ removes the $\sigma'$ value from each entry $(s',x',\sigma',r')$ of $d$. If $d$ is not splittable, then $\mathsf{Split}(\emptyset, d)$ leaves $d$ unchanged. In either case $\mathsf{Split}(\emptyset, d)$ leaves the $(s', x')$ values of each entry of $d$ unchanged. Therefore, we can remove the $U_\mathsf{Split}^\dag$ and $U_\mathsf{Split}$ operations from the SEQ check. \bhaskar{Make this argument more formal if there's time.}
    
    % First, in hybrid 6, at the start of the SEQ check, $C \times D$ are in the span of states of the form $\ket{c'}\ket{d'}$, where $(c',d') = \mathsf{Split}(\emptyset, d)$, and $d$ is a splittable database. \cref{thm:database-is-splittable}. 
    
    % Second, the SEQ check in hybrid 6 is equivalent to applying a bitflip to $B$ controlled on the value of $d$. This is because $U_\mathsf{Split}^\dag$ maps each computational-basis eigenstate of $C \times D$ to another computational-basis eigenstate. Then step \ref{hyb-6-step:bitflip} applies a bitflip to $B$ controlled on the computational basis value of $C \times D$. Finally, $U_\mathsf{Split}$ inverts $U_\mathsf{Split}^\dag$, mapping $C \times D$ back to its original computational-basis value.
    
    % More precisely, the SEQ check in hybrid 6 applies a bitflip to $B$ if $d$ contains no entry $(s')$ contain values $\ket{c'}\ket{d'}$ such that $(c',d') $
    
    % If $d$ contains no entry $(s',x',\sigma',r')$
    % \begin{align*}
    %     \ket{0}_B \otimes \ket{c'}_C \otimes \ket{d'}_D &\overset{U_\mathsf{Split}^\dag}{\longrightarrow} \ket{0}\ket{\emptyset} \ket{d}\\
    %     &\overset{\text{Step \ref{hyb-6-step:bitflip}}}{\longrightarrow} \ket{1}\ket{\emptyset} \ket{d}
    % \end{align*}
    
    % let $\Pi'_{s,x,\sigma}$ project the $C \times D$ registers onto states that do not break the strong unforgeability of the signature scheme. That is to say, $\Pi'_{s,x,\sigma}$ projects onto all eigenstates $\ket{c'}_C \otimes \ket{d'}_D$ such that 
    % \bhaskar{Make this argument more formal if there's time.}
\end{proof}

\paragraph{Hybrid 8.} Move the cache 1 and cache 2 steps to occur regardless of the value of $B$.
\begin{itemize} 
    \item The oracle has a database register $D$ implementing a compressed random oracle $H: \cS_\secp \times \cX \to \cR$. The oracle also has a cache register $C$, and control qubit $B$. $B \times C \times D$ are initialized to $\ket{0, \emptyset, \emptyset}$.
\end{itemize}
\begin{enumerate}
    \item \textbf{Inputs:} The oracle takes a query register $Q = Q_s \times Q_x \times Q_\sigma \times Q_u$ with eigenstates of the form $\ket{s, x, \sigma, u}$ where $(s, x, \sigma, u) \in \cS_\secp \times \cX \times \Sigma_\secp \times \cY$.
    \item Check that $\SigToken.\Ver(\pk, s, x, \sigma) = 1$. If so, then continue. If not, then abort and output register $Q$.\label{hyb-8-step:verify-signature}
    \item \textbf{SEQ Check:} Check that $D$ contains no entry $(s', x', r')$ such that $s' = s$ but $x' \neq x$. If the check passes, then flip $B$ to $\ket{1}$. \label{hyb-8-step:check-equality-of-signatures}
    \item \label{hyb-8-step:cache-1}\textcolor{red}{\textbf{Cache 1}: Check if $D$ contains an entry $(s', x', r')$ such that $s' = s$ and $x' \neq x \oplus 1$. If so, then do the following:
        \begin{enumerate}
            \item If $C$ contains the entry $(s, x, \sigma)$, then remove this entry from $C$.
            \item Else, add $(s, x, \sigma)$ to $C$.
        \end{enumerate}}
    \item If $B = \ket{1}$:
    \begin{enumerate}
        \item Query $H$ on $\ket{s,x}_{Q_s \times Q_x} \otimes \ket{0}_R \otimes \ket{d}_{D}$.\label{hyb-8-step:query-RO}
        \item Compute $y = f(s,x;r)$, and CNOT $y$ onto $Q_u$:
            \[\ket{s, x, \sigma, u}_Q \otimes \ket{r}_R \to \ket{s, x, \sigma, u \oplus f(s,x;r)}_Q \otimes \ket{r}_R\]
        \item Uncompute item \ref{hyb-8-step:query-RO}.\label{hyb-8-step:uncompute-RO-query}
    \end{enumerate}
    \item \textcolor{red}{\textbf{Cache 2}: Repeat step \ref{hyb-8-step:cache-1}.}\label{hyb-8-step:cache-2}
    \item Uncompute step \ref{hyb-8-step:check-equality-of-signatures}. \label{hyb-8-step:check-equality-of-signatures-2}
    \item Output register $Q$.
\end{enumerate}

\begin{lemma}
    Hybrids 7 and 8 are perfectly indistinguishable.
\end{lemma}
\begin{proof}
    The only difference between the hybrids is where the cache 1 and cache 2 steps occur. In hybrid 7, they only occur if $B = \ket{1}$. In hybrid 8, they occur regardless of $B$'s value.

    First, consider the case where $B = \ket{1}$ at the end of the SEQ check. Then the two hybrids act the same becuse they both apply cache 1 and cache 2.

    Second, consider the case where $B = \ket{0}$ at the end of the SEQ check. Then the two hybrids still compute equivalent functions. Hybrid 7 does not apply either cache 1 or cache 2, whereas hybrid 8 applies cache 1 followed immediately by cache 2. Note that Cache 1 and cache 2 are inverses because the cache operation is its own inverse (\cref{thm:cache-is-involution}). Therefore, applying the cache operation twice is equivalent to not applying it at all.

    We've shown that regardless of the value of $B$, hybrids 7 and 8 compute equivalent functions.
\end{proof}

\paragraph{Hybrid 9.} Change the order of the cache 1 step and the SEQ check.

\begin{itemize}
    \item The oracle has a database register $D$ implementing a compressed random oracle $H: \cS_\secp \times \cX \to \cR$. The oracle also has a cache register $C$, and control qubit $B$. $B \times C \times D$ are initialized to $\ket{0, \emptyset, \emptyset}$.
\end{itemize}
\begin{enumerate}
    \item \textbf{Inputs:} The oracle takes a query register $Q = Q_s \times Q_x \times Q_\sigma \times Q_u$ with eigenstates of the form $\ket{s, x, \sigma, u}$ where $(s, x, \sigma, u) \in \cS_\secp \times \cX \times \Sigma_\secp \times \cY$.
    \item Check that $\SigToken.\Ver(\pk, s, x, \sigma) = 1$. If so, then continue. If not, then abort and output register $Q$.\label{hyb-9-step:verify-signature}
    \item \label{hyb-9-step:cache-1}\textcolor{red}{\textbf{Cache 1:}} Check if $D$ contains an entry $(s', x', r')$ such that $s' = s$ and $x' \neq x \oplus 1$. If so, then do the following:
        \begin{enumerate}
            \item If $C$ contains the entry $(s, x, \sigma)$, then remove this entry from $C$.
            \item Else, add $(s, x, \sigma)$ to $C$.
        \end{enumerate}
    \item \textcolor{red}{\textbf{SEQ Check:}} Check that $D$ contains no entry $(s', x', r')$ such that $s' = s$ but $x' \neq x$. If the check passes, then flip $B$ to $\ket{1}$. \label{hyb-9-step:check-equality-of-signatures}
    \item If $B = \ket{1}$: \label{hyb-9-step:operation-if-B-equals-1}
    \begin{enumerate}
        \item Query $H$ on $\ket{s,x}_{Q_s \times Q_x} \otimes \ket{0}_R \otimes \ket{d}_{D}$.\label{hyb-9-step:query-RO}
        \item Compute $y = f(s,x;r)$, and CNOT $y$ onto $Q_u$:
            \[\ket{s, x, \sigma, u}_Q \otimes \ket{r}_R \to \ket{s, x, \sigma, u \oplus f(s,x;r)}_Q \otimes \ket{r}_R\]
        \item Uncompute item \ref{hyb-9-step:query-RO}.\label{hyb-9-step:uncompute-RO-query}
    \end{enumerate}
    \item \textcolor{red}{\textbf{SEQ Check:}} Uncompute step \ref{hyb-9-step:check-equality-of-signatures}.\label{hyb-9-step:check-equality-of-signatures-2}
    \item \textcolor{red}{\textbf{Cache 2:}} Repeat step \ref{hyb-9-step:cache-1}.\label{hyb-9-step:cache-2}
    \item Output register $Q$.
\end{enumerate}

\begin{lemma}
    Hybrids 8 and 9 are perfectly indistinguishable.
\end{lemma}
\begin{proof}
    The difference between the hybrids is the order in which they perform the cache and SEQ check operations. 
    
    However, the cache operation commputes with the SEQ check. Cache applies a unitary to $C$ controlled on the computational basis value of $Q \times D$. The SEQ check applies a unitary to $B$ controlled on the computational basis value of $Q \times D$. Such operations commute.

    Therefore, hybrids 8 and 9 compute equivalent functions.
\end{proof}

\paragraph{Hybrid 10.} Replace steps \ref{hyb-9-step:check-equality-of-signatures} - \ref{hyb-9-step:check-equality-of-signatures-2} from hybrid 9 with a call to the SEQ oracle.

\begin{itemize}
    \item \textcolor{red}{The oracle stores an instantiation of the SEQ oracle $O^\SEQ_f$ (\cref{fig:seq-oracle-with-serial-numbers}), which stores a database register $D$. The oracle also has a cache register $C$. $C \times D$ are initialized to $\ket{\emptyset, \emptyset}$}.
\end{itemize}
\begin{enumerate}
    \item \textbf{Inputs:} The oracle takes a query register $Q = Q_s \times Q_x \times Q_\sigma \times Q_u$ with eigenstates of the form $\ket{s, x, \sigma, u}$ where $(s, x, \sigma, u) \in \cS_\secp \times \cX \times \Sigma_\secp \times \cY$.
    \item Check that $\SigToken.\Ver(\pk, s, x, \sigma) = 1$. If so, then continue. If not, then abort and output register $Q$.\label{hyb-10-step:verify-signature}
    \item \label{hyb-10-step:cache-1}\textbf{Cache 1}: Check if $D$ contains an entry $(s', x', r')$ such that $s' = s$ and $x' \neq x \oplus 1$. If so, then do the following:
        \begin{enumerate}
            \item If $C$ contains the entry $(s, x, \sigma)$, then remove this entry from $C$.\label{hyb-10-step:add-to-C}
            \item Else, add $(s, x, \sigma)$ to $C$.\label{hyb-10-step:remove-from-C}
        \end{enumerate}
    \item \label{hyb-10-step:SEQ-query}\textcolor{red}{Initialize $\ket{+}_B$. Query $O^\SEQ_f$ on $\ket{s, x, u}_{Q_s \times Q_x \times Q_u} \otimes \ket{+}_B$.}
    \item \textbf{Cache 2}: Repeat step \ref{hyb-10-step:cache-1}.\label{hyb-10-step:cache-2}
    \item Output register $Q$.
\end{enumerate}

\begin{lemma}
    Hybrids 9 and 10 are perfectly indistinguishable.
\end{lemma}
\begin{proof}
    The differences between the hybrids are as follows. Hybrid 9 stores a database $D$ and includes an SEQ check (step \ref{hyb-9-step:check-equality-of-signatures}), a procedure for computing the answer (step \ref{hyb-9-step:operation-if-B-equals-1}), and an uncomputation of the SEQ check (step \ref{hyb-9-step:check-equality-of-signatures-2}). Hybrid 10 has query access to the SEQ oracle, which stores the database $D$ internally. Furthermore, steps \ref{hyb-9-step:check-equality-of-signatures} - \ref{hyb-9-step:check-equality-of-signatures-2} of hybrid 9 are replaced with a query to the SEQ oracle in step \ref{hyb-10-step:SEQ-query} of hybrid 10. Note that hybrid 10 queries the SEQ oracle on $B = \ket{+}$. That way, the state of $B$ is unchanged by the bitflip operation $\mathsf{X}$.

    The two hybrids are equivalent because the SEQ oracle stores the database $D$ internally, and the query to the SEQ oracle computes the same function as steps \ref{hyb-9-step:check-equality-of-signatures} - \ref{hyb-9-step:check-equality-of-signatures-2} of hybrid 9.
\end{proof}

\paragraph{Hybrid 11.} Modify the cache 1 and cache 2 steps to make queries to $O_f^\SEQ$ rather than operating on $D$ directly.

\begin{itemize}
    \item The oracle stores an instantiation of the SEQ oracle $O^\SEQ_f$ (\cref{fig:seq-oracle-with-serial-numbers}), which stores a database register $D$. The oracle also has a cache register $C$. $C \times D$ are initialized to $\ket{\emptyset,\emptyset}$.
\end{itemize}
\begin{enumerate}
    \item \textbf{Inputs:} The oracle takes a query register $Q = Q_s \times Q_x \times Q_\sigma \times Q_u$ with eigenstates of the form $\ket{s, x, \sigma, u}$ where $(s, x, \sigma, u) \in \cS_\secp \times \cX \times \Sigma_\secp \times \cY$.
    \item Check that $\SigToken.\Ver(\pk, s, x, \sigma) = 1$. If so, then continue. If not, then abort and output register $Q$.\label{hyb-11-step:verify-signature}
    \item \textbf{Cache 1}: \label{hyb-11-step:cache-1}
        \textcolor{red}{\begin{enumerate}
            \item Prepare $\ket{s, x \oplus 1, 0, 0}$ on an internal register $Q'$, and query $O^\SEQ_f$ on $Q'$. Let $\ket{s, x \oplus 1, y, b}$ be the final state of $Q'$. \label{hyb-11-step:query-SEQ-oracle}
            \item If $b = 0$, then do the following: \label{hyb-11-step:CNOT}
            \begin{enumerate}
                \item If $C$ contains entry $(s, x, \sigma)$, then remove this entry from $C$.
                \item Else, add $(s, x, \sigma)$ to $C$.
            \end{enumerate}
            \item Uncompute \cref{hyb-11-step:query-SEQ-oracle}.\label{hyb-11-step:Uncompute}
        \end{enumerate}
        }
    \item \label{hyb-11-step:SEQ-query}Initialize $\ket{+}_B$. Query $O^\SEQ_f$ on $\ket{s, x, u}_{Q_s \times Q_x \times Q_u} \otimes \ket{+}_B$.
    \item \textbf{Cache 2}: Repeat step \ref{hyb-11-step:cache-1}.\label{hyb-11-step:cache-2}
    \item Output register $Q$.
\end{enumerate}

\begin{lemma}
    Hybrids 10 and 11 are perfectly indistinguishable.
\end{lemma}
\begin{proof}
    The difference between the hybrids is how they implement cache 1 (and cache 2). Let us trace the operation of cache 1 in both hybrids and show that the implementations compute the same function. 
    
    First, the $C \times D$ registers have computational-basis eigenstates of the form $\ket{c}_C \otimes \ket{d}_D$. It suffices to consider the operation of cache 1 on each eigenstate $\ket{c} \otimes \ket{d}$.

    Second, the cache 1 step of hybrid 10 checks the following condition:
    \begin{equation}\label{eq:cache-condition}
    \begin{split}
        &\text{$d$ contains an entry $(s', x', r')$}\\
        &\text{such that $s' = s$ and $x' \neq x \oplus 1$}
    \end{split}    
    \end{equation}

    If and only if condition \ref{eq:cache-condition} is satisfied, then hybrid 10's cache 1 operation will execute steps \ref{hyb-10-step:add-to-C} and \ref{hyb-10-step:remove-from-C}.

    Third, let us consider hybrid 11's cache 1 operation. Step \ref{hyb-11-step:query-SEQ-oracle} queries $O_f^\SEQ$ on $\ket{s, x \oplus 1, 0, 0}$ to obtain $\ket{s,x \oplus 1,y, b}$. We claim that $b = 0$ if and only if condition \ref{eq:cache-condition} is satisfied. This is because $O_f^\SEQ$ checks in step \ref{step:seq-oracle-SEQ-check} whether condition \ref{eq:cache-condition} is satisfied. If the condition is satisfied, then $O_f^\SEQ$ outputs the query register unchanged, so $b = 0$. If the condition is not satisfied, then $O_f^\SEQ$ answers the query and applies a bitflip to $b$, resulting in $b = 1$.
    
    Fourth, step \ref{hyb-11-step:CNOT} of hybrid 11 will execute the same operation as hybrid 10's steps \ref{hyb-10-step:add-to-C} and \ref{hyb-10-step:remove-from-C}, if and only if $b=0$. This step does not change the state of the $D$ register.

    Fifth, step \ref{hyb-11-step:Uncompute} of hybrid 11 uncomputes step \ref{hyb-11-step:query-SEQ-oracle}. This returns the $Q'$ register to its initial state. 
    
    In total, the cache 1 operation of hybrid 11 checks whether condition \ref{eq:cache-condition} is satisfied, and if so, it modifies $C$ according to steps \ref{hyb-10-step:add-to-C} and \ref{hyb-10-step:remove-from-C} of hybrid 10. This shows that hybrids 10 and 11 implement equivalent cache 1 operations, so the two hybrids are equivalent.
 \end{proof}

Now we can finish the proof of security. Note that the oracle distribution in hybrid 1 is the real-world distribution of $O_{f,G,\pk}$ from program $P_f$ (\cref{fig:generalized-otp-compiler-construction}). Additionally, the oracle distribution in hybrid 11 is the ideal-world distribution of $O_\Sim^{O_f^\SEQ}$ implemented by $\Sim$ (\cref{fig:SEQ-simulator}). 

Next, we have shown that hybrid 1 and hybrid 11 are computationally indistinguishable. This means that any QPT distinguisher given $(\ket{T}, s, \pk, f, \aux_f)$ and oracle access to an oracle sampled from either hybrid 1 or hybrid 11 will have $\negl(\secp)$ advantage at distinguishing these two cases. This is exactly what the distinguisher in the definition of security (\cref{def:generalized-otp-compiler}) is asked to distinguish.

In summary, we have constructed a QPT simulator $\Sim^{O_f^\SEQ}(1^\secp, \pk, \aux_f)$ (\cref{fig:SEQ-simulator}) such that for all $f \in \cF$ and all QPT distinguishers $D$ that are given $(\ket{T}, s, \pk, f, \aux_f)$ and either $O_{f, G, \pk}$ or $O_\Sim^{O_f^\SEQ}(1^\secp, \pk, \aux_f)$, $D$ has $\negl(\secp)$ distinguishing advantage. This completes the proof of security. \fi
\section{Query-Limited RAM Programs}\label{sec:query-limited-RAM-programs}

A useful ability for programs is the ability to maintain and update a state. In a RAM program, the program may modify the memory contents as it is evaluated, resulting in a sequence of memory states $\st_0, \st_1, \dots$. In this section, we show how to limit queries to a RAM program so that any adversarial evaluator can only access the program on a program state that is consistent with the set of queries so far.

% RAM programs allow a natural extension of one time programs to \emph{query-limited programs}.\jiahui{We have removed the original section on query-limited programs.Are we keeping this name?}
As a motivating example, suppose that, instead of purchasing a single query, a customer purchased some budget $\$B$ worth of queries. A given query might cost more or less, depending on how complicated it is. The merchant can give out a program that limits the customer's queries to a total cost of $\$B$.

More formally, the merchant can give out a RAM program whose internal state is the total cost $\$C$ of all queries made so far (this can be much smaller than the list of all queries). When a query is made that costs $\$c_q$, the program updates the state to $\$(C + c_q)$. However, once the internal state goes above $\$B$, the program only ever outputs a helpful message ``Budget reached. Please contact us to buy more.''
Since the evaluator is limited to program states resulting from a consistent evaluation, they cannot evaluate a sequence of queries with budget exceeding $\$B$.

\subsection{Definition}

% \justin{RAM programs subsume QLPs by putting $\StateUpdate$, $\mathsf{State}$, and $\Ver$ into the program itself. This formulation also lets the program use $r$ to decide to output $\bot$ (which can still result in $r$ being recorded, but I think this allows a more graceful interpretation of that).}

\paragraph{RAM Programs.} A randomized RAM program consists of a function $P:\cM\times \cX \times \cR \rightarrow \cM\times \cY$ and an initial state $\st_1$. $P$ takes as input a memory state $\st_i$, an input $x$, and randomess $r$, then outputs an updated memory state $\st_{i+1}$ and the output $y$:
\[
    P:(\st_i, x; r) \mapsto (\st_{i+1}, y)
\]

\paragraph{Correctness and Intended Usage.} A RAM program is intended to be used as follows. Starting from $\st_1 = \bot$ and some initial input $x_1$, sample randomness $r_1$ and compute $(\st_{2}, y_1) = P(\st_1, x_1; r_1)$.\bhaskar{Perhaps change the indexing to start at $1$. For example, change $(x_0, r_0, y_0) \to (x_1, r_1, y_1)$. Later notation seems to start the indexing at $1$.} \justin{Good call.} Then, choose a new input $x_2$ and repeat the process: sample $r_2$, then compute $(\st_{3}, y_2) = P(\st_2, x_2; r_2)$. This continues until the user is done evaluating the RAM program. As a simplification, the user can sample their randomness $r_i = H(i\concat x_i)$ using a random oracle $H$. Including $i$ in the input to $H$, as opposed to $\st_i$, ensures that fresh randomness is used for every query, even if the state loops.
% We emphasize that 

\jiahui{I think we are still using LEQ and REQ oracles, QLP and RAM programs interchangeably}
\paragraph{RAM Programs in the Effective Query Model.} Next, we give the formal definition of a RAM program in the effective query model in \Cref{fig:REQ}, which we call the LEQ oracle. 
At a high level, the LEQ oracle maintains a coherent list of evaluations made so far and the program states associated with them. To enforce a consistent history of the computation so far, it only allows query in the form of a stack - the last $q$'th query must be uncomputed to access the $q-1$'th query.

\ifllncs
\begingroup
% Preserve the paragraph settings and font used by LNCS for figure contents.
\csname @parboxrestore\endcsname
\normalfont\small
% Keep the entire oracle in one frame that can continue across a page break.
\begin{mdframed}[nobreak=false,everyline=false]
\else
\begingroup
% Preserve the article figure font while allowing the frame to flow with the text.
\csname @parboxrestore\endcsname
\normalfont\normalsize
\begin{mdframed}[nobreak=false,everyline=false]
\fi
% \justin{This is a version of the REQ oracle which is closer to the updated SEQ oracle from our best-possible paper}
    \paragraph{Registers.} The oracle acts on the following registers:
    \begin{itemize}
        \item $\cQ$: The query register taken as input. $\cQ$ has standard basis states of the form $\ket{t, x, u, b}_\cQ$. $t$ denotes the tag, $x$ denotes the input, $u$ is the register where the output will be written, and $b$ serves as an indicator for whether the query was answered.
        
        \item $\cS$: A register containing the current state $\st$ of the program. $\cS$ is initialized to $\ket{\st_1}$.
        
        \item $\cA = \cA_1\otimes \dots \otimes \cA_k$: A variable-length register containing ancillas for the evaluations. Initially, $\cA$ has length $0$.
        
        \item $\cG = \cG_1 \otimes \dots \otimes \cG_k$: A variable-length register where each $\cG_i$ contains a bit $g_i$ indicating whether the oracle is ``open'' (the last operation on tag $i$ was $U_P$) or ``closed'' (the last operation on tag $i$ was $U_P^\dagger$) for tag $i$. Initially, $\cG$ has length $0$.
        
        \item $\cB$: A work register to validate the query. $\cB$ is initialized to $\ket{0}$.
    \end{itemize}

    \paragraph{Sub-Routine.} The oracle makes use of a unitary $U_P$ that starts from fresh ancillas $\ket{\mathbf{0}}_{\cA_i}$, samples randomness $r$, and computes $P$ on its input. Concretely, for any $x\in \cX$ and $u\in \cY$, the unitary $U_P$ maps
    \[
         \ket{x, u}_{\cQ} \otimes \ket{\st}_{\cS} \otimes \ket{\mathbf{0}}_{\cA_i} 
         \leftrightarrow
         \frac{1}{\sqrt{|\cR}|}\sum_{r} \ket{x, u \oplus y_r}_{\cQ} \otimes \ket{\st'_{x,r}}_{\cS} \otimes \ket{\st, x, r}_{\cA_i}
    \]
    where $P(\st, x; r) = (\st'_{x,r}, y_r)$. Furthermore, $U_P$ acts as the identity on orthogonal states.
    
    \paragraph{Queries.} To respond to a query, the oracle acts on $\cQ \times \cS \times \cA \times \cG \times \cB$ as the following unitary. Given some basis state $\ket{t, x, u, b}_\cQ \otimes \ket{\st}_{\cS} \otimes \ket{a}_\cA \otimes \ket{g}_{\cG} \otimes \ket{0}_{\cB}$, the oracle does the following:
    % To respond to a query contained in register $\cQ$, the oracle acts on $\cQ \times \cD \times \cG \times \cW \times \cB$ as the following unitary. Given some basis state $\ket{t, x, u, b, c}_\cQ \otimes \ket{D}_\cD \otimes \ket{g}_\cG \otimes \ket{0,0}_\cW \otimes \ket{0}_{\cB}$, the oracle does the following:
    \begin{enumerate}
        \item \label{REQ-step:expand-registers}\textbf{Expand $\cG$ and $\cA$ by One Register.} Append $\ket{0}_{\cG_{k+1}}$ to $\cG$, and append $\ket{\mathbf{0}}_{\cA_{k+1}}$ to $\cA$.
        \item \label{REQ-step:validate-query}\textbf{Validate Query.} Let $k_{\max}$ be the maximum index $i \in [k]$ such that $g_i = 1$, or if no such $i$ exists, then let $k_{\max} = 0$.

        If $t \in \{k_{\max}, k_{\max}+1\}$, then
        % $t =  \MaxEntry(D) \land x = x_{\MaxEntry(D)}$ or if $t = \MaxEntry(D) + 1 \land g_{t-1} = 1$, 
        apply $\mathsf{X}$ to register $\cB$ to obtain $\ket{1}_\cB$.
        \label{step:REQ-valid-query}

        \item \textbf{Answer Query.} \label{step:REQ-answer-query}
        If $\cB$ contains $1$, then answer the query as follows.
        \begin{enumerate}
            \item \textbf{Close $U_P$ if $g_t$ is Currently ``Open''.} \label{step:REQ-close}
            If $g_t = 1$, then apply $U_P^\dagger$ to registers $\cQ \otimes \cS \otimes \cA_t$. 

            \item \textbf{Switch Mode if Returned to Original State.} \label{step:REQ-switch-mode}
            If register $\cA_t$ contains $\ket{\mathbf{0}}$, then flip $g_t$ and $b$ by applying $\mathsf{X}\otimes \mathsf{X}$ to $\cG_t \times \cQ_b$. 
            In other words, apply
            \[
                \ketbra{\mathbf{0}}_{\cA_t} \otimes \mathsf{X}_{\cG_t}\otimes \mathsf{X}_{\cQ_b} + \left(I - \ketbra{\mathbf{0}}\right)_{\cA_t} \otimes I_{\cG_t} \otimes I_{\cQ_b}
            \]
            to registers $\cA_t \times \cG_t \times \cQ_b$.

            \item \textbf{Open $U_P$ if $g_t$ is Set to ``Open''.} \label{step:REQ-open}
            If $g_t = 1$, apply $U_P$ to registers $\cQ \otimes \cS \otimes \cA_t$. 
        \end{enumerate}

        \item \label{step:REQ-uncompute-valid-query}Uncompute step \ref{step:REQ-valid-query}.
    \end{enumerate}
\end{mdframed}
\ifllncs
% Set the caption type only after the frame, so mdframed remains breakable.
\makeatletter
\def\@captype{figure}
\makeatother
\else
% Set the caption type after the breakable frame in article mode as well.
\captionsetup{type=figure}
\fi
\caption{The Limited Effective Query (LEQ) Oracle $O^{\LEQ}_{P}$ for RAM Programs}\label{fig:REQ}
\ifllncs
\endgroup
\else
\endgroup
\fi

\vspace{3mm}
A useful fact about the LEQ oracle is that it enforces a consistent history of the computation so far. In particular, the ancilla registers contain a superposition over ordered sequences of queries, randomnesses, and resulting states (entangled with the evaluator's state). This follows from the fact that the oracle restricts access to only the top layer of queries.

% \justin{This version is for the updated SEQ oracle.}
\begin{lemma}
    Let $A$ be an oracle algorithm interacting with the REQ oracle $O_P^{\REQ}$ from \Cref{fig:REQ}.
    After every query in this interaction, the mode register $\cG$ is supported on strings $g = 1^{k_{\max}}\concat 0^*$ with an all-ones prefix for some $k_{\max}$.
    Furthermore, the joint ancilla register $\cA_1 \otimes \dots \otimes \cA_{k_{\max}}$ and current state register $\cS$ are supported only on computational basis states
    \[
        \ket{(\st_1, x_1, r_1), \dots, (\st_{k_{\max}}, x_{k_{\max}}, r_{k_{\max}})}_{\cA_{[k_{\max}]}} \otimes \ket{\st_{k_{\max}+1}}_\cS
    \]
    for which there exist outputs $y_1,\dots,y_{k_{\max}+1}$ satisfying
    \[
        P(\st_i, x_i; r_i) = (\st_{i+1}, y_{i}) \quad \forall i \in [k_{\max}].
    \]
\end{lemma}
\begin{proof}
    We proceed by induction over the sequence of oracle queries. This clearly holds when the oracle is initialized. Now fix an interaction step and consider any computational basis state in the support of $\cQ \otimes \cS \otimes \cA \otimes \cG \otimes \cB$ immediately before the oracle call.
    Let $k_{\max}$ be the maximum index with $g_{k_{\max}}=1$ (or $k_{\max}=0$ if no such index exists).
    By the validation step, any answered query has tag $t\in\{k_{\max},k_{\max}+1\}$.

    We first isolate the net effect of Step~\ref{step:REQ-answer-query} on the registers associated with tag $t$. 
    Consider any state where the ancillas are $\ket{\mathbf{0}}$ in step \ref{step:REQ-switch-mode}
    Observe that step \ref{step:REQ-answer-query} acts as the identity on any states where the ancillas are orthogonal to $\ket{\mathbf{0}}$ in step \ref{step:REQ-switch-mode}, since it either applies $I$ or $U_P^\dagger U_P = I$, depending on $g_t$. Otherwise, we can explicitly compute the behavior by pre-and-post-applying $U_P$ controlled on $g_t$. Thus, step \ref{step:REQ-answer-query} has the overall effect
    \[
         \ket{x, u}_{\cQ} \otimes \ket{\st}_{\cS} \otimes \ket{\mathbf{0}}_{\cA_t} \otimes \ket{0}_{\cG_t}
         \leftrightarrow
         \frac{1}{\sqrt{|\cR}|}\sum_{r} \ket{x, u \oplus y_r}_{\cQ} \otimes \ket{\st'_r}_{\cS} \otimes \ket{\st, x, r}_{\cA_t} \otimes \ket{1}_{\cG_t}
    \]
    for every $(x, u) \in \cX\times \cY$ and acts as the identity otherwise, where $P(\st, x;r) = (\st'_r, y_r)$.

    Therefore $\cA_{t-1, t} \otimes \cG_{t-1, t} \otimes \cS$ satisfy the requirements of the lemma. Finally, the fact that $t\in\{k_{\max},k_{\max}+1\}$ implies that 1) all indices before or after $t$ satisfy the requirements by the inductive hypothesis 2) all indices $t'>t$ are $g_{t'} = 1$, and 3) all indices after  $t'<t$ are $g_{t'} = 0$.
\end{proof}

\begin{definition}[Query-Limited RAM Program (QLP) Compiler]\label{def:RAM-program-compiler}
Let $\cP = \{P:\cM \times \cX \times \cR \rightarrow \cM \times \cY\}_{P}$ be a family of RAM programs. A \textbf{query-limited RAM program compiler} for $\cP$ comprises the efficient algorithms $(\PGen, \Eval)$ and has the following syntax, correctness, and security guarantees.

\begin{itemize}
    \item \textbf{Syntax:}  
    \begin{itemize}
        \item $\PGen(1^\secp, P) \to \tilde{P}$: 
        Take a security parameter $\secp$ and a functionality $P \in \cP$ then outputs a program $\tilde{P}$ that implements $P$.
        
        \item $\Eval(\tilde{P}, x) \to (y, \tilde{P}')$: 
        Takes a program $\tilde{P}$ and an input $x$, and evaluates the program on $x$ to produce output $y$ and a new program state $\tilde{P}'$. 
        % $\Eval$ may modify the program state.
    \end{itemize}
    
    \item \textbf{Correctness:} 
    For every $P\in \cP$, every sufficiently large $\secp$, and every polynomial-length (i.e. $n = \mathsf{poly}(\secp)$) query sequence $(x_1, \dots, x_n) \in \cX^*$, the following distributions are statistically close:
    \[
        \left\{ (y_1, \dots, y_n): \begin{array}{c}
             (r_1, \dots, r_n) \gets \cR^{n}  
             \\
             (\st_1, y_1) \gets P(\st_0, x_1; r_1) 
             \\
             \dots
             \\
             (\st_n, y_n) \gets P(\st_{n-1}, x_n; r_n)
        \end{array}\right\}
        \approx_s
        \left\{ 
        (y_1, \dots, y_n)
        :
        \begin{array}{c}
            \tilde{P}_0 \gets \PGen(1^\secp, P)
            \\
            (y_1, \tilde{P}_1) \gets \Eval(\tilde{P}_0, x_1) 
            \\
            \dots 
            \\
            (y_n, \tilde{P}_n) \gets \Eval(\tilde{P}_{n-1}, x_n)
        \end{array}
        \right\}
    \]
    where $\st_0 = \bot$.

    % \justin{Old below here.}

    % Let us be given a functionality $(f, \Ver) \in \cF$, a security parameter $\secp \in \bbN$, a number of queries $q \in \bbN$, and a sequence of $q$ distinct queries $(x_1, \dots, x_q) \subseteq \cX^q$.  In the real world, let us compute $P^{(0)} = \PGen(1^\secp, f, \Ver)$, and for each $i \in [q]$, let us compute 
    % \[\Eval(P^{(i-1)}, x_i) \to (y_i, b_i)\] 
    % and set $P^{i}$ to be the state of the program after the $i$-th query. In the ideal world, let us initialize the oracle $O^{LEQ}_{f, \Ver}$ and then for each $i \in [q]$, compute
    % \[\ket{x_i, 0, 0} \overset{O^{LEQ}_{f, \Ver}}{\longrightarrow} \ket{x_i, y'_i, b'_i}\]
    % The compiler is correct if for any $(f, \Ver) \in \cF$, there is a negligible function $\negl(\secp)$ such that for any $\secp, q, x_1, \dots, x_q$, the distributions of $[(y_1, b_1), \dots, (y_q, b_q)]$ and $[(y'_1, b'_1), \dots, (y'_q, b'_q)]$ are $\negl(\secp)$-close in statistical distance.

    \item \textbf{Security:} 
    The scheme satisfies LEQ security if there exists a QPT simulator $\cS$ such that for any randomized RAM program \jiahui{randomized RAM or QLP?} $P \in \cP$ and any distinguisher $D$ with (quantum) auxiliary input $\aux_{P, \cP}$ (which may depend non-uniformly on $P$ and $\cP$), there exists a negligible function $\negl(\secp)$ such that for all $\secp \in \bbN$,
    \[
        \abs{\Pr\left[D(1^\secp, \PGen(1^\secp, P), \aux_{P, \cP}) \to 1\right] 
        - 
        \Pr\left[D\left(1^\secp, \cS^{O^{\REQ}_{P}}(1^\secp), \aux_{P, \cP})\right) \to 1\right]} 
        \leq \negl(\secp)
    \]
\end{itemize}
\end{definition}

% \subsection{Properties of the REQ Oracle}\label{sec:REQ-props}

\subsection{Construction}

Here we construct a RAM program compiler (\cref{def:RAM-program-compiler}).
We use the following tools:
\begin{itemize}
    \item A one-shot program compiler (\cref{def:osp-compiler}) $\OSP.(\KeyGen, \PGen, \TokenGen, \Eval)$ with serial number space $\cS_\OSP$.
    
    %\proofauditloc{P06}{ The one-shot program compiler$\OSP.(\KeyGen,\PGen,\TokenGen,\Eval)$ of\Cref{theorem:OSP_from_OSS}, obtained from \Cref{fig:generalized-otp-compiler-construction} by instantiating$\SigToken$ with a one-shot signature scheme, with serial number space$\cS_\OSP$.}

    \item A blind-unforgeable signature scheme $\mathsf{Sig}.(\KeyGen, \Sign, \Ver)$ with signature space $\Sigma_\Sig$.
    \item A secret-key encryption scheme $\SKE.(\KeyGen, \Enc, \Dec)$ with ciphertext space $\cC_\SKE$ satisfying qIND-qCPA security (\Cref{def:qIND-qCPA}).
    \item A random oracle 
    $G_\Sign: \cS_\OSP \times \cC_\SKE \times \{0,1\}^{2\secp} \rightarrow \{0,1\}^{\secp}$.
    \item A random oracle $G': \left(\cX \times \cC_\SKE \times \{0,1\}^{\secp} \times \Sigma_\Sig \times \cS_\OSP \right) \times \{0,1\}^{\secp}
    \rightarrow 
    \{0,1\}^{2\secp}$.
    % \item A random oracle $G': \left(\cX \times \{0,1\}^{\secp} \times \Sigma_\Sig \times \cS_\OSP \right) \times \{0,1\}^{\secp}
    % \rightarrow 
    % \{0,1\}^{\secp}$.
\end{itemize}
The random oracles $G_\Sign$ and $G'$ can be replaced by psuedorandom functions. We present the construction as if they were oracles to slightly simplify the proof of security by removing an indistinguishable transition from PRF to random oracle.

\begin{figure}[t]
    \begin{mdframed}
            \begin{enumerate}
            \itemindent=-15pt
                \item[] \textbf{Hardcoded:} 
                % $(\sk, \pk, P)$ %, 
                $(\sk, \pk, \sk_\Enc, P)$.
                \item[] \textbf{Input:} $(x, \tilde{\st}, r_s, \sigma, s'; s; r)$. 
                $x$ is the chosen input, 
                % $\st$ is the current program state
                $\tilde{\st}$ is an encryption of the current program state, 
                $r_s$ is an arbitrary string, $\sigma$ is a signature, $s$ is the evaluation serial number for the one-shot program, $s'$ is an arbitrary serial number, and $r$ is the sampled randomness.
                \\
                For ease of notation, we denote $\mathsf{inp} = (x, \tilde{\st}, r_s, \sigma, s')$.
                % \proofauditloc{P05}{Are we supposed to use $\tilde{\st}$ in $\inp$? the input is an authenticated ciphertext, not the plaintext. The notations are a little confusing}
                \itemindent=0pt
                \item Compute $\Sig.\Ver(\pk, s\concat\st\concat r_s, \sigma)$. If the signature is valid, continue. Otherwise output $\bot$.
                \item Decrypt $\st \gets \SKE.\Dec(\sk_\Enc, \tilde{\st}; r_\Enc)$.\label{step:REQ-construction-decrypt-state}

                \item Parse $r = r_x\concat r_{s'}'$. 
                \label{step:REQ-construction-after-decrypt-state}
                \item Evaluate $(\st', y) \gets P(\st, x; r_x)$. 
                
                \item Compute randomness for signing 
                % $r_\Sign = G'(s\concat \inp)$.
                and encrypting $r_\Enc\concat r_\Sign = G'(s\concat \inp)$.
                \item Encrypt $\tilde{\st'} \gets \SKE.\Enc(\sk_\Enc, \st';\ r_\Enc)$.\label{step:REQ-construction-encrypt-next-state}
                \item Compute the following two values:\label{step:REQ-construction-sign-next-serial-and-state}
                \begin{align*}
                    r_{s'} &= r_{s'}'\concat G_\Sign\left(s'\big\concat \tilde{\st'}\big\concat r_{s'}'\right)
                    \\
                    % \sigma' &= \Sig.\Sign\left(\sk,\ s'\big\concat \st' \big\concat r_{s'};\ r_{\Sign}\right).
                    \sigma' &= \Sig.\Sign\left(\sk,\ s'\big\concat \tilde{\st'}\big\concat r_{s'};\ r_{\Sign}\right).
                \end{align*}
                \item Output $\left(y, (\tilde{\st'}, r_{s'}, \sigma')\right)$.
            \end{enumerate}
            \end{mdframed}
    \caption{Classical Sub-Program $P_\OSP$ for the Query-Limited RAM Compiler}
    \label{fig:P_OSP}
\end{figure}

\begin{figure}[t]
\begin{mdframed}
    % \begin{itemize}
    %     \item 
        \paragraph{$\PGen(1^\secp, P)$:}
        \begin{enumerate}
            \item Sample a signature key pair $(\sk, \pk)\gets \Sig.\KeyGen(1^\secp)$.
            % \item Sample a SKE encryption key pair $(\sk_\Enc, \sk_\Enc) \gets \SKE.\KeyGen(1^\secp)$.
            \item Generate a one-shot program common reference string $\crs \gets \OSP.\KeyGen(1^\secp)$.\footnote{Alternatively, the program may use a pre-existing OSP common reference string, since the common reference string is independent of the OSP functionality.}%\footnotemark[1]
            \item Generate a one-shot program $\tilde{P}_{\OSP} \gets \OSP.\PGen(\crs, P_\OSP)$ where $P_\OSP$ is the classical randomized functionality defined in \Cref{fig:P_OSP}.
            
            \item Generate a one-shot program token $(s_1, \ket{T_1}) \gets \OSP.\TokenGen(\crs)$.\footnote{If interaction is permissible, $s$ may be taken externally to make all communication classical.}
            \item Encrypt $\tilde{\st_1} \gets \SKE.\Enc(\sk_\Enc, \bot)$.
            \justin{SKE. Check other points.}
            \item Sign $\sigma_1 \gets \Sig.\Sign(\sk, s_0\concat \tilde{\st_1}\concat 0^{2\secp})$.
            \item Output $\tilde{P} = \left(\crs, \tilde{P}_{\OSP}, (\tilde{\st_1}, 0^{2\secp}, \sigma_1), s_1, \ket{T_1}\right)$.
        \end{enumerate}

        % \item 
        \paragraph{$\Eval(P_f, x)$:}
        \begin{enumerate}
            % \item Parse $P_f = \left(\crs, \tilde{P}_{\OSP}, (\tilde{\st}, r_s, \sigma), s, \ket{T}\right)$.
            \item Parse $P_f = \left(\crs, \tilde{P}_{\OSP}, (\st, r_s, \sigma), s, \ket{T}\right)$.
            \item Generate $(s', \ket{T'}) \gets \OSP.\TokenGen(\crs)$.
            % \item Evaluate $\left(y, (\tilde{\st'}, r_{s'}, \sigma')\right) \gets \OSP.\Eval(\tilde{P}_\OSP, s, (x,\tilde{\st}, r_s),\sigma, s'), \ket{T})$.
            \item Evaluate $\left(y, (\st', r_{s'}, \sigma')\right) \gets \OSP.\Eval(\tilde{P}_\OSP, s, (x,\st, r_s),\sigma, s'), \ket{T})$.
            % \item Update $\tilde{P}$ to $(\crs, \tilde{P}_{\OSP}, (\tilde{\st'}, r_{s'}, \sigma'), \ket{T'})$.
            \item Update $\tilde{P}$ to $(\crs, \tilde{P}_{\OSP}, (\st', r_{s'}, \sigma'), \ket{T'})$.
            \item Output $y$.
        \end{enumerate}
    % \end{itemize}
    % \footnotetext[1]{Alternatively, the program may use a pre-existing OSP common reference string, since the common reference string is independent of the OSP functionality.}
\end{mdframed}
\caption{Query-Limited RAM Program Compiler Construction}\label{fig:RAM-program-compiler-construction}
\end{figure}

\begin{theorem} \label{thm:QLP_main_them}
    The construction in \cref{fig:RAM-program-compiler-construction} is query-limited RAM program compiler (\Cref{def:RAM-program-compiler}) for any family of randomized RAM programs $\cF$.
\end{theorem}
\begin{proof}
    We need to show correctness and simulation security.
    \ifllncs
     \paragraph{Correctness} 
     Due to the limitation of space, we leave the proof of correctness to section~\ref{sec:LEQ-security}.
     \else
\jiahui{I think the following revised proof is a more rigorous proof. Our original proof already grasps the ideas and is simply not clear enough. Either keeping the original proof or use the new proof should be fine}
\justin{I think the original is fine. The main issue chatgpt is pointing out is that we don't explicitly state anywhere that the OSS program is correct on repeated adaptive queries.}
\jiahui{Agreed. Keeping the original proof}
    % BEGIN PROOF AUDIT P06 (original line 442)
    \paragraph{Correctness.}
    Consider the set of serial numbers and tokens generated for any polynomial number of evaluations. By correctness and union bound, none of the generated pairs are $(\bot, \bot)$, except with negligible probability. Conditioned on this, the output of every query $(s; (x, \tilde{\st}, r_s), \sigma, s'))$ made to the one-shot program $\tilde{P}_\OSP$ is distributed statistically close to $P_\OSP((x, \tilde{\st}, r_s), \sigma, s'); s; r)$ as defined in \Cref{fig:RAM-program-compiler-construction}, for uniform $r$ and a polynomial number of queries. 
    Now consider directly querying $P_\OSP$. Observe that if $\sigma$ is a valid signature on $s\concat \tilde{\st}\concat r_s$
    % \proofauditloc{P05}{The signature authenticates $s\concat\tilde{\st}\concat r_s$, not the decrypted state.}
    % $s\concat \tilde{\st}\concat r_s$, and $\tilde{\st}$ is an encryption of some state $\st$, 
    then $P_\OSP$ computes $(\st', y) \gets P(\st, x; r_x)$, for uniform $r_x$, and outputs $y$, an encryption $\tilde{\st'}$ of $\st'$, and a signature $\sigma'$ on $s'\concat \tilde{\st'}\concat r_{s'}$ along with a string $r_{s'}$. Starting from the initial state $\st_0 = \bot$
    % $\tilde{\st} = \Enc(0)$ 
    and a valid $\sigma$, the sequence of $y_1, \dots, y_n$ resulting from directly querying $P_\OSP$ a total of $n=\mathsf{poly}(\secp)$ is distributed as 
    \[
        \left\{ (y_1, \dots, y_n): \begin{array}{c}
             (r_1, \dots, r_n) \gets \cR^{n}  
             \\
             (\st_1, y_1) \gets P(\st_0, x_1; r_1) 
             \\
             \dots
             \\
             (\st_n, y_n) \gets P(\st_{n-1}, x_n; r_n)
        \end{array}\right\}
    \]
    % When the first query $(s_q; (x_1, \tilde{\st}_1, r_{s_1}), \sigma_1, s_2))$ is made, $P_\OSP$ 
    % \justin{Will revisit correctness after the OSP correctness is updated. Just need to show that each OSP query is correctly distributed with overwhelming probability and union bound.}
     \fi
    
    \paragraph{Security.}
   
    By the security of the one-shot program compiler, the OSP $\tilde{P}_\OSP$ can be indistinguishably replaced by the OSP simulator with oracle access to the OSP SEQ functionality $O_\SEQ^{P_{\OSP}}$ (\Cref{fig:seq-oracle-with-serial-numbers}) for the program $P_{\OSP}$ given in \Cref{fig:RAM-program-compiler-construction}. 
    Thus, it suffices to show that $O_\SEQ^{P_\OSP}$ can be indistinguishably simulated using oracle access to $O_\REQ^{P}$.

    % We show how to simulate
    We first introduce a hybrid oracle $O_\Hyb^{P}$ in \Cref{fig:REQ-hybrid-oracle} and show that it can be used to simulate the one-shot-program SEQ oracle $O_\SEQ^{P_{\OSP}}$ (\Cref{fig:seq-oracle-with-serial-numbers}) in \Cref{prop:REQ-sim-SEQ-to-hyb} (\Cref{sec:LEQ:proof:simulating-seq}).
    Then, we show how to use $O_\REQ^{P}$ to simulate oracle access to $O_\Hyb^{P}$ in \Cref{prop:REQ-hyb-to-REQ} (\Cref{sec:LEQ:proof:simulating-hybrid}).
\ifllncs
    The full proof of security is in section~\ref{sec:LEQ-security}.
    \else \fi
\end{proof}

\ifsubmit \else
Informally, transitioning to $O_\Hyb^{P}$ achieves two important steps in preparation for the final transition to $O_\REQ^{P}$. 
First, it moves the core functionality -- evaluating $P(\st_{t-1},x;r_1)$ and tracking the new $\st_{t}$ -- into the SEQ oracle while stripping out the cryptographic tools $P_\OSP$ uses to enforce good behavior. 
Second, it enforces the contiguous history of the computation by restricting the adversary's queries to either a new query or repeating the most recent query. Thus, the adversary cannot attempt to erase the oracle's memory of old queries without first erasing the newer ones.
\fi

\ifllncs
  % In LNCS mode, main.tex includes this proof in the appendix.
\else
  \ifsubmit
\section{Proof of Security for Query Limited RAM Programs}\label{sec:LEQ-security}
In this section, we prove the correctness and security of \Cref{fig:RAM-program-compiler-construction}.
\else
\subsection{Proof of Security}\label{sec:LEQ-security}
\fi

\ifllncs
\jiahui{I think the following revised proof is a more rigorous proof. Our original proof already grasps the ideas and is simply not clear enough. Either keeping the original proof or use the new proof should be fine}
\justin{I think the original is fine. The main issue chatgpt is pointing out is that we don't explicitly state anywhere that the OSS program is correct on repeated adaptive queries.}
\jiahui{Agreed. Keeping the original proof}
    % BEGIN PROOF AUDIT P06 (original line 442)
    \paragraph{Correctness.}
    Consider the set of serial numbers and tokens generated for any polynomial number of evaluations. By correctness and union bound, none of the generated pairs are $(\bot, \bot)$, except with negligible probability. Conditioned on this, the output of every query $(s; (x, \tilde{\st}, r_s), \sigma, s'))$ made to the one-shot program $\tilde{P}_\OSP$ is distributed statistically close to $P_\OSP((x, \tilde{\st}, r_s), \sigma, s'); s; r)$ as defined in \Cref{fig:RAM-program-compiler-construction}, for uniform $r$ and a polynomial number of queries. 
    Now consider directly querying $P_\OSP$. Observe that if $\sigma$ is a valid signature on $s\concat \tilde{\st}\concat r_s$
    % \proofauditloc{P05}{The signature authenticates $s\concat\tilde{\st}\concat r_s$, not the decrypted state.}
    % $s\concat \tilde{\st}\concat r_s$, and $\tilde{\st}$ is an encryption of some state $\st$, 
    then $P_\OSP$ computes $(\st', y) \gets P(\st, x; r_x)$, for uniform $r_x$, and outputs $y$, an encryption $\tilde{\st'}$ of $\st'$, and a signature $\sigma'$ on $s'\concat \tilde{\st'}\concat r_{s'}$ along with a string $r_{s'}$. Starting from the initial state $\st_0 = \bot$
    % $\tilde{\st} = \Enc(0)$ 
    and a valid $\sigma$, the sequence of $y_1, \dots, y_n$ resulting from directly querying $P_\OSP$ a total of $n=\mathsf{poly}(\secp)$ is distributed as 
    \[
        \left\{ (y_1, \dots, y_n): \begin{array}{c}
             (r_1, \dots, r_n) \gets \cR^{n}  
             \\
             (\st_1, y_1) \gets P(\st_0, x_1; r_1) 
             \\
             \dots
             \\
             (\st_n, y_n) \gets P(\st_{n-1}, x_n; r_n)
        \end{array}\right\}
    \]
    % When the first query $(s_q; (x_1, \tilde{\st}_1, r_{s_1}), \sigma_1, s_2))$ is made, $P_\OSP$ 
    % \justin{Will revisit correctness after the OSP correctness is updated. Just need to show that each OSP query is correctly distributed with overwhelming probability and union bound.}

    \else
    \fi

   \ifllncs
 \paragraph{Security.}

    By the security of the one-shot program compiler, the OSP $\tilde{P}_\OSP$ can be indistinguishably replaced by the OSP simulator with oracle access to the OSP SEQ functionality $O_\SEQ^{P_{\OSP}}$ (\Cref{fig:seq-oracle-with-serial-numbers}) for the program $P_{\OSP}$ given in \Cref{fig:RAM-program-compiler-construction}. 
    Thus, it suffices to show that $O_\SEQ^{P_\OSP}$ can be indistinguishably simulated using oracle access to $O_\REQ^{P}$.

    % We show how to simulate
    We first introduce a hybrid oracle $O_\Hyb^{P}$ in \Cref{fig:REQ-hybrid-oracle} and show that it can be used to simulate the one-shot-program SEQ oracle $O_\SEQ^{P_{\OSP}}$ (\Cref{fig:seq-oracle-with-serial-numbers}) in \Cref{prop:REQ-sim-SEQ-to-hyb} (\Cref{sec:LEQ:proof:simulating-seq}).
    Then, we show how to use $O_\REQ^{P}$ to simulate oracle access to $O_\Hyb^{P}$ in \Cref{prop:REQ-hyb-to-REQ} (\Cref{sec:LEQ:proof:simulating-hybrid}).
       \else
       \fi
    
In this section, we complete the proof of security by defining the hybrid oracle $O_\Hyb^{P}$ (\cref{fig:REQ-hybrid-oracle}), showing how to simulate oracle access to $O_\SEQ^{P_{\OSP}}$ using $O_\Hyb^{P}$ (\Cref{prop:REQ-sim-SEQ-to-hyb}) and in turn how to simulate oracle access to $O_\Hyb^{P}$ using $O^\REQ_{P}$ (\Cref{prop:REQ-hyb-to-REQ}).

\begin{figure}[H]
\begin{mdframed}
    \paragraph{Registers.} The oracle acts on the following registers:
    \begin{itemize}
        \item $\cQ$: The query register taken as input. $\cQ$ has standard basis states of the form $\ket{t, x, u, b}_\cQ$. $t$ denotes the tag, $x$ denotes the input, $u$ is the register where the output will be written, and $b$ serves as an indicator for whether the query was answered.
        \item $\cD$: An internal register that represents the database of a compressed oracle $H:[2^\secp] \times \cX \rightarrow \cR \times \{0,1\}^\secp$. $\cD$ is initialized to $\ket{\emptyset}$.%\bhaskar{Should this say $[2^\secp]$ instead of $[2^n]$?} \justin{In general $2^n$ could be any bound we set on the total number of queries, but it's fine to use $\secp$.}
        \item $\cW$: An internal register that is used to query the compressed oracle. $\cW$ is initialized to $\ket{0,0}$.
        \item $\cB$: An internal register to validate the query. $\cB$ is initialized to $\ket{0}$.
    \end{itemize}
    % They are initialized to $\ket{\emptyset}_\cD \otimes \ket{0,0}_\cW \otimes \ket{0}_{\cB}$. 
    % The oracle also takes a query register $\cQ$ with classical eigenstates of the form $\ket{s, x,u,b}_\cQ$.
    % \\

    % It uses $\cD$ to maintain a compressed oracle $H:[2^n] \times \cX \rightarrow \cR \times \{0,1\}^\secp$.\\

    \paragraph{Sub-Routine.} The oracle uses a subroutine $\StateD_{t}(D)$ that computes the state of the RAM program at the start of the $t$-th query. $\StateD_{t}(D)$ is parameterized by an index $t \in \bbN$ and takes as input a database $D = \{((i, x_{i}), r_{i,x}\concat r_{i,\Sign})\}_{i \in \{1, \dots, |D|\}}$. 
    % \begin{enumerate}
        % \item If $|D| < t-1$, then $\StateD_t(D)$ returns $\textsf{``too short''}$.
        % \item Otherwise (if $|D| \geq t-1$), then 
        $\StateD_{t}(D)$ computes the following values
        \begin{align*}
            \st_1 &= \bot\\
            (\st_{i+1}, y_i) &= P(\st_i, x_i; r_{i, x}), \quad \forall i \in [t-1]
        \end{align*}
        and returns $\st_{t}$.
    % \end{enumerate}
    % $\MaxEntry(D)$ is the maximum over $i$ such that $(i, x, r) \in D$.
    
    \paragraph{Queries.} To respond to a query, the oracle acts on $\cQ \times \cD \times \cW \times \cB$ as the following unitary. Given some basis state $\ket{t, x,u,b}_\cQ \otimes \ket{D}_\cD \otimes \ket{0,0}_\cW \otimes \ket{0}_{\cB}$, the oracle does the following:
    \begin{enumerate}
        \item \label{step:REQ-hybrid-valid-query}\textbf{Validate Query.} If either $t =  |D| \land x = x_{|D|}$ or if $t = |D| + 1$, then apply $\mathsf{X}$ to register $\cB$ to obtain $\ket{1}_\cB$.%\bhaskar{Is $\MaxEntry$ the same as $|D|-1$?} \justin{Effectively yes, but technically they are not the same. They differ if $D$ has multiple $i$ entries or if $D$ has an $i$ entry greater than its length, both of which will never occur in this oracle.}
        % If $\mode = 0$ and if either $k=0$ or $x_k \neq x$, return immediately. Otherwise, apply $X$ to register $\cB$ to flip the bit.
        \item \textbf{Answer Query.} If $\cB$ contains $1$, then answer the query as follows.%\bhaskar{Describe how to handle a database that is too short.}% If $\StateD_{t}(D) = \textsf{``too short''}$, then set $(y, r_2) = (0,0)$.
        \begin{enumerate}
            \item \label{step:REQ-hybrid-query-QRO}CNOT $(t, x)$ from $\cQ$ to $\cW$. Then query $\ket{(t, x), 0}_{\cW}$ to the compressed oracle to obtain a superposition over $\ket{(t, x), r}$.
            % Write $\ket{(t, x), 0}_{\cW}$ to $\cW$ using controlled NOT operations.
            % \label{step:REQ-prep-work}
            % \item Apply the compressed oracle to $\cW \otimes \cD$. The state on $\cW$ is now a superposition over states of the form $\ket{(t, x), r}$, where we may parse $r = r_1\concat r_2$.
            % \item Let $\st_{t} = \StateD_{t}(D)$ and let $(\st_{t+1}, y) = P_\OSP(\st_{t}, x;r_1)$.\footnote{This is a syntactic step for the sake of exposition. The computation can be absorbed into the next step.}
            % \item Using CNOTs, write the value $(y, r_2)$ to the sub-register of $\cQ$ containing $\ket{u}$ and flip $\ket{b}$:
            \item \label{step:REQ-hybrid-compute-y} Let $\st_{t} = \StateD_{t}(D)$ and parse $r = (r_1, r_2) \in \cR \times \bit^\secp$. Then compute $(\st_{t+1}, y) = P(\st_{t}, x;r_1)$.\footnote{This is a syntactic step for the sake of exposition. The computation can be absorbed into the next step.}
            \item \label{step:REQ-hybrid-copy-output}Using CNOTs, write the value $(y, r_2)$ to the sub-register of $\cQ$ containing $\ket{u}$ and flip $\ket{b}$:
            % Write, via a controlled NOT, the value $P(\st_{k-1+\mode}, x;r)$ to 
            % Compute the output as
            \begin{gather*}
                \ket{t, x, u, b}_\cQ \otimes \ket{(t, x), r}_{\cW} \\
                \mapsto \\
                \ket{t, x, u\oplus (y, r_2), b\oplus 1}_\cQ \otimes \ket{(t, x), r}_{\cW}
            \end{gather*}
            % In other words, flip $b$ and write $(P(\st_{k-1+\mode}, x;r_1), r_2)$ to the sub-register containing $u$ using controlled NOTs.
            \item \label{step:REQ-hybrid-uncompute-RO-query}Uncompute step \ref{step:REQ-hybrid-query-QRO}.%, then step \ref{step:REQ-prep-work}.
        \end{enumerate}
        \item \label{step:REQ-hybrid-uncompute-validate-query}Uncompute step \ref{step:REQ-hybrid-valid-query}.
    \end{enumerate}
\end{mdframed}
\caption{A Hybrid REQ Oracle $O_\Hyb^P$}\label{fig:REQ-hybrid-oracle}
\end{figure}

\subsubsection{Ordering the Database}\label{sec:LEQ-proof-ordering}

Before showing how to simulate $O_{\SEQ}^{P_{\OSP}}$ using $O_{\Hyb}^{P}$, we first establish a property of the SEQ oracle's database which will be useful for establishing a total ordering of the queries recorded in its database. 
% Intuitively, the claim says that for every encrypted program state $\tilde{\st_i}$ which is recorded by the SEQ oracle, either $\st_i = \st_0$ or the SEQ oracle \emph{also} records an encryption $\tilde{\st}_{i-1}$ of a previous program state.
Intuitively, the claim states that after interacting with the SEQ oracle for $P_{\OSP}$, it cannot sign any messages $s\concat \tilde{\st}\concat *$ which are inconsistent with either the starting state $(s_0, \bot)$ or some state $(s_{j}, \st_j)$ which is recorded by the SEQ oracle. 

In particular, this holds even if the adversary gets to see a measured copy of the SEQ oracle's database after interacting with it.
Since every entry in the SEQ database includes a valid signature on some $s\concat \tilde{\st}\concat *$, it must either use the first serial number $s_0$ or have a precursor entry also recorded in the SEQ database. 
This allows us to iteratively order the SEQ database starting from $s_0$ as
% \begin{align*}
%     ({\color{red}s_0},\ &(x_0, \tilde{\bot}, *, 0^{\secp}\concat 0^\secp, {\color{blue}s_1}), r_{0,x}\concat {\color{blue}r'_{1}})
%     \\
%     ({\color{blue}s_1},\ &(x_1, \tilde{\st_1}, *, {\color{blue}r'_{1}}\concat *, {\color{olive}s_2}), r_{1,x}\concat r'_{2})
%     \\
%     \dots
%     \\
%     (s_{i},\ &(x_{i}, \tilde{\st_i}, *, r'_{i}\concat *, s_{i+1}), r_{i,x}\concat r'_{i+1})
% \end{align*}
\[
    \begin{array}{rcl}
    ({\color{red}s_0}, 
    &(x_0, \tilde{\bot}, *, {\color{red}0^{\secp}}\concat 0^\secp, {\color{blue}s_1}), 
    &r_{0,x}\concat {\color{blue}r'_{1}})
    \\
    ({\color{blue}s_1}, 
    &(x_1, \tilde{\st_1}, *, {\color{blue}r'_{1}}\concat *, {\color{olive}s_2}), 
    &r_{1,x}\concat {\color{olive}r'_{2}})
    \\
    &\dots
    \\
    (s_{i}, 
    &(x_{i}, \tilde{\st_i}, *, r'_{i}\concat *, s_{i+1}), 
    &r_{i,x}\concat r'_{i+1})
    \end{array}
\]

\begin{claim}[Dynamically Constrained Signatures]
    \label{claim:REQ-proof-sEQ-ordering}
    Let $A^{(\cdot)}$ be a QPT oracle algorithm, let $A_2$ be a QPT algorithm, and let $P$ be a RAM program. Consider the following experiment $\mathsf{DC\text{-}Unforg}(A^{(\cdot)}, A_2, P, q, \secp)$, parameterized by a query number $q \in \bbN$ and a security parameter $\secp\in \bbN$.
    \begin{enumerate}
        \item Sample $(\sk, \pk) \gets\Sig.\KeyGen(1^\secp)$ and $\sk_\Enc \gets \SKE.\KeyGen(1^\secp)$.
        \item Sample $(s_0, \ket{T_0})\gets \OSP.\TokenGen(\crs)$, encrypt $\tilde{\st_0} = \SKE.\Enc(\sk_\Enc, \bot)$ and sign $\sigma_0 \gets \Sig.\Sign(\sk, s_0\concat \tilde{\st_0}\concat 0)$.
        \item Run $A^{(\cdot)}(s_0, \ket{T_0}, \tilde{\st_0}, \sigma_0)$ with oracle access to $O_{\SEQ}^{P_{\OSP}[\sk, \pk, \sk_\Enc]}$ until $q$ queries are made, where $(\sk, \pk, \sk_\Enc)$ are hard-coded into $P_\OSP$. Let register $\cA$ contain $A$'s state at the end of this step.
        \item Measure the SEQ oracle's compressed database in the computational basis to get $D_q$.
        \item Run $A_2(\cA, D_q)$ to obtain $\sigma'$ and $m$.
        \item Output ``$\mathsf{lose}$'' if $\Sig.\Ver(\pk, m, \sigma')$ rejects or if $m$ can be parsed as $m = s\concat \tilde{\st}\concat r_s$ satisfying either of the following two conditions.
        \begin{enumerate}
            \item $s = s_0$ and $\SKE.\Dec(\sk_\Enc, \tilde{\st}) = \bot$
            \item or there exists an entry $(*, (x_{j}, \tilde{\st}_{j}, *, *, s)), r_{j,x}\concat r'_s) \in D_q$ such that the program states $\st_j = \SKE.\Dec(\sk_\Enc, \tilde{\st_j})$ and $\st_{j+1} = \SKE.\Dec(\sk_\Enc, \tilde{\st})$ satisfy
            \[
                \begin{aligned}
                    &P(\st_j, x_j; r_{j,x}) = (\st_{j+1}, *)
                \\
                \text{and}\quad & r_s = r_s'\concat *
                \end{aligned}
            \]
            \label{step:REQ-proof-SEQ-ordering-condition-complicated}
        \end{enumerate} 
        Otherwise, output ``$\mathsf{win}$''.
    \end{enumerate}
    Assuming $\Sig$ is blind-unforgeable, then for all sufficiently large $\secp \in \bbN$, every $q = \mathsf{poly}(\secp)$, every QPT oracle algorithm $A^{(\cdot)}$, every QPT $A_2$, and every RAM program $P$,
    \[
        \Pr[\mathsf{win} \gets \mathsf{DC\text{-}Unforg}(A^{(\cdot)}, A_2, P, q, \secp)] = \negl(\secp)
    \]

    % $D_q$ satisfies the following: for every $((s, (x, \tilde{st}, r_s, \sigma, s')), r) \in D_q$, one of the following holds:
    % \begin{itemize}
    %     \item $s = s_0$
    %     \item or there exists an entry $(s_{i-1}, (x_{i-1}, \tilde{\st}_{i-1}, *, *, s)), r_{i-1}) \in D_q$ for some $s_{i-1}$.

    %     Futhermore, if this case occurs then $\SKE.\Dec(\sk_\Enc, \tilde{\st}_{i-1}) = \st_{i-1}$ and $\SKE.\Dec(\sk_\Enc, \tilde{\st}) = \st$ such that
    %     \[
    %         P(\st_{i-1}, x; r_{x, i-1}) = (\st, *)
    %     \]
    %     where $r_{i-1} = r_{x,i-1}\concat *$.
    % \end{itemize}
    % \justin{TODO: modify so that it includes an extra sig output from the adversary (maybe after seeing $D_q$?)}
\end{claim}

To aid in proving this claim, we introduce the following strengthening of the compressed oracle chaining lemma from \cite{GLRRV25}. Intuitively, \cite{GLRRV25}'s original version showed that if an adversary has access to a composition of random oracles $H\circ G$ and $H$'s compressed database records a value $H(y) = z$, then $G$'s compressed database records a matching value $G(x) = y$, for some $x$. The lemma below strengthens this in two ways. First, instead of directly evaluating $H$ on $G(x)$, we consider evaluating $H$ on some arbitrary function $f$ of $x$ and $G(x)$, along with half of $G(x)$.\footnote{Including half of $G(x)$ ensures that $H$'s input has a minimal dependence on $G(x)$. Otherwise, $f$ could ignore its inputs when choosing the input to $H$.} Second, it enforces that the \emph{input} $x_g$ recorded by $G$ is a preimage of the input $x_h$ recorded by $H$.

\begin{lemma}[Compressed Oracle Chaining]\label{lem:compressed-chaining}
    Let $G:\cX_G \rightarrow \cY_1\times \cY_2$ and $H:\cX_H \times \cY_2 \rightarrow \cZ$ be random oracles implemented by the compressed oracle technique.
    
    Let $f:\cX_G \times \cY_1 \times \cY_2 \rightarrow \cX_H$ be an arbitrary function. Define the function $F_f: \cX \rightarrow \cZ$ by the following computation for $F(x_g)$:
    \begin{enumerate}
        \item Compute $y_1\concat y_2 \coloneqq G(x_g)$.
        \item Compute $x_h \coloneqq f(x_g, y_1\concat y_2)$.
        \item Output $H(x_h, y_2)$.
    \end{enumerate}
    Consider running an interaction of an oracle algorithm with $F_f$ until query $t$, then measuring the internal state of $G$ and $H$ to obtain $D_G$ and $D_H$. 

    Let $E_t$ be the event that after the measurement at time $t$, for all $(x_h\concat y_2, z)\in D_H$, there exists an entry $(x_g, y_1\concat y_2) \in D_G$ such that $f(x_g, y_1\concat y_2) = x_h$.
    Then
    \[
        \Pr[E_t]
        \geq 
        1 - \frac{8t^2}{|\cY_2|}
    \]
\end{lemma}

The proof of the strengthened compressed oracle chaining lemma is similar to that of the original, but requires some careful modifications. We defer it to \Cref{sec:compressed-chaining} in favor of focusing on the direct analysis of RAM programs.

\proofaudit{P07}{Bound the joint bad event, not just its marginals}{%
The final inference in this argument needs a common joint experiment. As defined here, $E_q^*$ is a database/predecessor condition, not a subevent of successful signature output $E_q$; the two marginal bounds alone do not show $\Pr[E_q\land\neg E_q^*]$ is small. Apply the recording argument directly to a valid output with no required predecessor, keeping the measured outer database as side information, and then apply chaining to the jointly recorded witness. }
\begin{proof}[Proof of \Cref{claim:REQ-proof-sEQ-ordering}]
    % We first claim that with overwhelming probability, every entry $((s, (*, \tilde{st}, r_s, \sigma, s')), r) \in D_q$ satisfies $\sigma$ is a valid signature on $s\concat \tilde{\st} \concat r_s$.
    % \begin{itemize}
    %     \item 
    %     \item and $r_s = r_s' \concat G_\Sign(s\concat \tilde{\st} \concat r_s')$ or $s = s_0$.
    % \end{itemize}
    Observe that if $A$ submits a query $((s, (*, \tilde{st}, r_s, \sigma, *)), r)$ to the SEQ oracle where $\sigma$ were \emph{not} a valid signature on $s\concat \tilde{\st} \concat r_s$, then $P_\OSP$ deterministically outputs $\bot$. Since the SEQ oracle never records deterministic queries, $D_q$ never records queries failing this condition.
    
    Next, we show that either $s=s_0$ or $r_s = r_s' \concat G_\Sign(s\concat \tilde{\st} \concat r_s')$.
    Observe that $P_\OSP$ only ever signs messages $s\concat \tilde{\st} \concat r_s$ where $r_s = r_s' \concat G_\Sign(s\concat \tilde{\st} \concat r_s')$. Furthermore, the experiment can be internally simulated given query access to a signing oracle that only signs such messages. Therefore by the blind-unforgeability of $\Sig$, $A_2$ can only find signatures on either
    \begin{itemize}
        \item $s_0\concat \tilde{\st_0}\concat 0^{2\secp}$, which was signed as $\sigma_0$,
        \item or on $s\concat \tilde{\st} \concat r_s$ where $r_s = r_s' \concat G_\Sign(s\concat \tilde{\st} \concat r_s')$ for some $r_{s}'$.
    \end{itemize}

    Since the adversary loses in the former case, we restrict our attention to the latter. Let $E_{q}$ be the event that $A_2$ outputs a valid signature satisfying the latter after $q$ queries. Also consider the event $E^*_{q}$ where condition \ref{step:REQ-proof-SEQ-ordering-condition-complicated} holds for
    this output and the measured database $D_q$. We want to reason about $\Pr[E_q \land \neg E^*_q]$.
    \proofauditloc{P07}{$E_q^*$ does not include signature validity; it is not a subevent of $E_q$.}%

    Consider the experiment where $G_\Sign$ is implemented as a compressed oracle and let $E_{q,G_{\Sign}}$ be the event that $A_2$ successfully signs a message $s\concat \tilde{\st} \concat r_s$ where $G_\Sign$'s compressed database records an entry $(s\concat \tilde{\st}\concat r'_s, r_s'')$ where $r_s = r_s'\concat r_s''$. By \Cref{lem:zhandry_lemma5},
    \[
        \Pr[E_{q,G_\Sign}] \geq \left(\sqrt{\Pr[E_{q}]} - \sqrt{1/2^{\secp}}\right)^2
        = \Pr[E_{q}] - \negl(\secp)
    \]
    % Now consider the latter case. Let $E_{k, t}$ be the event where $k$ such tuples occur in $D_q$ when it is measured at time $t$. We want to reason about the probability of the event where each of these tuples has a recorded precursor according to the requirements of the lemma, which we call $E^*_{k,t}$, conditioned on $E_{k,t}$. Consider the experiment where $G_\Sign$ is implemented as a compressed oracle and let $E'_{k,t}$ be the event that $k$ tuples occur in $D_q$ and have matching entries in $G_\Sign$'s compressed database. By \Cref{lem:zhandry_lemma5}, for any $k=\mathsf{poly}(\secp)$,
    % \[
    %     \Pr[E'_{k,t}] \geq \left(\sqrt{\Pr[E_{k, t}]} - \sqrt{k/2^{-\secp}}\right)^2
    %     = \Pr[E_{k, t}] - \negl(\secp)
    % \]
    Furthermore, by the compressed oracle chaining lemma (\Cref{lem:compressed-chaining}), $D_q$ must have a corresponding entry $((s_{i-1}, (x_{i-1}, \tilde{\st}_{i-1}, *, *, s)), r_s'\concat *)$ satisfying condition \ref{step:REQ-proof-SEQ-ordering-condition-complicated} after any polynomial number of queries, except with negligible probability.
    \proofauditloc{P07}{Apply recording and chaining in the same joint experiment, retaining the measured outer database as side information.}%

  \iffalse  
\jiahui{Set
    $H=G_\Sign$ and let $G$ be the SEQ oracle's internal compressed
    oracle. Fix the keys and the independent oracle $G'$, so the
    function $f$ below is deterministic. For
    \[
        x_g=
        \bigl(s_{i-1},
          (x_{i-1},\tilde{\st}_{i-1},*,*,s)\bigr),
        \qquad
        G(x_g)=r_{i-1,x}\concat r_s',
    \]
    set $y_1=r_{i-1,x}$ and $y_2=r_s'$. Define
    $f(x_g,r_{i-1,x}\concat r_s')$ to be the function which
    1) decrypts $\tilde{\st}_{i-1}$,
    2) evaluates $P$ using input $x_{i-1}$ and randomness
    $r_{i-1,x}$,
    3) encrypts the new state to produce $\tilde{\st}$ using
    the coins specified by $G'$ in $P_\OSP$, and
    4) outputs $s\concat\tilde{\st}$.
    Thus the corresponding input to $H$ is
    \[
        f(x_g,r_{i-1,x}\concat r_s')\concat r_s'
        =s\concat\tilde{\st}\concat r_s'.
    \]}
    \fi

    In particular, the chaining lemma requires that the entry $(x_g, y_1\concat y_2) \in G$ matching $(x_h\concat y_2, *)\in H$ satisfies $x_h = f(x_g, G(x_g))$ and $y_2$ appears in both databases. 
    Set $H = G_\Sign$, $G$ to be the SEQ oracle's internal compressed oracle, $x_g = ((s_{i-1}, (x_{i-1}, \tilde{\st}_{i-1}, *, *, s))$, and $y_2 = r'_{s}$ where $H(x_g) = *\concat r'_{s}$. 
    Then, set $f(x_g, r_x\concat r'_{s})$ to be the function which 1) decrypts $\tilde{\st}_{i-1}$, then 2) evaluates $P$, then 3) encrypts the new state $\st$ to produce $\tilde{\st}$, and finally 4) outputs $s\concat \tilde{\st}\concat r'_{s}$. 
    Together with the correctness of $\SKE$, it must be the case that $P(\st_{i-1}, x; r_{x,i-1}) = (\st, *)$ where $\st_{i-1} = \Dec(\sk_\Enc, \tilde{\st}_{i-1})$ and $\st = \Dec(\sk_\Enc, \tilde{\st})$. Furthermore, $y_2=r'_s$ is recorded in both the input of $H$ (also known as $G_\Sign$) and the output of $G$ (also known as the SEQ oracle's internal compressed oracle). Therefore
    \[
        \Pr[E^*_{q}] \geq \Pr[E_{q,G_{\Sign}}] - \negl(\secp) \geq \Pr[E_q] - \negl(\secp).
    \]
    \proofauditloc{P07}{This marginal comparison alone does not bound $\Pr[E_q\land\neg E_q^*]$; prove a joint-event bound.}%
    Finally, $\Pr[E_{t} \land \neg E^*_{k,t}] = \negl(\secp)$ because $E^*_q$ is a sub-event of $E_q$.
    % which implies that $\Pr[E_{t} \land \neg E^*_{k,t}] = 1-\negl(\secp)$ for any $k=\mathsf{poly}(\secp)$ after any polynomial number of queries.
\end{proof}

\subsubsection{Simulating the One-Shot Program SEQ Oracle}
\label{sec:LEQ:proof:simulating-seq}

\begin{prop}\label{prop:REQ-sim-SEQ-to-hyb}
    There exists a simulator $\Sim_1$ such that for all RAM programs $P$, all QPT oracle algorithms $A^{(\cdot)}$ and all sufficiently large $\secp \in \bbN$,
    \[
        \left\{A^{O_{\SEQ}^{P_{\OSP}[\sk, \pk, \sk_\Enc]}}: \begin{array}{c}
             (\sk, \pk) \gets \Sig.\KeyGen(1^\secp) \\
             \sk_\Enc\gets \SKE.\KeyGen(1^\secp) 
        \end{array}\right\} 
        \approx_c
        \left\{A^{\Sim_1^{O_{\Hyb}^{P}}}\right\} 
    \]
    where $P_{\OSP}[\sk, \pk, \sk_\Enc]$ denotes the functionality $P_\OSP$ with $\sk$, $\pk$, and $\sk_\Enc$ hard-coded.
\end{prop}
\begin{proof}
    % The $\Reflect$ operation is specific to the SEQ oracle. It takes as input a serial number $s$ and a 
    The simulator $\Sim_1$ is given in \Cref{fig:REQ-hybrid-oracle-sim-workreg,fig:REQ-hybrid-oracle-sim}. 
    
    \begin{figure}[H]
    \begin{mdframed}
        $\Sim_1$ maintains a cache register $\cC = (\cC_{\serialindex}, \cC_{\inputcache}, \cC_{\outputcache})$ containing three key-value stores:
        \begin{itemize}
            \item $\serialindex = \{(s:i_s)\}_{s}$ is a mapping from serial numbers $s$ to indices $i_s$. It is initialized to $\{(0:s_0)\}$.
            \item $\inputcache = \{(i:\inp_i)\}_{i}$ is a mapping from indices $i$ to inputs $\inp_i = (x_i, \tilde{\st}_i, r_{s_i}, \sigma_i, s')$ for $P_\OSP$. It is initialized to $\emptyset$.
            \item $\outputcache = \{(i:\outp_i)\}$ is a mapping from indices $i$ to outputs $\outp_i = (y, (\tilde{\st'}, r_{s'}, \sigma'))$ resulting from evaluating $P_\OSP$.
        \end{itemize}

        Additionally, it uses a $\Reflect^{O_\Hyb^P}_{i}$ subroutine parameterized by an index $i$ and an input $x\in \cX$ which acts on a state $\ket{b}$ as follows.\footnote{Intuitively, this routine reflects around whether $O_\Hyb^P$ has recorded some query $(i, *)$.} Fix any $x\neq x' \in \cX$.
        % Intuitively, this routine outputs whether $O_\Hyb^P$ has recorded an entry $(i, x)$ by checking if it is willing to answer a query on a different input $x'$.
        \begin{enumerate}
            \item Prepare and query $\propto \ket{(i, x)}\otimes \sum_{u} \ket{u} \otimes \ket{0}$ to $O_\Hyb^P$ to obtain $\propto \ket{(i, x)}\otimes \sum_{u} \ket{u} \otimes \ket{b_1}$.\label{step:REQ-sim-hyb-reflect1}
            \item Prepare and query $\propto \ket{(i, x')}\otimes \sum_{u} \ket{u} \otimes \ket{b_2}$ to $O_\Hyb^P$ to obtain $\propto \ket{(i, x')}\otimes \sum_{u} \ket{u} \otimes \ket{b_2}$.\label{step:REQ-sim-hyb-reflect2}
            \item Compute $\ket{b \oplus (b_1\land b_2)}$.
            \item Uncompute steps \ref{step:REQ-sim-hyb-reflect2} and \ref{step:REQ-sim-hyb-reflect1}.
            % \item Prepare and query 
        \end{enumerate}

        % Additionally, it uses a $\Reflect^{O_\Hyb^P}_{(i, x)}$ subroutine parameterized by an index $i$ and an input $x\in \cX$ which acts on a state $\ket{b}$ as follows. 
        % Intuitively, this routine outputs whether $O_\Hyb^P$ has recorded an entry $(i, x)$ by checking if it is willing to answer a query on a different input $x'$.
        % \begin{enumerate}
        %     \item Set $x' = x\oplus 1$. Prepare and query $\propto \ket{(i, x')}\otimes \sum_{u} \ket{u} \otimes \ket{b}$ to $O_\Hyb^P$ to obtain $\propto \ket{(i, x')}\otimes \sum_{u} \ket{u} \otimes \ket{b'}$.
        %     \item Uncompute the first two registers then output $\ket{b'}$.
        %     % \item Prepare and query 
        % \end{enumerate}
        \end{mdframed}  
        \caption{Work Registers and Subroutines for $\Sim_1^{O_\Hyb^P}$}\label{fig:REQ-hybrid-oracle-sim-workreg}
    \end{figure}

    \begin{claim}\label{claim:REQ-sim-hyb-reflect-claim}
        $\Reflect^{O_\Hyb^P}_{i}$ is equivalent to the following procedure: flip $b$ if $\not\exists ((i, *), *))\in D$.
    \end{claim}
    \begin{proof}
        First, we show that step \ref{step:REQ-sim-hyb-reflect1} is equivalent to flipping $b_1 = 0$ controlled on $\not\exists ((i, x''), *))\in D$ for all $x''\neq x$. 
        If there is such an entry $((i, x''), *))\in D$, then the query to $O_\Hyb^P$ acts as the identity. 
        Note that in the case of $i < \MaxEntry(D)$, the former holds and $O_\Hyb^P$ also acts as the identity. 
        Furthermore, writing the output of the $P$ evaluation to $\sum_{u}\ket{u}$ does not change the state and so the evaluations of the compressed oracle cancel with each other. 
        Similarly, step \ref{step:REQ-sim-hyb-reflect2} is equivalent to flipping $b_2 = 0$ controlled on $\not\exists ((i, x''), *))\in D$ for all $x''\neq x'$.

        Now observe that since $x' \neq x$, both $b_1=1$ and $b_2=1$ if and only if $\not\exists ((i, *), *))\in D$. Finally, uncomputing $b_1$, $b_2$, and the work registers removes any effect other than the flip.
    \end{proof}

    \begin{figure}[H]
    \begin{mdframed}
         To answer a query $\ket{s, \inp, u, b}_\cQ$ where $\inp = (x, \tilde{\st}, r_{s}, \sigma, s')$ do the following.
        \begin{enumerate}
            \item \textbf{SEQ Check:} Check that the cache does not contain a conflicting $\inputcache[\serialindex[s]] = \inp'$ where $\inp' \notin \{\bot, \inp\}$.\footnote{We define $\inputcache[\bot] = \bot$ to handle the case where $\serialindex[s] = \bot$.} If it does, then skip the following steps.
            \item \textbf{Signature Validation:} Compute $\Sig.\Ver(\pk, s\concat\tilde{\st}\concat r_s, \sigma)$. If the signature is valid, then flip $b$, write $\bot$ to $u$, and skip the following steps.
            
            \item \textbf{Early Answer:} Let $i_{\max}$ be the highest index recorded in $\serialindex$. If $\serialindex[s] < i_{\max}-1$, compute 
            \[
                \ket{s, \inp, u\oplus \outputcache[\serialindex[s]], b\oplus 1}_{\cQ}
            \]
            from the output cache and skip the following steps.
            
            \item \textbf{Cache Clear:} \label{step:REQ-hybrid-oracle-sim-cache}
            \begin{enumerate}
                \item Initialize a temporary ancilla register $\ket{g} = \ket{1}$. Controlled on the input, evaluate $\Reflect^{O_\Hyb^P}_{\serialindex[s], x} \ket{g}$. \label{step:REQ-hybrid-oracle-sim-cache-reflect}
                % flip $g$ if there is an entry $((\serialindex[s], x), *)\in D$.
                \item \textbf{Serial and Input Cache:} If $g = 1$, controlled on the input register $\ket{s, \inp, *, *}$, set $i_s = \serialindex[s]$ and update the serial and input caches to
                \begin{align*}
                    \serialindex &\mapsto \serialindex \triangle \{(s': i_s + 1)\}
                    \\
                    \inputcache &\mapsto \inputcache \triangle \{(i_s: (s, \inp))\}
                \end{align*}
                in the cache registers $\cC_{\serialindex}$ and $\cC_{\inputcache}$.
                % set $i_s = \serialindex[s]$ and apply to $\cC$ the operation mapping
                % \[
                %     \ket{\serialindex, \inputcache}_\cC \otimes \ket{s, \inp, *, *}_\cQ 
                %     \ \leftrightarrow\ 
                %     \ket{\serialindex \triangle \{(s':i_s+1)\}, \inputcache \triangle \{(i_s: \inp)}_\cC \otimes \ket{s, \inp, *, *}_\cQ 
                % \]
                % and acting as the identity on all orthogonal states.
                % update $\serialindex \mapsto \serialindex\triangle \{(s': \serialindex[s]+1)\}$ and $\inputcache \mapsto \inputcache\triangle \{(\serialindex[s]: \inp)\}$ as before.

                \item \textbf{Output Cache:} 
                If $g=0$, controlled on the input register $\ket{s, \inp, *, *}$, set $i_s = \serialindex[s]$ and look up $(s_{i_s-1}, \inp_{i_s - 1}) = \inputcache[i_s - 1]$.
                Then, compute $\outp = P_\OSP(\inp_{i_s-1}; s;r)$ where $r = H(\serialindex[s], x)$ as in the \textbf{Evaluation} step, and update 
                \[
                    \outputcache 
                    \mapsto 
                    \outputcache \triangle \{(i_s - 1: \outp)\}
                \]
                in the cache register $\cC$. Note that this requires two queries to $O_{\Hyb}^{P}$.
   % \proofauditloc{}{This simulator is given only $O_\Hyb^P$; $O_{\SEQ}^{P_\OSP}$ is supposedly the one it tries to simulate?}
                
                \item Repeat step \ref{step:REQ-hybrid-oracle-sim-cache-reflect}, then discard the temporary register.
            \end{enumerate}
            
            \item \textbf{Evaluation:}
            \begin{enumerate}
                \item \textbf{Query $O_{\SEQ}^{P_{\OSP}}$}: Prepare and query $\ket{(\serialindex[s], x), 0}$ to $O_{\SEQ}^{P_{\OSP}}$ to obtain $\ket{(\serialindex[s], x), y, r_2}$.\label{step:REQ-hybrid-oracle-sim-query-oracle}
                
                \item Set $r_\Enc \concat r_\Sign = G'(s\concat \inp)$.
                \item \textbf{Compute Next Encrypted State.} Compute $\tilde{\st'} = \Enc(\sk_\Enc, 0; r_{\Enc})$.
                \item \textbf{Compute Next Signature.} Compute the following two values (as in step \ref{step:REQ-construction-sign-next-serial-and-state} of \Cref{fig:RAM-program-compiler-construction}):
                \begin{align*}
                    r_{s'} &= r_{s'}'\concat G_\Sign\left(s'\big\concat \tilde{\st'}\big\concat r_{s'}'\right)
                    \\
                    \sigma' &= \Sig.\Sign\left(\sk,\ s'\big\concat \tilde{\st'}\big\concat r_{s'};\ r_{\Sign}\right).
                \end{align*}
                
                \item \textbf{Answer Query.} Write the response to the query register containing $u$ as $\ket{u} \mapsto \ket{u \oplus (y, (\tilde{\st'}, r_{s'}, \sigma')}$ and flip the success indicator bit $b$.
% \proofauditloc{P09}{Should a successful response also flip the caller\textquotesingle s bit $b$?}
                \item \textbf{Unquery $O_{\SEQ}^{P_{\OSP}}$}: Repeat step \ref{step:REQ-hybrid-oracle-sim-query-oracle}
            \end{enumerate}
            \item \textbf{Cache Fill:} Repeat step \ref{step:REQ-hybrid-oracle-sim-cache}.
        \end{enumerate}
    \end{mdframed}
    \caption{Simulator $\Sim_1^{O_\Hyb^P}$ for the SEQ Oracle $O_{\SEQ}^{P_{\OSP}}$}\label{fig:REQ-hybrid-oracle-sim}
    \end{figure}

    % To answer a query $\ket{s, \inp, u, b}_\cQ$, do the following.
    %     \begin{enumerate}
    %         \item \textbf{SEQ Check:} as before.
    %         \item \textbf{Signature Validation:} as before.
    %         \item \textbf{History Check:} as before.
    %         \item \textbf{Early Answer {\color{red}(New)}:} Let $i_{\max}$ be the highest index recorded in $\serialindex$. If $\serialindex[s] < i_{\max}$, compute 
    %         \[
    %             \ket{s, \inp, u\oplus \outputcache[\serialindex[s]], b\oplus 1}_{\cQ}
    %         \]
    %         from the output cache and skip the following steps.
    %         \item \textbf{Cache Clear:} as before (if no early answer).
    %         \item \textbf{Evaluation:} as before (if no early answer).
    %         \item \textbf{Cache Fill:} as before (if no early answer).
    %     \end{enumerate}

    Consider the following sequence of hybrids.
    % \begin{itemize}
        % \item 
        \paragraph{\underline{$\Hyb_0$}} is $A^{O_{\SEQ}^{P_{\OSP}[\sk, \pk, \sk_\Enc]}}$'s view with freshly sampled $(\sk, \pk)$ and $\sk_\Enc$. It can generally be divided into two steps.
        \begin{enumerate}
            \item \textbf{SEQ Check:} Check that the database $D$ does not contain a conflicting entry $(s, \inp', *)$ where $\inp' \neq (x, \tilde{\st}, r_s,\sigma, s')$; if it does, skip the following steps.
            \item \textbf{Evaluation:} Compute $P_\OSP(x, \tilde{\st}, r_s,\sigma, s'; s; r)$ where $r = H(s, x, \tilde{\st}, r_s,\sigma, s')$. Write the result to $u$ and flip $b$. This involves two queries to the compressed oracle $H$: one to obtain $r$ and another to erase it.
        \end{enumerate}
        
        % \item 
        \paragraph{\underline{$\Hyb_1$ (Move Signature Validation)}} moves the signature check step in $P_{\OSP}$ to before the compressed oracle is queried. Explicitly, respond to oracle queries $\ket{s,(x, \tilde{\st}, r_s,\sigma, s'), u, b}$ as follows.
        \begin{enumerate}
            \item \textbf{SEQ Check:} as before.
            % Check that the database $D$ does not contain a conflicting entry $(s, \inp', *)$ where $\inp' \neq (x, \tilde{\st}, r_s,\sigma, s')$; if so, skip the following steps.
            \item \textbf{Signature Validation {\color{red} (Moved)}:} Compute $\Sig.\Ver(\pk, s\concat\tilde{\st}\concat r_s, \sigma)$. If the signature is valid, then flip $b$, write $\bot$ to $u$, and go straight to outputting, skipping the following steps. Otherwise, continue
            % \proofauditloc{}{Should this be "If the signature is invalid, then flip $b$, write $\bot$ to $u$,
% and skip the following steps. Otherwise, continue."}
            \item \textbf{Evaluation: {\color{red} (Modified)}}: Compute $P_\OSP(x, \tilde{\st}, r_s,\sigma, s'; s; r)$ where $r = H(s, x, \tilde{\st}, r_s,\sigma, s')$ starting from step \ref{step:REQ-construction-decrypt-state} in the construction (\Cref{fig:RAM-program-compiler-construction}). Write the result to $u$ and flip $b$. This includes 2 queries to $H$.
        \end{enumerate}

        $Hyb_0$ is equivalent to $\Hyb_1$ since the signature validation is independent of $r$.
        % We prove equivalence to $\Hyb_0$ in \Cref{claim:REQ-sim-SEQ-to-hyb_proof-h0-h1}.
        % \justin{Equivalence to $\Hyb_0$: signature validation doesn't use $r$ so it commutes with the rest of the evaluation.}

        % \item 
        \paragraph{\underline{$\Hyb_2$ (Add History Check)}} adds an additional ``history consistency'' check after signature validation. Specifically, respond to oracle queries $\ket{s,(x, \tilde{\st}, r_s,\sigma, s'), u, b}$ as follows.
        \begin{enumerate}
            \item \textbf{SEQ Check:} as before.
            % Check that the database $D$ does not contain a conflicting entry $(s, \inp', *)$ where $\inp' \neq (x, \tilde{\st}, r_s,\sigma, s')$; if so, skip the following steps.
            \item \textbf{Signature Validation:} as before.
            % Compute $\Sig.\Ver(\pk, s\concat\tilde{\st}\concat r_s, \sigma)$. If the signature is valid, then flip $b$, write $\bot$ to $u$, and output.

            \item \textbf{History Consistency Check {\color{red}(New)}:} 
            Check that either $s = s_0$ or that there exist entries 
            \[
                \begin{array}{rcl}
                     (s_0,\ &(x_0, \bot, *, 0^\secp\concat 0^\secp, s_1), &r_{0,x}\concat r'_{1})
                \\
                (s_1,\ &(x_1, \tilde{\st_1}, *, r'_{1}\concat *, s_2), &r_{1,x}\concat r'_{2})
                \\
                &\dots&
                \\
                (s_{i-1},\ &(x_{i-1}, \tilde{\st}_{i-1}, *, r'_{i-1}\concat *, s_i), &r_{i-1,x}\concat r'_{i})
                \end{array}
            \]
            % \begin{align*}
            %     (s_0,\ &(x_0, \bot, *, 0^\secp\concat 0^\secp, s_1), r_{0,x}\concat r'_{1})
            %     \\
            %     (s_1,\ &(x_1, \tilde{\st_1}, *, r'_{1}\concat *, s_2), r_{1,x}\concat r'_{2})
            %     \\
            %     \dots
            %     \\
            %     (s_{i-1},\ &(x_{i-1}, \tilde{\st}_{i-1}, *, r'_{i-1}\concat *, s_i), r_{i-1,x}\concat r'_{i})
            % \end{align*}
            in $D$ where $s_i = s$ and $r_{s} = r'_{i}\concat *$ such that for every $j < i$, the program states $\st_j = \SKE.\Dec(\sk_\Enc, \tilde{\st_j})$ and $\st_{j+1} = \SKE.\Dec(\sk_\Enc, \tilde{\st_{j+1}})$ satisfy
            \[
                P(\st_j, x_j; r_{j,x}) = (\st_{j+1}, *)
            \]
            where $r_j = r_{j,x}\concat r'_{j,s_{j+1}}$ and $\st_0 = \bot$. 

            \emph{Measure} whether this occurs and if not, abort the experiment completely.\footnote{This is the only physical measurement that occurs while answering a query -- all other computations are done coherently.}
            % Check that either $s = s_0$ or there exists an entry $(s', (*, *, *, s), *) \in D$ for some serial number $s'$, where $D$ is the compressed oracle database. If neither of those are the case, abort the experiment completely.
            % \justin{Modify the history check to also include $\tilde{\st}$ consistency, as in \Cref{claim:REQ-proof-sEQ-ordering}.}
            
            \item \textbf{Evaluation:} as before.
            % Compute $P_\OSP(x, \tilde{\st}, r_s,\sigma, s'; s; r)$ where $r = H(s, x, \tilde{\st}, r_s,\sigma, s')$ starting from step \ref{step:REQ-construction-decrypt-state} in the construction (\Cref{fig:RAM-program-compiler-construction}). Write the result to $u$ and flip $b$. This includes 2 queries to $H$.
        \end{enumerate}
        We prove computational indistinguishability to $\Hyb_1$ in \Cref{claim:REQ-hyb-oracle_history}.
        % \justin{Indistinguishability to $\Hyb_1$: ordering claim \Cref{claim:REQ-proof-sEQ-ordering} implies history is consistent if the signature validates, with overwhelming probability.}

        \paragraph{\underline{$\Hyb_3$ ($\st$ from Database)}} changes how $\st$ is computed in the \textbf{Evaluation} step. Instead of decrypting it from $\tilde{\st}$, $\st$ is computed using the compressed database $D$. To answer a query $\ket{s,(x, \tilde{\st}, r_s,\sigma, s'), u, b}_\cQ$, do the following.
        \begin{enumerate}
            \item \textbf{SEQ Check:} as before.
            % Check that the database $D$ does not contain a conflicting entry $(s, \inp', *)$ where $\inp' \neq (x, \tilde{\st}, r_s,\sigma, s')$; if so, skip the following steps.
            \item \textbf{Signature Validation:} as before.
            % Compute $\Sig.\Ver(\pk, s\concat\tilde{\st}\concat r_s, \sigma)$. If the signature is valid, then flip $b$, write $\bot$ to $u$, and output.

            \item \textbf{History Consistency Check:} as before.\label{step:REQ-sim-SEQ-to-hyb_proof-hyb4-history}
            % Check that either $s = s_0$ or there exists an entry $(s', (*, *, *, s), *) \in D$ for some serial number $s'$, where $D$ is the compressed oracle database. If neither of those are the case, abort the experiment completely.
            \item \textbf{Cache Clear:} as before.
            
            \item \textbf{Evaluation {\color{red}(Modified)}:} 
            Compute $P_\OSP(x, \tilde{\st}, r_s,\sigma, s'; s; r)$ where $r = H(s, x, \tilde{\st}, r_s,\sigma, s')$, starting from step \ref{step:REQ-construction-decrypt-state} in the construction (\Cref{fig:RAM-program-compiler-construction}).
            \begin{enumerate}
                \item \textbf{Compute $\st$ {\color{red}(Modified)}}: 
                Order the database, starting from $s_0$, as containing
                \begin{align*}
                    (s_0,\ &(x_0, \tilde{\bot}, *, *, s_1), r_{0,x}\concat *)
                    \\
                    (s_1,\ &(x_1, \tilde{\st_1}, *, *, s_2), r_{1,x}\concat *)
                    \\
                    \dots
                    \\
                    (s_{i},\ &(x_{i}, \tilde{\st_i}, *, *, s_{i+1}), r_{i,x}\concat *)
                \end{align*}
                Find $i_s$ such that $s_{i_s} = s$, then iteratively compute
                \[
                    P(\st_j, x_j; r_{j,x}) = (\st_{j+1}, *)
                \]
                starting from $j = 0$ and $\st_0 = \bot$ to $j = i_s-1$. Set $\st = \st_{i_s}$.
                % % Starting from $i=0$ to $i= i_s-1$, where $i_s$ is as computed in $\Hyb_3$:
                % % \begin{enumerate}
                % %     \item Find $s_i$ by looking for an entry $(s_{i}: i) \in \serialindex$.
                % %     \item Find an entry $(s_i, (x_{i}, *, *, *, s'_{i}), r_i) \in D$ and parse $r_i = r_{i,x}\concat r_{i,s'_{i}}$.
                % %     \item Compute $\st_{i+1}$ as $P(\st_{i}, x_i; r_{i,x}) = (\st_{i+1}, y_{i})$.
                % % \end{enumerate}
                % Set $\st = \st_{i_s - 1}$.
                % %\justin{The following directly finds $s'$ in the database.}
                % Starting from $i=0$ until $s_i = s$:
                % \begin{enumerate}
                %     \item Find an entry $(s_i, (x_{i}, *, *, *, s'_{i}), r_i) \in D$ and parse $r_i = r_{i,x}\concat r_{i,s'_{i}}$.
                %     \item Compute $\st_{i+1}$ as $P(\st_{i}, x_i; r_{i,x}) = (\st_{i+1}, y_{i})$.
                %     \item Set $s_{i+1} = s'_i$ and increment $i$.
                % \end{enumerate}
                % Set $\st_{i} = $

                \item Continue from step \ref{step:REQ-construction-after-decrypt-state} in \Cref{fig:RAM-program-compiler-construction}.
                \item Write the output to $u$ and flip $b$.
            \end{enumerate}

            \item \textbf{Cache Fill:} as before.
        \end{enumerate}
        We prove equivalence to $\Hyb_2$ in \Cref{claim:REQ-hyb-oracle_st-from-db}.

        \paragraph{\underline{$\Hyb_4$ (Simplified History Check)}} simplifies the history check by removing the condition on $\tilde{\st_j}$. Explicitly, answer queries by doing the following.
        \begin{enumerate}
            \item \textbf{SEQ Check:} as before.
            \item \textbf{Signature Validation:} as before.
            \item \textbf{History Consistency Check {\color{red}(Modified)}}:
            Check that either $s = s_0$ or that there exist entries 
            \[
                \begin{array}{rcl}
                ({\color{red}s_0}, 
                &(x_0, \tilde{\st}, *, {\color{red}0^\secp}\concat 0^\secp, {\color{blue}s_1}), 
                &r_{0,x}\concat {\color{blue}r'_{1}})
                \\
                ({\color{blue}s_1}, 
                    &(x_1, \tilde{\st}, *, {\color{blue}r'_1}\concat *, {\color{violet}s_2}), 
                    &r_{1,x}\concat {\color{violet}r'_{2}})
                \\
                \dots
                \\
                (s_{i-1}, 
                &(x_{i-1}, \tilde{\st}, *, r'_{i}\concat *, s_i), 
                &r_{i-1,x}\concat r'_{i})
                \end{array}
            \]
            in $D$ where $s_i = s$ and $r_s = r'_{i} \concat *$.
            \item \textbf{Evaluation:} as before.
        \end{enumerate}
        We prove computational indistinguishability to $\Hyb_3$ in \Cref{claim:REQ-hyb-oracle_simplified-history}.

        \paragraph{\underline{$\Hyb_5$ (Hide $\st$)}} changes how $\tilde{\st'}$ is computed in the \textbf{Evaluation} of $P_\OSP(\inp; s; r)$ (step \ref{step:REQ-sim-SEQ-to-hyb_proof-hyb6-evaluation}). Instead of computing it as an encryption of the next state $\st'$ (step \ref{step:REQ-construction-encrypt-next-state} of \Cref{fig:RAM-program-compiler-construction}), generate $\tilde{\st'} = \Enc(\sk_\Enc, 0; r_{\Enc})$ as an encryption of $0$. We prove computational indistingiushability to $\Hyb_4$ in \Cref{claim:REQ-hyb-oracle_hide_st}.

        % \item 
        \paragraph{\underline{$\Hyb_6$ (Input Cache)}} adds the serial-to-index cache $\serialindex$ and the input cache $\inputcache$, contained in a variable-length register $\cC$. $\serialindex = \{(s:i_s)\}_{s}$ is a mapping from serial numbers $s$ to indices $i_s$. $\inputcache = \{(i:\inp_i\}_{i}$ is a mapping from indices $i$ to inputs $\inp_i$. $\serialindex$ is initialized to $\{(0:s_0)\}$, while $\inputcache$ is initialized to $\emptyset$.

        To answer a query 
        % $\ket{s,(x, \tilde{\st}, r_s,\sigma, s'), u, b}_\cQ$
        $\ket{s,\inp, u, b}_\cQ$, do the following.
        \begin{enumerate}
            \item \textbf{SEQ Check:} as before.
            \item \textbf{Signature Validation:} as before.
            \item \textbf{History Check:} as before.
            
            \item \textbf{Cache Clear {\color{red}(New)}:} If there exists an entry $(s, (*, *, *, *, *), *) \in D$, where $D$ is the compressed database, do the following.
            \begin{itemize}
                \item Set $i_s = \serialindex[s]$ and update the caches to\footnote{A query using $s$ registers the \emph{next} serial number $s'$ into $\serialindex$ and the \emph{current} input $\inp$ into $\inputcache$ under $s$. \textbf{Cache Clear} prepares for answering each query by (if necessary) de-registering the next serial number $s'$ from $\serialindex$ and clearing the current input $\inp$ from $\inputcache$ for $s$. The symmetric difference $\triangle$ is used to ensure the operation is unitary. A new element is never added in \textbf{Cache Clear} because it is (indirectly) conditioned on an element already existing.}
                \begin{align*}
                    \serialindex &\mapsto \serialindex \triangle \{(s': i_s + 1)\}
                    \\
                    \inputcache &\mapsto \inputcache \triangle \{i_s: (s, \inp)\}
                \end{align*}
                
            \end{itemize}
            % \begin{itemize}
            %     \item Find an entry $(s'', (*, *, *, *, s), *) \in D$.\footnote{Such an entry is guaranteed to exist by the history check.} Set $i_s = \serialindex[s'']+1$ and apply to $\cC$ the operation mapping
            %     \[
            %         \ket{\serialindex, \inputcache}_\cC \otimes \ket{s, \inp, *, *}_\cQ 
            %         \ \leftrightarrow\ 
            %         \ket{\serialindex \triangle \{(s:i_s)\}, \inputcache \triangle \{(i_s: \inp)}_\cC \otimes \ket{s, \inp, *, *}_\cQ 
            %     \]
            %     and acting as the identity on all orthogonal states.\footnote{We maintain the invariant that at the start and end of every query, $\serialindex[s]\neq \bot$ and $\inputcache[i_s]\neq \bot$ if and only if there is an entry $(s, *, *)\in D$. Thus, after \textbf{Input Cache Clear}, $s$, $i_s$, and $\inp$ are erased from $\serialindex$ and $\inputcache$.}
            % \end{itemize}
            \label{step:REQ-sim-SEQ-to-hyb_proof-hyb3-inputcache}
            
            \item \textbf{Evaluation:} as before.
            
            \item \textbf{Cache Fill {\color{red}(New)}:} Repeat step \ref{step:REQ-sim-SEQ-to-hyb_proof-hyb3-inputcache}.
        \end{enumerate}
        We prove equivalence to $\Hyb_5$ in \Cref{claim:REQ-hyb-oracle_input-cache}.
        
        \paragraph{\underline{$\Hyb_7$ (SEQ Check using Cache)}} changes the \textbf{SEQ Check} to use the cache instead of looking in the database.
        To answer a query 
        % $\ket{s,(x, \tilde{\st}, r_s,\sigma, s'), u, b}_\cQ$
        $\ket{s,\inp, u, b}_\cQ$, do the following.
        \begin{enumerate}
            \item \textbf{SEQ Check {\color{red}(Modified)}:} Check that the cache does not contain a conflicting $\inputcache[\serialindex[s]] = (s', \inp')$ where $(s', \inp') \notin \{\bot, (s, \inp)\}$.\footnote{We define $\inputcache[\bot] = \bot$ to handle the case where $\serialindex[s] = \bot$.} If it does, then skip the following steps.
            
            \item \textbf{Signature Validation:} as before.
            \item \textbf{History Check:} as before.
            
            \item \textbf{Cache Clear :} as before.
            
            \item \textbf{Evaluation:} as before.
            
            \item \textbf{Cache Fill:} as before.
        \end{enumerate}
        We prove equivalence to $\Hyb_6$ in \Cref{claim:REQ-hyb-oracle_seq-cache}.
        \paragraph{\underline{$\Hyb_8$ ($H$ Input Renaming)}} renames the inputs to the compressed oracle $H$. $s$ is renamed to its index $\serialindex[s]$ and $\inp = (x, \tilde{\st}, r_s,\sigma, s')$ is renamed to just the chosen input $x$. For example, instead of querying $H$ on $(s, \inp)$, the simulator would query it on $(\serialindex[s], x)$.
        % Instead of querying it on $(s, (x, *, *, *, *))$, the simulator queries $H$ on $(\serialindex[s], x)$, where $i_s$ is computed as determined in $\Hyb_3$. Similarly, if it looks for an entry starting with $s$ in the database, it 
        % Each time the simulator would query $H$ on $(s,\inp)$ or otherwise look for an entry starting with $s$ in $H$'s compressed database, the simulator substitutes $(i_s, x)$ 

        To answer a query $\ket{s,(x, \tilde{\st}, r_s,\sigma, s'), u, b}_\cQ$, do the following.
        \begin{enumerate}
            \item \textbf{SEQ Check:} as before.
            \item \textbf{Signature Validation:} as before.
            \item \textbf{History Check {\color{red}(Modified)}:} Check that either $s = s_0$ or there exist entries
            \begin{gather*}
                ((0, x_0), r_0)
                \\
                ((1, x_1), r_1)
                \\
                \dots
                \\
                ((\serialindex[s]-1, x_{\serialindex[s]-1}), r_{\serialindex[s]-1})
            \end{gather*}
            in $D$.
            \item \textbf{Cache Clear {\color{red}(Modified)}:} If there exists an entry $((\serialindex[s], *), *) \in D$, where $D$ is the compressed database, clear the cache as before. \label{step:REQ-sim-SEQ-to-hyb_proof-hyb6-inputcache}
            \item \textbf{Evaluation {\color{red}(Modified)}:}
            Compute $P_\OSP(\inp; s; r)$ as before, except $r = H(\serialindex[s], x)$ \label{step:REQ-sim-SEQ-to-hyb_proof-hyb6-evaluation}
            \item \textbf{Cache Fill {\color{red}(Modified)}:} Repeat step \ref{step:REQ-sim-SEQ-to-hyb_proof-hyb6-inputcache}.
        \end{enumerate}
        We prove equivalence to $\Hyb_7$ in \Cref{claim:REQ-hyb-oracle_rename-oracle}.

        % We prove computational indistinguishability to $\Hyb_6$ in \Cref{claim:REQ-sim-SEQ-to-hyb_proof-h6-h7}.
        % \justin{Indistinguishability: decryption isn't done anywhere else now that $\st$ is computed directly from $D$ and the history check doesn't check $\tilde{\st}$. Might be easier to switch this to public key encryption instead of using an encryption oracle to get the other ciphertexts.}

        \paragraph{\underline{$\Hyb_9$ (Output Cache)}} adds an output cache $\outputcache$ to the cache register $\cC$.
        $\outputcache = \{(i:\outp_i)\}$ is a key-value store mapping indices $i$ to outputs $\outp_i = (y, (\tilde{\st'}, r_{s'}, \sigma'))$ resulting from querying $P_\OSP$ on $(s_i, \inputcache[i])$. 

        To answer a query $\ket{s,(x, \tilde{\st}, r_s,\sigma, s'), u, b}_\cQ$, do the following.
        \begin{enumerate}
            \item \textbf{SEQ Check:} as before.
            \item \textbf{Signature Validation:} as before.
            \item \textbf{History Check:} as before.
            \item \textbf{Cache Clear {\color{red}(Modified)}:}  \label{step:REQ-sim-SEQ-to-hyb_proof-hyb8-cache}
            \begin{enumerate}
                \item If there exists an entry $(\serialindex[s], x, *)\in D$, update the serial-to-index and input caches to
                \begin{align*}
                    \serialindex &\mapsto \serialindex \triangle \{(s': i_s + 1)\}
                    \\
                    \inputcache &\mapsto \inputcache \triangle \{i_s: \inp\}
                \end{align*}

                \item If there \emph{does not} exist an entry $(\serialindex[s], x, *) \in D$, look up $(s_{i-1}, \inp_{i-1}) = \inputcache[\serialindex[s] -1]$ and parse $\inp_{i-1} = (x_{i-1}, \dots)$ Then, compute $r_{i-1} = D(\serialindex[s]-1, x_{i-1})$ \emph{directly from the database register} and compute $\outp_{i-1} = P_{\OSP}(s_{i-1}, \inp_{i-1}; r_{i-1})$ as in the \textbf{evaluation} step.\footnote{We emphasize that this does not query the compressed oracle to obtain $r_{i-1}$.} Finally,
                update the output cache to
                \[
                  \outputcache \mapsto \outputcache \triangle \{(\serialindex[s]-1: \outp\}  
                \]

            \end{enumerate}
            % If there exists an entry $(\serialindex[s], x, *)\in D$, clear the cache as follows.
            % \begin{enumerate}
            %     \item \textbf{Input Cache:} Update $\serialindex \mapsto \serialindex\triangle \{(s': \serialindex[s]+1)\}$ and $\inputcache \mapsto \inputcache\triangle \{(\serialindex[s]: \inp)\}$ as before.
                
            %     \item \textbf{Output Cache {\color{red}(New)}:}
            %     compute $\outp = P_\OSP(\inp; s;r)$ as in the \textbf{Evaluation} step, except $r = D(\serialindex[s], x)$. \emph{We emphasize that this directly reads $r$ from $D$ and does not query $H$ to obtain it.}
                
            % \end{enumerate}

            \item \textbf{Evaluation:} as before.
            
            \item \textbf{Cache Fill {\color{red}(Modified)}:} Repeat step \ref{step:REQ-sim-SEQ-to-hyb_proof-hyb8-cache}.
            \label{step:REQ-sim-SEQ-to-hyb_proof-hyb8-outputcache}
        \end{enumerate}
        We prove equivalence to $\Hyb_8$ in \Cref{claim:REQ-hyb-oracle_output-cache}.
        % The output cache works similarly to the input cache by caching a value if the oracle database contains a matching entry. The difference is that it actually computes the output, which disturbs the compressed oracle database slightly because recording isn't perfect.}
        
        % \justin{indistinguishability: check the loss on recording this. Should argue fidelity/trace distance of the overall state of the system after every query vs the previous hybrid.}

        \paragraph{\underline{$\Hyb_{10}$ (Early Answer)}} uses the output cache to answer queries which use a serial number in the middle of the recorded serial numbers. 
        To answer a query $\ket{s, \inp, u, b}_\cQ$ where $\inp = (x, \tilde{\st}, r_s, \sigma, s')$, do the following.
        \begin{enumerate}
            \item \textbf{SEQ Check:} as before.
            \item \textbf{Signature Validation:} as before.
            \item \textbf{History Check:} as before.
            \item \textbf{Early Answer {\color{red}(New)}:} Let $i_{\max}$ be the highest index recorded in $\serialindex$. If $\serialindex[s] < i_{\max}$, compute 
            % \proofauditloc{P14}{The $+1$ admits uncached outputs, even initially; is the intended threshold $i_{\max}-1$?} %\Justin{Yes. $\SI$ is one ahead of the other caches because it records the *next* serial number. Good catch.}
            \[
                \ket{s, \inp, u\oplus \outputcache[\serialindex[s]], b\oplus 1}_{\cQ}
            \]
            from the output cache and skip the following steps.
            \item \textbf{Cache Clear:} as before (if no early answer).
            \item \textbf{Evaluation:} as before (if no early answer).
            \item \textbf{Cache Fill:} as before (if no early answer).
        \end{enumerate}
        We prove equivalence to $\Hyb_9$ in \Cref{claim:REQ-hyb-oracle_early-answer}.

        \paragraph{\underline{$\Hyb_{11}$ (Output Cache Queries $H$)}} modifies how the output cache works. Instead of directly using $r = D(i, x)$ from the database, it queries the random oracle $H$ on $(i, x)$ to obtain $r$. 

        To answer a query $\ket{s,(x, \tilde{\st}, r_s,\sigma, s'), u, b}_\cQ$, do the following.
        \begin{enumerate}
            \item \textbf{SEQ Check:} as before.
            \item \textbf{Signature Validation:} as before.
            \item \textbf{Early Answer:} as before.
            \item \textbf{Cache Clear {\color{red}(Modified)}:}  
            If there exists an entry $(\serialindex[s], x, *)\in D$, clear the cache as follows.
            \begin{enumerate}
                \item \textbf{Serial-to-Index and Input Caches:} Same as before.

                \item \textbf{Output Cache {\color{red}(Modified)}:} 
                If there \emph{does not} exist an entry $(\serialindex[s], x, *) \in D$, look up $(s_{i-1}, \inp_{i-1}) = \inputcache[\serialindex[s] -1]$ and parse $\inp_{i-1} = (x_{i-1}, \dots)$ Then, compute $r_{i-1} = H(\serialindex[s]-1, x_{i-1})$ \emph{by querying the compressed oracle $H$} and compute $\outp_{i-1} = P_{\OSP}(s_{i-1}, \inp_{i-1}; r_{i-1})$ as in the \textbf{evaluation} step.\footnote{The change is to compute $r$ by querying the compressed oracle.} Finally,
                update the output cache to
                \[
                  \outputcache \mapsto \outputcache \triangle \{(\serialindex[s]-1: \outp\}  
                \]
                \label{step:REQ-hybrid-oracle-sim-output-cache-H}
            
                % \item Initialize an ancilla register $\ket{g} = \ket{0}$ and flip $g$ if there is an entry $((\serialindex[s], *), *)\in D$.\footnote{This step implicitly occurs in the cache in the previous hybrids. We make this step explicit here because clearing/filling the output cache can affect $D$ and thereby the uncomputation of $g$.}
                % \item If $g = 1$, update $\serialindex \mapsto \serialindex\triangle \{(s': \serialindex[s]+1)\}$ and $\inputcache \mapsto \inputcache\triangle \{(\serialindex[s]: \inp)\}$ as before.
                
                % \item If $g=1$, compute $\outp = P_\OSP(\inp; s;r)$ where $r = H(\serialindex[s], x)$ as in the \textbf{Evaluation} step. This requires two queries to $H$: one to compute $r$ and another to erase it after computing $\outp$.
                
                % \item If $g=1$, update $\outputcache$ in the cache register $\cC$ using the map
                % \[
                %     \outputcache 
                %     \mapsto 
                %     \outputcache \triangle \{(\serialindex[s]: \outp)\}
                % \]
                % \label{step:REQ-hybrid-oracle-sim-output-cache-H}
                
                \item Flip $g$ if there is an entry $(s, x, *)\in D$. Measure whether $g=0$ and abort the experiment if not.
            \end{enumerate}
            \label{step:REQ-sim-SEQ-to-hyb_proof-outcache-H}
            
            \item \textbf{Evaluation:} as before.
            
            \item \textbf{Cache Fill {\color{red}(Modified)}:} Repeat step \ref{step:REQ-sim-SEQ-to-hyb_proof-outcache-H}.
            % \label{step:REQ-sim-SEQ-to-hyb_proof-hyb8-outputcache}
        \end{enumerate}
        We prove statistical closeness to $\Hyb_{10}$ in \Cref{claim:REQ-hyb-oracle_outcache-H}.

        \paragraph{\underline{$\Hyb_{12}$ (Remove History Check)}} removes the \textbf{history consistency check}. 
        
        We prove statistical closeness to $\Hyb_{11}$ in \Cref{claim:REQ-hyb-oracle_remove-history}.

        \paragraph{\underline{$\Hyb_{13}$ (Cache Uses Oracle)}} replaces the conditions for the cache to activate. Instead of the cache activating if $\exists ((\serialindex[s], *),*)\in D$ matching the input $(s, *)$, it uses $\Reflect^{O^P_\Hyb}_{\serialindex[s]}$ to decide whether to activate the cache.
        
        To answer a query $\ket{s, \inp, u, b}_\cQ$, do the following.
        \begin{enumerate}
            \item \textbf{SEQ Check:} as before.
            \item \textbf{Signature Validation:} as before.
            \item \textbf{Early Answer:} as before.
            \item \textbf{Cache Clear {\color{red}(Modified)}:} (if no early answer)\label{step:REQ-sim-SEQ-to-hyb_proof-hyb13-cache-oracle}
            \begin{enumerate}
                \item \textbf{\color{red}(Modified)} Run $\Reflect_{\serialindex[s]}^{O_\Hyb^P}$ on $\ket{1}$ to obtain $\ket{g}$.\footnote{The computation of $\ket{g}$ is implicit in prior hybrids. The modification is in \emph{how} it is computed, whether by directly checking the database $D$ or by using $\Reflect_{\serialindex[s]}^{O_\Hyb^P}$.}
                \label{step:REQ-sim-SEQ-to-hyb_proof-hyb13-activate-cache}
                \item \textbf{Input Cache:} If $g=1$, then update $\serialindex$ and $\inputcache$ as before.
                \item \textbf{Output Cache:} If $g=0$, then update $\outputcache$ as before.\label{step:REQ-sim-SEQ-to-hyb_proof-hyb13-cache-activate}
                
                \item \textbf{\color{red}(Modified)} Repeat step \ref{step:REQ-sim-SEQ-to-hyb_proof-hyb13-activate-cache} to erase $\ket{g}$.
            \end{enumerate}
            \item \textbf{Evaluation:} as before (if no early answer).
            \item \textbf{Cache Fill {\color{red}(Modified)}:} repeat step \ref{step:REQ-sim-SEQ-to-hyb_proof-hyb13-cache-oracle} (if no early answer).
        \end{enumerate}
        $\Hyb_{13}$ is equivalent to $\Hyb_{12}$ by \Cref{claim:REQ-sim-hyb-reflect-claim}, which states that $\Reflect_{\serialindex[s]}^{O_{\Hyb}^P}$ is equivalent to checking whether $\not\exists ((\serialindex[s], *), *)\in D$, and the fact that $g$ starts as $1$.

        \paragraph{\underline{$\Hyb_{14} = \Sim^{O_\Hyb^P}$ (Evaluation Uses Oracle)}} separates the computation of $\tilde{\st'}$, the encryption of the next state, and $\sigma$, the signature on the next serial number and encrypted next state, from the computation of $P(\st, x;r_x)$ inside the \textbf{Evaluation} step. This allows the computation of $P(\st, x;r_x)$ to be re-interpreted as a query to $O_\Hyb^{P}$.
        
        To answer a query $\ket{s,(x, \tilde{\st}, r_s,\sigma, s'), u, b}_\cQ$, do the following.
        \begin{enumerate}
            \item \textbf{SEQ Check:} as before.
            \item \textbf{Signature Validation:} as before.
            \item \textbf{Cache Clear:} as before.
            \item \textbf{Evaluation {\color{red}(Modified)}:}
            \begin{enumerate}
                \item \textbf{Query $O_{\SEQ}^{P_{\OSP}}$ {\color{red}(Separated)}}: 
                Prepare and query $\ket{(\serialindex[s], x), 0}$ to $O_{\Hyb}^{P}$ to obtain $\ket{(\serialindex[s], x), y, r_2}$.\label{step:REQ-hybrid-oracle-sim-hyb13-query-oracle}
                
                \item Set $r_\Enc \concat r_\Sign = G'(s\concat \inp)$.
                \item \textbf{Compute Next State.} Compute $\tilde{\st'} = \Enc(\sk_\Enc, 0; r_{\Enc})$.
                \item \textbf{Compute Next Signature.} Compute the following two values (as in step \ref{step:REQ-construction-sign-next-serial-and-state} of \Cref{fig:RAM-program-compiler-construction}):
                \begin{align*}
                    r_{s'} &= r_{s'}'\concat G_\Sign\left(s'\big\concat \tilde{\st'}\big\concat r_{s'}'\right)
                    \\
                    \sigma' &= \Sig.\Sign\left(\sk,\ s'\big\concat \tilde{\st'}\big\concat r_{s'};\ r_{\Sign}\right).
                \end{align*}
                
                \item \textbf{Answer Query.} Write the response to the query register containing $u$ as $\ket{u} \mapsto \ket{u \oplus (y, (\tilde{\st'}, r_{s'}, \sigma')}$

                \item \textbf{Unquery $O_{\SEQ}^{P_{\OSP}}$ {\color{red}(Separated)}}: Repeat step \ref{step:REQ-hybrid-oracle-sim-hyb13-query-oracle}
            \end{enumerate}
            \item \textbf{Cache Fill:} as before.
        \end{enumerate}
        We prove equivalence to $\Hyb_{13}$ in \Cref{claim:REQ-hyb-oracle_eval-oracle}.

        We show statistical or computational indistinguishability between each pair of subsequent hybrids in \cref{claim:REQ-hyb-oracle_h1-h2,claim:REQ-hyb-oracle_h2-h3,claim:REQ-hyb-oracle_h3-h4,claim:REQ-hyb-oracle_h4-h5,claim:REQ-hyb-oracle_h5-h6,claim:REQ-hyb-oracle_h6-h7,claim:REQ-hyb-oracle_h7-h8,claim:REQ-hyb-oracle_h8-h9,claim:REQ-hyb-oracle_h9-h10,claim:REQ-hyb-oracle_h10-h11,claim:REQ-hyb-oracle_h11-h12,claim:REQ-hyb-oracle_h12-h13,claim:REQ-hyb-oracle_h13-h14}. Therefore the real world $\Hyb_0$ is indistinguishable from $\Sim^{O_{\Hyb}^P} = \Hyb_{14}$.

        \begin{claim}[Hybrid: History Check]\label{claim:REQ-hyb-oracle_history}\label{claim:REQ-hyb-oracle_h1-h2}
            For all QPT adversaries,
            \[
                \Hyb_{1} \approx_c \Hyb_{2}
            \]
            Furthermore, the \textbf{history consistency check} fails with negligible probability in $\Hyb_2$.
        \end{claim}
        \begin{proof}
            It suffices to show that the \textbf{history consistency check} fails with negligible probability in $\Hyb_2$ and is therefore gentle. Suppose it failed with noticeable probability. Since the SEQ check restricts $D$ to record a single entry for each serial number $s$, either the new query $\ket{s, (x, \tilde{\st}, \sigma, r_s, s'), u, b}$ or an existing entry $((s, (x, \tilde{\st}, \sigma, r_s, s'), r) \in D$ does not have a precursor entry $(*, (*, *, *, *, s), *) \in D$ satisfying the state requirements. In either case, $\sigma$ is a valid signature on $s\concat \tilde{\st}\concat r_s$ due to the \textbf{signature validation} step.

            Therefore if this occurs with noticeable probability after any $q = \mathsf{poly}(\secp)$ queries, we can construct a QPT adversary that wins the $\mathsf{DC\text{-}Unforg}$ security game from \Cref{claim:REQ-proof-sEQ-ordering} with noticeable probability. Specifically, run $\Hyb_2$ and stop after random query $q' < q$, before it is answered. Find and output the signature $\sigma$ and message $s\concat \tilde{\st}\concat r_s$ corresponding to the query or entry which does not have a precursor in $D$.
        \end{proof}

        \begin{claim}[Hybrid: $\st$ from Database]\label{claim:REQ-hyb-oracle_st-from-db}\label{claim:REQ-hyb-oracle_h2-h3}
            For all adversaries,
            \[
                \Hyb_2 = \Hyb_3
            \]
            Furthermore, the \textbf{history consistency check} fails with negligible probability in $\Hyb_3$.
        \end{claim}
        \begin{proof}
            The only difference between $\Hyb_2$ and $\Hyb_3$ is how $\st$ is computed. In $\Hyb_2$ is is computed by decrypting $\tilde{\st}$, whereas in $\Hyb_3$ it is iteratively computed by decrypting $\st_{j}$ in the database and computing $P(\st_j, x_j; r_{j,x}) = (\st_{j+1}, *)$. The \textbf{history consistency check} requires that these are equivalent, and the experiment aborts otherwise. 
            Finally, since the two experiments are perfectly indistinguishable and the history check failing can be detected, it must happen with the same probability in $\Hyb_3$ as in $\Hyb_2$, i.e. negligible.
        \end{proof}

        \begin{claim}[Hybrid: Simplified History Check]\label{claim:REQ-hyb-oracle_simplified-history}\label{claim:REQ-hyb-oracle_h3-h4}
            For all QPT adversaries,
            \[ 
                \Hyb_{3} \approx_c \Hyb_4    
            \]
            Furthermore, the \textbf{history consistency check} fails with negligible probability in $\Hyb_4$.
        \end{claim}
        \begin{proof}
            The hybrids differ only in the \textbf{History Consistency Check} (Step~\ref{step:REQ-sim-SEQ-to-hyb_proof-hyb4-history}). $\Hyb_4$ performs a subset of the checks that $\Hyb_3$ does. Thus, it is equivalent in $\Hyb_3$ to first perform the full check, then additionally perform the simplified check from $\Hyb_4$. Furthermore, \ref{claim:REQ-hyb-oracle_st-from-db} states that the first check fails with negligible probability. Therefore it is gentle and the state in $\Hyb_3$ is negligibly far in trace distance from the state if the full check was not performed; in other words, the state in $\Hyb_4$.

            Finally, since the two experiments are computationally indistinguishable and the history check failing can be detected, it must happen in $\Hyb_3$ with probability negligibly different from in $\Hyb_2$, i.e. negligible.
        \end{proof}

        \begin{claim}[Hybrid: Hide $st$]\label{claim:REQ-hyb-oracle_hide_st}\label{claim:REQ-hyb-oracle_h4-h5}
            For all QPT adversaires,
            \[
                \Hyb_4 \approx_c \Hyb_5
            \]
            Furthermore, the \textbf{history consistency check} fails with negligible probability in $\Hyb_5$.
        \end{claim}
        \begin{proof}
            The difference between $\Hyb_4$ and $\Hyb_5$ is that each $\tilde{\st}$ is an encryption of either a program state $\st$ in $\Hyb_4$ or a fixed string $0$ in $\Hyb_5$. 
            Observe that $\Hyb_4$ and $\Hyb_5$ never decrypt any $\tilde{\st}$ and only use the key to encrypt. Therefore indistinguishability of $\Hyb_4$ and $\Hyb_5$ can be reduced to security of the encryption scheme under quantum queries in both the learning and challenge portions of the security game, i.e. qInd-qCPA security (\Cref{def:qIND-qCPA}).

            % \justin{Indistinguishability: decryption isn't done anywhere else now that $\st$ is computed directly from $D$ and the history check doesn't check $\tilde{\st}$. Might be easier to switch this to public key encryption instead of using an encryption oracle to get the other ciphertexts.}

            Finally, since the two experiments are computationally indistinguishable and the history check failing can be detected, it must happen in $\Hyb_5$ with probability negligibly different from in $\Hyb_4$, i.e. negligible.
        \end{proof}

        Before showing that $\Hyb_5$ and $\Hyb_6$ are perfectly indistinguishable in \Cref{claim:REQ-hyb-oracle_input-cache}, we first show a useful property of the cache in $\Hyb_6$.

        \begin{claim}[Input Cache Invariant]\label{claim:REQ-sim-SEQ-to-hyb_proof-inputcache-invariant}
            In $\Hyb_5$, after every query, there are entries
            \begin{align*}
                (s_{i+1}: i+1) &\in \serialindex
                \\
                \text{and } 
                (i: (s_{i}, (\inp', s_{i+1})) &\in \inputcache
            \end{align*}
            for some index $i>0$, some serial numbers $s_i$ and $s_{i+1}$, and some partial input $\inp' = (x, \tilde{\st}, r_s)$ if and only if there is an entry 
            \[
                ((s_{i}, (\inp', s_{i+1})), r_i) \in D
            \]
        \end{claim}
        \begin{proof}
            We induct over the number of queries. This property clearly holds at $0$ queries. Consider the state of the simulator upon receiving the $(q+1)$'th query. Before answering it, the invariant holds because it holds after answering the $q$'th query by the inductive hypothesis and the adversary's operations between queries do not affect the simulator's state.

            Consider how the simulator answers a query $\ket{s, (\inp', s')}$. If either the \textbf{SEQ check} or \textbf{history consistency check} fail, then the invariant holds after answering the query. 

            Now consider the effect of \textbf{cache clear}. 
            We claim that at the end of this step, $\serialindex[s'] = \bot$ and $\inputcache[\serialindex[s]] = \bot$, i.e. the cache does not contain entries corresponding to the new query.
            There are two cases at this stage, which may be in superposition or entangled with an external state: either there exists an entry $(s, (\inp', s'), *) \in D$ or there does not. 
            In the former case, by the invariant and the SEQ check, $\serialindex[s'] = \serialindex[s]+1$ and $\inputcache[\serialindex[s]] = (\inp', s')$. Furthermore in this case, \textbf{cache clear} computes 
            \begin{gather*}
                \serialindex \mapsto \serialindex\triangle \{(s': \serialindex[s]+1)\}
                \\
                \inputcache \mapsto \inputcache \triangle \{(\serialindex[s]: (\inp', s'))\}
            \end{gather*}
            which clears both entries.
            In the latter, $\serialindex[s'] = \bot$ and $\inputcache[\serialindex[s]] = \bot$ at the beginning of the step and \textbf{cache clear} acts as the identity.

            \textbf{Evaluation} only queries $H$ on , so the only possible difference between the database $D$ at the start of \textbf{Evaluation} and the updated $D'$ at the end it is at $D(s, (\inp', s'))$ and $D'(s, (\inp', s'))$. These $D$ and $D'$ are the same as at the start and end of answering the query, because no other operation modifies the database in the computational basis.

            Finally, consider the effect of \textbf{cache fill}. At the start of this step, $\serialindex[s'] = \bot$ and $\inputcache[\serialindex[s]] = \bot$. Again there are two cases (potentially in superposition or entangled externally): either $(s, (\inp', s'), *) \in D'$ or not. In the former case, applying the symmetric difference causes $\serialindex[s'] = \serialindex[s]+1$ and $\inputcache[\serialindex[s]] = (\inp', s')$ as required. In the latter case, the cache is unmodified and $\serialindex[s'] = \bot$ and $\inputcache[\serialindex[s]] = \bot$ as required.
        \end{proof}

        \begin{claim}[Hybrid: Input Cache]\label{claim:REQ-hyb-oracle_input-cache}\label{claim:REQ-hyb-oracle_h5-h6}
            For all adversaries,
            \[
                \Hyb_5 = \Hyb_6
            \]
            Furthermore, the \textbf{history consistency check} fails with negligible probability in $\Hyb_5$.
        \end{claim}
        \begin{proof}
            Consider the unitary $U$ which, controlled on the compressed database $D$, maps the cache to
            \begin{gather*}
                \serialindex \mapsto \serialindex\triangle \{(s': \serialindex[s]+1) \quad \forall (s, (\inp', s'), r) \in D\}
                \\
                \inputcache \mapsto \inputcache \triangle \{(\serialindex[s]: (\inp', s'))  \quad \forall (s, (\inp', s'), r) \in D\}
            \end{gather*}
            Using the invariant shown in \Cref{claim:REQ-sim-SEQ-to-hyb_proof-inputcache-invariant}, $U$ maps between the state of the simulator after each query in $\Hyb_5$ and its state in $\Hyb_4$ (with some additional $\ket{0}$ ancillas). Specifically, if applied at the start of a query in $\Hyb_5$, it completely empties the cache, and if applied at the end of a query in $\Hyb_4$, it completely fills the cache according to $D$. Therefore $\Hyb_5$ can be computed by applying $U$ at the start of a query, answering it as in $\Hyb_4$, then applying $U$ again after the query.
            
            Finally, since the two experiments are perfectly indistinguishable and the history check failing can be detected, it must happen with in $\Hyb_3$ as in $\Hyb_2$, i.e. negligible.
        \end{proof}

        \begin{claim}[Hybrid: SEQ Using Cache]\label{claim:REQ-hyb-oracle_seq-cache}\label{claim:REQ-hyb-oracle_h6-h7}
            For all adversaries,
            \[
                \Hyb_6 = \Hyb_7
            \]
            Furthermore, the \textbf{history consistency check} fails with negligible probability in $\Hyb_7$.
        \end{claim}
        \begin{proof}
            It suffices to show that the \textbf{SEQ check} from $\Hyb_6$ rejects if and only if the version from $\Hyb_7$ would reject. The $\Hyb_5$ version rejects if there is an entry $((s, \inp'), *) \in D$ where $\inp \neq \inp'$. The $\Hyb_7$ version rejects if there is a cache entry $\inputcache[\serialindex[s]] = \inp'$ where $\inp' \notin \inp$. By the cache invariant (\Cref{claim:REQ-sim-SEQ-to-hyb_proof-inputcache-invariant}), $((s, \inp'), *) \in D$ if and only if $\inputcache[\serialindex[s]] = \inp'$ (since $\inp'$ is recorded in $D$, $\inp' \neq \bot$.

            Finally, since the two experiments are perfectly indistinguishable and the history check failing can be detected, it must happen with the same probability in $\Hyb_7$ as in $\Hyb_6$, i.e. negligible.
        \end{proof}

        \begin{claim}[Hybrid: $H$ Input Renaming]\label{claim:REQ-hyb-oracle_rename-oracle}\label{claim:REQ-hyb-oracle_h7-h8}
            For all adversaries, 
            \[  
                \Hyb_7 = \Hyb_8
            \]
            Furthermore, the \textbf{history consistency check} fails with negligible probability in $\Hyb_8$.
        \end{claim}
        \begin{proof}
            Recall that the cache invariant (\Cref{claim:REQ-sim-SEQ-to-hyb_proof-inputcache-invariant}) requires that there is an entry $\inputcache[i] = \inp \neq \bot$ if and only if $((s_i, \inp), *) \in D$ for some $s_i$ such that $\serialindex[s_i] = i$. Since $D$ can only record one $\inp_i$ for each $s_i$, the caches $\serialindex$ and $\inputcache$ define a bijection 
            \[
                (s_i, \inp) \rightarrow (\serialindex[s_i], \inputcache[\serialindex[s_i]]
            \]
            covering every entry in $D$. Furthermore, compressed oracles are insensitive to bijective renaming (\Cref{claim:compressed-oracle-renaming}). 

            Finally, since the two experiments are perfectly indistinguishable and the history check failing can be detected, it must happen with the same probability in $\Hyb_6$ as in $\Hyb_5$, i.e. negligible.
        \end{proof}

        Before showing that $\Hyb_8$ and $\Hyb_9$, which introduces the output cache, are perfectly indistinguishable in \Cref{claim:REQ-hyb-oracle_h8-h9}, we first show a useful property of the output cache.
        
        \begin{claim}[Output Cache Invariant]\label{claim:REQ-hyb-oracle_output-cache-invariant}
            In $\Hyb_9$, after every query, there are entries
            \begin{align*}
                (s_{i+1}: i+1) &\in \serialindex
                \\
                \text{and }\quad (i: (s_{i}, ((x, \inp_i, s_{i})) &\in \inputcache
                \\
                \text{and }\quad (i-1: P_\OSP(x_{i-1}, \inp_{i-1}, s_{i-1}; s_{i}; r_{i-1})) &\in \outputcache,
            \end{align*}
            where $\inp_{i-1} = (\sigma_{i-1}, r_{\sigma, i-1})$ for some $\sigma_{i-1}$ and $r_{\sigma, i-1}$, if and only if there is an entry
            \[
               ((i, x), r) \in D
            \]
        \end{claim}
        \begin{proof}
            The conditions on $\serialindex$ and $\inputcache$ follow from the same argument as \Cref{claim:REQ-sim-SEQ-to-hyb_proof-inputcache-invariant}.
            It remains to analyze the output cache. 
            We show this by induction. We claim that at the end of the first cache step, $\outputcache[i-1] = P_\OSP(x, \inp', s_{i+1}; s; r)$ where $(i-1, x_{i-1}, r_{i-1}) \in D$. This is true because the inductive hypothesis requires that $\outputcache[i-1] = \bot$ if there are no entries $(i, *) *)$ in $D$ and the symmetric difference in the update swaps between $\bot$ and $P_\OSP(x, \inp', s_{i+1}; s; r)$.\footnote{If we had instead obtained $r$ by querying $H$, the database register \emph{can} change, which is more complicated to keep track of. We avoid this issue for now by directly reading $r$ off of the database.} 
            Then, after the second cache step, there are two cases: either there is an entry $(i, *) *) \in D$ or there is not. In the first case, $\outputcache[i-1]$ is not modified and remains filled, but $\serialindex[i+1]$ and $\inputcache[i]$ are filled. In the second case, $\outputcache[i-1]$ is emptied, but $\serialindex[i+1]$ and $\inputcache[i]$ are not modified and remain empty. In all cases, these are the only indices which are modified, so the inductive hypothesis holds.
        \end{proof}
        
        \begin{claim}[Hybrid: Output Cache]\label{claim:REQ-hyb-oracle_output-cache}\label{claim:REQ-hyb-oracle_h8-h9}
            For all adversaries,
            \[
                \Hyb_8 = \Hyb_9
            \]
            Furthermore, the \textbf{history consistency check} fails with negligible probability in $\Hyb_9$.
        \end{claim}
        \begin{proof}
            Consider the unitary $U$ which, controlled on the compressed database $D$, maps the output cache to 
            \begin{gather*}
                \outputcache \mapsto \outputcache \triangle \left\{\big(\serialindex[s]: P_{\OSP}(\StateD_{i}(D), x; r)\big)  \quad \forall ((i, x_i), r_i) \in D\right\}
            \end{gather*}
            
            Using the invariant shown in \Cref{claim:REQ-sim-SEQ-to-hyb_proof-inputcache-invariant}, $U$ maps between the state of the simulator after each query in $\Hyb_9$ and its state in $\Hyb_8$ (with some additional $\ket{0}$ ancillas). Specifically, if applied at the start of a query in $\Hyb_9$, it completely empties the output cache $\outputcache$, and if applied at the end of a query in $\Hyb_8$, it completely fills the output cache according to $D$. Therefore $\Hyb_9$ can be computed by applying $U$ at the start of a query, answering it as in $\Hyb_8$, then applying $U$ again after the query.
            
            Finally, since the two experiments are perfectly indistinguishable and the history check failing can be detected, it must happen with the same probability in $\Hyb_9$ as in $\Hyb_8$, i.e. negligible.
        \end{proof}

\jiahui{The original proof uses an equality where it deletes the term
$(\bra{\psi}\Hyb_9)(I-\Pi_{\early})
(\Hyb_{10}\ket{\psi})$ due to Hyb 9 and Hyb 10 being identical except the early answer.
Identity of the two projected components does not make this
inner product zero. Overall statement is correct though. 
     Fixed below.   }

\begin{claim}[Hybrid: Early Answer]
\label{claim:REQ-hyb-oracle_early-answer}
\label{claim:REQ-hyb-oracle_h9-h10}
    For all adversaries,
    \[
        \Hyb_9 \approx_s \Hyb_{10}
    \]
    Furthermore, the \textbf{history consistency check} fails with
    negligible probability in $\Hyb_{10}$.
\end{claim}
\begin{proof}
    Consider the projector $\Pi_{\mathsf{early}}$ onto
    {\color{black}joint states of the query, database, and cache registers}
    which trigger an early answer (including passing the SEQ check,
    {\color{black}signature validation,} and history check).
    {\color{black}Use the early-answer condition
    $\serialindex[s]<i_{\max}-1$.}
    Let $\ket{\psi}$ be any {\color{black}normalized} state resulting
    from the computation so far, including the database registers,
    the query registers, {\color{black}the cache registers,} and $A$'s
    internal registers. We abuse notation slightly by writing
    $\Hyb_9\ket{\psi}$ (respectively $\Hyb_{10}\ket{\psi}$) to be one
    query in $\Hyb_9$ (respectively $\Hyb_{10}$) applied to $\ket{\psi}$.

    % CHANGED: full-state comparison in vector norm.
    {\color{black}
    For this comparison, use the same purification of the common
    history-check measurement in both hybrids, and include its
    outcome and all work registers in the states. 
    %Write $U,V$ for the resulting complete query operations for $\Hyb_9,\Hyb_{10}$ respective,
    respectively. They agree whenever no early answer is triggered,
    so
    \[
        (\Hyb_9-\Hyb_{10})(I-\Pi_{\mathsf{early}})\ket{\psi}=0.
    \]
Since we have
\begin{align*}
    0
    &\leq
    \left|1-\bra{\psi}\Hyb_9^\dagger\Hyb_{10}\ket{\psi}\right|^2\\
    &=
    1-2\operatorname{Re}
    \left(\bra{\psi}\Hyb_9^\dagger\Hyb_{10}\ket{\psi}\right)
    +\left|\bra{\psi}\Hyb_9^\dagger\Hyb_{10}\ket{\psi}\right|^2.
\end{align*}
Rearranging this inequality and using the trace-distance formula
for normalized pure states gives
\begin{align*}
    \mathsf{TD}[\Hyb_9\ket{\psi},\Hyb_{10}\ket{\psi}]^2
    &=
    1-\left|\bra{\psi}\Hyb_9^\dagger\Hyb_{10}\ket{\psi}\right|^2\\
    &\leq
    2-2\operatorname{Re}
    \left(\bra{\psi}\Hyb_9^\dagger\Hyb_{10}\ket{\psi}\right)\\
    &=
    \bigl\|\Hyb_9\ket{\psi}\bigr\|^2
    +\bigl\|\Hyb_{10}\ket{\psi}\bigr\|^2
    -2\operatorname{Re}
    \left(\bra{\psi}\Hyb_9^\dagger\Hyb_{10}\ket{\psi}\right)\\
    &=
    \bigl\|\Hyb_9\ket{\psi}-\Hyb_{10}\ket{\psi}\bigr\|^2\\
    &=
    \bigl\|(\Hyb_9-\Hyb_{10})\ket{\psi}\bigr\|^2.
\end{align*}
    Therefore,
    \begin{align*}
        \mathsf{TD}[\Hyb_9\ket{\psi},\Hyb_{10}\ket{\psi}]
        &\leq \bigl\|(\Hyb_9-\Hyb_{10})\ket{\psi}\bigr\| \\
          &\leq \bigl\|(\Hyb_9-\Hyb_{10})\Pi_{\mathsf{early}}\ket{\psi} + (\Hyb_9-\Hyb_{10})(I-\Pi_{\mathsf{early}})\ket{\psi}\bigr\| \\
        &=\bigl\|(\Hyb_9-\Hyb_{10})\Pi_{\mathsf{early}}\ket{\psi}\bigr\|.
    \end{align*}
    If $\Pi_{\mathsf{early}}\ket{\psi}=0$, this distance is zero.
    Otherwise, set
    \[
        \ket{\eta}
        =\frac{\Pi_{\mathsf{early}}\ket{\psi}}
               {\|\Pi_{\mathsf{early}}\ket{\psi}\|}.
    \]
    It suffices to bound $\|(\Hyb_9-\Hyb_{10})\ket{\eta}\|$ uniformly for such
    normalized early-answer states satisfying the cache invariants.
    The bound for the original state is then multiplied by
    $\|\Pi_{\mathsf{early}}\ket{\psi}\|$.
    }

    Recall how a query $H(i,x)$ works. It uses an operation
    $\Decomp_{i,x}$, which swaps
    \[
        \ket{D'}\leftrightarrow
        {\color{black}\frac{1}{\sqrt{|\cR|2^\secp}}}
        \sum_{r\in\cR\times\{0,1\}^\secp}
        \ket{D'\cup\{((i,x),r)\}},
    \]
    where $D'(i,x)=\bot$
    {\color{black}and fixes the orthogonal complement of these two
    vectors, for each fixed residual database $D'$.}
    Then, it classical-copies the result $r$ to the output register
    and re-applies $\Decomp_{i,x}$.

    % CHANGED: match the actual cache operations in Hyb_9.
    {\color{black}
    On query $(s,(x,\inp',s_{i_s+1}))$, set
    $i_s=\serialindex[s]$. Hybrid $\Hyb_9$ first performs
    \textbf{Cache Clear}. It then computes
    $P_\OSP(x,\inp',s_{i_s+1};s;r)$ using two queries to $H(i_s,x)$
    to compute and uncompute $r$, and finally performs
    \textbf{Cache Fill}.
    }
    The middle \textbf{evaluation} step can be simplified by
    expanding how queries work.
    {\color{black}The two middle $\Decomp$ cancel: the intermediate
    computation uses $r$ in the work register and uses only
    predecessor database cells to compute the program state.}
    This results in the following procedure for $\Hyb_9$, conditioned
    on $\Pi_{\mathsf{early}}$.
    \begin{enumerate}
        \item {\color{black}Perform \textbf{Cache Clear} as defined in
        $\Hyb_9$. On an early-answer state, this clears
        $\serialindex[s_{i_s+1}]$ and $\inputcache[i_s]$; it leaves
        $\serialindex[s]$, $\inputcache[i_s+1]$, and
        $\outputcache[i_s]$ unchanged.}

        \item $\Decomp_{i_s,x}$.

        \item Compute and classical-copy
        $P_\OSP(x,\inp',s_{{\color{black}i_s}+1};s;r)$ to the output
        register, controlled on $r=D(i_s,x)$,
        {\color{black}and flip $b$. Uncompute the temporary registers.}

        \item $\Decomp_{i_s,x}$.

        \item {\color{black}Perform \textbf{Cache Fill} as defined in
        $\Hyb_9$.}
    \end{enumerate}

    \noindent Now consider the following sub-hybrid experiments
    where we gradually modify $\Hyb_9$ into $\Hyb_{10}$
    {\color{black}on the early-answer component}.
    \begin{itemize}
        \item $\Hyb_{9,a}$ removes the first $\Decomp$ operation.

        \item $\Hyb_{9,b}$ removes the second $\Decomp$ operation.

        \item $\Hyb_{9,c}$ swaps the order of clearing the cache
        and the computation of
        $P_\OSP(x,\inp',s_{{\color{black}i_s}+1};s;r)$.

        At this point, {\color{black}\textbf{Cache Clear} and
        \textbf{Cache Fill}} take place immediately one after another and therefore cancel.

        \item $\Hyb_{10}$ differs from $\Hyb_{9,c}$ only by answering
        the query using the pre-computed value of $P_\OSP$ stored in
        the output cache, instead of computing it from scratch using
        randomness $r=D(i_s,x)$. Using the cache invariant on
        $\Hyb_{10}$, this is equivalent.
    \end{itemize}
    {\color{black}On the early-answer component,} the fact that
    \[
        \Hyb_{9,b}=\Hyb_{9,c}
    \]
    follows from the observation that
    {\color{black}clearing the cache leaves the database and query
    inputs unchanged, while evaluating $P_\OSP$ does not use the
    cache entries being cleared: the current input is in the query
    register, and computing the program state uses only predecessor
    entries. Thus these two operations commute. With both
 $\Decomp$ removed, the queried database cell remains
    present, so the two cache operations restore precisely the
    entries they clear.}

    % CHANGED: identify the two intermediate states to be bounded.
    {\color{black}
    Next, we bound the vector distance between
    $\Hyb_9\ket{\eta}$ and $\Hyb_{9,a}\ket{\eta}$, and between
    $\Hyb_{9,a}\ket{\eta}$ and $\Hyb_{9,b}\ket{\eta}$.
    It suffices to show that $\Decomp$ is close to the identity
    on the state immediately before each removed operation.
    For the second comparison, this is the state in $\Hyb_{9,a}$,
    where the first decompression has already been removed.
    }

    An early answer is triggered for query
    {\color{black}$(s,\inp_x)$} if
    $\exists(s_{\max}:i_{\max})\in\serialindex$ such that
    $\serialindex[s]<i_{\max}{\color{black}-1}$.
    Here we denote $i_s\coloneqq\serialindex[s]$ and
    {\color{black}$\inp_x=(x,\inp',s_{i_s+1})$.}
    In particular, there must be an $s_{i_s+2}$ where
    $\serialindex[s_{i_s+2}]=i_s+2$.
    Combining this with the output cache invariant
    (\Cref{claim:REQ-hyb-oracle_output-cache-invariant}) and history
    check, there must be entries
    \begin{align*}
        ((i_s,x),r)&\in D,
        \qquad\text{where }r=*\concat r'_{i_s+1},\\
        (i_s+1:
        {\color{black}(*,*,r'_{i_s+1}\concat *,*,s_{i_s+2})})
        &\in\inputcache.
    \end{align*}
    Furthermore, these two properties also imply the existence of
    input and output cache entries for $s$ and $\inp_x$, which are
    constrained by the SEQ check to be
    \begin{align*}
        (i_s:\inp_x)&\in\inputcache,\\
        (i_s:{\color{black}P_\OSP(\inp_x;s;r)})&\in\outputcache.
    \end{align*}
    {\color{black}Here the input-cache entries display their input
    component, suppressing the accompanying serial number when
    it is stored as part of the entry.}
    In particular, the input cache entry $\inputcache[i_s+1]$
    {\color{black}records $r'_{i_s+1}$ and} is not modified by
    answering a query to $s$, and the database entry
    $((i_s,x),*\concat r'_{i_s+1})\in D$ can only be modified by
    $\Decomp$ operations.

    % CHANGED: bound both removed decompressions in vector norm.
    {\color{black}
    Let $\ket{\chi}$ be the normalized state immediately before
    either of the two decompressions being removed. Before the first,
    \textbf{Cache Clear} has not changed the queried database cell or
    the successor input-cache entry. Before the second in
    $\Hyb_{9,a}$, the first decompression is absent, and the
    intermediate evaluation also leaves both of these records
    unchanged. Thus the queried cell is non-$\bot$, and its last
    $\secp$ bits agree with $r'_{i_s+1}$ recorded in
    $\inputcache[i_s+1]$ in both cases. The queried address is also
    recorded outside the database by the query input and the
    unchanged entry $\serialindex[s]$.

    We apply the vector estimate from
    \Cref{lem:decomp-approx-identity}, with
    $\cR_1=\cR$ and $\cR_2=\{0,1\}^\secp$.
    For completeness, the same estimate holds with arbitrary
    entanglement among the remaining registers. Group the state
    according to the queried address $(i_s,x)$, the recorded value
    $r'_{i_s+1}$, and the residual database. Write such a group as
    $j=(i_j,x_j,z_j,D'_j)$, where $D'_j(i_j,x_j)=\bot$. Then
    \[
        \ket{\chi}
        =\sum_j\sum_{r\in\cR}\ket{\xi_{j,r}}
          \otimes
          \ket{D'_j\cup\{((i_j,x_j),r\concat z_j)\}},
        \qquad
        \sum_j\sum_{r\in\cR}\|\ket{\xi_{j,r}}\|^2=1.
    \]
    The vectors $\ket{\xi_{j,r}}$ contain all registers other than
    the database. Distinct values of $(i_j,x_j,z_j)$ have orthogonal
    external records; for fixed $(i_j,x_j,z_j)$, distinct residual
    databases remain orthogonal.

    Let $P_+$ project the queried cell onto its uniform non-$\bot$
    state, controlled on the queried address. By Cauchy--Schwarz,
    \begin{align*}
        \|P_+\ket{\chi}\|^2
        &=\frac{1}{|\cR|2^\secp}
          \sum_j\biggl\|\sum_{r\in\cR}\ket{\xi_{j,r}}\biggr\|^2\\
        &\leq\frac{1}{2^\secp}
          \sum_j\sum_{r\in\cR}\|\ket{\xi_{j,r}}\|^2
         =2^{-\secp}.
    \end{align*}
    There is no $\bot$ component in the queried cell. Since
    $\Decomp$ swaps its uniform non-$\bot$ state with $\ket{\bot}$
    and fixes their orthogonal complement, the removed uniform
    component and the newly created $\bot$ component are orthogonal.
    Let $\ket{v}:=P_+\ket{\chi}$. Hence
\begin{align*}
    \bigl\|(\Decomp-I)\ket{\chi}\bigr\|^2
    &= \bigl\|\Decomp\ket{v}-\ket{v}\bigr\|^2\\
    &= \bigl\|\Decomp\ket{v}\bigr\|^2
       +\bigl\|\ket{v}\bigr\|^2
       -2\operatorname{Re}\bigl(\bra{v}\Decomp\ket{v}\bigr)\\
    &= \bigl\|\ket{v}\bigr\|^2+\bigl\|\ket{v}\bigr\|^2\\
    &= 2\bigl\|P_+\ket{\chi}\bigr\|^2 \leq\frac{1}{2^{\secp-1}}.
\end{align*}

    The operations following each replacement are identical
    isometries in the two purified circuits, so they preserve
    vector norm. Therefore
    \begin{align*}
        \bigl\|(\Hyb_9-\Hyb_{9,a})\ket{\eta}\bigr\|
        &\leq\sqrt{1/2^{\secp-1}},\\
        \bigl\|(\Hyb_{9,a}-\Hyb_{9,b})\ket{\eta}\bigr\|
        &\leq\sqrt{1/2^{\secp-1}}.
    \end{align*}
    Combining these bounds with the exact remaining changes gives
    \[
        \bigl\|(\Hyb_9-\Hyb_{10})\ket{\eta}\bigr\|
        \leq2\sqrt{1/2^{\secp-1}}.
    \]
    By the initial reduction and homogeneity,
    \[
        \mathsf{TD}[\Hyb_9\ket{\psi},\Hyb_{10}\ket{\psi}]
        \leq\bigl\|(\Hyb_9-\Hyb_{10})\ket{\psi}\bigr\|
        \leq2\sqrt{1/2^{\secp-1}}\,
             \|\Pi_{\mathsf{early}}\ket{\psi}\|
        \leq2\sqrt{1/2^{\secp-1}}.
    \]

    Let $q$ be the number of oracle calls made by the adversary.
    %Summing over the $q$ complete query operations, including the
    adversary's operations.
    %Use $\Hyb_{10}$ for the prefix before each replaced query: its early answers leave the   database and caches unchanged, and its other queries preserve    the preceding cache invariants. Thus each state at which a    query is replaced satisfies the invariants used above. 
    Writing $\ket{\psi_t}$ for the state before query $t$ in   $\Hyb_{10}$ and $\ket{\Psi_9},\ket{\Psi_{10}}$ for the final  purified states, we have
    \[
        \bigl\|\ket{\Psi_9}-\ket{\Psi_{10}}\bigr\|
        \leq\sum_{t=1}^q\bigl\|(\Hyb_9-\Hyb_{10})\ket{\psi_t}\bigr\|
        \leq2q\sqrt{1/2^{\secp-1}}=\negl(\secp)
    \]
    for polynomial $q$. Their trace distance, including the
    recorded outcomes of the history checks, is at most the same
    bound. This proves $\Hyb_9\approx_s\Hyb_{10}$, i.e. statistically indistinguishable.
    }

    Finally, since the two experiments are statistically
    indistinguishable and the history check failing can be detected,
    it must happen with almost the same probability in $\Hyb_{10}$
    as in $\Hyb_9$, i.e. negligible.
\end{proof}

%\proofaudit{P15}{revised proof ends}

        \begin{claim}[Hybrid: Output Cache Queries $H$]\label{claim:REQ-hyb-oracle_outcache-H}\label{claim:REQ-hyb-oracle_h10-h11}
            For all adveraries,
            \[
                \Hyb_{10} \approx_s \Hyb_{11}
            \]
            Furthermore, the \textbf{history consistency check} fails with negligible probability in $\Hyb_9$.
        \end{claim}
        \begin{proof}
            The only difference between $\Hyb_{10}$ and $\Hyb_{11}$ is in the caching routine, specifically when updating the output cache $\outputcache[\serialindex[s]]$ in step \ref{step:REQ-hybrid-oracle-sim-output-cache-H}.\footnote{The usage of the bit $g$ is implicit in $\Hyb_{10}$.}
            In particular, in $\Hyb_{11}$ the randomness $r$ is read directly from $D$'s database, while in $\Hyb_{11}$ the random oracle $H$ is queried to obtain $r$. Simplifying the computation as before, the output cache update in the two hybrids can be written as
            \begin{enumerate}
                \item ($\Hyb_{11}$ only) Compute $\Decomp_{\serialindex[s],x}$ on the database register.
                \item Compute and classical-copy $P_{\OSP}(x,\inp', s_{i+1}; s; r)$ to the output cache register using a symmetric difference, controlled on $r = D(i_s, x)$.
                \item ($\Hyb_{11}$ only) Compute $\Decomp_{\serialindex[s],x}$ on the database register.
            \end{enumerate}
            Thus, we need to show that whenever the output cache would be either cleared or filled, the two $\Decomp$ operations act statistically close to the identity. 
            The output cache is only cleared or filled at index $i_s-1$ when a query is made to a to serial number $s$ where $\serialindex[s] = i_s$ and an early answer is \emph{not} triggered.\footnote{Unlike in the proof of \Cref{claim:REQ-hyb-oracle_early-answer}, we cannot rely on the existence of a ``later'' entry in the cache because the early answer does \emph{not} trigger, meaning there is no later entry recorded.}
            By the history check, there exists an entry $(i_s-1, x_{i_s-1}, r_{i_s-1}) \in D$ such the $\inp_{i_s}$ part of the query contains the last that the last $\secp$ bits of $r_{i_s-1}$. By the SEQ check and cache invariant, there exists an entry $\inputcache[i_s - 1] \neq \bot$ containing $x_{i_s -1}$. Therefore by \Cref{lem:decomp-approx-identity}, $\Decomp$ acts close to the identity on the query $(i_s-1, x_{i_s-1}$ which is made to modify the output cache.

            Finally, since the two experiments are statistically indistinguishable and the history check failing can be detected, it must happen with almost the same probability in $\Hyb_{10}$ as in $\Hyb_{9}$, i.e. negligible.
        \end{proof}

        \begin{claim}[Hybrid: Remove History Check]\label{claim:REQ-hyb-oracle_remove-history}\label{claim:REQ-hyb-oracle_h11-h12}
            For all QPT adversaries, 
            \[
                \Hyb_{11} \approx_c \Hyb_{12}
            \]
        \end{claim}
        \begin{proof}
            The only difference between $\Hyb_{11}$ and $\Hyb_{12}$ is $\Hyb_{11}$ measures the history check and aborts if it fails. By \Cref{claim:REQ-hyb-oracle_early-answer}, this happens with negligible probability and so is gentle.
        \end{proof}

        \begin{claim}[Hybrid: Cache Uses Oracle]\label{claim:REQ-hyb-oracle_reflect-cache}\label{claim:REQ-hyb-oracle_h12-h13}
            For all adversaries,
            \[
                \Hyb_{12} = \Hyb_{13}
            \]
        \end{claim}
        \begin{proof}
            By \Cref{claim:REQ-sim-hyb-reflect-claim}, $\Reflect_{i}^{O_\Hyb^{P}}$ is equivalent to flipping $g$ controlled on $D$ not containing any entries $((i, *), *)$. Since $g$ starts at $1$, the cache steps activate in $\Hyb_{12}$ if and only if there \emph{is} an entry $((i, *), *) \in D$, the same as in $\Hyb_{11}$.
        \end{proof}

        \begin{claim}[Hybrid: Evaluation uses Oracle]\label{claim:REQ-hyb-oracle_eval-oracle}\label{claim:REQ-hyb-oracle_h13-h14}
            For all adversaries,
            \[
                \Hyb_{13} = \Hyb_{14}
            \]
        \end{claim}
        \begin{proof}
            Expanding the definition of $O_\Hyb^{P}$, there are three differences between $\Hyb_{13}$ and $\Hyb_{14}$. 
            First, the computation of $P(\st, x; r_x)$ is separated from the computation of the signatures. This clearly is only a syntactic difference. 
            
            Second, the compressed oracle $H$ is evaluated a second time to uncompute $r_x$ (note that $r_2$ is provided as part of the output of $O_\Hyb^{P}$, and so is not uncomputed) at the end of querying $O_\Hyb^{P}$ in step \ref{step:REQ-hybrid-oracle-sim-query-oracle} and start of unquerying $O_\Hyb^{P}$ in the last step of \textbf{Evaluation}.
            Observe that the steps in between querying $O_\Hyb^{P}$ do not operate on the database register $\cD$ or the work register used for querying $H$ (the output $(y, r_2)$ is moved to a separate register). Therefore the extra two queries to $H$ commute with the operations in between querying  $O_\Hyb^P$ and cancel with each other.

            Third, $O_\Hyb^P$ only answers queries $\ket{i, x, u, b}$ if $\exists ((i, x), *)\in D$ and no other entry has a higher index $i'$, or if $i$ is one higher than the highest index recorded in $D$.
            By the cache invariant (\Cref{claim:REQ-sim-SEQ-to-hyb_proof-inputcache-invariant}) and the \textbf{early answer} step, $\Hyb_{12}$ does not do \textbf{Evaluation} if there is an entry $((i', *), *) \in D$ where $i' > i$. Furthermore, if $i$ matches the maximum entry $((i, x'), *) \in D$, then the \textbf{SEQ check} guarantees that $x' = x$.
        \end{proof}

        Combining these claims, $A^{O_{\SEQ}^{P_\OSP[\sk, \pk, \sk_{\SKE}]}} = \Hyb_0$ is computationally indistinguishable from $\Hyb_{12} = A^{\Sim_1^{O_{\Hyb}^P}}$
\end{proof}

\subsubsection{Simulating the Hybrid Oracle}
\label{sec:LEQ:proof:simulating-hybrid}

Here we show that the hybrid oracle $O_{\Hyb}^{P}$ (\cref{fig:REQ-hybrid-oracle}) can be simulated with the REQ oracle $O_P^{\REQ}$ (\cref{fig:REQ}). The simulation is indistinguishable from the real hybrid oracle to any distinguisher that makes polynomially-many queries. Note that the distinguisher implicitly knows the program $P$ because the distinguisher is quantified after $P$, so the distinguisher may have $P$ hardwired.

\begin{prop}\label{prop:REQ-hyb-to-REQ}
    There exists a simulator $\Sim$ such that for all RAM programs $P$ and all distinguishers $D(1^\secp)$ that make $\mathsf{poly}(\secp)$-many queries to a given oracle, there is a negligible function $\negl(\cdot)$ such that for all $\secp \in \bbN$,
    \[
        \abs{\Pr[1 \gets D^{O_{\Hyb}^{P}(1^\secp)}(1^\secp)] - \Pr[1 \gets D^{\Sim^{O_P^{\REQ}}(1^\secp)}(1^\secp)]} \leq \negl(\secp)
    \]
    where $O_P^{\REQ}$ is the oracle given in \cref{fig:REQ}, and $O_{\Hyb}^{P}$ is the oracle given in \cref{fig:REQ-hybrid-oracle}.
\end{prop}
\begin{proof}
$ $
\subsubsection*{The simulator.} First, we describe the simulator $\Sim$ in \cref{fig:simulator-REQ-hybrid-oracle}.

\begin{figure}[H]
\begin{mdframed}
    \paragraph{Registers:} The simulator acts on the following registers:
    \begin{itemize}
        \item $\cQ$: The query register taken as input. $\cQ$ has standard basis states of the form $\ket{t, x, u, b}_\cQ$. $t$ denotes the tag, $x$ denotes the input, $u$ is the register where the output will be written, and $b$ serves as an indicator for whether the query was answered.
        
        \item $\cC = \cC_1 \otimes \dots \otimes \cC_{k'}$: An internal register of variable length that stores a cache $C = [(t_i, x'_i, y_i, g'_i)]_{i \in [k']}$.
        $t_i$ is the tag, $x'_i$ is the input, $y_i$ is the output, and $g'_i$ is a bit indicating whether this cache location is in use ($g'_i = 0$) or not ($g'_i = 1$) .
        Initially, $\cC$ has length $0$.
        
        \item $\cD$: An internal register that represents the database of a compressed oracle $G: [2^\secp] \times \cX \to \bit^\secp$. $\cD$ is initialized to $\ket{\emptyset}$.

        % \item $\cW$: An internal work register that is used to query the compressed oracle $G$. $\cW$ is initialized to $\ket{0}$, and $\cW = \ket{0}$ at the start of every query.

        \item $\cB'$: An internal register that decides whether to answer the query from the user. $\cB'$ is initialized to $\ket{0}$, and additionally $\cB' = \ket{0}$ at the start of every query.
    \end{itemize}
\end{mdframed}
\caption{Registers for The Simulator $\Sim^{O^\LEQ_P}$ for $O_\Hyb^P$}\label{fig:simulator-REQ-hybrid-oracle-registers}
\end{figure}

\begin{figure}[H]
\begin{mdframed}
    \justin{Just formatting and organizational tweaks so far.}
    \paragraph{Queries.} To respond to a query, the simulator acts on $\cQ \times \cC \times \cD \times \cW \times \cB'$ as the following unitary. Given some basis state $\ket{t, x,u,b}_\cQ \otimes \ket{C}_\cC \otimes \ket{D}_{\cD} \otimes \ket{\mathbf{0}}_\cW \otimes \ket{0}_{\cB'}$, the simulator does the following:

\begin{enumerate}
        \item \label{sim-step:expand-registers}\textbf{Expand $\cC$ by Two Registers.} Append two new entries $\ket{0, 0, 0, 0}_{\cC_{k'+1}} \otimes \ket{0, 0, 0, 0}_{\cC_{k'+2}}$ to $\cC$.
        
        \item \textbf{Validate Query.} \label{sim-step:compute-B} Let $k_{\max}$ be the maximum index $i \in [k']$ such that $g'_i = 1$, or if no such $i$ exists, then let $k_{\max} = 0$. 
        
        If $\cQ$ contains $t = k_{\max}$ and $x = x'_{k_{\max}}$, or if $\cQ$ contains $t = k_{\max}+1$, then apply $\mathsf{X}$ to $\cB'$.
        % \item \textbf{Answer Query.} If $\cB'$ contains $1$, then answer the query as follows.
        % \begin{enumerate}
            % \item If $t = k_{\max} + 1$, then append a new entry $\ket{0, 0, 0, 0}_{\cC_t}$ to $\cC$.
        \item \textbf{Manage Cache.}\label{step:sim-for-hybrid-oracle-manage-cache}
        \begin{enumerate}
            \item \label{sim-step:refl-and-copy-1}$\mathsf{ReflectAndCopy}$: If $g'_{t} = 0$% and $\cD$ does not contain an entry of the form $(t,x,*)$
            , then CNOT $(t,x)$ from $\cQ$ to $\cC_{t}$:
                \[\ket{t_{t},x'_{t},y_{t}, 0}_{\cC_{t}} \mapsto \ket{t_{t} \oplus t,x'_{t} \oplus x,y_{t}, 0}_{\cC_{t}}\]
            % \item \label{sim-step:eval-and-copy}$\mathsf{EvalAndCopy}$:
            % \begin{enumerate}
                % \item \label{sim-step:query-random-oracle}Apply $\Decomp_{t,x}$ to $\cD$.
            \item \label{sim-step:decomp} If $g_i = 1$, $\Decomp_{t,x}$ to $\cD$.
            \justin{Added this step as part of the fix.}
            
            \item \label{sim-step:query-REQ-oracle}\textbf{Query $O^\REQ_P$:} If $\cD$ is in the span of states of the form
            \begin{align*}
                \ket{D} \quad \text{or} \quad \frac{1}{\sqrt{2^{\secp}}} \cdot \sum_{r_2 \in \bit^\secp} \ket{D \cup \{(t,x,r_2)\}}
            \end{align*}
            for all databases $D$ that do not contain $(t,x,*)$, 
            then query $O^\REQ_P$ on $\cC_t$.\footnote{The last bit $g_t$ in $\cC_t$ serves as the response indicator bit for this query. It flips if the query is answered.}
            Otherwise, query $O^\REQ_P$ on a dummy query with $t = -1$.
            \item \label{sim-step:decomp-undo} Repeat step \ref{sim-step:decomp}.
            % \item \label{sim-step:query-random-oracle-2}Apply $\Decomp_{t,x}$ to $\cD$. Then apply $\CO'$ to $\cQ_u \times \cD$:
            % \[\ket{u}_{\cQ_u} \leftrightarrow \ket{u \oplus (0, r_2)}_{\cQ_u}\]
            \item \label{sim-step:refl-and-copy-2}$\mathsf{ReflectAndCopy}$: Repeat step \ref{sim-step:refl-and-copy-1}.
        \end{enumerate}
        \item \label{sim-step:copy-answer} \textbf{Answer Query:} If $\cB'$ contains $1$, then answer the query by copying $(y, r_2)$ from $\cC_{t} \times \cD$ onto $\cQ$ and flipping $b$:
        \[\ket{t,x,u,b}_\cQ \mapsto \ket{t, x, u \oplus (y, r_2), b\oplus 1}_\cQ\]
        \proofauditloc{P18}{This answer flips $b$ using the precomputed acceptance flag, even on the intermediate blank-history component. There is some later inconsistency in the writing on whether to flip $b$}%
        % If $(t,x,*) \notin D$, then set $r_2 = 0$.
        \item \textbf{Manage Cache.} Repeat step \ref{step:sim-for-hybrid-oracle-manage-cache}.
        % as follows:
        \label{sim-step:refl-and-copy-3}
        \label{sim-step:refl-and-copy-4}
        % \begin{enumerate}
            %     \item \label{sim-step:refl-and-copy-3}$\mathsf{ReflectAndCopy}$: Repeat step \ref{sim-step:refl-and-copy-2}.
            %     \item ($\Decomp_{t,x}$) Repeat step \ref{sim-step:decomp}.
            %     \item (Query $O^\REQ_P$) Repeat step \ref{sim-step:query-REQ-oracle}.
            % \item ($\Decomp_{t,x}$) Repeat step \ref{sim-step:decomp}.
            %     % \item Apply $\Decomp_{t,x}$ to $\cD$.
            %     % \item \label{sim-step:uncompute-oracle-queries} Repeat step \ref{sim-step:query-random-oracle} (querying $G$) and then step \ref{sim-step:query-REQ-oracle} (querying $O^\REQ_P$).
            % % \end{enumerate}
            % \item \label{sim-step:refl-and-copy-4}$\mathsf{ReflectAndCopy}$: Repeat step \ref{sim-step:refl-and-copy-1}.
            % % \item If $D$ does not contain $(t,x,*)$, then query $O^\REQ_P$ on $\cC_{k_{\max}}$
            % % \item If $b_t = 0$, then CNOT $(t,x)$ from $\cQ$ to $\cC_t$:
            % % \[\ket{t_{t},x_{t},y_{t}, 0}_{\cC_{t}} \mapsto \ket{t_{t} \oplus t,x_{t} \oplus x,y_{t}, 0}_{\cC_{t}}\]
            % % \item If $\cC_t = \ket{0,0,0,0}$, then remove register $\cC_t$.
        % \end{enumerate}
        \item (Validate Query) Repeat step \ref{sim-step:compute-B}.
\end{enumerate}
% \justin{Cases:
% \begin{itemize}
%     \item $r_2 \notin D$ and $g_t = 0$: Between steps 5-7, $g_t = 1$. Attempt to erase in step 8 iff $r_2\notin D$.
%     \item $r_2 \notin D$ and $g_t = 1$: Between steps 5-7, $g_t = 1$. Attempt to erase in step 8 iff $r_2\notin D$.
%     \item $r_2 \in D$ and $g_t = 0$: Not possible
%     \item $r_2 \in D$ and $g_t = 1$: Between steps 5-7, $g_t = 1$. Attempt to erase in step 8 iff $r_2\notin D$.
% \end{itemize}
% }
\end{mdframed}
\caption{The Simulator $\Sim^{O^\LEQ_P}$ for $O_\Hyb^P$}\label{fig:simulator-REQ-hybrid-oracle}
\end{figure}

\subsubsection*{State Transition Function of $O^\REQ_P$.}
We show that step \ref{step:REQ-answer-query} of $O^\REQ_P$ can be replaced with an operation that simply swaps two states (\cref{thm:state-transitions-of-REQ}).

\begin{definition}
    For any query $(t,x,u,b) \in [2^\secp] \times \cX \times \cY \times \bit$ to $O^\REQ_P$ and any program state $\st$, let
    \[\ket{\phi^\REQ_{t,x,u,b,\st}} = \frac{1}{\sqrt{\abs{\cR}}} \cdot \sum_{r_1 \in \cR} \ket{t,x,u \oplus y_{r_1},b \oplus 1}_\cQ \otimes \ket{\st'_{x,{r_1}}}_\cS \otimes \ket{\st, x, {r_1}}_{\cA_t} \otimes \ket{1}_{\cG_t}\]
    where $(\st'_{x,r_1}, y_{r_1}) = P(\st, x; r_1)$.
\end{definition}

\begin{definition}[Reachable States of $O^\REQ_P$]\label{def:reachable-states-of-REQ}
    Let the state of $S^\REQ$ be the space of states on $O^\REQ_P$'s internal registers for which $\cG_i = \ket{1}$ or $\cG_i \times \cA_i = \ket{0} \otimes \ket{\mathbf{0}}$ for all $i \in [k]$.
\end{definition}

Then \cref{thm:state-transitions-of-REQ} describes the state transition function of step \ref{step:REQ-answer-query} on any state in $S^\REQ$. 
\begin{lemma}\label{thm:state-transitions-of-REQ}
    For any query $(t,x,u,b) \in [2^\secp] \times \cX \times \cY \times \bit$ to $O^\REQ_P$ and any program state $\st$, step \ref{step:REQ-answer-query} of $O^\REQ_P$ swaps the following states if $\cB = \ket{1}$:
    \[\ket{t,x,u,b}_\cQ \otimes \ket{\st}_\cS \otimes \ket{\mathbf{0}}_{\cA_t} \otimes \ket{0}_{\cG_t} \leftrightarrow \ket{\phi^\REQ_{t,x,u,b,\st}}\]
    and acts as the identity on all states in $S^\REQ$ that are orthogonal to the ones above.
\end{lemma}
\begin{proof}
We will consider an orthonormal basis for $S^\REQ$ and show that for every basis state, step \ref{step:REQ-answer-query} acts the way that \cref{thm:state-transitions-of-REQ} claims.

    \paragraph{Case 1:} The input to step \ref{step:REQ-answer-query} has $\cB = \ket{0}$.

    Then step \ref{step:REQ-answer-query} aborts and acts as the identity on this state.
    
    \paragraph{Case 2:} The input to step \ref{step:REQ-answer-query} is has $\cA_t \times \cG_t \times \cB = \ket{\mathbf{0}} \otimes \ket{0} \otimes \ket{1}$. 
    
    Then the input is in the span of states of the form: $\ket{t,x,u,b}_\cQ \otimes \ket{\st}_\cS \otimes \ket{\mathbf{0}}_{\cA_t} \otimes \ket{0}_{\cG_t} \otimes \ket{1}_\cB$.
    
    Let us step through the action of step \ref{step:REQ-answer-query} on this state. Since $\cB = \ket{1}$, step \ref{step:REQ-answer-query} does not abort. Next, step \ref{step:REQ-close} does nothing because $\cG_t = \ket{0}$. Step \ref{step:REQ-switch-mode} verifies that $\cA_t = \ket{\mathbf{0}}$ and then maps:
    \[\ket{0}_{\cG_t} \otimes \ket{b}_{\cQ_b} \mapsto \ket{1}_{\cG_t}  \otimes \ket{b \oplus 1}_{\cQ_b}\]
    Finally, step \ref{step:REQ-open} verifies that $\cG_t = \ket{1}$ and then applies $U_P$, which acts as follows:
    \begin{align*}
        &\ket{t, x, u, b \oplus 1}_{\cQ} \otimes \ket{\st}_{\cS} \otimes \ket{\mathbf{0}}_{\cA_i} \otimes \ket{1}_{\cG_t}\\
        &\mapsto \frac{1}{\sqrt{|\cR}|}\sum_{r_1 \in \cR} \ket{t, x, u \oplus y_{r_1}, b \oplus 1}_{\cQ} \otimes \ket{\st'_{x,r_1}}_{\cS} \otimes \ket{\st, x, r_1}_{\cA_i} \otimes \ket{1}_{\cG_t}\\
        &= \ket{\phi^\REQ_{t,x,u,b,\st}}
    \end{align*}
    This shows that step \ref{step:REQ-answer-query} maps
    \[\ket{t,x,u,b}_\cQ \otimes \ket{\st}_\cS \otimes \ket{\mathbf{0}}_{\cA_t} \otimes \ket{0}_{\cG_t} \mapsto \ket{\phi^\REQ_{t,x,u,b,\st}}\]

    \paragraph{Case 3:} The input to step \ref{step:REQ-answer-query} has $\cA_t \times \cG_t \times \cB = \ket{a} \otimes \ket{0} \otimes \ket{1}$, where $a \neq \mathbf{0}$.

    This state is not in $S^\REQ$, so we don't need to consider it.

    \paragraph{Case 4:} The input $\ket{\psi}$ to step \ref{step:REQ-answer-query} has $\cG_t \times \cB = \ket{1} \otimes \ket{1}$ and satisfies $\ket{\psi} = U_P \cdot \ketbra{\mathbf{0}}_{\cA_t} \cdot U_P^\dag \cdot \ket{\psi}$. 
    
    We claim that in this case, $\ket{\psi}$ is in the span of states of the form $\ket{\phi^\REQ_{t,x,u,b,\st}}$. First, $U_P^\dag \cdot \ket{\psi}$ is in the span of $\ketbra{\mathbf{0}}_{\cA_t} \otimes \ketbra{1}_{\cG_t} \otimes \ketbra{1}_\cB$. Then
    \[U_P^\dag \cdot \ket{\psi} = \sum_{t,x,u,b',\st} \alpha(t,x,u,b',\st) \cdot \ket{t,x,u,b'}_\cQ \otimes \ket{\st}_\cS \otimes \ket{\mathbf{0}}_{\cA_t} \otimes \ket{1}_{\cG_t} \otimes \ket{1}_\cB\]
    for some amplitude function $\alpha$. Next, we set $b = b' \oplus 1$:
    \[U_P^\dag \cdot \ket{\psi} = \sum_{t,x,u,b,\st} \alpha(t,x,u,b \oplus 1,\st) \cdot \ket{t,x,u,b \oplus 1}_\cQ \otimes \ket{\st}_\cS \otimes \ket{\mathbf{0}}_{\cA_t} \otimes \ket{1}_{\cG_t} \otimes \ket{1}_\cB\]
    Finally, we apply $U_P$:
    \[\ket{\psi} = U_P \cdot U_P^\dag \cdot \ket{\psi} = \sum_{t,x,u,b,\st} \alpha(t,x,u,b \oplus 1,\st) \cdot \ket{\phi^\REQ_{t,x,u,b,\st}}\]

    This shows that $\ket{\psi}$ is in the span of states of the form $\ket{\phi^\REQ_{t,x,u,b,\st}}$.

    Next, let us step through the action of step \ref{step:REQ-answer-query} on each basis state $\ket{\phi^\REQ_{t,x,u,b,\st}}$. Since $\cG_t \times \cB = \ket{1} \otimes \ket{1}$, step \ref{step:REQ-answer-query} does not abort, and then step \ref{step:REQ-close} applies $U_P^\dag$, which acts as follows:
    \begin{align*}
        \ket{\phi^\REQ_{t,x,u,b,\st}} &= \frac{1}{\sqrt{\abs{\cR}}} \cdot \sum_{r_1 \in \cR} \ket{t,x,u \oplus y_{r_1},b \oplus 1}_\cQ \otimes \ket{\st'_{x,{r_1}}}_\cS \otimes \ket{\st, x, {r_1}}_{\cA_t} \otimes \ket{1}_{\cG_t} \otimes \ket{1}_\cB\\
        &\mapsto \ket{t,x,u,b \oplus 1}_\cQ \otimes \ket{\st}_\cS \otimes \ket{\mathbf{0}}_{\cA_t} \otimes \ket{1}_{\cG_t} \otimes \ket{1}_\cB
    \end{align*}

    Next, step \ref{step:REQ-switch-mode} verifies that $\cA_t = \ket{\mathbf{0}}$ and then does the following:
    \[\ket{1}_{\cG_t} \otimes \ket{b \oplus 1}_{\cQ_b} \mapsto \ket{0}_{\cG_t} \otimes \ket{b}_{\cQ_b}\]

    Finally, step \ref{step:REQ-open} does nothing because $\cG_t = \ket{0}$. In total, this shows that step \ref{step:REQ-answer-query} maps:
    \[\ket{\phi^\REQ_{t,x,u,b,\st}} \mapsto \ket{t,x,u,b}_\cQ \otimes \ket{\st}_\cS \otimes \ket{\mathbf{0}}_{\cA_t} \otimes \ket{0}_{\cG_t} \otimes \ket{1}_\cB\]

    \paragraph{Case 5:} The input $\ket{\psi}$ to step \ref{step:REQ-answer-query} has $\cG_t \times \cB = \ket{1} \otimes \ket{1}$ and satisfies $\ket{\psi} = U_P \cdot \left(I - \ketbra{\mathbf{0}}\right)_{\cA_t} \cdot U_P^\dag \cdot \ket{\psi}$.

    Let us step through the action of step \ref{step:REQ-answer-query} on $\ket{\psi}$. Since $\cG_t \times \cB = \ket{1} \otimes \ket{1}$, step \ref{step:REQ-answer-query} does not abort, and then step \ref{step:REQ-close} applies $U_P^\dag$ to $\ket{\psi}$. Note that $U_P^\dag \cdot \ket{\psi}$ is in the span of $\left(I - \ketbra{\mathbf{0}}\right)_{\cA_t}$. Next, step \ref{step:REQ-switch-mode} acts as the identity because the state is in the span of $\left(I - \ketbra{\mathbf{0}}\right)_{\cA_t}$. Finally, step \ref{step:REQ-open} applies $U_P$ because $\cG_t = \ket{1}$, which restores the state to
    \[U_P \cdot U_P^\dag \cdot \ket{\psi} = \ket{\psi}\]
    In total, this shows that step \ref{step:REQ-answer-query} acts as the identity in this case.
\end{proof}

The following corollary says that $S^\REQ$ contains all reachable states of $O^\REQ_P$.
\begin{corollary}
    After any number of quantum queries to $O^\REQ_P$, the state of $O^\REQ_P$'s internal registers is in $S^\REQ$ (\cref{def:reachable-states-of-REQ}).
\end{corollary}
\begin{proof}
    We will prove this inductively. The initial state of $O^\REQ_P$ is in $S^\REQ$ because $\cA$ and $\cG$ have length $k = 0$. Next, let us assume that at the start of a given query, the state of $O^\REQ_P$ is in $S^\REQ$. Then we will show that at the end of the query, the state is still in $S^\REQ$.
    
    During a query to $O^\REQ_P$, step \ref{REQ-step:expand-registers} appends registers $\ket{0}_{\cG_{k+1}} \otimes \ket{\mathbf{0}}_{\cA_{k+1}}$, which satisfy the condition to be in $S^\REQ$. Step \ref{REQ-step:validate-query} does not change the computational-basis values of $\cG \times \cA$, so the state after step \ref{REQ-step:validate-query} is still in $S^\REQ$. 
    
    According to \cref{thm:state-transitions-of-REQ}, step \ref{step:REQ-answer-query} swaps 
    \[\ket{t,x,u,b}_\cQ \otimes \ket{\st}_\cS \otimes \ket{\mathbf{0}}_{\cA_t} \otimes \ket{0}_{\cG_t} \otimes \ket{1}_\cB \leftrightarrow \ket{\phi^\REQ_{t,x,u,b,\st}}\] 
    and acts as the identity on all states in $S^\REQ$ that are orthogonal to the ones above. Note that both states above are in $S^\REQ$: $\ket{t,x,u,b}_\cQ \otimes \ket{\st}_\cS \otimes \ket{\mathbf{0}}_{\cA_t} \otimes \ket{0}_{\cG_t} \otimes \ket{1}_\cB$ satisfies $\cG_t \times \cA_t = \ket{0} \otimes \ket{\mathbf{0}}$ and $\ket{\phi^\REQ_{t,x,u,b,\st}}$ satisfies $\cG_t = \ket{1}$. Therefore, step \ref{step:REQ-answer-query} maps any state in $S^\REQ$ to a state in $S^\REQ$.
    
    Finally, step \ref{step:REQ-uncompute-valid-query} does not change the computational-basis values of $\cG \times \cA$, so the state at the end of the query is still in $S^\REQ$.
\end{proof}

\newif\ifNewProof
\NewProoftrue

\ifNewProof
    \subsubsection*{Hybrids}
\paragraph{Hybrid 1:} The simulator $\Sim^{O^\REQ_P}(1^\secp)$ in \cref{fig:simulator-REQ-hybrid-oracle}.

\paragraph{Hybrid 2:} This hybrid opens up the queries to $O^\REQ_P$ and replaces it with its state transition function.

Given some basis state $\ket{t, x,u,b}_\cQ \otimes \ket{C}_\cC \otimes \ket{D}_{\cD} \otimes \ket{\mathbf{0}}_\cW \otimes \ket{0}_{\cB'}$, the simulator does the following:

\begin{enumerate}
        \item \label{sim-2-step:expand-registers}\textbf{Expand $\cC \times \cG \times \cA$ by Two Registers.} Append the following registers to $\cC \times \cG \times \cA$:
        \[\ket{\mathbf{0}}_{\cC_{k+1}} \otimes \ket{\mathbf{0}}_{\cC_{k+2}} \otimes \ket{0}_{\cG_{k+1}} \otimes \ket{0}_{\cG_{k+2}} \otimes \ket{\mathbf{0}}_{\cA_{k+1}} \otimes \ket{\mathbf{0}}_{\cA_{k+2}}\]
        
        \item \textbf{Validate Query.} \label{sim-2-step:compute-B} Let $k_{\max}$ be the maximum index $i \in [k]$ such that $g'_i = 1$, or if no such $i$ exists, then let $k_{\max} = 0$. 
        
        If $\cQ$ contains $t = k_{\max}$ and $x = x'_{k_{\max}}$, or if $\cQ$ contains $t = k_{\max}+1$, then apply $\mathsf{X}$ to $\cB'$.

        \item \textbf{Manage Cache.}\label{sim-2-step:hyb2:manage-cache}
        \begin{enumerate}
            \item \label{sim-2-step:refl-and-copy-1}$\mathsf{ReflectAndCopy}$: If $g'_{t} = 0$, then CNOT $(t,x)$ from $\cQ$ to $\cC_{t}$:
            \[ 
                \ket{t_{t},x'_{t},u_{t}, 0}_{\cC_{t}} \mapsto \ket{t_{t} \oplus t,x'_{t} \oplus x,u_{t}, 0}_{\cC_{t}}
            \]
            \item \label{sim-2-step:decomp1} If $g_t' = 1$, apply $\Decomp_{t,x}$ to $\cD$.
            \item \label{sim-2-step:query-REQ-oracle}(Query $O^\REQ_P$) If $\cD$ is in the span of states of the form
            \begin{align*}
                \ket{D} \quad \text{or} \quad \frac{1}{\sqrt{2^{\secp}}} \cdot \sum_{r_2 \in \bit^\secp} \ket{D \cup \{(t,x,r_2)\}}
            \end{align*}
            for all databases $D$ that do not contain $(t,x,*)$, 
            then swap states of the following form:
            \[
                \ket{t_t,x'_t,u_t,g'_t}_{\cC_t} \otimes \ket{\st}_\cS \otimes \ket{\mathbf{0}}_{\cA_t} \otimes \ket{0}_{\cG_t} \leftrightarrow \ket{\phi^\REQ_{t_t,x'_t,u_t,g'_t,\st}}
            \]
            \item \label{sim-2-step:decomp}If $g_t' = 1$, apply $\Decomp_{t,x}$ to $\cD$.
            
            \item \label{sim-2-step:refl-and-copy-2}$\mathsf{ReflectAndCopy}$: Repeat step \ref{sim-2-step:refl-and-copy-1}.
        \end{enumerate}
        
        \item \label{sim-2-step:copy-answer} \textbf{Answer Query:} If $\cB'$ contains $1$, then answer the query by copying $(y, r_2)$ from $\cC_{t} \times \cD$ onto $\cQ$ and flipping $b$:
        \[\ket{t,x,u,b}_\cQ \mapsto \ket{t, x, u \oplus (y, r_2), b\oplus 1}_\cQ\]
        \item \textbf{Manage Cache.} Repeat step \ref{sim-2-step:hyb2:manage-cache}.
        \label{sim-2-step:refl-and-copy-3}
        \label{sim-2-step:uncompute-query-REQ-oracle}
        \label{sim-2-step:refl-and-copy-4}
        %     % If $(t,x,*) \notin D$, then set $r_2 = 0$.
        %     \item \label{sim-2-step:refl-and-copy-3}$\mathsf{ReflectAndCopy}$: Repeat step \ref{sim-2-step:refl-and-copy-2}.
        %     \item ($\Decomp_{t,x}$) Repeat step \ref{sim-2-step:decomp}.
        %     \item \label{sim-2-step:uncompute-query-REQ-oracle}(Query $O^\REQ$) Repeat step \ref{sim-2-step:query-REQ-oracle}.
        % \item \label{sim-2-step:refl-and-copy-4}$\mathsf{ReflectAndCopy}$: Repeat step \ref{sim-2-step:refl-and-copy-1}.
        \item \label{sim-2-step:uncompute-B}(Validate Query) Repeat step \ref{sim-2-step:compute-B}.
\end{enumerate}

\begin{lemma}
    Hybrids 1 and 2 are perfectly indistinguishable after any number of quantum queries.
\end{lemma}
\begin{proof}
    The first difference between hybrids 1 and 2 is that step \ref{sim-step:expand-registers} of hybrid 1 appends two registers to $\cC$ whereas step \ref{sim-2-step:expand-registers} of hybrid 2 appends two registers to each of $\cC$, $\cG$, and $\cA$. Note that hybrid 1 always makes two queries to $O^\REQ_P$, and each query appends one register to $\cG$ and $\cA$ each. So Hybrid 1 also appends two registers to each of $\cC$, $\cG$, and $\cA$, but some of these append operations occur inside $O^\REQ_P$. Hybrid 2 simply moves all these append operations to step \ref{sim-2-step:expand-registers}.

    The second difference between hybrids 1 and 2 is that in hybrid 1, step \ref{step:REQ-answer-query} of $O^\REQ_P$ has been replaced by the state transition function given in \cref{thm:state-transitions-of-REQ}. These operations are equivalent according to \cref{thm:state-transitions-of-REQ}.

    The third and final difference between hybrids 1 and 2 is that hybrid 2 omits the query validation steps of $O^\REQ_P$ (steps \ref{REQ-step:validate-query} and \ref{step:REQ-uncompute-valid-query} of $O^\REQ_P$). This change is undetectable because hybrid 1 implements its own query validation step in step \ref{sim-step:compute-B}. If step \ref{sim-step:compute-B} accepts the query, then the subsequent calls to $O^\REQ_P$ will also accept the query and flip $\cB$. Furthermore, if step \ref{sim-step:compute-B} rejects the query, then the hybrid acts as the identity (other than appending new registers). If step \ref{sim-step:compute-B} rejects the query, then step \ref{sim-2-step:copy-answer} acts as the identity, and the remaining steps cancel out: steps \ref{sim-2-step:compute-B} - \ref{sim-2-step:refl-and-copy-2} cancel out with steps \ref{sim-2-step:refl-and-copy-3} - \ref{sim-2-step:uncompute-B}.
\end{proof}

% It will also be useful to characterize the effect of the \textbf{manage cache} step in Hybrid 2.

% \begin{lemma}
%     On computational basis query $\ket{t, x, u, b}$,
%     \justin{TODO}
% \end{lemma}
% \begin{proof}
%     First, we establish an invariant for the cache $\cC_t$ at the start and end of every \textbf{manage cache} step: it contains $t\neq 0$ if and only if $g_t' = 0$, and furthermore $x_t = 0$ if $g_t' = 0$. These parts of the cache register are initially $0$, establishing the invariant at time $0$ (when no queries have been made), and are modified only by $\mathsf{ReflectAndCopy}$. Now consider time greater than $0$. After the first $\mathsf{ReflectAndCopy}$, $\calC_t$ contains nonzero $t$ because $t < 1$ is not answered.
%     \justin{Is this true? We might have a $0$-indexing issue....}
%     The next time $\cC_t$ is potentially modified is when $\mathsf{ReflectAndCopy}$ is applied again conditioned on $g_t' = 0$. If $g_t' = 0$, this returned to $t = 0$ and $x_t = 0$ because register $\cQ$ has not changed, and otherwise continues to contain $t\neq 0$ and $x_t$ potentially not $0$.

%     Next, we establish a similar invariant for $\Decomp_{t,x}$: outside of the \textbf{manage cache} step, 
    
%     \justin{in progress}
    
% \end{proof}

\paragraph{Hybrid 3:} 
Modify when the $\Decomp_{t,x}$ operations trigger. 
Recall that in Hybrid 2, the first $\mathsf{ReflectAndCopy}$ operation is followed by the three steps: apply $\Decomp_{t,x}$ if $g'_t=1$; perform the RAM query only on components for which the database has no $(t,x,*)$ entry; and again apply $\Decomp_{t,x}$ if $g'_t=1$. The reverse sequence is used after
\textbf{Answer Query}, between the third and fourth $\mathsf{ReflectAndCopy}$ operations.

In step \ref{sim-step:decomp}, instead of checking $g_t'$, trigger $\Decomp_{t, x}$ if $\cC_t \times \cS \times \cG_t \times \cA_t$ is orthogonal to $\ket{\phi^\REQ_{t,x,u,b,\st}}$. Then, in step \ref{sim-step:decomp-undo}, trigger $\Decomp_{t, x}$ unconditionally.
Their reverse sequence after answering the query is also replaced accordingly.

% Add the following operation: If $\cC_t \times \cS \times \cG_t \times \cA_t$ is orthogonal to $\ket{\phi^\REQ_{t,x,u,b,\st}}$, then apply $\Decomp_{t,x}$ to $\cD$.

%This hybrid uses the procedure below and is compared with the revised Hybrid 2. 
% Recall that, in the revised Hybrid 2, the first
% $\mathsf{ReflectAndCopy}$ operation is followed by the three steps:
% apply $\Decomp_{t,x}$ if $g'_t=1$; perform the RAM query only on
% components for which the database has no $(t,x,*)$ entry; and again
% apply $\Decomp_{t,x}$ if $g'_t=1$. The reverse sequence is used after
% \textbf{Answer Query}, between the third and fourth
% $\mathsf{ReflectAndCopy}$ operations.
% In Hybrid 3, these three steps are replaced by steps
% \ref{sim-3-step:conditionally-apply-decomp}--\ref{sim-3-step:decomp}
% below, and their reverse sequence is replaced accordingly.

Given some basis state $\ket{t, x,u,b}_\cQ \otimes \ket{C}_\cC \otimes \ket{D}_{\cD} \otimes \ket{\mathbf{0}}_\cW \otimes \ket{0}_{\cB'}$, the simulator does the following:

\begin{enumerate}
    \item \label{sim-3-step:expand-registers}\textbf{Expand $\cC \times \cG \times \cA$ by Two Registers.} Append the following registers to $\cC \times \cG \times \cA$:
    \[\ket{\mathbf{0}}_{\cC_{k+1}} \otimes \ket{\mathbf{0}}_{\cC_{k+2}} \otimes \ket{0}_{\cG_{k+1}} \otimes \ket{0}_{\cG_{k+2}} \otimes \ket{\mathbf{0}}_{\cA_{k+1}} \otimes \ket{\mathbf{0}}_{\cA_{k+2}}\]
    
    \item \textbf{Validate Query.} \label{sim-3-step:compute-B} Let $k_{\max}$ be the maximum index $i \in [k]$ such that $g'_i = 1$, or if no such $i$ exists, then let $k_{\max} = 0$. 
    
    If $\cQ$ contains $t = k_{\max}$ and $x = x'_{k_{\max}}$, or if $\cQ$ contains $t = k_{\max}+1$, then apply $\mathsf{X}$ to $\cB'$.

    \item \textbf{Manage Cache.}\label{sim-3-step:manage-cache}
    \begin{enumerate} 
        \item \label{sim-3-step:refl-and-copy-1}$\mathsf{ReflectAndCopy}$: If $g'_{t} = 0$
        , then CNOT $(t,x)$ from $\cQ$ to $\cC_{t}$:
            \[\ket{t_{t},x'_{t},u_{t}, 0}_{\cC_{t}} \mapsto \ket{t_{t} \oplus t,x'_{t} \oplus x,u_{t}, 0}_{\cC_{t}}\]
  
        \item \label{sim-3-step:conditionally-apply-decomp}
    
        Apply $\Decomp_{t,x}$ to $\cD$, {\color{red}controlled on the component
        of $\cC_t\times\cS\times\cG_t\times\cA_t$ orthogonal to
        all the following states:
        \[
            \begin{split}
                &\ket{t,x,u,b}_{\cC_t}
                 \otimes\ket{\st}_{\cS}
                 \otimes\ket{0}_{\cG_t}
                 \otimes\ket{\mathbf{0}}_{\cA_t},
                 \qquad \forall(u,b,\st),\\
                &\ket{\phi^\REQ_{t,x,u,b,\st}},
                 \qquad \forall(u,b,\st).
            \end{split}
        \]
        }

        \item \label{sim-3-step:query-REQ-oracle}(Query $O^\REQ_P$) If $\cD$ is in the span of states of the form
        \begin{align*}
            \ket{D} \quad \text{or} \quad \frac{1}{\sqrt{2^{\secp}}} \cdot \sum_{r_2 \in \bit^\secp} \ket{D \cup \{(t,x,r_2)\}}
        \end{align*}
        for all databases $D$ that do not contain $(t,x,*)$, 
        then swap states of the following form:
            \[\ket{t_t,x'_t,u_t,g'_t}_{\cC_t} \otimes \ket{\st}_\cS \otimes \ket{\mathbf{0}}_{\cA_t} \otimes \ket{0}_{\cG_t} \leftrightarrow \ket{\phi^\REQ_{t_t,x'_t,u_t,g'_t,\st}}\]
        \item \label{sim-3-step:decomp}Apply $\Decomp_{t,x}$ to $\cD$ {\color{red} unconditionally}.
        \item \label{sim-3-step:refl-and-copy-2}$\mathsf{ReflectAndCopy}$: Repeat step \ref{sim-3-step:refl-and-copy-1}.
    \end{enumerate}

    \item \label{sim-3-step:copy-answer} \textbf{Answer Query:} If $\cB'$ contains $1$, then answer the query by copying $(y, r_2)$ from $\cC_{t} \times \cD$ onto $\cQ$ and flipping $b$:
    \[\ket{t,x,u,b}_\cQ \mapsto \ket{t, x, u \oplus (y, r_2), b\oplus 1}_\cQ\]

    \item \textbf{Manage Cache.}
    \begin{enumerate}
       
        \item ($\mathsf{ReflectAndCopy}$) \label{sim-3-step:refl-and-copy-3} Repeat step \ref{sim-3-step:refl-and-copy-2}.
        \item \label{sim-3-step:decomp-2}({\color{red} Unconditional} $\Decomp_{t,x}$) Repeat step \ref{sim-3-step:decomp}.
        \item \label{sim-3-step:uncompute-query-REQ-oracle}(Query $O^\REQ_P$) Repeat step \ref{sim-3-step:query-REQ-oracle}.
        \item \label{sim-3-step:conditionally-apply-decomp-2}\textcolor{red}{Repeat step \ref{sim-3-step:conditionally-apply-decomp}.}
        \item \label{sim-3-step:refl-and-copy-4}($\mathsf{ReflectAndCopy}$) Repeat step \ref{sim-3-step:refl-and-copy-1}.
        \item \label{sim-3-step:compute-B-2}(Validate Query) Repeat step \ref{sim-3-step:compute-B}.
    \end{enumerate}
\end{enumerate}

\begin{lemma}
    Hybrids 2 and 3 are perfectly indistinguishable after any number of quantum queries.
\end{lemma}
\begin{proof}
    The only changes are during the \textbf{manage cache} step, in particular steps \ref{sim-3-step:conditionally-apply-decomp} to \ref{sim-3-step:decomp} (and the reverse in the repetition of \textbf{manage cache}). We claim that outside of these steps, the state of the simulator is identical between Hybrids 2 and 3.
    By linearity, it suffices to show this for computational basis queries $\ket{t, x', u, 0}$. 

    Let $V_{t,x}$ be the span of states of the form
    \begin{align*}
            \ket{D} \quad \text{or} \quad \frac{1}{\sqrt{2^{\secp}}} \cdot \sum_{r_2 \in \bit^\secp} \ket{D \cup \{(t,x,r_2)\}}
        \end{align*}
    for all databases $D$ that do not contain $(t,x,*)$. Observe that $\Decomp_{t,x}$ acts trivially on the orthogonal space $V_{t,x}^\perp$.
    and denote 
    \[
        \ket{\xi^\REQ_{t_t,x'_t,u_t,g'_t,\st}} \coloneqq 
    \ket{t_t,x'_t,u_t,g'_t}_{\cC_t} \otimes \ket{\st}_\cS \otimes \ket{\mathbf{0}}_{\cA_t} \otimes \ket{0}_{\cG_t} 
    \]
    so that the query operation swaps
    \[
        \ket{\xi^\REQ_{t_t,x'_t,u_t,g'_t,\st}} \leftrightarrow \ket{\phi^\REQ_{t_t,x'_t,u_t,g'_t,\st}}
    \]
    Let $W_{\LEQ}$ be the span of these two states.

    Consider the following sequence of hybrid experiments which modify only the three steps \ref{sim-3-step:conditionally-apply-decomp} to \ref{sim-3-step:decomp}. First, Hybrid 2.1 inserts two $\Decomp_{t,x}$ operations after the conditional query:
    \begin{enumerate}
        \item If $g_t' = 1$, apply $\Decomp_{t,x}$.
        \item Conditional Query: If $\cD$ is supported on $V_{t,x}$, swap $\ket{\xi^\REQ_{t_t,x'_t,u_t,g'_t,\st}} \leftrightarrow \ket{\phi^\REQ_{t_t,x'_t,u_t,g'_t,\st}}$.
        \item\label{step:sim-h2h3-conditional-g0-decomp} {\color{red}\textbf{New:}} If $g_t' = 0$, apply $\Decomp_{t,x}$.
        \item {\color{red}\textbf{New:}} If $g_t' = 0$, apply $\Decomp_{t,x}$.
        \item If $g_t' = 1$, apply $\Decomp_{t,x}$.
    \end{enumerate}
    Since $\Decomp_{t,x}$ is an involution, the new steps are together the identity. Observe that the last two steps can be combined together as \emph{always} do $\Decomp_{t,x}$.

    Next, Hybrid 2.2 moves the new step \ref{step:sim-h2h3-conditional-g0-decomp} to \emph{before} the conditional LEQ evaluation:
    \begin{enumerate}
        \item If $g_t' = 1$, apply $\Decomp_{t,x}$.
        \item {\color{red} \textbf{Moved and Modified:}} 
        \begin{itemize}
            \item If $(\cC_t, \cS, \cA_t, \cG_t)$ are supported on $W_{\LEQ}$ and $g_t' = 0$, apply $\Decomp_{t,x}$.
            \item If $(\calC_t, \cS, \cA_t, \cG_t)$ are \emph{not} supported on $W_{\LEQ}$ and $g_t' = 1$, apply $\Decomp_{t,x}$.
        \end{itemize}
        \item \label{step:sim-h2h3-h2.2-conditionalquery}Conditional Query: If $\cD$ is supported on $V_{t,x}$, swap $\ket{\xi^\REQ_{t_t,x'_t,u_t,g'_t,\st}} \leftrightarrow \ket{\phi^\REQ_{t_t,x'_t,u_t,g'_t,\st}}$.
        \item Apply $\Decomp_{t,x}$.
    \end{enumerate}
    To see equivalence, first observe that since $V_{t,x}$ is invariant under $\Decomp_{t,x}$, an unconditional application of $\Decomp_{t,x}$ commutes with with the conditional query in step \ref{step:sim-h2h3-h2.2-conditionalquery}. When $(\cC_t, \cS, \cA_t, \cG_t)$ are supported on $W_{\LEQ}$, step \ref{step:sim-h2h3-h2.2-conditionalquery} flips the control bit $g'_t$. Therefore switching the ordering is equivalent with the control flipped when $(\cC_t, \cS, \cA_t, \cG_t)$ are supported on $W_{\LEQ}$.

    Finally, we do a case analysis to combine the three conditional applications of $\Decomp_{t,x}$ before step \ref{step:sim-h2h3-h2.2-conditionalquery}. The first and third step cancel whenever $(\calC_t, \cS, \cA_t, \cG_t)$ are \emph{not} supported on $W_{\LEQ}$. Otherwise, $\Decomp_{t,x}$ is applied when $g_t' = 0$ \emph{and} when $g_t' = 1$. In other words, the aggregate effect of these three steps is to apply $\Decomp_{t,x}$ when $(\calC_t, \cS, \cA_t, \cG_t)$ are supported on $W_{\LEQ}$. Thus Hybrid 2.2 is equivalent to Hybrid 3.

\end{proof}

\subsubsection*{State Transition Function of $\Sim$.}
\begin{definition}[States of $\Sim$ in Hybrid 4]\label{def:basis-of-Sim}
    Let $s \in (\cX \times \cR \times \bit^\secp)^*$ be a (potentially empty) tuple, and for each $i \in [|s|]$, let 
    \[s_i = (x_i, r_{1,i}, r_{2, i}) \in \cX \times \cR \times \bit^\secp.\]
    
    Let $k \in \{0\} \cup \bbN$ satisfy $k \geq |s|$. Let $\st_1 = \bot$, and for each $i \in [|s|]$, let
    \begin{align*}
        (\st_{i+1}, y_i) &= P(\st_i, x_i; r_{1,i})\\
        a_i &= (\st_i, x_i, r_{1,i})\\
        c_i &= (i, x_i, y_i, 1)\\
        d_i &= (i, x_i, r_{2,i})\\
        g_i &= 1
    \end{align*}
    Then, for each $i \in (|s|, k]$, let
    \begin{align*}
        a_i &= (0, 0, 0)\\
        c_i &= (0, 0, 0, 0)\\
        g_i &= 0
    \end{align*}
    For each such $k, s$, let us define the following computational-basis eigenstate:
    \begin{align*}
        \ket{\psi^\Sim_{k, s}} = &\ket{0}_\cW \otimes \ket{0}_{\cB'}\\
        &\otimes \ket{c_1}_{\cC_1} \otimes \dots \otimes \ket{c_k}_{\cC_{k}} \otimes \ket{\{d_i\}_{i \in [|s|]}}_\cD \\
        &\otimes \ket{a_1}_{\cA_1} \otimes \dots \otimes \ket{a_k}_{\cA_k} \otimes \ket{g_1}_{\cG_1} \otimes \dots \otimes \ket{g_k}_{\cG_k}\\
        &\otimes \ket{\st_{|s|+1}}_\cS \otimes \ket{0}_\cB
    \end{align*}

    Let us define a Fourier basis state as follows. For any such $k, s$ and any $x \in \cX$, $\tilde{r}_1 \in \cR$ and $\tilde{r}_2 \in \bit^\secp$, let:

    \begin{align*}
        \ket{\phi^\Sim_{k, s, x, \tilde{r}_1, \tilde{r}_2}} &= \frac{1}{\sqrt{\abs{R} \cdot 2^\secp}} \cdot \sum_{(r_1, r_2) \in \cR \times \bit^\secp} (-1)^{\langle r_1, \tilde{r}_1\rangle + \langle r_2, \tilde{r}_2\rangle} \cdot \ket{\psi^\Sim_{k, s||(x, r_1, r_2)}}
    \end{align*}
\end{definition}

Note that the initial state of $\Sim$'s internal registers is $\ket{\psi^\Sim_{0, \emptyset}}$:
\[\ket{\psi^\Sim_{0, \emptyset}} = \ket{0}_\cW \otimes \ket{0}_{\cB'} \otimes \ket{\emptyset}_\cD \otimes \ket{\bot}_\cS \otimes \ket{0}_\cB\]

\begin{claim}
    For any given $k \in \{0\} \cup \bbN$, the states $\left\{\ket{\psi^\Sim_{k, s}}\right\}_{s}$ of \cref{def:basis-of-Sim} are orthonormal.
\end{claim}
\begin{proof}
    First, each $\ket{\psi^\Sim_{k, s}}$ is a computational-basis eigenstate, so it has unit norm. 
    
    Next, let us consider two different $s$-values $s \neq s'$ that have the same length $k_{\max}$. We will show that $\braket{\psi^\Sim_{k, s}}{\psi^\Sim_{k, s'}} = 0$. First, $k_{\max} \geq 1$ because there is only one $s$-value ($s = \emptyset$) of length $0$. Second, every component of $s$ or $s'$ is written on some register of $\ket{\psi^\Sim_{k, s}}$ or $\psi^\Sim_{k, s'}$ respectively. For each $i \in [k_{\max}]$, $(x_i, r_{1,i})$ are written on register $\cA_i$, and $r_{2,i}$ is written in the $i$-th entry $d_i$ of register $\cD$. If $s \neq s'$, then $\ket{\psi^\Sim_{k, s}}$ and $\psi^\Sim_{k, s'}$ will disagree on one of the registers listed above. Therefore, $\braket{\psi^\Sim_{k, s}}{\psi^\Sim_{k, s'}} = 0$.

    Finally, let us consider two $s$-values $(s, s')$ of different lengths. We will show that $\braket{\psi^\Sim_{k, s}}{\psi^\Sim_{k, s'}} = 0$. Let $k_{\max}$ be the length of $s$, and let $k'_{\max}$ be the length of $s'$. Without loss of generality, let us assume that $k_{\max} > k'_{\max}$. Then the register $\cG_{k_{\max}}$ will store $\ket{1}$ in $\ket{\psi^\Sim_{k, s}}$ and will store $\ket{0}$ in $\ket{\psi^\Sim_{k, s'}}$. Therefore, $\braket{\psi^\Sim_{k, s}}{\psi^\Sim_{k, s'}} = 0$.

    In summary, we've shown that if $s \neq s'$, then $\braket{\psi^\Sim_{k, s}}{\psi^\Sim_{k, s'}} = 0$.
\end{proof}

% \paragraph{Unreachable States of $\Sim$.}
% At the start of any query, for any $(t,x,u,b,\st)$, the following states are unreachable:
% \begin{itemize}
%     \item $(t,x) \notin D$ and $\cC_t \times \cS \times \cG_t \times \cA_t \times \cB = \ket{\phi^\REQ_{t,x,u,b,\st}}$ ($r_{2,|s|} = \bot$ and $\tilde{r}_1 = 0$).
%     \item $(t, x) \in D$ and $\cC_t \times \cS \times \cG_t \times \cA_t \times \cB = \ket{\mathbf{0}}$.
%     \item $D_{t,x} = \ket{+}$ ($\tilde{r}_2 = 0$).
% \end{itemize}

% At the start of step 6 (Answer Query), it is always true that $g_t = 1$ and $(t,x) \in D$.

\paragraph{Hybrid 4:}

The initial state of the simulator is $\ket{\psi^\Sim_{0,\emptyset}}$. Next, given any query of the form $\ket{t, x,u,b}_\cQ$, the simulator does the following:
%\justin{TODO: transition from 3-4}\bhaskar{This transition is done.}
\begin{enumerate}
    \item \label{sim-4-step:increase}$\mathsf{Increase}$: Map
    \[\ket{\psi^\Sim_{k,s}} \mapsto \ket{\psi^\Sim_{k+2,s}}\]
    \item \label{sim-4-step:decomp-sim-1}$\Decomp^\Sim$: For any  sequence $s \in \left(\cX \times \cR \times \bit^\secp\right)^*$ of length $t-1$, swap the following states:
    \[\ket{\psi^\Sim_{k,s}} \leftrightarrow \ket{\phi^\Sim_{k,s, x, 0, 0}}\]
    and act as the identity on all states orthogonal to the ones above.
    
    \item \label{sim-4-step:copy-sim}$\mathsf{Copy}^\Sim$: 
    \begin{itemize}
        \item For any sequence $s \in \left(\cX \times \cR \times \bit^\secp\right)^*$ of length $t$ for which $s_t = (x, *, *)$, swap the following states:
        \[\ket{t,x,u,b}_\cQ \otimes \ket{\psi^\Sim_{k,s}} \leftrightarrow \ket{t,x,u \oplus (y_t, r_{2,t}),b \oplus 1}_\cQ \otimes \ket{\psi^\Sim_{k,s}}\]
        where $y_t, r_{2,t}$ are computed from $s$ as described in \cref{def:basis-of-Sim}. 
        % \item For any sequence $s \in \left(\cX \times \cR \times \bit^\secp\right)^*$ of length $t-1$, swap the following states:
        % \[\ket{t,x,u,b}_\cQ \otimes \ket{\psi^\Sim_{k,s}} \leftrightarrow \ket{t,x,u,b \oplus 1}_\cQ \otimes \ket{\psi^\Sim_{k,s}}\]
        % where $(*, y_t) = P(\st_t, x; 0)$, and $\st_t$ is computed from $s$ as described in \cref{def:basis-of-Hyb}.
        % \item 
        
        Act as the identity on all states orthogonal to the ones above.
    \end{itemize}
    \item \label{sim-4-step:decomp-sim-2}$(\Decomp^\Sim)^\dag$: Uncompute step \ref{sim-4-step:decomp-sim-1}.
\end{enumerate}

The following definition defines the space of reachable states of hybrid 3 and 4's internal registers. 
\begin{definition}[Reachable States of the Simulator]\label{def:reachable-states-of-Sim}
    Let $S^\Sim$ be the span of states of the form $\ket{\psi^\Sim_{k,s}}$ for all (potentially empty) sequences $s \in (\cX \times \cR \times \bit^\secp)^*$ and all $k \geq |s|$.

    % Let $S^\Sim$ be the span of $\ket{\psi^\Sim_{0, \emptyset}}$ and states of the form $\ket{\phi^\Sim_{k,s, x, \tilde{r}_1, \tilde{r}_2}}$ for all (potentially empty) sequences $s \in (\cX \times \cR \times \bit^\secp)^*$, all $k \geq |s|$, all $x$, and all $(\tilde{r}_1, \tilde{r}_2) \neq (0,0)$.

    % Let $S'^\Sim$ be the span of states of the form $\ket{\phi^\Sim_{k,s, x, \tilde{r}_1, \tilde{r}_2}}$ for all (potentially empty) sequences $s \in (\cX \times \cR \times \bit^\secp)^*$, all $k \geq |s|$, and all $x,\tilde{r}_1, \tilde{r}_2$.
\end{definition}

The following will help us show that step \ref{sim-3-step:manage-cache} of hybrid 3 can be replaced with $\Decomp^\Sim$, step \ref{sim-4-step:decomp-sim-1} of hybrid 4.
\begin{lemma}\label{thm:decomp-sim-equivalence}
    For any state of the form $\ket{t,x,u,b}_\cQ \otimes \ket{\psi^\Sim_{k,s}}$, where $(t,x) = (|s|, x_{|s|})$ or $t = |s| + 1$, step \ref{sim-3-step:manage-cache} of hybrid 3 acts on this state equivalently to step \ref{sim-4-step:decomp-sim-1} of hybrid 4 ($\Decomp^\Sim$) and produces a state in $S^\Sim$.
\end{lemma}
\begin{proof}
$ $
    \paragraph{Case 1:} The input is $\ket{t,x,u,b}_\cQ \otimes \ket{\psi^\Sim_{k,s}}$, where $|s| = t-1$.

    Let us step through the action of steps \ref{sim-3-step:refl-and-copy-1} - \ref{sim-3-step:refl-and-copy-2} of hybrid 3 on this state. Initially, $\cC_t \times \cG_t \times \cA_t = \ket{\mathbf{0}}_{\cC_t} \times \ket{0}_{\cG_t} \times \ket{\mathbf{0}}_{\cA_t}$, and $(t,x,*) \notin D$. Step \ref{sim-3-step:refl-and-copy-1} ($\mathsf{ReflectAndCopy}$) maps:
    \[\ket{\mathbf{0}}_{\cC_t} \times \ket{0}_{\cG_t} \times \ket{\mathbf{0}}_{\cA_t} \mapsto \ket{t,x,0,0}_{\cC_t} \times \ket{0}_{\cG_t} \times \ket{\mathbf{0}}_{\cA_t}\]
    Step \ref{sim-3-step:conditionally-apply-decomp} acts as the identity because $\cC_t \times \cS \times \cG_t \times \cA_t$ is in the span of the states
    \[\ket{t,x,u,b}_{\cC_t} \otimes \ket{\st}_{\cS} \otimes \ket{0}_{\cG_t} \otimes \ket{\mathbf{0}}_{\cA_t}, \quad \forall (u,b,\st).\]
    Step \ref{sim-3-step:query-REQ-oracle} (Query $O^\REQ_P$) verifies that $(t,x,*) \notin D$ and then maps:
    \[\ket{t,x,0,0}_{\cC_t} \otimes \ket{\st_{|s|+1}}_\cS \otimes \ket{0}_{\cG_t} \otimes \ket{\mathbf{0}}_{\cA_t} \mapsto \ket{\phi^\REQ_{t,x,0,0,\st_{|s|+1}}}\]
    At the start of step \ref{sim-3-step:decomp} ($\Decomp_{t,x}$), $\cD$ is in the span of databases $D$ such that $(t,x,*) \notin D$. Step \ref{sim-3-step:decomp} maps:
    \[\ket{D}_{\cD} \mapsto \frac{1}{\sqrt{2^{\secp}}} \cdot \sum_{r_2 \in \bit^\secp} \ket{D \cup \{(t,x,r_2)\}}\]

    Step \ref{sim-3-step:refl-and-copy-2} acts as the identity because $g'_t = 1$.
    
    The final state of the simulator's internal registers at the end of step \ref{sim-3-step:decomp} is:
    \begin{align*}
        \frac{1}{\sqrt{\abs{R} \cdot 2^\secp}} \cdot \sum_{(r_1, r_2) \in \cR \times \bit^\secp} \ket{\psi^\Sim_{k, s||(x, r_1, r_2)}} &= \ket{\phi^\Sim_{k, s, x, 0,0}}
    \end{align*}

    Next, let us examine the action of $\Decomp^\Sim$ (step \ref{sim-4-step:decomp-sim-1} of hybrid 4) on $\ket{t,x,u,b}_\cQ \otimes \ket{\psi^\Sim_{k,s}}$. $\Decomp^\Sim$ maps:
    \[\ket{\psi^\Sim_{k,s}} \mapsto \ket{\phi^\Sim_{k,s,x,0,0}}\]
    In this case, $\Decomp^\Sim$ acts equivalently to steps \ref{sim-3-step:refl-and-copy-1} - \ref{sim-3-step:refl-and-copy-2} of hybrid 3, and the final state is in $S^\Hyb$.

    \paragraph{Case 2:} The input is $\ket{t,x,u,b}_\cQ \otimes  \ket{\phi^\Sim_{k,s, x, \tilde{r}_1, \tilde{r}_2}}$, where \underline{$\tilde{r}_1 = 0$, $\tilde{r}_2 = 0$}, and $|s| = t-1$.

    Let us step through the action of steps \ref{sim-3-step:refl-and-copy-1} - \ref{sim-3-step:refl-and-copy-2} of hybrid 3 on this state. Initially, since $\tilde{r}_1 = 0$, 
    \[\cC_t \times \cS \times \cG_t \times \cA_t = \ket{\phi^\REQ_{t,x,0,0,\st_{|s|+1}}},\]
    and since $\tilde{r}_2 = 0$, $\cD$ is in the span of states of the form:
    \[\frac{1}{\sqrt{2^{\secp}}} \cdot \sum_{r_2 \in \bit^\secp} \ket{D \cup \{(t,x,r_2)\}}\]
    for all databases $D$ that do not contain $(t,x,*)$.
    
    Step \ref{sim-3-step:refl-and-copy-1} ($\mathsf{ReflectAndCopy}$) verifies that $g'_t = 1$ and then acts as the identity.
    
    Step \ref{sim-3-step:conditionally-apply-decomp} acts as the identity because $\cC_t \times \cS \times \cG_t \times \cA_t$ is in the span of the states
    \[\ket{\phi^\REQ_{t,x,u,b,\st}}, \quad \forall (u,b,\st)\]
    
    Step \ref{sim-3-step:query-REQ-oracle} (Query $O^\REQ_P$) maps:
    \[\ket{\phi^\REQ_{t,x,0,0,\st_{|s|+1}}} \mapsto \ket{t,x,0,0}_{\cC_t} \otimes \ket{\st_{|s|+1}}_\cS \otimes \ket{0}_{\cG_t} \otimes \ket{\mathbf{0}}_{\cA_t}\]
    
    Step \ref{sim-3-step:decomp} ($\Decomp_{t,x}$) maps:
    \[\frac{1}{\sqrt{2^{\secp}}} \cdot \sum_{r_2 \in \bit^\secp} \ket{D \cup \{(t,x,r_2)\}} \mapsto \ket{D}_{\cD}\]
    where $(t,x,*) \notin D$.

    Step \ref{sim-3-step:refl-and-copy-2} verifies that $g'_t = 0$ and then maps:
    \[\ket{t,x,0,0}_{\cC_t} \mapsto \ket{0,0,0,0}_{\cC_t}\]
    
    The final state of the simulator's internal registers at the end of step \ref{sim-3-step:refl-and-copy-2} is $\ket{\psi^\Sim_{k,s}}$.

    Next, let us examine the action of $\Decomp^\Sim$ (step \ref{sim-4-step:decomp-sim-1} of hybrid 4) on the initial state $\ket{t,x,u,b}_\cQ \otimes \ket{\phi^\Sim_{k,s, x, 0, 0}}$. $\Decomp^\Sim$ maps:
    \[\ket{\phi^\Sim_{k,s, x, 0, 0}} \mapsto \ket{\psi^\Sim_{k,s}}\]
    In this case, $\Decomp^\Sim$ acts equivalently to steps \ref{sim-3-step:refl-and-copy-1} - \ref{sim-3-step:refl-and-copy-2} of hybrid 3, and the final state is in $S^\Hyb$.

\iffalse
    \paragraph{Case 3:} The input is $\ket{t,x,u,b}_\cQ \otimes  \ket{\phi^\Sim_{k, s, x, \tilde{r}_1, \tilde{r}_2}}$, where \underline{$\tilde{r}_1 = 0$, $\tilde{r}_2 \neq 0$}, and $|s| = t-1$.

    This state is actually not reachable because the phases on the $t$-th entry of the sequence are $(\tilde{r}_1, \tilde{r}_2) = (0,0)$. If at the end of a previous $\mathsf{Copy}^\Sim$ operation, the state were $\ket{\phi^\Sim_{k, s, x, \tilde{r}_1, \tilde{r}_2}}$, then the subsequent $\Decomp^\Sim$ operation would have removed the final entry of the sequence. 
\fi
\bhaskar{More justification needed.}
%\iffalse
    \proofaudit{P17}{Case 3 must be checked rather than discarded}{%
Case 3 assumes $\tilde r_1=0$ and $\tilde r_2\ne0$. Thus its database
component has the coefficients $(-1)^{\langle r_2,\tilde r_2\rangle}$
from \Cref{def:basis-of-Sim}; they are not all $1$. This component is
orthogonal both to a database with no $(t,x,*)$ entry and to the uniform
superposition over the values of that entry. Consequently,
$\Decomp_{t,x}$ fixes it instead of removing the entry. The state can
arise when an accepted query copies $r_2$ into a response register
prepared as
$2^{-\secp/2}\sum_v(-1)^{\langle v,\tilde r_2\rangle}\ket v$.
Replace the assertion that this case is unreachable by checking steps
\ref{sim-3-step:refl-and-copy-1}--\ref{sim-3-step:refl-and-copy-2}.
Each of these steps fixes the state in Case 3, as does $\Decomp^\Sim$.
}
%\fi

\proofaudit{17}{}{Revised proof of case 3 below}

\paragraph{Case 3:} The input is
$\ket{t,x,u,b}_\cQ\otimes
\ket{\phi^\Sim_{k,s,x,\tilde r_1,\tilde r_2}}$,
where \underline{$\tilde r_1=0$, $\tilde r_2\ne0$},
and $|s|=t-1$.

{\color{black}
    Let $D=\{d_i\}_{i\in[|s|]}$, where $d_i$ is defined in
    \Cref{def:basis-of-Sim}; in particular, $D$ does not contain
    $(t,x,*)$. By that definition, the state on
    $\cC_t\times\cS\times\cG_t\times\cA_t\times\cD$ factors as
    \[
        \ket{\phi^\REQ_{t,x,0,0,\st_t}}
        \otimes
        \frac{1}{\sqrt{2^\secp}}
        \sum_{r_2\in\bit^\secp}
        (-1)^{\langle r_2,\tilde r_2\rangle}
        \ket{D\cup\{(t,x,r_2)\}}.
    \]
    All the other internal registers are fixed by $k,s$.
    The database component is orthogonal to every database
    state with no $(t,x,*)$ entry. Its inner product with
    \[
        \frac{1}{\sqrt{2^\secp}}
        \sum_{r_2\in\bit^\secp}
        \ket{D\cup\{(t,x,r_2)\}}
    \]
    is
    \[
        \frac{1}{2^\secp}
        \sum_{r_2\in\bit^\secp}
        (-1)^{\langle r_2,\tilde r_2\rangle}=0,
    \]
    since $\tilde r_2\ne0$. The uniform states associated with
    different residual databases are orthogonal as well.

    Step \ref{sim-3-step:refl-and-copy-1}
    ($\mathsf{ReflectAndCopy}$) acts as the identity because
    $g'_t=1$.

    Step \ref{sim-3-step:conditionally-apply-decomp} acts as the
    identity because the RAM component is
    $\ket{\phi^\REQ_{t,x,0,0,\st_t}}$, one of the states excluded
    by its control condition.

    Step \ref{sim-3-step:query-REQ-oracle} acts as the identity
    because, by the two orthogonality calculations above, the
    database component is orthogonal to the entire subspace
    specified in its control condition.

    Step \ref{sim-3-step:decomp} ($\Decomp_{t,x}$) acts as the
    identity because the database component is orthogonal to
    both states that this decompression swaps, for every fixed
    residual database.

    Step \ref{sim-3-step:refl-and-copy-2}
    ($\mathsf{ReflectAndCopy}$) acts as the identity because
    $g'_t$ is still $1$.

    Therefore steps
    \ref{sim-3-step:refl-and-copy-1}--\ref{sim-3-step:refl-and-copy-2}
    leave $\ket{\phi^\Sim_{k,s,x,0,\tilde r_2}}$ unchanged.
    The operator $\Decomp^\Sim$ also leaves it unchanged:
    by \Cref{def:basis-of-Sim},
    \[
        \left\langle\psi^\Sim_{k,s}\middle|
          \phi^\Sim_{k,s,x,0,\tilde r_2}\right\rangle=0,
        \qquad
        \left\langle\phi^\Sim_{k,s,x,0,0}\middle|
          \phi^\Sim_{k,s,x,0,\tilde r_2}\right\rangle=0,
    \]
    and the states belonging to different predecessor
    sequences are orthogonal. Thus the input is orthogonal to
    every state swapped by $\Decomp^\Sim$ for this query.
    The two operations agree in Case 3, and the final state
    remains in $S^\Sim$.
}

    \paragraph{Case 4:} The input is $\ket{t,x,u,b}_\cQ \otimes  \ket{\phi^\Sim_{k, s, x, \tilde{r}_1, \tilde{r}_2}}$, where \underline{$\tilde{r}_1 \neq 0$, $\tilde{r}_2 = 0$}, and $|s| = t-1$.

    Let us step through the action of steps \ref{sim-3-step:refl-and-copy-1} - \ref{sim-3-step:refl-and-copy-2} of hybrid 3 on this state. Initially, since $\tilde{r}_1 \neq 0$, $\cC_t \times \cS \times \cG_t \times \cA_t$ is orthogonal to states of the form:
    \begin{align*}
        &\ket{t,x,u,b}_{\cC_t} \otimes \ket{\st}_{\cS} \otimes \ket{0}_{\cG_t} \otimes \ket{\mathbf{0}}_{\cA_t}, \quad \forall (u,b,\st)\\
        &\ket{\phi^\REQ_{t,x,u,b,\st}}, \quad \forall (u,b,\st),
    \end{align*}
    and since $\tilde{r}_2 = 0$, $\cD$ is in the span of states of the form:
    \begin{align*}
        &\frac{1}{\sqrt{2^{\secp}}} \cdot \sum_{r_2 \in \bit^\secp} \ket{D \cup \{(t,x,r_2)\}}
    \end{align*}
    for all databases $D$ that do not contain $(t,x,*)$.
    
    Step \ref{sim-3-step:refl-and-copy-1} ($\mathsf{ReflectAndCopy}$) verifies that $g'_t = 1$ and then acts as the identity.
    
    Step \ref{sim-3-step:conditionally-apply-decomp} applies $\Decomp_{t,x}$ to $\cD$, which maps:
    \[\frac{1}{\sqrt{2^{\secp}}} \cdot \sum_{r_2 \in \bit^\secp} \ket{D \cup \{(t,x,r_2)\}} \mapsto \ket{D}\]
    
    Step \ref{sim-3-step:query-REQ-oracle} (Query $O^\REQ_P$) verifies that $\cD$ is in the span of databases $D$ that do not contain $(t,x,*)$. Then the step acts on $\cC_t \times \cS \times \cG_t \times \cA_t$ as the identity because $\cC_t \times \cS \times \cG_t \times \cA_t$ is orthogonal to the states that this step swaps.
    
    Step \ref{sim-3-step:decomp} ($\Decomp_{t,x}$) maps:
    \[\ket{D} \mapsto \frac{1}{\sqrt{2^{\secp}}} \cdot \sum_{r_2 \in \bit^\secp} \ket{D \cup \{(t,x,r_2)\}}\]

    Step \ref{sim-3-step:refl-and-copy-2} verifies that $g'_t = 1$ and then acts as the identity.
    
    The final state of the simulator's internal registers at the end of step \ref{sim-3-step:refl-and-copy-2} is the initial state $\ket{\phi^\Sim_{k, s, x, \tilde{r}_1, \tilde{r}_2}}$.

    Next, let us examine the action of $\Decomp^\Sim$ (step \ref{sim-4-step:decomp-sim-1} of hybrid 4) on the initial state $\ket{t,x,u,b}_\cQ \otimes \ket{\phi^\Sim_{k, s, x, \tilde{r}_1, \tilde{r}_2}}$. $\Decomp^\Sim$ also acts as the identity. Since $\tilde{r}_1 \neq 0$, then $\ket{\phi^\Sim_{k, s, x, \tilde{r}_1, \tilde{r}_2}}$ is orthogonal to the states that $\Decomp^\Sim$ swaps.
    
    In this case, $\Decomp^\Sim$ acts equivalently to steps \ref{sim-3-step:refl-and-copy-1} - \ref{sim-3-step:refl-and-copy-2} of hybrid 3, and the final state is in $S^\Hyb$.

    \paragraph{Case 5:} The input is $\ket{t,x,u,b}_\cQ \otimes  \ket{\phi^\Sim_{k, s, x, \tilde{r}_1, \tilde{r}_2}}$, where \underline{$\tilde{r}_1 \neq 0$, $\tilde{r}_2 \neq 0$}, and $|s| = t-1$.

    Let us step through the action of steps \ref{sim-3-step:refl-and-copy-1} - \ref{sim-3-step:refl-and-copy-2} of hybrid 3 on this state. Initially, since $\tilde{r}_1 \neq 0$, $\cC_t \times \cS \times \cG_t \times \cA_t$ is orthogonal to states of the form:
    \begin{align*}
        &\ket{t,x,u,b}_{\cC_t} \otimes \ket{\st}_{\cS} \otimes \ket{0}_{\cG_t} \otimes \ket{\mathbf{0}}_{\cA_t}, \quad \forall (u,b,\st)\\
        &\ket{\phi^\REQ_{t,x,u,b,\st}}, \quad \forall (u,b,\st),
    \end{align*}
    and since $\tilde{r}_2 \neq 0$, $\cD$ is orthogonal to states of the form:
    \begin{align*}
        &\ket{D}\\
        &\frac{1}{\sqrt{2^{\secp}}} \cdot \sum_{r_2 \in \bit^\secp} \ket{D \cup \{(t,x,r_2)\}}
    \end{align*}
    for all databases $D$ that do not contain $(t,x,*)$.
    
    Step \ref{sim-3-step:refl-and-copy-1} ($\mathsf{ReflectAndCopy}$) verifies that $g'_t = 1$ and then acts as the identity.
    
    Step \ref{sim-3-step:conditionally-apply-decomp} applies $\Decomp_{t,x}$ to $\cD$, which also acts as the identity because $\cD$ is orthogonal to the states that $\Decomp_{t,x}$ swaps.
    
    Step \ref{sim-3-step:query-REQ-oracle} (Query $O^\REQ_P$) acts as the identity because $\cD$ is orthogonal to the states that this step checks.
    
    Step \ref{sim-3-step:decomp} ($\Decomp_{t,x}$) acts as the identity because $\cD$ is orthogonal to the states that $\Decomp_{t,x}$ swaps.

    Step \ref{sim-3-step:refl-and-copy-2} verifies that $g'_t = 1$ and then acts as the identity.
    
    The final state of the simulator's internal registers at the end of step \ref{sim-3-step:refl-and-copy-2} is the initial state $\ket{\phi^\Sim_{k, s, x, \tilde{r}_1, \tilde{r}_2}}$.

    Next, let us examine the action of $\Decomp^\Sim$ (step \ref{sim-4-step:decomp-sim-1} of hybrid 4) on the initial state $\ket{t,x,u,b}_\cQ \otimes \ket{\phi^\Sim_{k, s, x, \tilde{r}_1, \tilde{r}_2}}$. $\Decomp^\Sim$ also acts as the identity. Since $\tilde{r}_1 \neq 0$ and $\tilde{r}_2 \neq 0$, then $\ket{\phi^\Sim_{k, s, x, \tilde{r}_1, \tilde{r}_2}}$ is orthogonal to the states that $\Decomp^\Sim$ swaps.
    
    In this case, $\Decomp^\Sim$ acts equivalently to steps \ref{sim-3-step:refl-and-copy-1} - \ref{sim-3-step:refl-and-copy-2} of hybrid 3, and the final state is in $S^\Hyb$.
\end{proof}

\begin{lemma}
    At the start of step \ref{sim-3-step:copy-answer} of hybrid 3, the state of the hybrid
\end{lemma}

\begin{lemma}
    Hybrids 3 and 4 are perfectly indistinguishable after any number of quantum queries.
\end{lemma}
\begin{proof}
    We will prove that (1) at the start of any query, the state of hybrid 3 or 4's internal registers is in $S^\Sim$, and (2) that a query to hybrids 3 or 4 acts equivalently if the initial state is in $S^\Sim$. We will prove these claims inductively. For the base case, note that the initial state of the simulator's registers in hybrids 3 or 4 is $\ket{\psi^\Sim_{0, \emptyset}} \in S^\Sim$. Next, let us assume as the inductive hypothesis that at the start of any query, the state of the simulator's internal registers is in $S^\Sim$. We will show that then hybrids 3 and 4 act equivalently on this state, and the state at the end of the query is also in $S^\Sim$.

    % First, note that several steps of hybrids 3 and 4 are equivalent. Step \ref{sim-3-step:expand-registers} of hybrid 3 is equivalent to step \ref{sim-4-step:increase} of hybrid 4. They both append registers to the oracle state. Next, step \ref{sim-3-step:manage-cache} of hybrid 3 is equivalent to step \ref{sim-4-step:decomp-sim-1} of hybrid 4 ($\Decomp^\Sim$).

    It suffices to consider the action of each hybrid on every state of the form: $\ket{t,x,u,b}_\cQ \otimes \ket{\psi^\Sim_{k,s}}$.

    \paragraph{Case 1:} The input is $\ket{t,x,u,b}_\cQ \otimes  \ket{\psi^\Sim_{k,s}}$, where $(t, x) \neq (|s|, x_{|s|})$ and $t \neq |s|+1$.

    Let us step through the action of hybrid 3 on this state. Step \ref{sim-3-step:expand-registers} appends registers to the state, mapping:
    \[\ket{\psi^\Sim_{k,s}} \mapsto \ket{\psi^\Sim_{k+2,s}}\]
    Step \ref{sim-3-step:compute-B} (Validate Query) computes $(k_{\max}, x'_{k_{\max}}) = (|s|, x_{|s|})$ and then leaves $\cB' = \ket{0}$ because $(t, x) \neq (k_{\max}, x'_{k_{\max}})$ and $t \neq k_{\max}+1$.

    Next, each step from \ref{sim-3-step:refl-and-copy-1} to \ref{sim-3-step:refl-and-copy-2} is an involution (it is its own inverse). Step \ref{sim-3-step:copy-answer} (Answer Query) acts as the identity because $\cB' = \ket{0}$. Then steps \ref{sim-3-step:refl-and-copy-3} - \ref{sim-3-step:refl-and-copy-4} uncompute steps \ref{sim-3-step:refl-and-copy-1} - \ref{sim-3-step:refl-and-copy-2} because they apply steps \ref{sim-3-step:refl-and-copy-1} - \ref{sim-3-step:refl-and-copy-2} in reverse order. 
    
    Finally, step \ref{sim-3-step:compute-B-2} acts as the identity, the same as step \ref{sim-3-step:compute-B}. The final state is $\ket{t,x,u,b}_\cQ \otimes  \ket{\psi^\Sim_{k+2,s}}$.

    Now let us step through the action of hybrid 4 on $\ket{t,x,u,b}_\cQ \otimes  \ket{\psi^\Sim_{k,s}}$. Step \ref{sim-4-step:increase} appends registers, mapping:
    \[\ket{\psi^\Sim_{k,s}} \mapsto \ket{\psi^\Sim_{k+2,s}}\]
    Step \ref{sim-4-step:decomp-sim-1} ($\Decomp^\Sim$) acts as the identity because $|s| \neq t - 1$. Step \ref{sim-4-step:copy-sim} ($\mathsf{Copy}^\Sim$) acts as the identity because $(t, x) \neq (|s|, x_{|s|})$ and $|s| \neq t-1$. Step \ref{sim-4-step:decomp-sim-2} ($\Decomp^\Sim$) acts as the identity as well. The final state is $\ket{t,x,u,b}_\cQ \otimes  \ket{\psi^\Sim_{k+2,s}}$, which is in $S^\Sim$. Furthermore, we've shown that hybrids 3 and 4 act equivalently in this case.

    \paragraph{Case 2:} The input is $\ket{t,x,u,b}_\cQ \otimes  \ket{\psi^\Sim_{k,s}}$, where $t = |s|+1$.%$(t, x) = (|s|, x_{|s|})$

    Let us step through the action of hybrid 3 on this state. Step \ref{sim-3-step:expand-registers} appends registers to the state, mapping:
    \[\ket{\psi^\Sim_{k,s}} \mapsto \ket{\psi^\Sim_{k+2,s}}\]
    which acts equivalently to step \ref{sim-4-step:increase} of hybrid 4.
    Step \ref{sim-3-step:compute-B} (Validate Query) computes $(k_{\max}, x'_{k_{\max}}) = (|s|, x_{|s|})$ and then maps 
    \[\ket{0}_{\cB'} \mapsto \ket{1}_{\cB'}\]
    because $t = k_{\max}+1 = |s| + 1$.%$(t, x) = (k_{\max}, x'_{k_{\max}}) = (|s|, x_{|s|})$

    Next, steps \ref{sim-3-step:refl-and-copy-1} - \ref{sim-3-step:refl-and-copy-2} act equivalently to $\Decomp^\Sim$. This follows from \cref{thm:decomp-sim-equivalence}. At the end of step \ref{sim-3-step:refl-and-copy-2}, the state is:
    \[\ket{t,x,u,b}_\cQ \otimes  \ket{\phi^\Sim_{k+2,s,x,0,0}}\]
    Note that $\ket{\phi^\Sim_{k+2,s,x,0,0}}$ is a superposition over states of the form $\ket{\psi^\Sim_{k+2,s||(x,r_1,r_2)}} \in S^\Sim$.

    Step \ref{sim-3-step:copy-answer} (Answer Query) verifies that $(t,x,*) \in D$, then reads $(y, r_2)$ from $\cC_t \times D_{t}$, and finally maps:
    \[\ket{t,x,u,b}_\cQ \otimes  \ket{\psi^\Sim_{k+2,s||(x,r_1,r_2)}} \mapsto \ket{t,x,u \oplus (y, r_2),b \oplus 1}_\cQ \otimes  \ket{\psi^\Sim_{k+2,s||(x,r_1,r_2)}}\]
    Note that step \ref{sim-3-step:copy-answer} acts the same as $\mathsf{Copy}^\Sim$.

    Steps \ref{sim-3-step:refl-and-copy-3} - \ref{sim-3-step:refl-and-copy-4} uncompute steps \ref{sim-3-step:refl-and-copy-1} - \ref{sim-3-step:refl-and-copy-2} and act equivalently to $(\Decomp^\Sim)^\dag = \Decomp^\Sim$.

    Finally, step \ref{sim-3-step:compute-B-2} (Validate Query) computes $(k_{\max}, x'_{k_{\max}}) = (|s|+1, x) = (t,x)$ and then maps:
    \[\ket{1}_{\cB'} \mapsto \ket{0}_{\cB'}\]
    This shows that in this case, hybrid 3 acts equivalently to hybrid 4, and the final state is in $S^\Sim$.

    \paragraph{Case 3:} The input is $\ket{t,x,u,b}_\cQ \otimes  \ket{\psi^\Sim_{k,s}}$, where $(t, x) = (|s|, x_{|s|})$.

    Let us step through the action of hybrid 3 on this state. Step \ref{sim-3-step:expand-registers} appends registers to the state, mapping:
    \[\ket{\psi^\Sim_{k,s}} \mapsto \ket{\psi^\Sim_{k+2,s}}\]
    which acts equivalently to step \ref{sim-4-step:increase} of hybrid 4.
    Step \ref{sim-3-step:compute-B} (Validate Query) computes $(k_{\max}, x'_{k_{\max}}) = (|s|, x_{|s|})$ and then maps 
    \[\ket{0}_{\cB'} \mapsto \ket{1}_{\cB'}\]
    because $(t, x) = (|s|, x_{|s|}) = (k_{\max}, x'_{k_{\max}})$.

    At the start of step \ref{sim-3-step:refl-and-copy-1}, the state of the simulator is $\ket{\psi^\Sim_{k+2, s}}$, which is a superposition over states of the form
    \[\ket{\phi^\Sim_{k+2,s_{[t-1]}, x, \tilde{r}_1, \tilde{r}_2}}, \quad \forall (\tilde{r}_1, \tilde{r}_2) \in \cR \times \bit^\secp\]
    where $s_{[t-1]} = (s_1, \dots, s_{t-1})$.
    
    Next, steps \ref{sim-3-step:refl-and-copy-1} - \ref{sim-3-step:refl-and-copy-2} act equivalently to $\Decomp^\Sim$. This follows from \cref{thm:decomp-sim-equivalence}. At the end of step \ref{sim-3-step:refl-and-copy-2}, the state is in a superposition over states of the form:
    \begin{align*}
        &\ket{\psi^\Sim_{k+2,s_{[t-1]}}}\\
        &\ket{\psi^\Sim_{k+2,s_{[t-1]}||(x, r_1, r_2)}}, \quad \forall (r_1, r_2)
    \end{align*}

\proofauditloc{P18}{%
This first row has only $t-1$ history entries. Step
\ref{sim-4-step:copy-sim} of Hybrid 4 in
\Cref{sec:LEQ:proof:simulating-hybrid} therefore leaves $b$
unchanged on it, whereas the next paragraph claims that the copy
operation flips $b$ on this same row.
}

\jiahui{Take a stab after the deadline.}

    Step \ref{sim-3-step:copy-answer} (Answer Query) acts equivalently to $\mathsf{Copy}^\Sim$. If the simulator's state is $\ket{\psi^\Sim_{k+2,s_{[t-1]}}}$, then $(t,x,*) \notin D$, so step \ref{sim-3-step:copy-answer} maps
    \begin{align*}
        \ket{t,x,u,b}_{\cQ} &\mapsto \ket{t,x,u,b \oplus 1}_{\cQ},
    \end{align*}
    which $\mathsf{Copy}^\Sim$ does as well. \proofauditloc{P18}{%
This contradicts step \ref{sim-4-step:copy-sim} of Hybrid 4 in
\Cref{sec:LEQ:proof:simulating-hybrid}: its identity clause leaves
$b$ unchanged on a history of length $t-1$. The source's separate
rule that would flip $b$ on this history is commented out.
} If the simulator's state is $\ket{\psi^\Sim_{k+2,s_{[t-1]}||(x, r_1, r_2)}}$, then $(t,x,*) \in D$, so step \ref{sim-3-step:copy-answer} acts the same as it did in case 2 above.

    Steps \ref{sim-3-step:refl-and-copy-3} - \ref{sim-3-step:refl-and-copy-4} uncompute steps \ref{sim-3-step:refl-and-copy-1} - \ref{sim-3-step:refl-and-copy-2} and act equivalently to $(\Decomp^\Sim)^\dag = \Decomp^\Sim$.

    Finally, step \ref{sim-3-step:compute-B-2} (Validate Query) computes either $(k_{\max}, x'_{k_{\max}}) = (|s|, x) = (t,x)$ or $k_{\max} = t-1$. In either case, the step maps:
    \[\ket{1}_{\cB'} \mapsto \ket{0}_{\cB'}\]
    This shows that in this case, hybrid 3 acts equivalently to hybrid 4, and the final state is in $S^\Sim$.
\end{proof}
\else
    \input{old-proof-for-hybrid-REQ-oracle}
\fi

\subsubsection*{State Transition Function of $O_\Hyb^P$.}
Let us define the reachable states of $O_\Hyb^P$'s internal registers.
\begin{definition}[States of $\Sim$ in Hybrid 5]\label{def:basis-of-Hyb}
    Let $s \in (\cX \times \cR \times \bit^\secp)^*$ be a (potentially empty) tuple, and for each $i \in [|s|]$, let 
    \[s_i = (x_i, r_{1,i}, r_{2, i}) \in \cX \times \cR \times \bit^\secp.\]
    
    Next, let $\st_1 = \bot$, and for each $i \in [|s|]$, let
    \begin{align*}
        d_i &= (i, s_i) = (i, x_i, r_{1,i}, r_{2,i})\\
        (\st_{i+1}, y_i) &= P(\st_i, x_i; r_{1,i})
    \end{align*}
    
    Let us define a computational-basis state as follows:
    \begin{align*}
        \ket{\psi^\Hyb_{s}} &= \ket{\{d_i\}_{i \in [|s|]}}_\cD \otimes \ket{0,0}_\cW \otimes \ket{0}_{\cB}
    \end{align*}

    Let us define a Fourier basis state as follows. For any $x \in \cX$, $\tilde{r}_1 \in \cR$ and $\tilde{r}_2 \in \bit^\secp$, let:
    \begin{align*}
        \ket{\phi^\Hyb_{s, x, \tilde{r}_1, \tilde{r}_2}} &= \frac{1}{\sqrt{\abs{R} \cdot 2^\secp}} \cdot \sum_{(r_1, r_2) \in \cR \times \bit^\secp} (-1)^{\langle r_1, \tilde{r}_1 \rangle + \langle r_2, \tilde{r}_2 \rangle} \ket{\psi^\Hyb_{s||(x, r_1, r_2)}}\\
        &= \frac{1}{\sqrt{\abs{R} \cdot 2^\secp}} \cdot \sum_{(r_1, r_2) \in \cR \times \bit^\secp} (-1)^{\langle r_1, \tilde{r}_1 \rangle + \langle r_2, \tilde{r}_2 \rangle} \ket{\{d_i\}_{i \in [|s|]} \cup \{(|s|+1, x, r_1, r_2)\}}_\cD \otimes \ket{0,0}_\cW \otimes \ket{0}_{\cB}
    \end{align*}
\end{definition}

% The following claim says that the computational-basis states $\ket{\psi^\Hyb_{s|| (x, r_1, r_2)}}$ span the same space as the Fourier-basis states $\ket{\phi^\Hyb_{s, x, \tilde{r}_1, \tilde{r}_2}}$.
% \begin{lemma}\label{thm:comp-to-fourier-bases}
%     For any sequence $s \in \left(\cX \times \cR \times \bit^\secp\right)^*$ and any $x \in \cX$,
%     \[\left\{\ket{\psi^\Hyb_{s|| (x, r_1, r_2)}}\right\}_{(r_1, r_2) \in \cR \times \bit^\secp} \quad \text{and} \quad \left\{\ket{\phi^\Hyb_{s, x, \tilde{r}_1, \tilde{r}_2}}\right\}_{(\tilde{r}_1, \tilde{r}_2) \in \cR \times \bit^\secp}\]
%     are orthonormal bases for the same space.
% \end{lemma}
% \begin{proof}
%     \bhaskar{TBD}
% \end{proof}

\paragraph{Hybrid 5:}
The initial state of the simulator is $\ket{\psi^\Hyb_\emptyset}$. Next, given any query of the form $\ket{t, x,u,b}_\cQ$, the simulator does the following:
\begin{enumerate}
    \item \label{sim-5-step:increase}$\mathsf{Increase}$: \textcolor{red}{(omitted)}
    \item \label{sim-5-step:decomp-hyb-1}$\Decomp^{\textcolor{red}{\Hyb}}$: For any sequence $s \in \left(\cX \times \cR \times \bit^\secp\right)^*$ of length $t-1$, swap the following states:
    \[\ket{\psi\textcolor{red}{^\Hyb_s}} \leftrightarrow \ket{\phi\textcolor{red}{^\Hyb_{s, x, 0, 0}}}\]
    and act as the identity on all states orthogonal to the ones above.
    
    \item \label{sim-5-step:copy-hyb}$\mathsf{Copy}^{\textcolor{red}{\Hyb}}$: 
    \begin{itemize}
        \item For any sequence $s \in \left(\cX \times \cR \times \bit^\secp\right)^*$ of length $t$ for which $s_t = (x, *, *)$, swap the following states:
        \[\ket{t,x,u,b}_\cQ \otimes \ket{\psi\textcolor{red}{^\Hyb_s}} \leftrightarrow \ket{t,x,u \oplus (y_t, r_{2,t}),b \oplus 1}_\cQ \otimes \ket{\psi\textcolor{red}{^\Hyb_s}}\]
        where $y_t, r_{2,t}$ are computed from $s$ as described in \textcolor{red}{\cref{def:basis-of-Hyb}}. 
        
        % \item For any sequence $s \in \left(\cX \times \cR \times \bit^\secp\right)^*$ of length $t-1$, swap the following states:
        % \[\ket{t,x,u,b}_\cQ \otimes \ket{\psi\textcolor{red}{^\Hyb_s}} \leftrightarrow \ket{t,x,u,b \oplus 1}_\cQ \otimes \ket{\psi\textcolor{red}{^\Hyb_s}}\]
        % \item 
        Act as the identity on all states orthogonal to the ones above.
    \end{itemize}
    \item \label{sim-5-step:decomp-hyb-2} $(\Decomp^{\textcolor{red}{\Hyb}})^\dag$: Uncompute step \ref{sim-5-step:decomp-hyb-1}.
\end{enumerate}

\begin{lemma}
    Hybrids 4 and 5 are perfectly indistinguishable after any number of quantum queries.
\end{lemma}
\begin{proof}
    Hybrids 4 and 5 implement essentially the same state transition function with different eignenbases. We will construct a unitary that maps between the bases of hybrids 4 and 5. 
    
    Hybrid 4 works with states of the form $\ket{\psi^\Sim_{k,s}}$ (\cref{def:basis-of-Sim}). $k$ refers to the length of registers $\cC, \cG, \cA$. Note that after any number of queries $q \in \{0\} \cup \bbN$, $k = 2 q$. This is because before any queries have been made, $k = 0$, and $\mathsf{Increase}$ (step \ref{sim-4-step:increase}) is the only step where $k$ changes, and this step increases $k$ by $2$ during each query.

    Hybrid 5 works with states of the form $\ket{\psi^\Hyb_{s}}$ (\cref{def:basis-of-Hyb}). These states do not depend on $k$.\\

    For any query number $q \in \{0\} \cup \bbN$, let us construct a unitary $U_q$ that maps the state in hybrid 4 after $q$ queries to the state in hybrid 5 after $q$ queries. For each sequence $s$ of length $|s| \leq 2q$, let $U_q$ map:
    \[\ket{\psi^\Sim_{2q,s}} \mapsto \ket{\psi^\Hyb_{s}}\]
    Let $U_q$ act as the identity on all states orthogonal to the ones above.

    Note that 
    \[U_0 \ket{\psi^\Sim_{0,\emptyset}} = \ket{\psi^\Hyb_{\emptyset}}\]
    So $U_0$ maps the initial state of hybrid 4 to the initial state of hybrid 5. Additionally, $U_q$ acts on the simulator's internal registers only.\\

    Let us use $\{U_q\}_{q \in \{0\} \cup \bbN}$ to transform hybrid 4 into hybrid 5. Let us start with hybrid 4. Then change the initial state to be $\ket{\psi^\Hyb_{\emptyset}}$. Then on each query $q \in \bbN$, the simulator applies $U_{q-1}^\dag$ at the beginning and applies $U_q$ at the end. The state of the simulator after $q$ queries is computed as follows (from right to left):
    \begin{equation}\label{eq:hybrid-4-with-conjugation}
        \left(\textcolor{red}{U_q} \cdot \text{Hybrid 4, Query q} \cdot \textcolor{red}{U_{q-1}}^\dag\right) \cdot \ldots \cdot \left(\textcolor{red}{U_2} \cdot \text{Hybrid 4, Query 2} \cdot \textcolor{red}{U_1^\dag}\right) \cdot \left(\textcolor{red}{U_1} \cdot \text{Hybrid 4, Query 1} \cdot \textcolor{red}{U_0^\dag}\right) \cdot \textcolor{red}{\ket{\psi^\Hyb_{\emptyset}}}
    \end{equation}
    
    Procedure \ref{eq:hybrid-4-with-conjugation} is perfectly indistinguishable from hybrid 4. The procedure computes the same state as the following procedure:
    \begin{equation}\label{eq:hybrid-4-with-extra-unitary}
        \textcolor{red}{U_q} \cdot \text{Hybrid 4, Query q} \cdot \ldots \cdot \text{Hybrid 4, Query 2} \cdot \text{Hybrid 4, Query 1} \cdot \ket{\psi^\Sim_{0,\emptyset}}
    \end{equation}
    This is because the unitaries between queries, such as $U_{1}^\dag \cdot U_{1}$, cancel out, and $U_0^\dag \cdot \ket{\psi^\Hyb_{\emptyset}} = \ket{\psi^\Sim_{0,\emptyset}}$. Procedure \ref{eq:hybrid-4-with-extra-unitary} computes hybrid 4 and then applies $U_q$ after the final query. The application of $U_q$ is undetectable to the party that queries the simulator because $U_q$ acts on the simulator's internal registers only.
    
    Additionally, procedure \ref{eq:hybrid-4-with-conjugation} is equivalent to hybrid 5. Just like hybrid 5, the procedure starts with the state $\ket{\psi^\Hyb_{\emptyset}}$. Additionally, each operation of the form $\left(\textcolor{red}{U_q} \cdot \text{Hybrid 4, Query q} \cdot \textcolor{red}{U_{q-1}}^\dag\right)$ is equivalent to hybrid 5 because hybrids 4 and 5 implement the same state transition function, just on different eigenbases. This shows that hybrids 4 and 5 are perfectly indistinguishable.
\end{proof}

Next, we show that hybrid 5 is equivalent to the hybrid oracle $O^P_\Hyb(1^\secp)$. Hybrid 5 simply describes the abstract state transition function of $O^P_\Hyb(1^\secp)$.

\paragraph{Hybrid 6:} The oracle $O^P_\Hyb(1^\secp)$ in \cref{fig:REQ-hybrid-oracle}.

Given some basis state $\ket{t, x,u,b}_\cQ \otimes \ket{D}_\cD \otimes \ket{0,0}_\cW \otimes \ket{0}_{\cB}$, the oracle does the following:
    \begin{enumerate}
        \item \label{sim-6-step:REQ-hybrid-valid-query}\textbf{Validate Query.} If either $t =  |D| \land x = x_{|D|}$ or if $t = |D| + 1$, then apply $\mathsf{X}$ to register $\cB$ to obtain $\ket{1}_\cB$.
        \item \textbf{Answer Query.} If $\cB$ contains $1$, then answer the query as follows.
        \begin{enumerate}
            \item \label{sim-6-step:REQ-hybrid-query-QRO}CNOT $(t, x)$ from $\cQ$ to $\cW$. 
            \[\ket{0,0,0}_\cW \mapsto \ket{t,x,0}_\cW\]
            Then query $\ket{(t, x), 0}_{\cW}$ to the compressed oracle by applying
            \[\Decomp_{t,x} \cdot \CO' \cdot \Decomp_{t,x}\]
            to registers $\cW \times \cD$.
            \item \label{sim-6-step:REQ-hybrid-compute-y} Let $\st_{t} = \StateD_{t}(D)$ and parse $r = (r_1, r_2) \in \cR \times \bit^\secp$ from register $\cW$. Then compute $(\st_{t+1}, y) = P(\st_{t}, x;r_1)$.
            \item \label{sim-6-step:REQ-hybrid-copy-output}Using CNOTs, write the value $(y, r_2)$ to the sub-register of $\cQ$ containing $\ket{u}$ and flip $\ket{b}$:
            \begin{gather*}
                \ket{t, x, u, b}_\cQ \otimes \ket{(t, x), r}_{\cW} \\
                \mapsto \\
                \ket{t, x, u\oplus (y, r_2), b\oplus 1}_\cQ \otimes \ket{(t, x), r}_{\cW}
            \end{gather*}
            \item \label{sim-6-step:REQ-hybrid-uncompute-RO-query}Uncompute step \ref{sim-6-step:REQ-hybrid-query-QRO}. First apply
            \[\Decomp_{t,x} \cdot \CO' \cdot \Decomp_{t,x}\]
            to registers $\cW \times \cD$. Then map
            \[\ket{t,x,0}_\cW \mapsto \ket{0,0,0}_\cW\]
        \end{enumerate}
        \item \label{sim-6-step:REQ-hybrid-uncompute-validate-query}Uncompute step \ref{sim-6-step:REQ-hybrid-valid-query}.
    \end{enumerate}

% The following definition defines the space of reachable states of $O^P_\Hyb$'s internal registers. 
% \begin{definition}[Reachable States of $O^P_\Hyb$]\label{def:reachable-states-of-hyb}
%     Let $S^\Hyb$ be the span of $\ket{\psi^\Hyb_{\emptyset}}$ and all states of the form $\ket{\phi^\Hyb_{s,x',\tilde{r}_1,\tilde{r}_2}}$ for all (potentially empty) sequences $s \in (\cX \times \cR \times \bit^\secp)^*$, all $x' \in \cX$, all $\tilde{r}_1 \in \cR$, and all $\tilde{r}_2 \in \bit^\secp$ such that $(\tilde{r}_1, \tilde{r}_2) \neq (0,0)$.
% \end{definition}

\begin{lemma}
    Hybrids 5 and 6 are perfectly indistinguishable after any number of quantum queries.
\end{lemma}
% \begin{lemma}
%     At the start of a query, if the state of $\cD \times \cW \times \cB$ is in the span of $S^\Hyb$, then the query to $O^P_\Hyb(1^\secp)$ acts on $\cQ \times \cD \times \cW \times \cB$ equivalently to:
%     \[\Decomp^\Hyb \cdot \mathsf{Copy}^\Hyb \cdot \Decomp^\Hyb\]
%     and the state at the end of the query is also in $S^\Hyb$.
% \end{lemma}
\begin{proof}
    Let us first simplify hybrid 6, steps \ref{sim-6-step:REQ-hybrid-query-QRO} - \ref{sim-6-step:REQ-hybrid-uncompute-RO-query}, which query the compressed oracle and copy the output to $\cQ$. Hybrid 6 is equivalent to the following:

    \paragraph{Hybrid 5.c:}
    Given some basis state $\ket{t, x,u,b}_\cQ \otimes \ket{D}_\cD \otimes \ket{0,0}_\cW \otimes \ket{0}_{\cB}$, the oracle does the following:
    \begin{enumerate}
        \item \label{sim-6-1-step:REQ-hybrid-valid-query}\textbf{Validate Query.} If either $t =  |D| \land x = x_{|D|}$ or if $t = |D| + 1$, then apply $\mathsf{X}$ to register $\cB$ to obtain $\ket{1}_\cB$.
        \item \textbf{Answer Query.} If $\cB$ contains $1$, then answer the query as follows.
        \begin{enumerate}
            \item \label{sim-6-1-step:Decomp-1}Apply $\Decomp_{t,x}$ to register $\cD$.
            \item \label{sim-6-1-step:CNOT-onto-W}CNOT $(t, x)$ from $\cQ$ to $\cW$. 
            \[\ket{0,0,0}_\cW \mapsto \ket{t,x,0}_\cW\]
            \item \label{sim-6-1-step:CNOT-D-to-W}Apply $\CO'$ to registers $\cW \times \cD$.
            \item \label{sim-6-1-step:REQ-hybrid-compute-y} Let $\st_{t} = \StateD_{t}(D)$ and parse $r = (r_1, r_2) \in \cR \times \bit^\secp$ from register $\cW$. Then compute $(\st_{t+1}, y) = P(\st_{t}, x;r_1)$.
            \item \label{sim-6-1-step:REQ-hybrid-copy-output}Using CNOTs, write the value $(y, r_2)$ to the sub-register of $\cQ$ containing $\ket{u}$ and flip $\ket{b}$:
            \begin{gather*}
                \ket{t, x, u, b}_\cQ \otimes \ket{(t, x), r}_{\cW} \\
                \mapsto \\
                \ket{t, x, u\oplus (y, r_2), b\oplus 1}_\cQ \otimes \ket{(t, x), r}_{\cW}
            \end{gather*}
            \item \label{sim-6-1-step:REQ-hybrid-uncompute-RO-query}Apply $\CO'$ to registers $\cW \times \cD$. 
            \item \label{sim-6-1-step:CNOT-onto-W-2}CNOT $(t, x)$ from $\cQ$ to $\cW$:
            \[\ket{t,x,0}_\cW \mapsto \ket{0,0,0}_\cW\]
            \item \label{sim-6-1-step:Decomp-2}Apply $(\Decomp_{t,x})^\dag$ to register $\cD$.
        \end{enumerate}
        \item \label{sim-6-1-step:REQ-hybrid-uncompute-validate-query}Uncompute step \ref{sim-6-1-step:REQ-hybrid-valid-query}.
    \end{enumerate}
        The only changes we have made are to the $\Decomp_{t,x}$ operations. The first $\Decomp_{t,x}$ operation has been moved to step \ref{sim-6-1-step:Decomp-1} because it commutes with step \ref{sim-6-1-step:CNOT-onto-W}. Likewise, the fourth $\Decomp_{t,x}$ operation has been moved to step \ref{sim-6-1-step:Decomp-2} because it commutes with step \ref{sim-6-1-step:CNOT-onto-W-2}. Note that we've also replaced the fourth $\Decomp_{t,x}$ with $(\Decomp_{t,x})^\dag$. This is just a notational change because $\Decomp_{t,x} = (\Decomp_{t,x})^\dag$. Finally, the second and third $\Decomp_{t,x}$ operations have been removed from steps \ref{sim-6-step:REQ-hybrid-query-QRO} and \ref{sim-6-step:REQ-hybrid-uncompute-RO-query} of hybrid 6. We have cancelled them out. First, $\Decomp_{t,x}$ commutes with the steps in the middle (\ref{sim-6-step:REQ-hybrid-compute-y} and \ref{sim-6-step:REQ-hybrid-copy-output} of hybrid 6). These steps compute $\st_t$ by reading the first $t-1$ entries of $\cD$, and they read $r = (r_1, r_2)$ from $\cW$. In contrast, $\Decomp_{t,x}$ acts only on the $t$-th entry of $\cD$, so it commutes with steps \ref{sim-6-step:REQ-hybrid-compute-y} and \ref{sim-6-step:REQ-hybrid-copy-output} of hybrid 6. Next, $\Decomp_{t,x}$ is its own inverse, so the second and third applications of $\Decomp_{t,x}$ in hybrid 6 cancel out with each other.\\

        Next, hybrid 5.c is equivalent to the following hybrid:
    \paragraph{Hybrid 5.b:}
    Given some basis state $\ket{t, x,u,b}_\cQ \otimes \ket{D}_\cD \otimes \ket{0,0}_\cW \otimes \ket{0}_{\cB}$, the oracle does the following:
    \begin{enumerate}
        \item \label{sim-6-2-step:REQ-hybrid-valid-query}\textbf{Validate Query.} If either $t =  |D| \land x = x_{|D|}$ or if $t = |D| + 1$, then apply $\mathsf{X}$ to register $\cB$ to obtain $\ket{1}_\cB$.
        \item \textbf{Answer Query.} If $\cB$ contains $1$, then answer the query as follows.
        \begin{enumerate}
            \item \label{sim-6-2-step:decomp-1}Apply $\Decomp_{t,x}$ to $\cD$.
            \item \label{sim-6-2-step:copy-prep}Let $\st_{t} = \StateD_{t}(D)$, read $r$ from the entry $(t,x,r) \in D$, and parse $r = (r_1, r_2) \in \cR \times \bit^\secp$. Then compute $(\st_{t+1}, y) = P(\st_{t}, x;r_1)$.
            \item \label{sim-6-2-step:copy}Using CNOTs, write the value $(y, r_2)$ to the sub-register of $\cQ$ containing $\ket{u}$ and flip $\ket{b}$:
            \begin{gather*}
                \ket{t, x, u, b}_\cQ
                \mapsto \\
                \ket{t, x, u\oplus (y, r_2), b\oplus 1}_\cQ
            \end{gather*}
            \item \label{sim-6-2-step:decomp-2}Apply $(\Decomp_{t,x})^\dag$
            to register $\cD$.
        \end{enumerate}
        \item \label{sim-6-2-step:REQ-hybrid-uncompute-validate-query}Uncompute step \ref{sim-6-2-step:REQ-hybrid-valid-query}.
    \end{enumerate}
        The only difference is that we have eliminated all actions on $\cW$ (we've cut out the middle man). Now we read $r$ directly from the $t$-th entry of $\cD$. It is clear by inspection that hybrid 5.c is equivalent to hybrid 5.b.

        Next, hybrid 5.b is equivalent to the following hybrid:
    \paragraph{Hybrid 5.a:}
    Given some basis state $\ket{t, x,u,b}_\cQ \otimes \ket{D}_\cD \otimes \ket{0,0}_\cW \otimes \ket{0}_{\cB}$, the oracle does the following:
    \begin{enumerate}
        \item \label{sim-6-3-step:REQ-hybrid-valid-query}\textbf{Validate Query.} If either $t =  |D| \land x = x_{|D|}$ or if $t = |D| + 1$, then apply $\mathsf{X}$ to register $\cB$ to obtain $\ket{1}_\cB$.
        \item \textbf{Answer Query.} If $\cB$ contains $1$, then answer the query as follows.
        \begin{enumerate}
            \item \label{sim-6-3-step:decomp-1}Apply $\Decomp^\Hyb$.
            \item \label{sim-6-3-step:copy}
            Apply $\mathsf{Copy}^\Sim$.
            \item \label{sim-6-3-step:decomp-2}Apply $(\Decomp^\Hyb)^\dag$.
        \end{enumerate}
        \item \label{sim-6-3-step:REQ-hybrid-uncompute-validate-query}Uncompute step \ref{sim-6-3-step:REQ-hybrid-valid-query}.
    \end{enumerate}

    We have replaced $\Decomp_{t,x}$ with $\Decomp^\Hyb$. These two operations compute the same function. We have also replaced steps \ref{sim-6-2-step:copy-prep} and \ref{sim-6-2-step:copy} with $\mathsf{Copy}^\Sim$. These two operations compute the same function as long as the state at the start of the operation is a superposition over databases $D$ with $(|D|, x_{|D|}) = (t, x)$. This property is enforced by the query validation in step \ref{sim-6-2-step:REQ-hybrid-valid-query} and the properties of the compressed oracle.\bhaskar{More justification needed.}
    \proofauditloc{P18}{Validation before decompression does not exclude a blank-history component at this copy step.}%

    Finally, we claim that hybrid 5.a is equivalent to hybrid 5. The only difference is that hybrid 5 always executes $(\Decomp^\Hyb)^\dag \cdot \mathsf{Copy^\Sim} \cdot \Decomp^\Hyb$, whereas hybrid 5.a executes these steps only when the initial state passes query validation in step \ref{sim-6-3-step:REQ-hybrid-valid-query}. Hybrids 5.a and 5 are equivalent because $(\Decomp^\Hyb)^\dag \cdot \mathsf{Copy^\Sim} \cdot \Decomp^\Hyb$ acts as the identity on any state that fails query validation.

\end{proof}

In summary, we have shown that $\Sim^{O_P^{\REQ}}(1^\secp)$ (hybrid 1) and $O_{\Hyb}^{P}(1^\secp)$ (hybrid 6) are perfectly indistinguishable. This completes the proof of \cref{prop:REQ-hyb-to-REQ}.
\end{proof}
\fi

\ifllncs
\else
\section{AI Disclosure}
LLM tools have been used to check correctness of the proofs and provide feedback on our writing. 
\iffalse
    GPT helped detect minor correctness issues in the first version of proofs for helper lemmas \Cref{claim:REQ-hyb-oracle_early-answer} and \Cref{lem:decomp-approx-identity} from incorrect use of inequalities and suggested fixes. The statements and claims are correct.
    
    The following are the main prompts we used:
    \begin{itemize}
        \item Read this paper draft and make the following audits, by adding comments in red color besides these locations: 1. detect any proof writing that may have caveats and suggest how to fix it; 2. point out any proof writing that is not clear/rigorous and how to make it clearer. Only point out issues that is obviously unclear or potentially wrong and is important to the results of the paper.
     
    Only audit the files used in the main file "main.tex". Ignore the commented out sections and un-used files.
    Do not change the structure and directories of the project. Add commands and Make audits on the original files. Output the entire zip file.

    \item Explain your audit comment number XXX in more details, in intuitive languages. Do not use terms/abstractions that have not appeared in the draft.
    
    \item Please review the following revised proof for audit XXX. Do not be nitpicky. Simply let me know if there is any substantial gap. If not then it's good.
    \end{itemize}
\fi
\fi

\bibliographystyle{alpha}
\bibliography{references,abbrev3,crypto}

\newcommand{\etalchar}[1]{$^{#1}$}
\begin{thebibliography}{HHWW19}

\bibitem[AB26]{ananth2026quantum}
Prabhanjan Ananth and Divyanshu Bhardwaj.
\newblock Quantum one time programs: Less assumptions, more feasibility and one
  message 2pc.
\newblock {\em Cryptology ePrint Archive}, 2026.

\bibitem[ABG26]{cananth2026noninteractive}
Prabhanjan Ananth, Divyanshu Bhardwaj, and Aparna Gupte.
\newblock Non interactive {MPC}, (quantumly) revisited.
\newblock Cryptology {ePrint} Archive, Paper 2026/302, 2026.

\bibitem[AGKZ20]{AGKZ20}
Ryan Amos, Marios Georgiou, Aggelos Kiayias, and Mark Zhandry.
\newblock One-shot signatures and applications to hybrid quantum/classical
  authentication.
\newblock In {\em Proceedings of the 52nd Annual ACM SIGACT Symposium on Theory
  of Computing}, STOC 2020, page 255–268. Association for Computing
  Machinery, 2020.

\bibitem[AMRS20]{EC:AMRS20}
Gorjan Alagic, Christian Majenz, Alexander Russell, and Fang Song.
\newblock Quantum-access-secure message authentication via
  blind-unforgeability.
\newblock In Anne Canteaut and Yuval Ishai, editors, {\em EUROCRYPT~2020,
  Part~III}, volume 12107 of {\em {LNCS}}, pages 788--817. Springer, Cham, May
  2020.

\bibitem[BDS23]{BDS23}
Shalev Ben-David and Or~Sattath.
\newblock Quantum tokens for digital signatures.
\newblock {\em Quantum}, 7:901, jan 2023.

\bibitem[BGS13]{BGS13}
Anne Broadbent, Gus Gutoski, and Douglas Stebila.
\newblock Quantum one-time programs.
\newblock In {\em Advances in Cryptology {\textendash} {CRYPTO} 2013}, pages
  344--360. Springer Berlin Heidelberg, 2013.

\bibitem[CHJV15]{STOC:CHJV15}
Ran Canetti, Justin Holmgren, Abhishek Jain, and Vinod Vaikuntanathan.
\newblock Succinct garbling and indistinguishability obfuscation for {RAM}
  programs.
\newblock In Rocco~A. Servedio and Ronitt Rubinfeld, editors, {\em 47th ACM
  STOC}, pages 429--437. {ACM} Press, June 2015.

\bibitem[CLLZ21]{coladangelo2021hidden}
Andrea Coladangelo, Jiahui Liu, Qipeng Liu, and Mark Zhandry.
\newblock Hidden cosets and applications to unclonable cryptography.
\newblock In {\em Advances in Cryptology--CRYPTO 2021: 41st Annual
  International Cryptology Conference, CRYPTO 2021, Virtual Event, August
  16--20, 2021, Proceedings, Part I 41}, pages 556--584. Springer, 2021.

\bibitem[{Gav}12]{gavinsky2014classicalqm}
D.~{Gavinsky}.
\newblock Quantum money with classical verification.
\newblock In {\em 2012 IEEE 27th Conference on Computational Complexity}, pages
  42--52, June 2012.

\bibitem[GGH{\etalchar{+}}16]{garg2016candidate}
Sanjam Garg, Craig Gentry, Shai Halevi, Mariana Raykova, Amit Sahai, and Brent
  Waters.
\newblock Candidate indistinguishability obfuscation and functional encryption
  for all circuits.
\newblock {\em SIAM Journal on Computing}, 45(3):882--929, 2016.

\bibitem[GHS16]{gagliardoni2016semantic}
Tommaso Gagliardoni, Andreas H{\"u}lsing, and Christian Schaffner.
\newblock Semantic security and indistinguishability in the quantum world.
\newblock In {\em Annual international cryptology conference}, pages 60--89.
  Springer, 2016.

\bibitem[GKR08]{goldwasser2008one}
Shafi Goldwasser, Yael~Tauman Kalai, and Guy~N Rothblum.
\newblock One-time programs.
\newblock In {\em Advances in Cryptology--CRYPTO 2008: 28th Annual
  International Cryptology Conference, Santa Barbara, CA, USA, August 17-21,
  2008. Proceedings 28}, pages 39--56. Springer, 2008.

\bibitem[GLQ{\etalchar{+}}26]{eprint:GLQRRZ26}
Aparna Gupte, Jiahui Liu, Luowen Qian, Justin Raizes, Bhaskar Roberts, and Mark
  Zhandry.
\newblock On best-possible one-time programs.
\newblock {\em {IACR} Cryptol. ePrint Arch.}, 2026:413, 2026.

\bibitem[GLR{\etalchar{+}}25]{GLRRV25}
Aparna Gupte, Jiahui Liu, Justin Raizes, Bhaskar Roberts, and Vinod
  Vaikuntanathan.
\newblock Quantum one-time programs, revisited.
\newblock In {\em Proceedings of the 57th Annual ACM Symposium on Theory of
  Computing}, pages 213--221, 2025.

\bibitem[GM24]{gunn2024quantum}
Sam Gunn and Ramis Movassagh.
\newblock Quantum one-time protection of any randomized algorithm.
\newblock {\em private communication}, 2024.

\bibitem[HHWW19]{C:HHWW19}
Ariel Hamlin, Justin Holmgren, Mor Weiss, and Daniel Wichs.
\newblock On the plausibility of fully homomorphic encryption for {RAMs}.
\newblock In Alexandra Boldyreva and Daniele Micciancio, editors, {\em
  CRYPTO~2019, Part~I}, volume 11692 of {\em {LNCS}}, pages 589--619. Springer,
  Cham, August 2019.

\bibitem[HV25]{HV25}
Andrew Huang and Vinod Vaikuntanathan.
\newblock A simple and efficient one-shot signature scheme, 2025.

\bibitem[KL20]{KatLin}
J.~Katz and Y.~Lindell.
\newblock {\em Introduction to Modern Cryptography}.
\newblock Chapman \& Hall/CRC Cryptography and Network Security Series. CRC
  Press, 2020.

\bibitem[Liu23]{liu2023depth}
Qipeng Liu.
\newblock Depth-bounded quantum cryptography with applications to one-time
  memory and more.
\newblock In {\em 14th Innovations in Theoretical Computer Science Conference
  (ITCS 2023)}, 2023.

\bibitem[LMW23]{STOC:LinMooWic23}
Wei-Kai Lin, Ethan Mook, and Daniel Wichs.
\newblock Doubly efficient private information retrieval and fully homomorphic
  {RAM} computation from ring {LWE}.
\newblock In Barna Saha and Rocco~A. Servedio, editors, {\em 55th ACM STOC},
  pages 595--608. {ACM} Press, June 2023.

\bibitem[LSZ20]{liu2020quantum}
Qipeng Liu, Amit Sahai, and Mark Zhandry.
\newblock Quantum immune one-time memories.
\newblock {\em Cryptology ePrint Archive}, 2020.

\bibitem[RS19]{radian2019semi}
Roy Radian and Or~Sattath.
\newblock Semi-quantum money.
\newblock In {\em Proceedings of the 1st ACM Conference on Advances in
  Financial Technologies}, AFT '19, page 132–146. Association for Computing
  Machinery, 2019.

\bibitem[Shm22a]{shmueli2022public}
Omri Shmueli.
\newblock Public-key quantum money with a classical bank.
\newblock In {\em Proceedings of the 54th Annual ACM SIGACT Symposium on Theory
  of Computing}, pages 790--803, 2022.

\bibitem[Shm22b]{shmueli2022semi}
Omri Shmueli.
\newblock Semi-quantum tokenized signatures.
\newblock In {\em Annual International Cryptology Conference}, pages 296--319.
  Springer, 2022.

\bibitem[Sta25a]{stambler2025cryptography}
Lev Stambler.
\newblock Cryptography without long-term quantum memory and global
  entanglement: Classical setups for one-time programs, copy protection, and
  stateful obfuscation.
\newblock {\em arXiv preprint arXiv:2504.21842}, 2025.

\bibitem[Sta25b]{stambler2025information}
Lev Stambler.
\newblock Information theoretic one-time programs from geometrically local
  $\text{QNC}_0$ adversaries.
\newblock {\em arXiv preprint arXiv:2503.22016}, 2025.

\bibitem[SZ25a]{SZ25}
Omri Shmueli and Mark Zhandry.
\newblock On one-shot signatures, quantum vs classical binding, and obfuscating
  permutations.
\newblock Cryptology {ePrint} Archive, Paper 2025/486, 2025.

\bibitem[SZ25b]{SZ25b}
Omri Shmueli and Mark Zhandry.
\newblock Unclonable cryptography in linear quantum memory.
\newblock Cryptology {ePrint} Archive, Paper 2025/2056, 2025.

\bibitem[WCS15]{CCS:WanChaShi15}
Xiao Wang, T.-H.~Hubert Chan, and Elaine Shi.
\newblock Circuit {ORAM}: On tightness of the {Goldreich}-{Ostrovsky} lower
  bound.
\newblock In Indrajit Ray, Ninghui Li, and Christopher Kruegel, editors, {\em
  ACM CCS 2015}, pages 850--861. {ACM} Press, October 2015.

\bibitem[Zha18]{Zha18}
Mark Zhandry.
\newblock How to record quantum queries, and applications to quantum
  indifferentiability.
\newblock Cryptology {ePrint} Archive, Paper 2018/276, 2018.

\bibitem[Zha19]{zhandry19compressed}
Mark Zhandry.
\newblock How to record quantum queries, and applications to quantum
  indifferentiability.
\newblock In Alexandra Boldyreva and Daniele Micciancio, editors, {\em Advances
  in Cryptology -- CRYPTO 2019}, pages 239--268, Cham, 2019. Springer
  International Publishing.

\end{thebibliography}
\appendix

\section{Compressed Oracle Chaining}\label{sec:compressed-chaining}

In this section, we prove the strengthened variant of \cite{GLRRV25}'s compressed oracle chaining lemma. We recall the statement of the lemma here for convenience.

Intuitively, the lemma considers an adversary which access to a general composition of random oracles $G$ and $H$, where $G$ is used to determine the input $x_h$ to $H$ in a somewhat structured manner. It states that, when $G$ and $H$ are implemented as compressed oracles, if $H$ records some input/output $H(x_h) = z$ pair, then $G$ also records a matching input/output pair that leads to $x_h$.

\begin{lemma}[Compressed Oracle Chaining]
    Let $G:\cX_G \rightarrow \cY_1\times \cY_2$ and $H:\cX_H \times \cY_2 \rightarrow \cZ$ be random oracles implemented by the compressed oracle technique.
    
    Let $f:\cX_G \times \cY_1 \times \cY_2 \rightarrow \cX_H$ be an arbitrary function. Define the function $F_f: \cX \rightarrow \cZ$ by the following computation for $F(x_g)$:
    \begin{enumerate}
        \item Compute $y_1\concat y_2 \coloneqq G(x_g)$.
        \item Compute $x_h \coloneqq f(x_g, y_1\concat y_2)$.
        \item Output $H(x_h, y_2)$.
    \end{enumerate}
    Consider running an interaction of an oracle algorithm with $F_f$ until query $t$, then measuring the internal state of $G$ and $H$ to obtain $D_G$ and $D_H$. 

    Let $E_t$ be the event that after the measurement at time $t$, for all $(x_h\concat y_2, z)\in D_H$, there exists an entry $(x_g, y_1\concat y_2) \in D_G$ such that $f(x_g, y_1\concat y_2) = x_h$.
    Then
    \[
        \Pr[E_t]
        \geq 
        1 - \frac{8t^2}{|\cY_2|}
    \]
\end{lemma}

\subsection{Alternative Representation of the Decompression Function}\label{sec:decompression-alternative-form}

We first derive a useful alternative representation of the $\Decomp$ operation used for compressed oracles.

Let $D$ be a database for a random function $H$ such that $D(x) = \bot$. If one were to query $H$ and receive result $H(x) = y$, the state of the oracle before the final decompression operation is $\ket{D \cup (x, y)}$. The effect of the $x$ decompression operation $\mathsf{Decomp}_x$ on a state $\ket{D \cup (x, y)}$ is
\begin{align*}
    &\mathsf{Decomp}_x \ket{D\cup (x, y)}
    \\
    &= \mathsf{Decomp}_x \frac{1}{|\cY|} \sum_{y'} \ket{D\cup (x, y')} \sum_{r\in \cY}(-1)^{r\cdot (y'-y)} 
    \\
    &= \mathsf{Decomp}_x  \frac{1}{|\cY|}\left(\sum_{r\neq 0} (-1)^{r\cdot -y} \sum_{y'}(-1)^{r\cdot y'} \ket{D\cup (x, y')} + \sum_{y'} \ket{D\cup (x, y')} \right)
    \\
    &=  \frac{1}{|\cY|} \left( \sum_{r\neq 0} (-1)^{r\cdot -y} \sum_{y'}(-1)^{r\cdot y'} \ket{D\cup (x, y')} + \sqrt{|\cY|} \ket{D} \right)
    \\
    &=  \left(\ket{D\cup (x, y)} - \frac{1}{|\cY|} \sum_{y'} \ket{D\cup (x, y')} + \frac{1}{\sqrt{|\cY|}} \ket{D} \right)
    \\
    &=  \left(\left(1-\frac{1}{|\cY|} \right)\ket{D\cup (x, y)} - \frac{1}{|\cY|} \sum_{y'\neq y} \ket{D\cup (x, y')} + \frac{1}{\sqrt{|\cY|}} \ket{D} \right)
\end{align*}

We note that this only describes the state \emph{in between} queries to $x$. A second query to $x$ will apply $\mathsf{Decomp}_x$ to the database register before determining the response, which maps the database register back to $\ket{D\cup (x,y)}$. Thus, despite the support of the state on other databases, $G(x)$ cannot actually change in between queries to $x$. However, this form is relevant when looking at the database register without knowledge of $x$.

\subsection{Proof}

We first explicitly describe how $F$ works. The internal states of $H$ and $G$ are stored in registers $\cD_H$ and $\cD_G$, respectively.
In general, we can consider a basis query $\ket{x_g, u}_{\cQ}$ in register $\cQ = (\cQ_{\cX_G}, \cQ_{\cZ})$. For convenience, we insert into $\cQ$ a additional work registers $\cQ_{\cX_H}$ and $\cQ_\cY = (\cQ_{\cY_1}, \cQ_{\cY_2})$ which are initialized to $\ket{0}$, resulting in $\ket{x_g, 0, 0, u}_{\cQ}$. To compute $F$ on this query, do the following.
\begin{enumerate}
    \item Query $(\cQ_{\cX_G}, \cQ_\cY)$ to $G$ to obtain $\ket{x_g, 0, G(x_g), u}_{\cQ}$, i.e. apply $\Decomp_G\circ \CO'_G \circ \Decomp_G$ on $(\cQ_{\cX_G}, \cQ_\cY, \cD_G)$.\footnote{Since $G$ and $H$ have different domains and ranges, we differentiate their decompression operations, which depend on the domain/range. We also differentiate the $\CO'_G$ operation to clarify that it acts on registers corresponding to a query to $G$.}
    \item Compute $f$ on registers $(\cX_G, \cY)$ and write the output to register $\cX_H$ to obtain $\ket{x_g, x_h, G(x_g), u}_{\cQ}$. Let $U_f$ be the corresponding unitary.
    \item Query $(\cQ_{\cX_H}, \cQ_{\cY_2}, \cQ_{\cZ})$ to $H$ to obtain $\ket{x_g, x_h, G(x_g), u\oplus H(x_h, G(x_g))}_{\cQ}$.
    \item Compute $f$ on registers $(\cX_G, \cY)$ and write the output to register $\cX_H$ to obtain $\ket{x_g, 0, G(x_g), u\oplus F(x_g)}_{\cQ}$.
    \item Query $(\cQ_{\cX_G}, \cQ_\cY)$ to $G$ again to obtain $\ket{x_g, 0, 0, u\oplus F(x_g)}_{\cQ}$.
    \item Return registers $(\cQ_{\cX_G}, \cQ_{\cZ})$
\end{enumerate}

% To compute $F$ on this query, 
% Then compute $f$ on registers $(\cX_G, \cY)$ and write the output to register $\cX_H$ to obtain $\ket{x_g, x_h, G(x_g), u}_{\cQ}$
% Then query $(\cQ_{\cX_H}, \cQ_\cY, \cQ_{\cZ})$ to $H$ to obtain $\ket{x_g, x_h, G(x_g), u\oplus H(x_h, G(x_g))}_{\cQ}$.
% Finally, query $(\cQ_{\cX_G}, \cQ_\cY)$ to $G$ again to obtain $\ket{x_g, x_h, 0, u\oplus H(x_h, G(x_g))}_{\cQ}$ and return registers $(\cQ_{\cX_G}, \cQ_{\cX_H}, \cQ_{\cZ})$.

Observe that $\Decomp_G$ commutes with the computation of $f$, since $\Decomp_G$ acts on registers $(\cQ_{\cX_G}, \cD_G)$ while $U_f$ acts on registers $(\cQ_{\cX_G}, \cY, \cX_H)$ and both are diagonal in the computational basis on the only overlapping register, $\cQ_{\cX_G}$.
Furthermore, $\Decomp_G$ commutes with the query to $H$, since they operate on disjoint registers $(\cQ_{\cX_G}, \cD_G)$ and $(\cQ_{\cX_H}, \cQ_\cY, \cQ_{\cZ}, \cD_H)$, respectively. Recall that $\Decomp_G \circ \Decomp_G = I$.
Thus, if we write the operation of $H$ as $U_H$, we may write the implementation of a query to $F$ as
\[
    U_F
    \coloneqq
    (\Decomp_G \circ \CO'_G) \circ (U_f \circ U_H \circ U_f) \circ (\CO'_G \circ \Decomp_G)
\]
where $\Decomp_G$ and $\CO'_G$ act on registers $(\cQ_{\cX_G}, \cQ_\cY, \cD_G)$, while $U_H$ acts on registers $(\cQ_{\cX_H}, \cQ_\cY, \cQ_{\cZ}, \cD_H)$ and $U_f$ operates on registers $(\cQ_{\cX_G}, \cY, \cX_H)$ .

Now consider the interaction of an oracle algorithm with $F$, where the algorithm maintains an additional internal register $\cA$.
Define the projector $E$ onto states 
\[
    \ket{a}_{\cA}\otimes \ket{x_g,*,*, *}_{\cQ}\otimes \ket{D_G, D_H}_{\cD}
\]
where $D_G$ and $D_H$ satisfy the requirements of event $E_t$ and define $\overline{E} = I - E$.
We will upper bound the norm $\norm{\overline{E} U_F \ket{\psi}}$ from after the query in terms of the norm $\norm{\overline{E}\ket{\psi}}$ from before the query.

We first bound the intermediate operations, in between the two $\Decomp_G$ operations.

\begin{claim}\label{claim:chaining-co-H}
    \[
        \norm{\overline{E} (\CO'_G \circ U_f) \circ U_H \circ (U_f \circ \CO'_G) \circ \Decomp_G \ket{\psi}}
        =
        \norm{\overline{E} (\Decomp_G) \ket{\psi}}
    \]
\end{claim}
\begin{proof}
    First, observe that for all states $\ket{\psi'}$,
    \[
        \norm{\overline{E} \CO'_G \ket{\psi'}}
        =
        \norm{\overline{E} U_f \ket{\psi'}}
        =
        \norm{\overline{E}  \ket{\psi'}}
    \]
    since neither $\CO'_G$ nor $U_f$ modify the database registers.

    Now consider the operation of $U_H$ on $(\CO'_G \cdot \Decomp_G) \ket{\psi}$.
    Observe that by the definitions of $\CO'_G$ and $f$, the state of the query register and oracle databases when $H$ is queried is always supported on basis states
    \[
        \ket{x_g, x_h, y, u}_{\cQ} \otimes \ket{D_G\cup (x_g, y)}_{\cD_G} \otimes \ket{D_H}_{\cD_H}
    \]
    where $f(x_g, y) = x_h$ and $y = y_1\concat y_2$.
    A query of $(x_h, y_2)$ to $H$ can only result in a new database $D_H'$ where the only potential difference between $D_H$ and $D_H'$ is $D_H(x_h\concat y_2) \neq D_H'(x_h\concat y_2)$.
    Since $(x_g, y)$ is already recorded in the contents of register $\cD_G$ and $f(x_g, y) = x_h$, applying $H$ always results in a valid database state with respect to the event $E$.
    In particular, if $D_H$ and $D_G$ already satisfied $E$ before the query to $H$, they continue to do so afterwards. Thus
    \[
        \norm{\overline{E}  U_H \circ (U_f \circ \CO'_G) \circ \Decomp_G \ket{\psi}}
        =
        \norm{\overline{E}  (U_f \circ \CO'_G) \circ \Decomp_G) \ket{\psi}}.
    \]
    Putting this together with the prior bounds on the effect of $\CO'_G$ and $U_f$ yields the claim.
\end{proof}

Next, we bound the effect of $\Decomp_G$ on a general state $\ket{\psi'}$. In general, both the first and the last $\Decomp_G$ operation will operate on a state of the form
\[
    \sum_{a,x_g,u,D_G,D_H} \alpha_{a,x_g, u,D_G,D_H} \ket{a}_{\cA} \otimes \ket{x_g, 0, 0, u}_{\cQ} \otimes \ket{D_G, D_H}_{\cD}
\]
where $\cA$ is the adversary's internal register, $\cQ$ contains the (expanded) query, and $\cD$ contains the oracle databases.
This is clearly true for the first application; it holds for the second because $\CO'_G$ and $U_f$ are applied twice, and $U_H$ does not modify the registers that either on.

\jiahui{Original proof had an issue with $ \norm{\overline{E} \circ \Decomp_G \circ E_{Y+,Z}\ket{\psi'}}
        =
        0.$ not true for general $f$. Fixed. }

\begin{claim}\label{claim:chaining-decomp}
    For every normalized state $\ket{\psi'}$,
    \[
        \norm{\overline{E} \Decomp_G \ket{\psi'}}
        \leq
        \norm{\overline{E} \ket{\psi'}}
        +\sqrt{\frac{2}{|\cY_2|}}.
    \]
\end{claim}
\begin{proof}
We first bound $\norm{\overline{E}\Decomp_G E\ket{\psi'}}$.
Fix the queried input $x_g$, the database $D_H$, and all cells of
$D_G$ other than $x_g$. Write $D'_G$ for these fixed cells, with
$D'_G(x_g)=\bot$. Also fix computational-basis values of the other
registers. These choices define mutually orthogonal subspaces, each
preserved by both $E$ and $\Decomp_G$. This is a decomposition for
the calculation; no intermediate measurement is performed.

For these fixed values, let $T$ be the set of input pairs recorded
in $D_H$ that have no matching entry in $D'_G$:
\[
    T:=\left\{(h,v)\in\cX_H\times\cY_2:
    \begin{array}{l}
        D_H(h\concat v)\neq\bot,\ \text{and there is no}\\
        (x'_g,y_1\concat v)\in D'_G
        \text{ with } f(x'_g,y_1\concat v)=h
    \end{array}\right\}.
\]
Every pair in $T$ must therefore be matched by the queried cell
$D_G(x_g)$ for the complete databases to satisfy $E$.

If $T=\emptyset$, the fixed cells already provide all required
matching entries. Every value of the queried cell, including
$\bot$, satisfies $E$, so
$\overline{E}\Decomp_G E=0$ on this subspace.
If $|T|\geq2$, no value of the queried cell can make the databases
satisfy $E$: a non-$\bot$ value $y_1\concat y_2$ matches only the
single input pair $(f(x_g,y_1\concat y_2),y_2)$.
Thus $E=0$ on this subspace, so
$\overline{E}\Decomp_G E=0$ here too.

It remains to consider $T=\{(h,v)\}$. Define
\[
    \cY:=\cY_1\times\cY_2,\qquad N:=|\cY|,
    \qquad
    A:=\{y_1\concat v:f(x_g,y_1\concat v)=h\},
    \qquad a:=|A|\leq|\cY_1|.
\]
Here $y_1$ ranges over $\cY_1$. In this subspace, the complete
databases satisfy $E$ exactly when $D_G(x_g)\in A$.
For the following calculation, abbreviate
\[
    \ket{\bot}:=\ket{D'_G},\qquad
    \ket{y}:=\ket{D'_G\cup(x_g,y)}
    \quad(y\in\cY),
\]
with all other fixed registers suppressed. Thus $E$ acts as
$\sum_{y\in A}\ket{y}\!\bra{y}$ on the queried cell.
By \Cref{sec:decompression-alternative-form},
\[
    \Decomp_{G,x_g}\ket{y}
    =\ket{y}-\frac1N\sum_{w\in\cY}\ket{w}
      +\frac1{\sqrt N}\ket{\bot}
    \qquad(y\in\cY).
\]
For any, possibly subnormalized, vector
$\ket{\eta}=\sum_{y\in A}c_y\ket{y}$, it follows that
\[
    \overline{E}\Decomp_{G,x_g}\ket{\eta}
    =\left(\sum_{y\in A}c_y\right)
      \left(
          \frac1{\sqrt N}\ket{\bot}
          -\frac1N\sum_{w\in\cY\setminus A}\ket{w}
      \right).
\]
Orthogonality of the database basis and Cauchy--Schwarz give
\begin{align*}
    \norm{\overline{E}\Decomp_{G,x_g}\ket{\eta}}^2
    &=\left(\frac1N+\frac{N-a}{N^2}\right)
        \left|\sum_{y\in A}c_y\right|^2\\
    &\leq\left(\frac{2a}{N}-\frac{a^2}{N^2}\right)
        \sum_{y\in A}|c_y|^2\\
    &\leq\frac{2}{|\cY_2|}\norm{\ket{\eta}}^2.
\end{align*}
In fact, for $a>0$, taking $c_y=1/\sqrt a$ attains the first
inequality, so the squared operator norm of
$\overline{E}\Decomp_{G,x_g}E$ on this subspace is exactly
$2a/N-(a/N)^2$. For $a=0$, the operator is zero.

The bound is uniform over all fixed choices of $x_g$, $D'_G$,
$D_H$, and the other registers. Since these subspaces and their
images are mutually orthogonal, summing their squared contributions
gives
\begin{equation}\label{eq:compressed-chaining-good-to-bad}
    \norm{\overline{E}\Decomp_G E\ket{\psi'}}
    \leq
    \sqrt{\frac{2}{|\cY_2|}}\norm{E\ket{\psi'}}.
\end{equation}

The contribution from $\overline{E}\ket{\psi'}$ satisfies
\begin{equation}\label{eq:compressed-chaining-event-not-e}
    \norm{\overline{E}\cdot\Decomp_G\cdot\overline{E}
          \ket{\psi'}}
    \leq\norm{\overline{E}\ket{\psi'}}.
\end{equation}

Putting together
\Cref{eq:compressed-chaining-good-to-bad,eq:compressed-chaining-event-not-e}
with $\ket{\psi'}=E\ket{\psi'}+\overline{E}\ket{\psi'}$, we have
\begin{align*}
    \norm{\overline{E}\Decomp_G\ket{\psi'}}
    &\leq
      \norm{\overline{E}\Decomp_G\overline{E}\ket{\psi'}}
      +\norm{\overline{E}\Decomp_G E\ket{\psi'}}\\
    &\leq
      \norm{\overline{E}\ket{\psi'}}
      +\sqrt{\frac{2}{|\cY_2|}}\norm{E\ket{\psi'}}\\
    &\leq
      \norm{\overline{E}\ket{\psi'}}
      +\sqrt{\frac{2}{|\cY_2|}},
\end{align*}
where the last inequality uses $\norm{\ket{\psi'}}=1$.
\end{proof}

Putting together \Cref{claim:chaining-co-H} and
\Cref{claim:chaining-decomp}, the norm after any single query to $F$
is bounded as
\begin{align*}
    \norm{\overline{E}\cdot U_F\ket{\psi}}
    &\leq
    \norm{\overline{E}\cdot(\CO'_G\circ U_f)\circ U_H
          \circ(U_f\circ\CO'_G)\circ\Decomp_G\ket{\psi}}
      +\sqrt{\frac{2}{|\cY_2|}}\\
    &=\norm{\overline{E}\Decomp_G\ket{\psi}}
      +\sqrt{\frac{2}{|\cY_2|}}\\
    &\leq\norm{\overline{E}\ket{\psi}}
      +2\sqrt{\frac{2}{|\cY_2|}}.
\end{align*}

Here $\ket{\psi}$ is normalized. The first and last inequalities
account for the two $\Decomp_G$ operations remaining in $U_F$;
the equality uses \Cref{claim:chaining-co-H} for the intervening
operations. We may purify the adversary's randomness and measurements
in $\cA$. Its operations between complete queries then act only on
$\cA$ and $\cQ$, not on the database registers, so they commute
with $E$ and preserve $\norm{\overline{E}\ket{\psi}}$.

The norm starts at $0$, so after $t$ queries to $F$ it is at most
$2t\sqrt{\frac{2}{|\cY_2|}}$.
The probability of seeing the event corresponding to $\overline{E}$
when we measure $\cD$ is the square of the norm, which is at most
$\frac{8t^2}{|\cY_2|}$.
Therefore the probability of seeing the complementary event $E$
is at least $1-\frac{8t^2}{|\cY_2|}$, as claimed.

\section{Applications}
Here we show a variety of applications of one-shot programs (\cref{sec:OSP}) and query-limited programs (\cref{sec:query-limited-RAM-programs}).

\subsection{Pay-Per-Use Programs}

A pay-per-use program can be evaluated many times, after which the user proves the total cost of the queries that they effectively made. The central subtlety is that, in the quantum setting, it is unclear how to charge for a query that is later coherently uncomputed. As in the rest of the paper, we therefore measure usage through the \emph{effective-query} viewpoint: only queries that remain recorded in the ideal oracle count toward the final bill.

Let $\cP = \{P:\cM \times \cX \times \cR \rightarrow \cM \times \cY\}_{P}$ be a family of randomized RAM programs. Throughout this subsection we assume that every output $y \in \cY$ contains a designated cumulative-cost field, written $\mathsf{cost}(y) \in \bbR$, and that the hidden RAM state likewise stores the same cumulative cost. We write ``the current cost of $\st$'' for this designated value.

Our construction extends a query-limited program with a special $\Prove$-mode. In $\Eval$-mode, the program behaves exactly like the underlying RAM program $P$. In $\Prove$-mode, it outputs the current cost together with a signature certifying that cost, and then irreversibly switches to a distinguished halt state. Intuitively, once a valid proof is produced, no additional effective evaluations should be possible.

\begin{definition}[Pay-Per-Use Program Compiler]\label{def:pay-per-use-program-compiler}
    Let $\cP = \{ P : \cM \times \cX \times \cR \rightarrow \cM \times \cY \}_{ P }$ be a family of RAM programs whose outputs include a designated cumulative cost value $\mathsf{cost}(y) \in \bbR$, initialized to $0$ when the program starts from $\st_0 = \bot$.

    A \textbf{pay-per-use (PPU) program compiler} for $\cP$ comprises the following QPT algorithms.
    \begin{itemize}
        \item \textbf{Syntax.}
        \begin{itemize}
            \item $\PGen(1^\secp, P) \to (\tilde{P}, \pk)$: takes a security parameter $\secp$ and a program $P \in \cP$, and outputs a compiled program $\tilde{P}$ together with a public verification key $\pk$.
            \item $\Eval(\tilde{P}, x) \to (y, \tilde{P}')$: evaluates $\tilde{P}$ on input $x$ and returns an output $y$ and an updated program state $\tilde{P}'$.
            \item $\Prove(\tilde{P}, c, \pk) \to b \in \bit$: a QPT interactive protocol between a prover and a verifier. The prover's inputs are a program $\tilde{P}$ and a claimed cost $c$. The verifier's inputs are $c$ and $\pk$. The verifier outputs $1$ iff it accepts that the effective queries encoded by $\tilde{P}$ have total cost $c$.
        \end{itemize}

        \item \textbf{Correctness.} The compiler satisfies the correctness property of \cref{def:RAM-program-compiler} with respect to the underlying family $\cP$.\bhaskar{The simulator in the definition of security should simulate both outputs of $\PGen$: both $\tilde{P}$ and $\pk$.}

        \proofaudit{P20}{Pay-per-use needs completeness definition and proof for honest proofs}{%
The correctness clause only requires the $\PGen,\Eval$ interface to be correct; it does not require an honest $\Prove$ execution to be accepted. Add the definition and proof
} 
\jiahui{added}

\item \textbf{Completeness.}
For every $P\in\cP$, every polynomially bounded $n\geq 0$, and every
input sequence $(x_1,\ldots,x_n)\in\cX^n$, the following experiment
accepts with probability at least $1-\negl(\secp)$:
\[
    \Pr\!\left[
        b=1:
        \begin{array}{l}
            (\tilde P_0,\pk)\gets\PGen(1^\secp,P),\\
            (y_i,\tilde P_i)\gets\Eval(\tilde P_{i-1},x_i)
                \quad\text{for }i=1,\ldots,n,\\
            b\gets\Prove(\tilde P_n,c,\pk)
        \end{array}
    \right]\geq 1-\negl(\secp),
\]
where both parties follow the protocol honestly and the claimed cost
is computed before running $\Prove$ as
\[
    c:=
    \begin{cases}
        0, & n=0,\\
        \mathsf{cost}(y_n), & n\geq 1.
    \end{cases}
\]
Any failure to produce a well-formed output in this experiment is
counted as rejection. The value $c$ is the cumulative cost of the
preceding evaluations; running $\Prove$ does not add another evaluation
of $P$ to this cost. Both parties obtain $c$ after the protocol, so its
value is not read from the program state after proving.

        \item \textbf{Security.} There exists a QPT simulator $\cS$ such that for every $P \in \cP$ and every non-uniform QPT distinguisher $D$ with auxiliary input $\aux_{P,\cP}$, there is a negligible function $\negl$ such that for all $\secp \in \bbN$,
        \[
            \abs{\Pr\!\left[D\!\left(1^\secp, \PGen(1^\secp,P), \aux_{P,\cP}\right) \to 1\right]
            -
            \Pr\!\left[D\!\left(1^\secp, \cS^{O_P^{\REQ}}(1^\secp), \aux_{P,\cP}\right) \to 1\right]} \leq \negl(\secp),
        \]
        where $\cS^{O_P^{\REQ}}(1^\secp)$ outputs both the simulated program and the public key, and $O_P^\REQ$ is the REQ oracle for $P$ from \cref{fig:REQ}.

        \item \textbf{Soundness.} For a fixed $P \in \cP$, let $\cS^{O_P^{\REQ}}$ be the simulator from the security item above. Consider the following experiment.
        \begin{enumerate}
            \item The challenger sends $(\tilde{P},\pk) \gets \cS^{O_P^{\REQ}}(1^\secp)$ to the adversary.
            \item The adversary outputs a claimed cost $c \in \bbR$.
            \item The adversary and the challenger engage in $\Prove(\tilde{P},c,\pk)$, with the adversary acting as prover and the challenger as verifier. Let $b$ be the verifier's output.
            \item The challenger reads the true effective cost $c^\star$ from the internal state of $O_P^{\REQ}$. Concretely, if there is no index $k$ with $g_k=1$, then $c^\star := 0$. Otherwise, let $k_{\max}$ be the maximum index with $g_{k_{\max}}=1$, read $(\st,x,r)$ from register $\cA_{k_{\max}}$, compute $(*,y)=P(\st,x;r)$, and set $c^\star := \mathsf{cost}(y)$.
            \item The adversary wins if $b=1$ and $c \neq c^\star$.
        \end{enumerate}
        The compiler is \emph{sound} if every QPT adversary wins with at most negligible probability.
    \end{itemize}
\end{definition}

\paragraph{Construction.}
Let $\QLP.(\PGen,\Eval)$ be the QLP compiler from \cref{fig:RAM-program-compiler-construction}. Let $\BUS.(\KeyGen,\Sign,\Ver)$ be a blind-unforgeable signature scheme, and let $\PRF$ be a pseudorandom function (\cref{def:PRF}).

The compiler samples a signature key pair $(\sk,\pk)$ and two PRF keys $k_1,k_2$. It then defines the following two-mode RAM program $P_{\mathsf{PPU}}$.
\begin{figure}
\begin{mdframed}
    \paragraph{$\PGen(1^\secp,P)$:}
    \begin{enumerate}
        \item Sample
        \begin{equation}\label{eq:sample-PPU-parameters}
        \begin{split}
            (\sk,\pk) &\gets \BUS.\KeyGen(1^\secp),\\
            k_1 &\getsr \bit^\secp,\\
            k_2 &\getsr \bit^\secp.
        \end{split}
        \end{equation}

        \item Using $(P,\sk,\pk,k_1,k_2)$, define the RAM program $P_{\mathsf{PPU}}$:
        \begin{mdframed}
        \begin{enumerate}
            \itemindent=-15pt
            \item[] \textbf{Inputs:} $[\st,(x,\mode);r]$, where $\st \in \cM$ is the current RAM state, $x \in \cX$ is the chosen input, $\mode \in \{\Eval,\Prove\}$, and $r \in \cR$ is fresh randomness.
            \itemindent=0pt
            \item If $\st = \mathsf{Halt}$, output $(\st',y) = (\mathsf{Halt},0)$.
            \item If $\mode=\Eval$ and $\st \neq \mathsf{Halt}$, output
            \[
                (\st',y) = P(\st,x;r).
            \]
            \item If $\mode=\Prove$ and $\st \neq \mathsf{Halt}$, let $c' := \mathsf{cost}(\st)$, set $\st' := \mathsf{Halt}$, and compute
            \begin{equation}\label{eq:sign-cost}
            \begin{split}
                r_1 &= \PRF(k_1,(c',r)),\\
                r_2 &= \PRF(k_2,(c',r,r_1)),\\
                \sigma &= \BUS.\Sign(\sk,(c',r,r_1);r_2),\\
                y &= (c',r,r_1,\sigma).
            \end{split}
            \end{equation}
            Then output $(\st',y)$.
        \end{enumerate}
        \end{mdframed}

        \item Generate a QLP for $P_{\mathsf{PPU}}$:
        \[
            \tilde{P} \gets \QLP.\PGen(1^\secp, P_{\mathsf{PPU}}).
        \]
        \item Output $(\tilde{P},\pk)$.
    \end{enumerate}

    \paragraph{$\Eval(\tilde{P},x)$:}
    \begin{enumerate}
        \item Query the QLP in $\Eval$-mode:
        \begin{align*}
            \mode &= \Eval,\\
            (y,\tilde{P}') &\gets \QLP.\Eval(\tilde{P},(x,\mode)).
        \end{align*}
        \item Output $(y,\tilde{P}')$.
    \end{enumerate}
\end{mdframed}
\caption{Pay-Per-Use Program Compiler Construction, Part 1}\label{fig:PPU-program-compiler-construction-1}
\end{figure}

\begin{figure}
\begin{mdframed}
    \paragraph{$\Prove(\tilde{P},c,\pk)$:}
    \begin{enumerate}
        \item \textbf{Input:} The prover has a QLP $\tilde{P}$ and a claimed cost $c$. The verifier has the same cost $c$ and the public key $\pk$.
        \item The prover queries the QLP in $\Prove$-mode:
        \begin{align*}
            (x,\mode) &= (0,\Prove),\\
            (y,\tilde{P}') &\gets \QLP.\Eval(\tilde{P},(x,\mode)).
        \end{align*}
        The prover sends $y$ to the verifier.
        \item The verifier parses $y=(c',r,r_1,\sigma)$ and outputs
        \[
            b \gets (c'=c) \land \left(1 = \BUS.\Ver(\pk,(c',r,r_1),\sigma)\right).
        \]
    \end{enumerate}
\end{mdframed}
\caption{Pay-Per-Use Program Compiler Construction, Part 2}\label{fig:PPU-program-compiler-construction-2}
\end{figure}

\begin{theorem}[PPU Program Compiler]
    The construction in \cref{fig:PPU-program-compiler-construction-1,fig:PPU-program-compiler-construction-2} is a pay-per-use program compiler for $\cP$ in the sense of \cref{def:pay-per-use-program-compiler}.
\end{theorem}
\begin{proof}
    The syntax is immediate. Correctness is proved in \cref{thm:ppu-program-correctness}, security in \cref{thm:ppu-program-security}, and soundness in \cref{thm:ppu-program-soundness}.
\end{proof}

\begin{lemma}[PPU Program Correctness]\label{thm:ppu-program-correctness}
    The construction in \cref{fig:PPU-program-compiler-construction-1,fig:PPU-program-compiler-construction-2} satisfies the correctness property of \cref{def:pay-per-use-program-compiler}.
\end{lemma}
\begin{proof}
    The compiled program is simply a QLP for the two-mode RAM program $P_{\mathsf{PPU}}$:
    \[
        \tilde{P} \gets \QLP.\PGen(1^\secp, P_{\mathsf{PPU}}).
    \]
    Moreover, $\Eval(\tilde{P},x)$ invokes that QLP only on inputs of the form $(x,\Eval)$. On such inputs, $P_{\mathsf{PPU}}$ behaves exactly like the underlying program $P$.

    Therefore, by the correctness guarantee of the QLP compiler, for every polynomial-length query sequence $(x_1,\ldots,x_n) \in \cX^*$, the distributions
    \begin{align*}
        &\left\{(y_1,\ldots,y_n):\begin{array}{c}
            \st_0 = \bot,\\
            (r_1,\ldots,r_n) \gets \cR^n,\\
            (\st_i,y_i) \gets P(\st_{i-1},x_i;r_i) \quad \forall i\in[n]
        \end{array}\right\}
        \\
        &\qquad\text{and}\qquad
        \left\{(y_1,\ldots,y_n):\begin{array}{c}
            (\tilde{P}_0,\pk) \gets \PGen(1^\secp,P),\\
            (y_i,\tilde{P}_i) \gets \Eval(\tilde{P}_{i-1},x_i) \quad \forall i\in[n]
        \end{array}\right\}
    \end{align*}
    are statistically close. This is exactly the required correctness statement.
\end{proof}

\jiahui{added completeness proof}
\begin{lemma}[PPU Program Completeness]
\label{thm:ppu-program-completeness}
    The construction in
    \cref{fig:PPU-program-compiler-construction-1,fig:PPU-program-compiler-construction-2}
    satisfies the completeness property of
    \cref{def:pay-per-use-program-compiler}.
\end{lemma}
\begin{proof}
    Fix $P\in\cP$ and a polynomial-length input sequence
    $(x_1,\ldots,x_n)$, allowing $n=0$.
    The honest execution consists of $n$ evaluations followed by
    one execution of $\Prove$. By the construction, these operations
    invoke the same QLP for $P_{\mathsf{PPU}}$ on the sequence
    \[
        (x_1,\Eval),\ldots,(x_n,\Eval),(0,\Prove).
    \]
    In particular, the final $\Prove$  uses the remaining
    compiled program $\tilde P_n$, not a newly generated program.

    The correctness guarantee of the QLP compiler
    (\cref{def:RAM-program-compiler}) applies to this entire sequence of
    $n+1$ queries. Denote the output of the last query by $y$.
    The joint distribution of $(\pk,y_1,\ldots,y_n,y)$ is
    statistically close to the corresponding distribution obtained
    by directly evaluating $P_{\mathsf{PPU}}$ with independent
    uniform randomness at each step.  $\pk$ is obtained by first fixing the parameters sampled in
    \eqref{eq:sample-PPU-parameters}, applying QLP correctness to the
    resulting program $P_{\mathsf{PPU}}$ averaged over
    these parameters.

    Consider the following execution. The first $n$ queries are all
    in $\Eval$-mode, so they evaluate the underlying RAM program
    $P$ starting from $\st_0=\bot$. Let $\st_n$ be its state after
    these evaluations. By the designated cumulative-cost field
    assumed for $P$, the claimed cost satisfies
    \[
        c=\mathsf{cost}(\st_n)
        =
        \begin{cases}
            0, & n=0,\\
            \mathsf{cost}(y_n), & n\geq 1.
        \end{cases}
    \]
    Since no preceding query used $\Prove$-mode, the final query
    executes the prove-branch in
    \cref{fig:PPU-program-compiler-construction-1}, reads
    $c':=\mathsf{cost}(\st_n)$, and produces
    \begin{align*}
        r_1&=\PRF(k_1,(c',r)),\\
        r_2&=\PRF(k_2,(c',r,r_1)),\\
        \sigma&=\BUS.\Sign(\sk,(c',r,r_1);r_2),\\
        y&=(c',r,r_1,\sigma).
    \end{align*}
    Here $r$ is the randomness of the final query. In particular,
    $c'=c$. By perfect correctness of the signature scheme
    (\cref{def:classical-signature-scheme}),
    \[
        \BUS.\Ver(\pk,(c',r,r_1),\sigma)=1.
    \]
    Perfect correctness applies to the specified signing coins
    $r_2$ as well.
    %; no pseudorandomness or unforgeability argument is needed for this step. 
    Therefore, the verifier in
    \cref{fig:PPU-program-compiler-construction-2} accepts this
    direct-execution transcript with probability one.

    The claimed cost $c$ is a function of the preceding outputs
    (or the constant $0$ when $n=0$), and the verifier's decision
    is a function of $(\pk,c,y)$. Applying these functions to the
    two statistically close joint output distributions cannot
    increase their statistical distance. Consequently, the
    verifier accepts the honest compiled execution with
    probability at least $1-\negl(\secp)$.

    When $n=0$, the only query is $(0,\Prove)$, which reads the
    initial cost $\mathsf{cost}(\bot)=0$ and signs it, so the same
    argument applies to this case. The prove-branch subsequently
    sets the state to $\mathsf{Halt}$, but the certified cost
    remains in the output $y$ and in the parties' remaining
    value $c$. %no cost field of $\mathsf{Halt}$ is needed for this completeness statement.
\end{proof}

\begin{lemma}[PPU Program Security]\label{thm:ppu-program-security}
    The construction in \cref{fig:PPU-program-compiler-construction-1,fig:PPU-program-compiler-construction-2} satisfies the security property of \cref{def:pay-per-use-program-compiler}.
\end{lemma}
\begin{proof}
    Let $P_{\mathsf{PPU}}$ be the two-mode RAM program defined in \cref{fig:PPU-program-compiler-construction-1}. The real compiler outputs
    \[
        (\tilde{P},\pk) = \big(\QLP.\PGen(1^\secp,P_{\mathsf{PPU}}),\pk\big).
    \]
    By the security of the QLP compiler (\cref{def:RAM-program-compiler}), there exists a QPT simulator $\cS'$ such that
    \begin{equation}\label{eq:ppu-security-step1}
    \abs{\Pr\!\left[D\!\left(1^\secp,\QLP.\PGen(1^\secp,P_{\mathsf{PPU}}),\aux_{P,\cP}\right)\to 1\right]
    -
    \Pr\!\left[D\!\left(1^\secp,\cS'^{O_{P_{\mathsf{PPU}}}^{\REQ}}(1^\secp),\aux_{P,\cP}\right)\to 1\right]} \le \negl(\secp).
    \end{equation}
    Thus it remains to simulate $O_{P_{\mathsf{PPU}}}^{\REQ}$ given only $O_P^{\REQ}$.

    \begin{lemma}\label{thm:simulate-one-REQ-oracle-with-another}
        For every $P \in \cP$, there exists a QPT oracle simulator $\cS_{P\to P_{\mathsf{PPU}}}$ with quantum query access to $O_P^{\REQ}$ that perfectly simulates $O_{P_{\mathsf{PPU}}}^{\REQ}$.
        Moreover, after every query, the cost register maintained by $\cS_{P\to P_{\mathsf{PPU}}}$ equals the cumulative cost of the current effective state recorded by $O_P^{\REQ}$ (or $0$ if no effective evaluation has yet occurred).
    \end{lemma}
    \begin{proof}
        The simulator follows the REQ procedure of \cref{fig:REQ}, with its own bookkeeping registers $\cA',\cG',\cB'$, and augments them with a local state register $\cS'$ storing a pair $(c,h)$, where $c$ is the current cumulative cost and $h\in\bit$ indicates whether the program is halted. Initially $\cS'$ stores $(0,0)$.

        The only nontrivial task is to implement the REQ subroutine $U_{P_{\mathsf{PPU}}}$.
        \begin{itemize}
            \item In $\Eval$-mode and with $h=0$, the program $P_{\mathsf{PPU}}$ simply evaluates $P$. The simulator therefore delegates this branch to its oracle access to $O_P^{\REQ}$ on the same tag and input $x$, obtains the output $y$, XORs $y$ into the outer query register exactly as $U_{P_{\mathsf{PPU}}}$ would, and updates the cost component of $\cS'$ to $\mathsf{cost}(y)$.
            \item In $\Prove$-mode and with $h=0$, the hidden state of $P$ is not needed. The simulator computes the output locally by coherently sampling $r\in\cR$, setting $r_1=\PRF(k_1,(c,r))$, $r_2=\PRF(k_2,(c,r,r_1))$, and $\sigma=\BUS.\Sign(\sk,(c,r,r_1);r_2)$, XORing $(c,r,r_1,\sigma)$ into the output register, and flipping the halt bit to $h=1$.
            \item If $h=1$, then $P_{\mathsf{PPU}}$ always outputs $(\mathsf{Halt},0)$, so the simulator can implement this branch locally as well.
        \end{itemize}

        On every computational-basis branch of the interaction, these are exactly the three branches of $P_{\mathsf{PPU}}$. Hence the simulator applies the same unitary transformation as $O_{P_{\mathsf{PPU}}}^{\REQ}$, including the inverse transformation on repeated queries. By linearity, the induced channel on arbitrary superpositions is identical, so the simulation is perfect.

        The invariant on the cost register follows by induction over the sequence of effective queries. Initially both costs are $0$. Each effective $\Eval$-query updates the simulator's cost register to $\mathsf{cost}(y)$, which is precisely the cost stored by the next effective state of $P$ in $O_P^{\REQ}$. A $\Prove$-query does not change the effective state of $P$ and only freezes further evolution, so the invariant continues to hold.
    \end{proof}

    Now define the final simulator $\cS$ by composition:
    \[
        \cS^{O_P^{\REQ}}(1^\secp) := \big(\cS'^{\cS_{P\to P_{\mathsf{PPU}}}^{O_P^{\REQ}}}(1^\secp),\pk\big),
    \]
    where $\pk$ is the public key sampled by $\cS_{P\to P_{\mathsf{PPU}}}$.
    By perfect emulation of $O_{P_{\mathsf{PPU}}}^{\REQ}$ we have
    \begin{equation}\label{eq:ppu-security-step2}
        \Pr\!\left[D\!\left(1^\secp,\cS'^{O_{P_{\mathsf{PPU}}}^{\REQ}}(1^\secp),\aux_{P,\cP}\right)\to 1\right]
        =
        \Pr\!\left[D\!\left(1^\secp,\cS^{O_P^{\REQ}}(1^\secp),\aux_{P,\cP}\right)\to 1\right].
    \end{equation}
    Combining \eqref{eq:ppu-security-step1} and \eqref{eq:ppu-security-step2} proves the claim.
\end{proof}

\begin{lemma}[PPU Program Soundness]\label{thm:ppu-program-soundness}
    The construction in \cref{fig:PPU-program-compiler-construction-1,fig:PPU-program-compiler-construction-2} satisfies the soundness property of \cref{def:pay-per-use-program-compiler}.
\end{lemma}
\begin{proof}
    Let $A$ be any QPT adversary in the soundness game. By \cref{thm:simulate-one-REQ-oracle-with-another}, the game in \cref{def:pay-per-use-program-compiler} is perfectly equivalent to a game in which the prover interacts with $\cS'^{O_{P_{\mathsf{PPU}}}^{\REQ}}$ instead of $\cS^{O_P^{\REQ}}$, and the challenger checks the true cost using the cost register of the simulated oracle $O_{P_{\mathsf{PPU}}}^{\REQ}$. Thus we may treat the composition of $A$ with $\cS'$ as a single QPT oracle algorithm $\widehat{A}$ interacting with $O_{P_{\mathsf{PPU}}}^{\REQ}$.

    We now modify the prove-branch of $O_{P_{\mathsf{PPU}}}^{\REQ}$ through a sequence of hybrids.

    \paragraph{Hybrid 0.}
    This is the game above, i.e. $\widehat{A}$ interacts with $O_{P_{\mathsf{PPU}}}^{\REQ}$, where the prove-branch uses the PRFs $\PRF(k_1,\cdot)$ and $\PRF(k_2,\cdot)$.

    \paragraph{Hybrid 1.}
    Replace $\PRF(k_2,\cdot)$ in the prove-branch with a truly random oracle $H_2$. Standard PRF security implies that Hybrids 0 and 1 are computationally indistinguishable.

    \paragraph{Hybrid 2.}
    Let
    \[
        B := \{(c,r,r_1) : r_1 \neq \PRF(k_1,(c,r))\}.
    \]
    Replace the signing procedure in the prove-branch with a \emph{memoized blinded signing oracle} $\widehat{\Sign}_B$. On a query $m=(c,r,r_1)$, the oracle behaves as follows: if $m\in B$ it outputs $\bot$; if $m\notin B$ and $m$ has been seen before, it returns the memoized signature for $m$; otherwise it queries the blind signing oracle $\Sign_B(\sk,m)$ once, stores the resulting signature, and returns it. Since the prove-branch of $P_{\mathsf{PPU}}$ only ever queries the signing procedure on messages with $r_1=\PRF(k_1,(c,r))$, the blinded branch is never taken, and Hybrids 1 and 2 are perfectly indistinguishable.

    \paragraph{Hybrid 3.}
    Replace $\PRF(k_1,\cdot)$ with a truly random oracle $H_1$ both in the generation of $r_1$ and in the definition of the blinding set:
    \[
        B := \{(c,r,r_1) : r_1 \neq H_1(c,r)\}.
    \]
    Because the prove-branch and the blinding test access $\PRF(k_1,\cdot)$ only as an oracle, Hybrids 2 and 3 are computationally indistinguishable by PRF security.

    \paragraph{Hybrid 4.}
    Implement $H_1$ as a compressed oracle rather than as a lazy-sampled random oracle. This does not change the adversary's view.

    Fix Hybrid 4. Let $E_{\mathsf{succ}}$ be the event that the verifier accepts a proof for a claimed cost $c$ while the true effective cost at the end of the game is some $c^\star \neq c$. We show that $\Pr[E_{\mathsf{succ}}]$ is negligible.

    Let the verifier-accepted message be $m=(c,r,r_1)$ with signature $\sigma$.
    There are two cases.

    \medskip
    \noindent\textbf{Case 1:} $m \in B$, i.e. $r_1 \neq H_1(c,r)$.
    In this case, the signature $(m,\sigma)$ is valid (because the verifier accepted), yet $m$ lies in the blinding set of the oracle $\widehat{\Sign}_B$. By construction, $\widehat{\Sign}_B$ is implemented using the blind signing oracle $\Sign_B(\sk,\cdot)$ together with local memorization. Therefore a successful prover in this case yields a successful blind-unforgeability adversary for $\BUS$. Hence this case occurs with negligible probability.

    \medskip
    \noindent\textbf{Case 2:} $m \notin B$, i.e. $r_1 = H_1(c,r)$.
    Since Hybrid 4 implements $H_1$ as a compressed oracle, we may apply \cref{lem:zhandry_lemma5} with $k=1$. It follows that, except with negligible probability, whenever $\widehat{A}$ outputs a tuple $(c,r,r_1)$ satisfying $r_1 = H_1(c,r)$, the compressed database of $H_1$ contains the recorded entry $((c,r),r_1)$ at the end of the experiment.

    But $H_1$ is queried \emph{only} inside the $\Prove$-branch of $P_{\mathsf{PPU}}$. Consequently, a recorded entry for $(c,r)$ implies that a $\Prove$-query with randomness $r$ remains effective at the end of the interaction. By the definition of $P_{\mathsf{PPU}}$, that query reads the current cost $c$ from the program state, outputs a signature on $m=(c,r,r_1)$, and sets the program state to $\mathsf{Halt}$. Once this happens, every later effective query leaves the program in the halt state and cannot change the cost. Therefore the final true effective cost must equal $c$, i.e. $c^\star = c$, contradicting the event $E_{\mathsf{succ}}$.

    We have shown that both cases occur only with negligible probability in Hybrid 4, so the same is true in Hybrid 4 itself. Undoing the hybrids changes the success probability by at most a negligible amount. Hence the soundness error in the original game is negligible.
\end{proof}

\subsection{Semi-Quantum Query-Limited Programs}\label{sec:semi-quantum-QLP}
% \subsection{Semi-Quantum Query-Limited Programs}\label{sec:semi-quantum-QLP}
We construct query-limited programs that can be sent over classical channels. This is possible because in our construction, the power to evaluate the program is conferred by a one-shot program token. As with one-shot signatures, the power conferred by this token can be transferred over classical channels by using the token to sign the serial number of a new token.

\paragraph{Definition.} A program is \textit{semi-quantum} if the ability to evaluate the program can be transferred from a sender to a receiver using only classical communication. In \cref{def:semi-quantum-query-limited-program-compiler} below, we augment the definition of a QLP compiler to include an interactive protocol $\Transfer$ that allows control of a program to be transferred using classical communication. 

Our correctness property is similar to the one for regular QLPs (\cref{def:RAM-program-compiler}) except that the program evaluations can be interleaved with program transfers. Recall that an honestly generated QLP
\[\tilde{P} \gets \PGen(1^\secp, P)\]
allows the user to make a sequence of evaluations (executions of $\Eval(\tilde{P},x)$), and the correctness property of \cref{def:RAM-program-compiler} ensures that $\tilde{P}$'s responses will be statistically close to the responses that $P$ would have given. For semi-quantum QLPs, we allow the user to interleave the program evaluations with program transfers (executions of $\Transfer$), and we require that correctness of the evaluations still holds in this setting.
\bhaskar{Should we also provide a way to verify the transferred program?}
% Additionally, we add a function $\mathsf{PVer}$ that verifies whether a program is well-formed. We require the following properties:
% \begin{itemize}
%     \item \textbf{Validity of honest programs:} A program generated honestly with $\mathsf{PGen}$ will be accepted by $\mathsf{PVer}$ with overwhelming probability.
%     \item \textbf{Gentleness of $\mathsf{PVer}$:} If a state is accepted by $\mathsf{PVer}$, then we can apply $\mathsf{PVer}$ again and this changes 
% \end{itemize}
% For correctness, we have the guarantee that  Additionally, if the program sent during $\Pi_\mathsf{Transfer}$ passes $\mathsf{PVer}$ at the start of the protocol, then the program received at the end of the protocol will still pass $\mathsf{PVer}$ with overwhelming probability.

\begin{definition}[Semi-Quantum Query-Limited Program Compiler]\label{def:semi-quantum-query-limited-program-compiler}
    Let $\cP = \{P:\cM \times \cX \times \cR \rightarrow \cM \times \cY\}_{P}$ be a family of RAM programs. A \textbf{semi-quantum query-limited program compiler} for $\cP$ is a query-limited program compiler for $\cP$ (\cref{def:RAM-program-compiler}) with the following additional properties. 
    \begin{itemize}
    
    \item \textbf{Syntax:} In addition to $\PGen$ and $\Eval$ from \cref{def:RAM-program-compiler}, the compiler also includes the following protocol:
    \begin{itemize}
        % \item $\mathsf{PVer}(\tilde{P}) \to \tilde{P}', b$: A QPT algorithm that verifies a given program $\tilde{P}$. $\mathsf{PVer}$ outputs a new program $\tilde{P}'$ and a bit $b$ to indicate acceptance ($b=1$) or rejection ($b=0$).
% 
        \item $\Transfer$: An interactive protocol between a QPT sender $S$ and a QPT receiver $R$. $S$ takes as input a program $\tilde{P}$. Next, the parties may only send classical messages. If we are working in an oracle model, then the parties may also send access to classical oracles. Finally, as output $R$ has a new program $\tilde{P}'$. We notate this protocol as $\Transfer(\tilde{P}) \to \tilde{P}'$.
    \end{itemize}

    % \item \textbf{Validity of Honest Programs:} For any $P \in \cP$, there is a negligible function $\negl(\secp)$ such that for any $\secp \in \bbN$,
    % \begin{align*}
    % \Pr\left[b = 1 : \substack{
    %     \tilde{P} \gets \mathsf{PGen}(1^\secp, P)\\
    %     (\tilde{P}', b) \gets \mathsf{PVer}(\tilde{P})
    %     }\right] \geq 1 - \negl(\secp)
    % \end{align*}
    % Additionally, for any 

    % \item \textbf{Gentleness of $\mathsf{PVer}$:} For any program $\tilde{P}$, let 
    % \begin{align*}
    %     (\tilde{P}', b) &\gets \mathsf{PVer}(\tilde{P})\\
    %     (\tilde{P}'', b') &\gets \mathsf{PVer}(\tilde{P}')
    % \end{align*}
    % Given that $b = 1$, $\tilde{P}'$ and $\tilde{P}''$ are statistically indistinguishable.

    % \item \textbf{Transfer Security:} 

    \item \textbf{Correctness:} Let an operation sequence of length $n$ be a sequence $[(\mathsf{mode}_i, x_i)]_{i \in [n]}$, where for each $i \in [n]$, $\mathsf{mode}_i \in \{\mathsf{Eval}, \mathsf{Transfer}\}$ and $x_i \in \cX$. 
    % Given a program $P$, a security parameter $\secp$, and an operation sequence, let $\mathsf{Real}$ and $\mathsf{Ideal}$ be the following distributions over $y$-values: 
    % \begin{itemize}
    %     \item $\mathsf{Real}(P, \secp, [(\mathsf{mode}_i, x_i)]_{i \in [n]})$:
    %     \begin{enumerate}
    %         \item Compute $\tilde{P}_0 \gets \PGen(1^\secp, P)$.
    %         \item For each $i \in [n]$, if $\mathsf{mode}_i = \mathsf{Eval}$, then compute
    %         \[(y_i, \tilde{P}_i) \gets \Eval(\tilde{P}_{i-1}, x_i).\]
    %         If $\mathsf{mode}_i = \mathsf{Transfer}$, then compute $y_i = \bot$ and
    %         \[\tilde{P}_i \gets \Pi_\mathsf{Transfer}(\tilde{P}_{i-1}).\]
    %         \item Output $(y_1, \dots, y_n)$.
    %     \end{enumerate}

    %     \item $\mathsf{Ideal}(P, \secp, [(\mathsf{mode}_i, x_i)]_{i \in [n]})$:
    %     \begin{enumerate}
    %         \item Compute $\st_0 = \bot$.
    %         \item For each $i \in [n]$, if $\mathsf{mode}_i = \mathsf{Eval}$, then sample $r_i \getsr \cR$ and compute
    %         \[(\st_i, y_i) \gets P(\st_{i-1}, x_i; r_i).\]
    %         If $\mathsf{mode}_i = \mathsf{Transfer}$, then compute $y_i = \bot$ and $\st_i = \st_{i-1}$.
    %         \item Output $(y_1, \dots, y_n)$.
    %     \end{enumerate}
    % \end{itemize}
    
    The compiler satisfies correctness if for every $P\in \cP$ and every polynomial function $n(\secp) = \mathsf{poly}(\secp)$, there exists a negligible function $\negl(\secp)$, such that for every $\secp \in \bbN$ and every operation sequence $[(\mathsf{mode}_i, x_i)]_{i \in [n]}$ of length $n = n(\secp)$, 
    % the distributions
    % \[\mathsf{Real}(P, \secp, [(\mathsf{mode}_i, x_i)]_{i \in [n]}) \quad \text{ and } \quad \mathsf{Ideal}(P, \secp, [(\mathsf{mode}_i, x_i)]_{i \in [n]})\]
% 
    the ideal-world distribution
    \[\left\{ (y_1, \dots, y_n): \begin{array}{c}
             (r_1, \dots, r_n) \gets \cR^{n}  
             \\
             \st_0 = \bot\\
             (\st_i, y_i) \gets \begin{cases}
                 P(\st_{i-1}, x_i; r_i), & \mathsf{mode}_i = \mathsf{Eval}\\
                 (\st_{i-1}, \bot), & \mathsf{mode}_i = \mathsf{Transfer}
             \end{cases}, \forall i \in [n]
             \end{array}\right\}\]

             and the real-world distribution
             \[\left\{ 
        (y_1, \dots, y_n)
        :
        \begin{array}{c}
            \tilde{P}_0 \gets \PGen(1^\secp, P)
            \\
            (y_i, \tilde{P}_i) \gets \begin{cases}
                 \Eval(\tilde{P}_{i-1}, x_i), & \mathsf{mode}_i = \mathsf{Eval}\\
                 (\bot, \Transfer(\tilde{P}_{i-1}), & \mathsf{mode}_i = \mathsf{Transfer}
             \end{cases}, \forall i \in [n]
        \end{array}
        \right\}\]
    are $\negl(\secp)$-close in statistical distance.
    
    \item \textbf{Security} The compiler essentially satisfies the security property of \cref{def:RAM-program-compiler}.  We will view all the senders and receivers involved in the protocol as one colluding adversary. Let the quantum distinguisher $D$ be a non-uniform quantum algorithm that can be divided into the following stages $((S_1, R_1), (S_2, R_2),\cdots, (S_L, R_L), D')$ for an arbitrary polynomial $L= L(\lambda)$, where each $(S_i, R_i)$ consists of possibly several evaluations and then a sender-receiver $\Transfer$ protocol. $D'$ is a final distinguisher that outputs a bit.
    
    The scheme satisfies LEQ security if there exists a QPT simulator $\cS$ such that for any semi-quantum query-limited program $P \in \cP$ and any distinguisher $D$ with (quantum) auxiliary input $\aux_{P, \cP}$ (which may depend non-uniformly on $P$ and $\cP$), there exists a negligible function $\negl(\secp)$ such that for all $\secp \in \bbN$,
    \[
        \abs{\Pr\left[D(1^\secp, \PGen(1^\secp, P), \aux_{P, \cP}) \to 1\right] 
        - 
        \Pr\left[D\left(1^\secp, \cS^{O^{\REQ}_{P}}(1^\secp), \aux_{P, \cP})\right) \to 1\right]} 
        \leq \negl(\secp)
    \]
\end{itemize}
 
\end{definition}

\jiahui{formalize security definition}

\paragraph{Construction.}

To construct semi-quantum QLPs, we will start with the construction of (regular) QLPs given in \cref{fig:RAM-program-compiler-construction} (our construction is not black-boxly from any QLPs, but a simple tweak of our QLP construction). The program $\tilde{P}$ comprises $(\crs, \tilde{P}_\OSP, (\st, r_s, \sigma), s, \ket{T})$. The values $(\crs, \tilde{P}_\OSP, (\st, r_s, \sigma), s)$ are classical strings, which can be sent over classical channels. (Recall that one-shot programs, such as $\tilde{P}_{\OSP}$, are classical programs.) Only $\ket{T}$, the one-shot program token, is a quantum state and cannot be sent over classical channels.

Next, to transfer the program, we will make a dummy evaluation of the program. To do so, we evaluate the program in $\mathsf{Transfer}$ mode, in which the program outputs $y = \bot$ on every input and leaves the RAM state $\st$ unchanged. The point of running $\Eval$ in $\mathsf{Transfer}$ mode is to consume the program token (held by the sender) and generate a new token (held by the receiver) that can be used on the next query.

$\Eval$ includes the following steps.
\begin{enumerate}
    \item Sample $(s', \ket{T'}) \gets \OSP.\TokenGen(\crs)$.\label{Eval-step:sample-new-token}
    \item Evaluate $\left(y, (\st', r_{s'}, \sigma')\right) \gets \OSP.\Eval(\tilde{P}_\OSP, s, (x,\st, r_s),\sigma, s'), \ket{T})$.\label{Eval-step:Evaluate-program}
    \item Update $\tilde{P}$ to $(\crs, \tilde{P}_{\OSP}, (\st', r_{s'}, \sigma'), \ket{T'})$. \label{Eval-step:output-new-program}
\end{enumerate}
Let us have the receiver execute step \ref{Eval-step:sample-new-token} to generate the new program token. Then the receiver sends $s'$ to the sender. Next, the sender executes step \ref{Eval-step:Evaluate-program} to evaluate the program. This step also generates a signature $\sigma'$ on $s'$, which ensures that $\ket{T'}$ can be used to make the next query. Then the sender sends $(y, (\st', r_{s'}, \sigma'))$ to the receiver, and the receiver executes step \ref{Eval-step:output-new-program}. 

At the end of this protocol, the parties have evaluated the program in $\mathsf{Transfer}$ mode, which does not change the RAM state. %\jiahui{Is the sender allowed to use part of the program and then send the program witj "remaining queries" to receiver? }
Additionally, the sender's program token has been consumed, and the receiver's program token now has the power to evaluate the program. In this way, the power to evaluate the program has been transferred from the sender to the receiver.\\

Now, \cref{fig:semi-quantum-QLP-compiler-construction} describes the construction of a semi-quantum QLP compiler in detail. Let $\QLP.(\PGen,\Eval)$ be the QLP compiler construction from \cref{fig:RAM-program-compiler-construction}. Let $\OSP.(\KeyGen,\TokenGen,\PGen,\Eval)$ be a one-shot program compiler (\cref{def:osp-compiler}).

\begin{figure}
\begin{mdframed}
    % \begin{itemize}
    %     \item 
        \paragraph{$\PGen(1^\secp, P)$:}
        \begin{enumerate}
            \item Use $P$ to construct the following RAM program $P'$:
            \begin{mdframed}
            \begin{enumerate}
            \itemindent=-15pt
                \item[] \textbf{Inputs:} $[\st, (x, \mode); r]$. 
                $\st \in \cM$ is the current program state, $x \in \cX$ is the chosen input, $\mode \in \{\Eval, \Transfer\}$ is the mode to run the program in, and $r \in \cR$ is the sampled randomness.
                \itemindent=0pt
                \item If $\mode = \Eval$, then compute and output $(\st', y) = P(\st, x; r)$.
                \item If $\mode = \Transfer$, then output $(\st', y) = (\st, \bot)$.
            \end{enumerate}
            \end{mdframed}
            \item Generate a QLP for $P'$: $\tilde{P} \gets \QLP.\PGen(1^\secp, P')$. Here, $\QLP.\PGen$ is the algorithms described in Figure ~\ref{fig:RAM-program-compiler-construction}.
            \item Output $\tilde{P}$.
        \end{enumerate}

        \paragraph{$\Eval(\tilde{P}, x)$:}
        \begin{enumerate}
            \item Query the QLP in $\Eval$ mode:
            \begin{align*}
                \mode &= \Eval\\
                (y, \tilde{P}') &\gets \QLP.\Eval(\tilde{P}, (x, \mode))
            \end{align*}
            \item Output $(y, \tilde{P}')$.
        \end{enumerate}
        
        % \item 
        \paragraph{$\Transfer(\tilde{P})$:}
        \begin{enumerate}
            \item \textbf{Input:} The sender $S$ has a QLP $\tilde{P}$, which is parsed as $\tilde{P} = (\crs, \tilde{P}_{\OSP}, (\st, r_s, \sigma), s, \ket{T})$. The receiver $R$ has no inputs.
            \item $S$ sends $\crs$ to $R$.
            \item $R$ generates a token
            $$(s', \ket{T'}) \gets \OSP.\TokenGen(\crs)$$
            and sends $s'$ to $S$.
            \item $S$ evaluates the program as follows:
            \begin{align*}
                (x, \mode) &= (0, \Transfer)\\
                (y, (\st', r_{s'}, \sigma')) &\gets \QLP.\Eval(\tilde{P}_\OSP, s, [(x, \mode),\st, r_s],\sigma, s'), \ket{T})
            \end{align*}
        
            \item $S$ sends $(\tilde{P}_\OSP, \st', r_{s'}, \sigma')$ to $R$.
            \item $R$ updates $\tilde{P}$ to 
            $$\tilde{P}' \gets (\crs, \tilde{P}_{\OSP}, (\st', r_{s'}, \sigma'), \ket{T'}).$$
            $R$'s output is $\tilde{P}'$.
        \end{enumerate}
    % \end{itemize}
\end{mdframed}
\caption{Construction of a Semi-Quantum QLP Compiler for $\cP$}\label{fig:semi-quantum-QLP-compiler-construction}
\end{figure}

\begin{theorem}
    The construction in \cref{fig:semi-quantum-QLP-compiler-construction} is a semi-quantum query-limited program compiler for $\cP$.
\end{theorem}
\begin{proof}
$ $
\paragraph{Syntax.} It is clear by inspection that the construction has the syntax of a semi-quantum QLP.

\paragraph{Correctness.}
First, let $[(\mathsf{mode}_i, x_i)]_{i \in [n]}$ be a given operation sequence. It suffices to consider the case where $x_i = 0$ if $\mode_i = \Transfer$, for all $i \in [n]$. This is because when $\mode_i = \Transfer$, the value of $x_i$ does not influence the output of the real- or ideal-world distributions, and $y_i = \bot$.

Second, the program $\tilde{P}$ output by $\PGen$ is a regular QLP for the functionality $P'$:
    \[\tilde{P} \gets \QLP.\PGen(1^\secp, P')\]

$\Eval(\tilde{P}, x)$ simply evaluates the program with $\mode = \Eval$:
    \[(y, \tilde{P}') \gets \QLP.\Eval(\tilde{P}, (x, \mode = \Eval))\]

Likewise, $\Transfer(\tilde{P})$ is equivalent to evaluating the program with $\mode = \Transfer$ and $x = 0$. That is to say: if we treat the sender and receiver as one joint evaluator, then $\Transfer(\tilde{P})$ is equivalent to this evaluator computing
    \[(y, \tilde{P}') \gets \QLP.\Eval(\tilde{P}, (x = 0, \mode = \Transfer))\]
and outputting $\tilde{P}'$. To see why, note that the $\Transfer$ protocol follows the same steps as $\QLP.\Eval$ in \cref{fig:RAM-program-compiler-construction}.

Third, the correctness of $\QLP$ (\cref{def:RAM-program-compiler}) implies that for any operation sequence $[(\mathsf{mode}_i, x_i)]_{i \in [n]}$, the following distributions are statistically close:
\begin{equation}\label{eq:ideal-world-dist-P-prime}
    \left\{ (y_1, \dots, y_n): \begin{array}{c}
             (r_1, \dots, r_n) \gets \cR^{n}  
             \\
             \st_0 = \bot\\
             (\st_i, y_i) \gets P'(\st_{i-1}, (x_i, \mode_i); r_i), \forall i \in [n]
             \end{array}\right\}
\end{equation}
and
\begin{equation}\label{eq:real-world-dist-P-prime}
    \left\{ 
        (y_1, \dots, y_n)
        :
        \begin{array}{c}
            \tilde{P}_0 \gets \PGen(1^\secp, P)
            \\
            (y_i, \tilde{P}_i) \gets \QLP.\Eval(\tilde{P}_{i-1}, (x_i, \mode_i)), \forall i \in [n]
        \end{array}
        \right\}
\end{equation}

Fourth, \cref{eq:ideal-world-dist-P-prime} is equivalent to the ideal-world distribution in \cref{def:semi-quantum-query-limited-program-compiler}'s correctness notion:
\begin{align*}
    &\left\{ (y_1, \dots, y_n): \begin{array}{c}
             (r_1, \dots, r_n) \gets \cR^{n}  
             \\
             \st_0 = \bot\\
             (\st_i, y_i) \gets P'(\st_{i-1}, (x_i, \mode_i); r_i), \forall i \in [n]
             \end{array}\right\}\\
    &= \left\{ (y_1, \dots, y_n): \begin{array}{c}
             (r_1, \dots, r_n) \gets \cR^{n}  
             \\
             \st_0 = \bot\\
             (\st_i, y_i) \gets \begin{cases}
                 P(\st_{i-1}, x_i; r_i), & \mathsf{mode}_i = \mathsf{Eval}\\
                 (\st_{i-1}, \bot), & \mathsf{mode}_i = \mathsf{Transfer}
             \end{cases}, \forall i \in [n]
             \end{array}\right\}
\end{align*}
This follows from the definition of $P'$.

Fifth, \cref{eq:real-world-dist-P-prime} is statistically close to the real-world distribution in \cref{def:semi-quantum-query-limited-program-compiler}'s correctness notion:
\begin{align*}
    &\left\{ 
        (y_1, \dots, y_n)
        :
        \begin{array}{c}
            \tilde{P}_0 \gets \PGen(1^\secp, P)
            \\
            (y_i, \tilde{P}_i) \gets \QLP.\Eval(\tilde{P}_{i-1}, (x_i, \mode_i)), \forall i \in [n]
        \end{array}
        \right\}\\
    &\approx_s \left\{ 
        (y_1, \dots, y_n)
        :
        \begin{array}{c}
            \tilde{P}_0 \gets \PGen(1^\secp, P)
            \\
            (\tilde{y}_i, \tilde{P}_i) \gets \QLP.\Eval(\tilde{P}_{i-1}, (x_i, \mode_i)), \forall i \in [n]\\
            y_i = \begin{cases}
                \tilde{y}_i,& \mode_i = \Eval\\
                \bot,& \mode_i = \Transfer
            \end{cases}, \forall i \in [n]
        \end{array}
        \right\}\\
    &= \left\{ 
        (y_1, \dots, y_n)
        :
        \begin{array}{c}
            \tilde{P}_0 \gets \PGen(1^\secp, P)
            \\
            (y_i, \tilde{P}_i) \gets \begin{cases}
                 \Eval(\tilde{P}_{i-1}, x_i), & \mathsf{mode}_i = \mathsf{Eval}\\
                 (\bot, \Transfer(\tilde{P}_{i-1}), & \mathsf{mode}_i = \mathsf{Transfer}
             \end{cases}, \forall i \in [n]
        \end{array}
        \right\}
\end{align*}
The second line describes the case where we evaluate $\QLP.\Eval(\tilde{P}_{i-1}, (x_i, \mode_i))$, and if $\mode_i = \Transfer$, we overwrite $y_i \gets \bot$. The first and second lines are statistically close because when $\mode_i = \Transfer$, $\QLP.\Eval(\tilde{P}_{i-1}, (x_i, \mode_i))$ produces $y = \bot$ with overwhelming probability, due to the correctness of $\QLP$.

In total, we have shown that the real- and ideal-world distributions in \cref{def:semi-quantum-query-limited-program-compiler}'s correctness notion are statistically close for any operation sequence. This completes the proof of correctness.

\paragraph{Security.} \bhaskar{TBD} The security notion of \cref{def:semi-quantum-query-limited-program-compiler} is the same as the one for query-limited RAM programs (\cref{def:RAM-program-compiler}). Furthermore, the program $\tilde{P}$ output by $\PGen$ is a regular QLP for the functionality $P'$:
    \[\tilde{P} \gets \QLP.\PGen(1^\secp, P')\]
Then the security of $\QLP$ (\cref{def:RAM-program-compiler}) implies the following: there exists a QPT simulator $\cS'$ such that for any randomized RAM program $P \in \cP$ and any distinguisher $D$ with (quantum) auxiliary input $\aux_{P, \cP}$ (which may depend non-uniformly on $P$ and $\cP$), there exists a negligible function $\negl(\secp)$ such that for all $\secp \in \bbN$,
    \[
        \abs{\Pr\left[D(1^\secp, \PGen(1^\secp, P), \aux_{P, \cP}) \to 1\right] 
        - 
        \Pr\left[D\left(1^\secp, \cS'^{O^{\REQ}_{\textcolor{red}{P'}}}(1^\secp), \aux_{P, \cP})\right) \to 1\right]} 
        \leq \negl(\secp)
    \]
This is very close to what we have to prove. The only difference is that the simulator $\cS'$ has access to $O^{\REQ}_{\textcolor{red}{P'}}$, but we need a simulator simulates $\PGen(1^\secp, P)$ with access to $O^{\REQ}_{P}$.

\Cref{thm:sim-P-to-P-prime} proves that there exists a simulator $\cS_{P \to P'}$ that receives oracle access to $O^{\REQ}_{P}$ and perfectly simulates the oracle $O^{\REQ}_{P'}$. Then we can compose $\cS'$ with $\cS_{P \to P'}$ to obtain a simulator $\cS$ that simulates $\PGen(1^\secp, P)$ given access to $O^{\REQ}_{P}$. 

% Queries on $\mode = \Transfer$ do not change the state of the RAM program $P'$, and they always produce output $\bot$. So on $\Eval$ queries, $O^{\REQ}_{P'}$ functions the same as $O^{\REQ}_{P}$. Whenever $\cS_{P \to P'}$ receives a query with $\mode = \Eval$, it forwards the query to $O^{\REQ}_{P}$, and whenever it receives a query with $\mode = \Transfer$, it simulates this query in its own internal registers. The joint state of the $O^{\REQ}_{P}$ oracle and the internal registers of $\cS_{P \to P'}$ accurately reflect the state that $O^{\REQ}_{P'}$ would hold at this point.
\bhaskar{This proof needs more detail.}
\jiahui{Should we change all REQ to LEQ?}\jiahui{Seems to be all changed}
\end{proof}

\begin{lemma}\label{thm:sim-P-to-P-prime}
    There exists a simulator $\cS_{P \to P'}$ that receives oracle access to $O^{\REQ}_{P}$ and perfectly simulates the oracle $O^{\REQ}_{P'}$.    
\end{lemma}
\begin{proof}
First, let us construct of $\cS_{P \to P'}$. 

At a high level, $\cS_{P \to P'}$ simulates $O^{\REQ}_{P'}$ by following the procedure in \cref{fig:REQ}. Whenever $O^{\REQ}_{P'}$ needs to compute $P(\st, x)$, the simulator queries $O^{\REQ}_{P}$ on $\ket{t, x, u, +}$. The $b$-value is in the $\ket{+}$ state so that any bitflip applied to $\ket{+}$ will leave $\ket{+}$ unchanged. Additionally, the simulator does not see the state of $P$, so the simulator's $\cS$ register stores nothing.\\

\noindent \underline{$\cS_{P \to P'}^{O^{\REQ}_{P}}$:}
\begin{enumerate}
    \item The simulator samples $(\sk, \pk, k_1, k_2)$ according to \cref{eq:sample-PPU-parameters}. Let $P'$ be the program defined in \cref{fig:semi-quantum-QLP-compiler-construction}.
    \item The simulator implements $O^\REQ_{P'}$ by mostly following the procedure in \cref{fig:REQ}. Any deviations from \cref{fig:REQ} are listed below.
    \item \textbf{Work Registers.} The simulator stores the following registers:
    \begin{itemize}
        \item $\cS'$: A register that stores $\ket{c, h}$, where $c$ is the cost of all the queries so far, and $h$ is a bit that equals $1$ if and only if the program should halt. $\cS'$ is initialized to $\ket{0,0}$.
        
        \item $\cA', \cG', \cB'$. Same as $\cA, \cG, \cB$ in \cref{fig:REQ}.
    \end{itemize}
    \item \textbf{Sub-Routine.} The simulator makes use of subroutine $U_{P'}$. For any $x\in \cX$, $\mode \in \{\Eval, \Transfer\}$, and $u\in \cY$ on register $\cQ$, and any $\ket{c,h}$ on register $\cS'$, $U_{P'}$ acts as follows:
    \begin{enumerate}
        \item If $\mode = \Eval$, then:
        \begin{enumerate}
            \item Map
            \[
            \ket{x, \mode, u, b}_{\cQ} \otimes \ket{0}_{\cA'_i} 
            \leftrightarrow
            \ket{x, \mode, u, b}_{\cQ} \otimes \ket{i, x, \mode, 0, 0}_{\cA'_i}
            \]
            \item Query $O^\REQ_P$ on $\cA'_i$ to map:
            \[\ket{i, x, \mode, 0, 0}_{\cA'_i} \to \ket{i, x, \mode, y, b'}_{\cA'_i}\]
            \item Then map:
            \begin{align*}
                &\ket{x, \mode, u, b}_{\cQ} \otimes \ket{i, x, \mode, y, b'}_{\cA'_i}\\
                \longrightarrow &\ket{x, \mode, u \oplus y, b \oplus b'}_{\cQ} \otimes \ket{i, x, \mode, y, b'}_{\cA'_i}
            \end{align*}
        \end{enumerate}
        
        \item If $\mode = \Transfer$, then map:
        \[
         \ket{x, \mode, u, b}_{\cQ} \otimes \ket{0}_{\cA'_i} 
         \leftrightarrow
         \frac{1}{\sqrt{|\cR}|}\sum_{r} \ket{x, \mode, u \oplus \bot, b}_{\cQ} \otimes \ket{x, \mode, r}_{\cA'_i}
    \]
    \bhaskar{This proof is unfinished} \jiahui{added the following}
    When $\mode = \Transfer$, then $U_{P'}$ does \emph{not} query $\O^\REQ_{P}$ and instead
implements a reversible no-op RAM step that only outputs $\bot$ and records a dummy
(transcript) coin $r\in\cR$ in $\cA'_i$.
Concretely, it performs the following reversible maps (controlled on the
$\mode$ register): 
\begin{enumerate}
    \item Map
    \[
        \ket{x,\mode,u,b}_{\cQ}\otimes \ket{0}_{\cA'_i}
        \;\leftrightarrow\;
        \ket{x,\mode,u,b}_{\cQ}\otimes \ket{x,\mode,0}_{\cA'_i}\,,
    \]
    i.e., it writes the pair $(x,\mode)$ into the transcript block $\cA'_i$.

    \item Prepare a uniform superposition over coins in the last component of $\cA'_i$:
    \[
        \ket{x,\mode,0}_{\cA'_i}
        \;\mapsto\;
        \frac{1}{\sqrt{|\cR|}}\sum_{r\in\cR}\ket{x,\mode,r}_{\cA'_i}\,.
    \]

    \item Finally, XOR the constant output $\bot$ into the output accumulator (leaving $b$
    unchanged):
    \[
        \ket{x,\mode,u,b}_{\cQ}\otimes \ket{x,\mode,r}_{\cA'_i}
        \;\mapsto\;
        \ket{x,\mode,u\oplus \bot,b}_{\cQ}\otimes \ket{x,\mode,r}_{\cA'_i}\,.
    \]
\end{enumerate}

Equivalently, for $\mode=\Transfer$, we have $U_{P'}$ apply: 
\[
\ket{x,\mode,u,b}_{\cQ}\otimes \ket{0}_{\cA'_i}
\;\leftrightarrow\;
\frac{1}{\sqrt{|\cR|}}\sum_{r\in\cR}
\ket{x,\mode,u\oplus \bot,b}_{\cQ}\otimes \ket{x,\mode,r}_{\cA'_i}\,.
\]
    \end{enumerate}
\end{enumerate}

To see why $\cS_{P\to P'}$ is a \emph{perfect} simulator, it suffices to compare its action with the ideal oracle $O_{P'}^\REQ$ on each computational-basis branch of the query history. In $\Eval$ mode, the only nontrivial step of $P'$ is to run the underlying program $P$ on input $x$, and $\cS_{P\to P'}$ does exactly this by delegating that step to its oracle access to $O_P^\REQ$, while keeping the same query-history, garbage, and bookkeeping registers that \cref{fig:REQ} would maintain for $O_{P'}^\REQ$. In $\Transfer$ mode, $P'$ does not access $P$ at all: by definition it leaves the hidden RAM state unchanged and outputs the distinguished symbol $\bot$, so $\cS_{P\to P'}$ can implement this branch entirely locally by coherently writing $\bot$ to the output register and storing the sampled randomness in its ancilla register exactly as the honest REQ oracle would. Therefore, on every basis state, $\cS_{P\to P'}$ applies exactly the same unitary transformation as $O_{P'}^\REQ$ (including the corresponding inverse on repeated-query branches), and hence by linearity it induces exactly the same quantum channel on arbitrary superpositions. Thus $\cS_{P\to P'}$ perfectly simulates $O_{P'}^\REQ$.

\end{proof}
The security notion of \cref{def:semi-quantum-query-limited-program-compiler}
is the same as the one for query-limited RAM programs
(\cref{def:RAM-program-compiler}), except that the distinguisher $D$ is allowed
to internally run several stages of $\Eval$ and $\Transfer$. However, this does
not change the structure of the proof: the entire view of all senders and
receivers is generated from the initial program state, and hence it suffices to
simulate that initial state.

Fix any $P\in\cP$ and any (non-uniform) QPT distinguisher $D$ with auxiliary input $\aux_{P,\cP}$.
Let $P'$ be the two-mode RAM program defined in \Cref{fig:semi-quantum-QLP-compiler-construction}, and observe that the real semi-quantum compiler on input $P$ simply outputs a regular $\QLP$ for $P'$, i.e.,
\[
\PGen(1^\secp,P)\equiv \QLP.\PGen(1^\secp,P').
\]
By the $\REQ$-security of $\QLP$ from \Cref{def:RAM-program-compiler}, there exists a QPT simulator $\cS'$ such that
\begin{equation}\label{eq:sq-qlp-sec-step1}
\abs{\Pr\!\left[D\!\left(1^\secp,\QLP.\PGen(1^\secp,P'),\aux_{P,\cP}\right)\to 1\right]
-
\Pr\!\left[D\!\left(1^\secp,\cS'^{\,O^{\REQ}_{P'}}(1^\secp),\aux_{P,\cP}\right)\to 1\right]}
\le \negl(\secp).
\end{equation}
Next, by \Cref{thm:sim-P-to-P-prime}, there exists an oracle simulator $\cS_{P\to P'}$ that, given oracle access to $O^{\REQ}_{P}$, perfectly emulates $O^{\REQ}_{P'}$; in particular, for every QPT oracle algorithm $\adv$, the final joint state (i.e., the full ``view'') of $\adv$ when interacting with $O^{\REQ}_{P'}$ is identical to its final joint state when interacting with $\cS_{P\to P'}^{\,O^{\REQ}_{P}}$.
Define the desired simulator $\cS$ by composing $\cS'$ with this oracle emulation:
\[
\cS^{\,O^{\REQ}_{P}}(1^\secp)\ :=\ \cS'^{\,\cS_{P\to P'}^{\,O^{\REQ}_{P}}}(1^\secp),
\]
meaning that $\cS$ runs $\cS'$ and answers each oracle query that $\cS'$ makes to $O^{\REQ}_{P'}$ by invoking $\cS_{P\to P'}$, which in turn may query $O^{\REQ}_{P}$.
By perfect emulation we have
\begin{equation}\label{eq:sq-qlp-sec-step2}
\Pr\!\left[D\!\left(1^\secp,\cS'^{\,O^{\REQ}_{P'}}(1^\secp),\aux_{P,\cP}\right)\to 1\right]
=
\Pr\!\left[D\!\left(1^\secp,\cS^{\,O^{\REQ}_{P}}(1^\secp),\aux_{P,\cP}\right)\to 1\right].
\end{equation}
Combining \eqref{eq:sq-qlp-sec-step1} and \eqref{eq:sq-qlp-sec-step2}, and using $\PGen(1^\secp,P)\equiv \QLP.\PGen(1^\secp,P')$, we obtain
\[
\abs{\Pr\!\left[D\!\left(1^\secp,\PGen(1^\secp,P),\aux_{P,\cP}\right)\to 1\right]
-
\Pr\!\left[D\!\left(1^\secp,\cS^{\,O^{\REQ}_{P}}(1^\secp),\aux_{P,\cP}\right)\to 1\right]}
\le \negl(\secp),
\]
which is exactly the security requirement of \Cref{def:RAM-program-compiler} (and hence of \Cref{def:semi-quantum-query-limited-program-compiler}).

\subsection{One-Shot Signatures Based On Any Blind-Unforgeable Signature Scheme}\label{sec:OSS-from-BUS}

The only constructions we have of one-shot signatures (\cite{SZ25,SZ25b,HV25}) produce signatures that are vectors in a subspace coset. \bhaskar{double check this claim} 

This limitation is not essential. We provide a generic construction of one-shot signatures with a variety of signature formats. Given any signature scheme that satisfies blind unforgeability (\cref{def:blind-unforge}), we construct a one-shot signature scheme whose verification function simply applies the verification function of the blind-unforgeable scheme. The proof is similar to that of \cite{GLRRV25}, who constructed \textit{one-time} signatures from any blind-unforgeable signature scheme.
\bhaskar{Can we construct one-shot signatures in the plain model if we replace the oracle with an iO program?}

% Namely, we construct one-shot signature schemes from any blind-unforgeable signature scheme $\mathsf{BUS}$, and the OSS verification function simply applies $\mathsf{BUS}.\Ver$. %We leave a formal security proof for future version of this paper.
% \bhaskar{Does the new signature scheme still satisfy blind unforgeability.}

\paragraph{Definition.} In \cref{def:OSS-weak} below, we define the one-shot signature primitive that we construct. It is a little weaker than the primitive defined in \cref{def:signature-token,def:one-shot-signature}. 

First, the correctness property says that an honestly generated signature will pass verification. \Cref{def:OSS-weak} guarantees this property with overwhelming probability over the randomness of $\KeyGen$ and $\TokenGen$, whereas \cref{def:signature-token}'s notion guarantees this property for any well-formed $(\sk, \pk, \ket{T}, s)$ output by $\KeyGen$ or $\TokenGen$.

Second, \cref{def:OSS-weak} satisfies (regular) unforgeability, which says that it is hard to produce valid signatures on two different messages with the same serial number. In contrast, \cref{def:signature-token} guarantees strong unforgeability, which says that it is also hard to find two different valid signatures on a single $(s, m)$ pair.

\begin{definition}[One-Shot Signature Scheme]\label{def:OSS-weak}
$ $
\begin{itemize}
    \item \textbf{Syntax:} A one-shot signature scheme comprises the following QPT algorithms:
    \begin{itemize}
        \item $\KeyGen(1^\secp) \to (\sk, \pk)$: Takes a security parameter $\secp \in \bbN$ and generates two equal keys $\sk = \pk$.
        \item $\TokenGen(\sk) \to (\ket{T}, s)$: Generates a quantum signing token $\ket{T}$ and a classical serial number $s \in \cS_\secp \cup \{\bot\}$ associated with $\ket{T}$.
        \item $\Sign(\ket{T}, m) \to \sigma$: Takes a signing token $\ket{T}$ and a message $m$ from the message space $\cM_\secp$ and outputs a signature $\sigma \in \Sigma_\secp$.
        \item $\Ver(\pk, s, m, \sigma) \to \bit$: Verifies the signature $\sigma$ on message $m$ with serial number $s$. $\Ver$ outputs $1$ for valid signatures and $0$ for invalid signatures.
    \end{itemize}
    The variables $(\secp, \sk, \pk, s, m, \sigma)$ are classical, and $\ket{T}$ is a quantum state.

    \item \textbf{Correctness:} The scheme is \textit{correct} if there is a negligible function $\negl(\cdot)$ such that for all $\secp \in \bbN$ and all $m \in \cM_\secp$,
    \[\Pr\left[\substack{
    b = 1
    }
    : \substack{
        (\sk, \pk) \gets \KeyGen(1^\secp)\\
        (\ket{T}, s) \gets \TokenGen(\sk)\\
        \sigma \gets \Sign(\ket{T}, m)\\
        b \gets \Ver(\pk, s, m, \sigma)}\right] \geq 1 - \negl(\secp)\]
    
    \item \textbf{(Regular) Unforgeability:} The scheme is \textit{unforgeable} if for any QPT adversary $\cA$, there exists a negligible function $\negl(\cdot)$ such that for any $\secp \in \bbN$,
    \[\Pr\Bigg[
    \substack{
        m_0 \neq m_1 \land\\
        1 \gets \Ver(\pk, s, m_0, \sigma_0) \land\\
        1 \gets \Ver(\pk, s, m_1, \sigma_1)
    }:
    \substack{
        (\sk, \pk) \gets \KeyGen(1^\secp)\\
        (m_0, m_1, \sigma_0, \sigma_1, s) \gets \cA(1^\secp, \pk) 
    }
    \Bigg] \leq \negl(\secp)\]
\end{itemize}
\end{definition}

\paragraph{Construction.} The construction uses a one-shot program for the $\mathsf{BUS}.\Sign$ functionality. Specifically, the functionality takes a message $m$ and serial number $s$ and outputs signature $$\sigma = (r, r_1, \sigma_\BUS),$$
where
$$\sigma_\BUS = \mathsf{BUS}.\Sign(\sk, (s, m, r, r_1)),$$
$r$ is random, and $r_1$ is pseudorandom. Next, we verify this signature by simply running 
$$\mathsf{BUS}.\Ver(\pk, (s,m,r,r_1), \sigma_{\mathsf{BUS}}).$$

Intuitively the construction is secure for the following reasons. If the adversary produces a valid message-signature tuple $(s, m, \sigma)$, then $(s,m)$ must have been queried to the BUS signing oracle. Furthermore, $\sigma$ contains $r$, which is highly entangled with the SEQ oracle's compressed oracle. Therefore, $(s,m)$ will be recorded in the SEQ oracle database with overwhelming probability. Next, the security of the one-shot program says that the SEQ oracle will not record multiple queries with the same serial number $s$. This allows us to show that an adversary cannot produce two valid message-signature pairs that are valid with respect to the same serial number $s$.\\

% This is similar to a one-time program, except that the program only outputs one valid signature for each serial number $s$.

Now we describe the construction in detail. First, let us define some parameters. Given a security parameter $\secp \in \bbN$, let $\cS_\secp$ be the space of serial numbers of the OSS scheme, and let $\cM_\secp$ be the space of messages. Let $\ell$ be the bitlength of a tuple $(s, m, r) \in \cS_\secp \times \cM_\secp \times \bit^\secp$.

Second, we use the following tools: 
\begin{itemize}
    \item Let $\mathsf{OSP}.(\KeyGen, \TokenGen, \PGen, \Eval)$ be a one-shot program compiler (\cref{def:generalized-otp-compiler,def:osp-compiler}).
    \item Let $\BUS.(\KeyGen, \Sign, \Ver)$ be a blind-unforgeable signature scheme (\cref{def:classical-signature-scheme,def:blind-unforge}). Let us assume that the message space $\cM_\secp^{\BUS}$ of $\BUS$ includes the set $\bit^{2 \ell}$, and $\BUS.\Sign$ uses a random tape of length $2 \ell$.
    \item Let $\PRF$ be a pseudorandom function (\cref{def:PRF}).
\end{itemize}

Third, the construction in \cref{fig:OSS-construction} is a one-shot signature scheme based on $\BUS$. Note that the scheme's $\Ver$ function simply applies $\BUS.\Ver$ to the message $(s, m, r, r_1)$.

\begin{figure}
\begin{mdframed}
\begin{itemize}
    \item $\KeyGen(1^\secp)$:
    \begin{enumerate}
        \item Sample:
        \begin{align*}
            (k_1, k_2) &\gets \bit^{\ell} \times \bit^{2 \ell}\\
            (\sk_\mathsf{BUS}, \pk_\mathsf{BUS}) &\gets \mathsf{BUS}.\KeyGen(1^\secp)\\
            \crs_\OSP &\gets \OSP.\KeyGen(1^\secp)
        \end{align*}
        % \item Sample a random oracle $H$.
        \item Define $f$ to be the following randomized function, parametrized by $(k_1, k_2, \sk_\BUS)$:
        \begin{mdframed}
            \begin{enumerate}
            \itemindent=-15pt
                \item[] \textbf{Inputs:} $(s,m;r)$. $s \in \cS_\secp$ is a serial number, $m \in \cM_\secp$ is a message, and $r \getsr \bit^\secp$ is a random string.
                \itemindent=0pt
                \item Compute
                \begin{align*}
                    r_1 &= \PRF(k_1, (s, m, r))\\
                    r_2 &= \PRF(k_2, (s, m, r, r_1))\\
                    \sigma_\BUS &= \BUS.\Sign(\sk_\BUS, (s, m, r, r_1);r_2)\\
                    \sigma &= (r, r_1, \sigma_\BUS).
                \end{align*}
                \item \textbf{Output:} $\sigma$
            \end{enumerate}
            \end{mdframed}
        
        \item Compute 
        \begin{align*}
            P &\gets \OSP.\PGen(\crs_\OSP, f)\\
            \sk = \pk &= (\crs_\OSP, \pk_\BUS, P)
        \end{align*}
        \item Output $(\sk, \pk)$.
    \end{enumerate}
    \item $\TokenGen(\sk)$:
    \begin{enumerate}
        \item Parse $\sk = (\crs_\OSP, \pk_\BUS, P)$.
        \item Compute $(\ket{T}_\OSP, s) \gets \mathsf{OSP}.\TokenGen(\crs_\OSP)$.
        % \item If $(\ket{T}_\OSP, s) = (\bot, \bot)$, then set $\ket{T} = \bot$. Otherwise, 
        \item Set $\ket{T} = (\ket{T}_\OSP, P, s)$.
        \item Output $(\ket{T}, s)$.
    \end{enumerate}
    \item $\Sign(\ket{T}, m)$: 
    \begin{enumerate}
        % \item If $\ket{T} = \bot$, then output $\sigma = \bot$ and halt. Otherwise, continue.
        \item Parse $\ket{T} = (\ket{T}_\OSP, P, s)$.
        \item Compute and output $\sigma \gets \OSP.\Eval(P, s, m, \ket{T}_\OSP)$.
    \end{enumerate}
    \item $\Ver(\pk, s, m, \sigma)$: 
    \begin{enumerate}
        \item Parse $\pk = (\crs_\OSP, \pk_\BUS, P)$ and $\sigma = (r, r_1, \sigma_\mathsf{BUS})$.
        \item Compute and output $b = \mathsf{BUS}.\Ver(\pk_\mathsf{BUS}, (s, m, r, r_1), \sigma_\mathsf{BUS})$.
    \end{enumerate}
\end{itemize}
\end{mdframed}
\caption{One-Shot Signature Scheme}\label{fig:OSS-construction}
\end{figure}

\begin{theorem}
    The construction in \cref{fig:OSS-construction} is a one-shot signature scheme (as defined in \cref{def:OSS-weak}).
\end{theorem}
\begin{proof}
    First, it is clear by inspection that the construction satisfies the syntax of a one-shot signature scheme. Note that for any $\secp \in \bbN$ and any $(\sk, \pk)$ in the support of $\KeyGen(1^\secp)$, $\sk = \pk$. Additionally, $\OSP.\PGen$ outputs a classical program $P$, so $\sk = \pk = (\crs_\OSP, \pk_\BUS, P)$ is a classical string.

    Second, \cref{thm:correctness-of-OSS} proves that the scheme satisfies correctness. Third, \cref{thm:regular-unforgeability-of-OSS} shows that the scheme satisfies regular unforgeability.
\end{proof}

\begin{lemma}\label{thm:correctness-of-OSS}
    The scheme constructed in \cref{fig:OSS-construction} satisfies correctness (\cref{def:OSS-weak}).
\end{lemma}
\begin{proof}
    This follows from the correctness of $\OSP$ and $\BUS$ and the pseudorandomness of $\PRF$. 
    
    % First, for a given $\secp \in \bbN$ and a given message $m \in \cM_\secp$, let 
    % \begin{align*}
    %     (\sk, \pk) &\gets \KeyGen(1^\secp)\\
    %     (\ket{T}, s) &\gets \TokenGen(\sk)\\
    %     \sigma &\gets \Sign(\ket{T}, m)\\
    %     b &\gets \Ver(\pk, s, m, \sigma)
    % \end{align*}
    
    % Second, $(\ket{T}, s) = (\bot, \bot)$ if and only if $\OSP.\TokenGen(\crs_\OSP)$ outputs $(\ket{T}_\OSP, s) = (\bot, \bot)$. This follows from the construction of $\TokenGen$. If $(\ket{T}_\OSP, s) = (\bot, \bot)$, then $\TokenGen$ outputs $(\ket{T}, s) = (\bot, \bot)$, and if $(\ket{T}_\OSP, s) \neq (\bot, \bot)$, then $\TokenGen$ outputs $(\ket{T}, s) = ((\ket{T}_\OSP, P, s), s) \neq (\bot, \bot)$. The probability that $(\ket{T}, s) = (\bot, \bot)$ equals the probability that $(\ket{T}_\OSP, s) = (\bot, \bot)$, which is $\negl(\secp)$, according to the correctness of $\OSP$ (\cref{def:generalized-otp-compiler}).

    % Third, let us consider the case where $(\ket{T}, s) \neq (\bot, \bot)$. This implies that $(\ket{T}_\OSP, s) \neq (\bot, \bot)$, which in turn implies that $\ket{T} = (\ket{T}_\OSP, P, s) \neq \bot$. Let us consider a sequence of hybrid distributions:
    For any given $\secp \in \bbN$ and message $m \in \cM_\secp$, let us define a sequence of hybrids.
    
    \paragraph{Hybrid 1.} This hybrid is the correctness experiment:
        \begin{align*}
            (\sk, \pk) &\gets \KeyGen(1^\secp)\\
            (\ket{T}, s) &\gets \TokenGen(\sk)\\
            \sigma &\gets \Sign(\ket{T}, m)\\
            b &= \Ver(\pk, s, m, \sigma)
        \end{align*}

    \paragraph{Hybrid 2.} This hybrid is equivalent to hybrid 1. We have just unfolded the notation.
        \begin{align*}
            (k_1, k_2) &\gets \bit^{\ell} \times \bit^{2 \ell}\\
            (\sk_\mathsf{BUS}, \pk_\mathsf{BUS}) &\gets \mathsf{BUS}.\KeyGen(1^\secp)\\
            \crs_\OSP &\gets \OSP.\KeyGen(1^\secp)\\
            P &\gets \OSP.\PGen(\crs_\OSP, f)\\
            (\ket{T}_\OSP, s) &\gets \mathsf{OSP}.\TokenGen(\crs_\OSP)\\
            \sigma &\gets \OSP.\Eval(P, s, m, \ket{T}_\OSP)\\
            b &= \mathsf{BUS}.\Ver(\pk_\mathsf{BUS}, (s, m, r, r_1), \sigma_\mathsf{BUS})
        \end{align*}

        \paragraph{Hybrid 3.} This hybrid is the same as hybrid 2, except that hybrid 3 does not generate a program $P$, and it computes $\sigma = f(s,m;r)$ for a uniformly random $r$.
        \begin{align*}
            (k_1, k_2) &\gets \bit^{\ell} \times \bit^{2 \ell}\\
            (\sk_\mathsf{BUS}, \pk_\mathsf{BUS}) &\gets \mathsf{BUS}.\KeyGen(1^\secp)\\
            \crs_\OSP &\gets \OSP.\KeyGen(1^\secp)\\
            (\ket{T}_\OSP, s) &\gets \mathsf{OSP}.\TokenGen(\crs_\OSP)\\
            r &\getsr \bit^\secp\\
            \sigma &= f(s, m; r)\\
            b &= \BUS.\Ver(\pk_\BUS, (s,m,r,r_1),\sigma_\BUS)
        \end{align*}

    \begin{claim}
        The probability that $b=1$ is negligibly close in hybrids 2 and 3. 
    \end{claim}
    \begin{proof}
        Hybrid 2 computes $\sigma$ as follows:
        \begin{align*}
            P &\gets \OSP.\PGen(\crs_\OSP, f)\\
            \sigma &\gets \OSP.\Eval(P, s, m, \ket{T}_\OSP),
        \end{align*}
    whereas hybrid 3 computes $\sigma$ as follows:
    \begin{align*}
        r &\getsr \bit^\secp\\
        \sigma &= f(s, m; r)
    \end{align*} 
    These two distributions over $\sigma$ are statistically close, by the correctness of $\OSP$ (\cref{def:generalized-otp-compiler}). This is the only difference between the two hybrids, so the distribution of $b$ is also statistically close in hybrids 2 and 3.
    \end{proof}
    
    \paragraph{Hybrid 4.} This hybrid is equivalent to hybrid 3. We have just unfolded the notation of $f(s, m; r)$.
        \begin{align*}
            (k_1, k_2) &\gets \bit^{\ell} \times \bit^{2 \ell}\\
            (\sk_\mathsf{BUS}, \pk_\mathsf{BUS}) &\gets \mathsf{BUS}.\KeyGen(1^\secp)\\
            \crs_\OSP &\gets \OSP.\KeyGen(1^\secp)\\
            (\ket{T}_\OSP, s) &\gets \mathsf{OSP}.\TokenGen(\crs_\OSP)\\
            r &\getsr \bit^\secp\\
            r_1 &= \PRF(k_1, (s, m, r))\\
            r_2 &= \PRF(k_2, (s, m, r, r_1))\\
            \sigma_\BUS &= \BUS.\Sign(\sk_\BUS, (s, m, r, r_1);r_2)\\
            \sigma &= (r, r_1, \sigma_\BUS)\\
            b &= \BUS.\Ver(\pk_\BUS, (s,m,r,r_1),\sigma_\BUS)
        \end{align*}
    
    \paragraph{Hybrid 5.} This hybrid is the same as hybrid 4, except that $r_1, r_2$ are truly random instead of pseudorandom.
        \begin{align*}
            (\sk_\mathsf{BUS}, \pk_\mathsf{BUS}) &\gets \mathsf{BUS}.\KeyGen(1^\secp)\\
            \crs_\OSP &\gets \OSP.\KeyGen(1^\secp)\\
            (\ket{T}_\OSP, s) &\gets \mathsf{OSP}.\TokenGen(\crs_\OSP)\\
            (r, r_1, r_2) &\getsr \bit^\secp \times \bit^{\ell} \times \bit^{2 \ell}\\
            \sigma_\BUS &= \BUS.\Sign(\sk_\BUS, (s, m, r, r_1);r_2)\\
            \sigma &= (r, r_1, \sigma_\BUS)\\
            b &= \BUS.\Ver(\pk_\BUS, (s,m,r,r_1),\sigma_\BUS)
        \end{align*}
    \begin{claim}
        The probability that $b=1$ is negligibly close in hybrids 4 and 5.
    \end{claim}
    \begin{proof}
        This follows from the security of $\PRF$ (\cref{def:PRF}).
    \end{proof}

    % The only difference between hybrids 2 and 3 is that hybrid 2 computes $(r_1, r_2)$ using $\PRF$ whereas hybrid 3 samples $(r_1, r_2)$ uniformly at random. This difference is indistinguishable due to the security of $\PRF$ (\cref{def:PRF}). The probability that $b=1$ in hybrids 2 and 3 are negligibly close. If they differed by a non-negligible amount, then we could construct an efficient distinguisher to distinguish $\PRF(k_1, \cdot)$ or $\PRF(k_2, \cdot)$ from a truly random function. We would simply run hybrid 2 or 3 and query the oracle to compute $(r_1, r_2)$. Then the distinguisher outputs the value $b$ that the hybrid computes.

    \begin{claim}
        In hybrid 5, $\Pr[b=1] = 1$.
    \end{claim}
    \begin{proof}
        This follows from the correctness of $\BUS$ (\cref{def:classical-signature-scheme}).
    \end{proof}

    The previous analysis implies that in hybrid 1, $\Pr[b=1]$ is overwhelming. This completes the proof of correctness.
\end{proof}

\begin{lemma}[Unforgeability]\label{thm:regular-unforgeability-of-OSS}
    The scheme constructed in \cref{fig:OSS-construction} satisfies (regular) unforgeability (\cref{def:OSS-weak}).
\end{lemma}
\begin{proof}
    Let us consider the following sequence of hybrids that transforms the unforgeability security game (hybrid 1) into a hybrid (hybrid 10) where the adversary has negligible probability of winning. We will show that in each pair of adjacent hybrids, the adversary's success probability is negligibly close. This shows that the adversary in the unforgeability game has negligible probability of winning.

\paragraph{Hybrid 1.} This hybrid is the unforgeability game (\cref{def:OSS-weak}) for the OSS scheme in \cref{fig:OSS-construction}. The adversary is given $(1^\secp, \pk, \ket{T}, s)$ and produces two messages with the same serial number $s$, and the challenger checks that the signatures on those messages are valid.\\

\begin{enumerate}
    \item \textbf{Inputs:} A security parameter $\secp \in \bbN$ and a QPT adverary $\cA$.
    \item The challenger samples 
    \begin{align*}
        (\sk, \pk) &\gets \KeyGen(1^\secp)
    \end{align*}
    and sends to a QPT adversary $\cA$ the values $(1^\secp, \pk)$.
    \item The adversary outputs $(m_0, m_1, \sigma_0, \sigma_1, s)$.
    \item The challenger checks that the following conditions are satisfied:\label{unforgeability-game-step:checks}
    \begin{align*}
        m_0 &\neq m_1\\
        1 &\gets \Ver(\pk, s, m_0, \sigma_0)\\
        1 &\gets \Ver(\pk, s, m_1, \sigma_1)
    \end{align*}
    \item \textbf{Output:} If the checks above all pass, then the output of the hybrid is $1$. Otherwise, the output is $0$.
\end{enumerate}

Note that during $\KeyGen(1^\secp)$, the challenger computes $P \gets \OSP.\PGen(\crs_\OSP, f)$.

\paragraph{Hybrid 2.} This hybrid is the same as hybrid 1, except that during $\KeyGen(1^\secp)$, the OSP $P$ is computed as 
$$P = \Sim^{O^\SEQ_{f}}(1^\secp, \crs_\OSP, \pk_\BUS),$$ 
where $\Sim$ is the simulator guaranteed by the SEQ security of $\OSP$ (\cref{def:generalized-otp-compiler}), and $O^\SEQ_f$ is the oracle defined in \cref{fig:seq-oracle-with-serial-numbers}.

\begin{claim}
    For any QPT adversary $\cA$, there is a negligible function $\negl(\cdot)$ such that for any $\secp \in \bbN$, $\abs{\Pr[\text{Hybrid }1(\secp, \cA) \to 1] - \Pr[\text{Hybrid }2(\secp, \cA) \to 1]} \leq \negl(\secp)$.
\end{claim}
\begin{proof}
    This follows from the SEQ security of $\OSP$ (\cref{def:generalized-otp-compiler}). 
    \iffalse
    SEQ security says that for any $f$ and any QPT distinguisher $D$, there is a negligible function $\negl(\cdot)$ such that for all $\secp \in \bbN$,
    \begin{align*}
        \Bigg|\Pr\Bigg[b = 1 &: \substack{
            \crs_\OSP \gets \OSP.\KeyGen(1^\secp)\\
            (\ket{T}, s) \gets \OSP.\TokenGen(\crs_\OSP)\\
            P \gets \OSP.\PGen(\crs_\OSP, f)\\
            b \gets D\left(P, \ket{T}, s, \crs_\OSP, f, \aux_f\right)
        }\Bigg]
        - \Pr\Bigg[b = 1 &: \substack{
            \crs_\OSP \gets \OSP.\KeyGen(1^\secp)\\
            (\ket{T}, s) \gets \OSP.\TokenGen(\crs_\OSP)\\
            P \gets \Sim^{O^\SEQ_{f}}(1^\secp, \pk, \aux_f)\\
            b \gets D\left(P, \ket{T}, s, \crs_\OSP, f, \aux_f\right)
        }\Bigg]\Bigg| \leq \negl(\secp)
    \end{align*}
    \fi
    \bhaskar{Add more detail.}
\end{proof}

\paragraph{Hybrid 3.} This hybrid is the same as hybrid 2, except that $O_f^\SEQ$ computes $\PRF(k_1, \cdot)$ by querying an oracle $\cO_1$, and computes $\BUS.\Sign(\sk, \cdot; \PRF(k_2, \cdot))$ by querying an oracle $\cO_\sigma$.

Let the oracle $\mathcal{O}_1$ be parametrized by $k_1$. $\cO_1$ takes inputs $(s, m, r) \in \cS_\secp \times \cM_\secp \times \bit^\secp$ and outputs
    \[r_1 = \PRF(k_1, (s, m, r))\] 
Let the oracle $\mathcal{O}_\sigma$ be parametrized by $(k_2, \sk_\BUS)$. $\cO_\sigma$ takes inputs $(s, m, r, r_1) \in \cS_\secp \times \cM_\secp \times \bit^\secp \times \bit^\ell$ and outputs
    \[\sigma_\BUS = \BUS.\Sign(\sk_\BUS, (s, m, r, r_1); \PRF(k_2, (s, m, r, r_1))\]

Next, when $O^\SEQ_f$ computes $f$, this entails computing 
\begin{align*}
    r_1 &= \PRF(k_1, (s, m, r))\\
    \sigma_\BUS &= \BUS.\Sign(\sk_\BUS, (s, m, r, r_1); \PRF(k_2, (s, m, r, r_1))
\end{align*}
In hybrid 3, $O^\SEQ_f$ computes $r_1$ by querying $\cO_1$ and computes $\sigma_\BUS$ by querying $\cO_\sigma$. After $O^\SEQ_f$ has computed $\sigma = f(s, m; r)$, $O^\SEQ_f$ uncomputes the values of $r_1$ and $\sigma_\BUS$ from its work registers.

\begin{claim}
    For any QPT adversary $\cA$ and any $\secp \in \bbN$, $\Pr[\text{Hybrid }2(\secp, \cA) \to 1] = \Pr[\text{Hybrid }3(\secp, \cA) \to 1]$.
\end{claim}
\begin{proof}
$O^\SEQ_f$ must use a compressed oracle to choose the random input $r$ for $f$. However, $\cO_1$ and $\cO_\sigma$ compute deterministic functions of their inputs, so it is syntactically valid for $O^\SEQ_f$ to query these external oracles to compute $(r_1, \sigma_\BUS)$. Furthermore, the final value of $\sigma$ that $O^\SEQ_f$ computes has the same distribution in hybrids 2 and 3. Therefore, the implementations of $O_f^\SEQ$ in hybrids 2 and 3 are equivalent.
\end{proof}

\paragraph{Hybrid 4.} This hybrid is the same as hybrid 3, except that $\PRF(k_2, \cdot)$ is replaced with a (truly) random oracle $H_2$. 

Let $H_2$ be a random oracle mapping $\cS_\secp \times \cM_\secp \times \bit^\secp \times \bit^\ell \to \bit^{2 \ell}$. Now $\mathcal{O}_\sigma$ is parametrized by $(H_2, \sk_\BUS)$. On input $(s, m, r, r_1)$, $\cO_\sigma$ outputs
    \[\sigma_\BUS = \BUS.\Sign(\sk_\BUS, (s, m, r, r_1); H_2(s, m, r, r_1))\]

\begin{claim}
    For any QPT adversary $\cA$, there is a negligible function $\negl(\cdot)$ such that for any $\secp \in \bbN$, $\abs{\Pr[\text{Hybrid }3(\secp, \cA) \to 1] - \Pr[\text{Hybrid }4(\secp, \cA) \to 1]} \leq \negl(\secp)$.
\end{claim}
\begin{proof}
This follows from the security of $\PRF$.
\end{proof}
    
\paragraph{Hybrid 5.} This hybrid is the same as hybrid 4, except that $\mathcal{O}_\sigma$ is blinded on a set $B$, as in the $\BUS$ security game.
    
Let the blinding set $B$ be all inputs $(s, m, r, r_1) \in \cS_\secp \times \cM_\secp \times \bit^\secp \times \bit^\ell$ such that $r_1 \neq \PRF(k_1, (s, m, r))$. Equivalently, $B$ is all inputs such that $r_1 \neq \cO_1(s, m, r)$. 

Next, on each input $(s, m, r, r_1)$, $\cO_\sigma$ queries $\cO_1$ to check that $r_1 = \cO_1(s, m, r)$. If the check passes (i.e. the input is $\notin B$), then $\cO_\sigma$ answers the query with $\sigma_\BUS = \BUS.\Sign(\sk_\BUS, (s, m, r, r_1); H_2(s, m, r, r_1))$. If the check fails (i.e. the input is $\in B$), then $\cO_\sigma$ responds with $\sigma_\BUS = \bot$. Finally, $\cO_\sigma$ acts coherently on superpositions of classical inputs.

\begin{claim}\label{thm:OSS-security-proof-hybrids-4-5}
    For any QPT adversary $\cA$ and any $\secp \in \bbN$, $\Pr[\text{Hybrid }4(\secp, \cA) \to 1] = \Pr[\text{Hybrid }5(\secp, \cA) \to 1]$.
\end{claim}
\begin{proof}
The only difference between the hybrids is how $\cO_\sigma$ acts when its input satisfies $(s, m, r, r_1) \in B$. In hybrid 4, $\cO_\sigma$ computes $\BUS.\Sign$ on such an input, whereas in hybrid 5, $\cO_\sigma$ outputs $\bot$.

In hybrids 4 and 5, it is not possible to query $\cO_\sigma$ on an input $(s, m, r, r_1) \in B$. $\cO_\sigma$ is only queried by $O^\SEQ_f$, which computes $r_1$ as $r_1 = \cO_1(s, m, r) = \PRF(k_1, (s, m, r))$. Therefore $\cO_\sigma$ is only queried on superpositions of inputs $\notin B$.

Therefore, the $O^\SEQ_f$ oracle in hybrids 4 and 5 are perfectly indistinguishable, so the two hybrids have the same probability of outputting $1$.
\end{proof}

\paragraph{Hybrid 6.} This hybrid is the same as hybrid 5, except that the challenger additionally checks that that $m, m' \notin B$ when deciding what to output.

More specifically, after $\cA$ outputs $(m_0, m_1, \sigma_0, \sigma_1, s)$, the challenger makes the following checks:
\begin{enumerate}
    \item Check that\label{unforgeability-game-step:old-checks}
    \begin{align*}
        m_0 &\neq m_1\\
        1 &\gets \Ver(\pk, s, m_0, \sigma_0)\\
        1 &\gets \Ver(\pk, s, m_1, \sigma_1)
    \end{align*}
    \item Parse $\sigma_0 = (r^0, r_1^0, \sigma_\BUS^0)$ and $\sigma_1 = (r^1, r_1^1, \sigma_\BUS^1)$. Check that\label{unforgeability-game-step:new-checks}
    \begin{align*}
        r_1^0 &= \cO_1(s, m_0, r^0)\\
        r_1^1 &= \cO_1(s, m_1, r^1)
    \end{align*}
    \item \textbf{Output:} If the checks above all pass, then the output of the hybrid is $1$. Otherwise, the output is $0$.
\end{enumerate}

Note that the checks in step \ref{unforgeability-game-step:new-checks} are new and were not present in hybrid 5.

\begin{lemma}
    For any QPT adversary $\cA$, there is a negligible function $\negl(\cdot)$ such that for any $\secp \in \bbN$, $\abs{\Pr[\text{Hybrid }5(\secp, \cA) \to 1] - \Pr[\text{Hybrid }6(\secp, \cA) \to 1]} \leq \negl(\secp)$.
\end{lemma}
\begin{proof}
    First, the only case when hybrids 5 and 6 output different values is when the checks in step \ref{unforgeability-game-step:old-checks} (verifying the signature) pass, but the checks in step \ref{unforgeability-game-step:new-checks} (checking that the signature is not in the blinded stet $B$) fail. In this case, hybrid 5 outputs $1$, but hybrid 6 outputs $0$.

    Second, the case described above occurs with only negligible probability due to the blind unforgeability of $\BUS$ (\cref{def:blind-unforge}). Note that the oracle $O_\sigma$ in hybrids 5 and 6 is equivalent to the oracle $\BUS.\Sign_B(\sk_\BUS, \cdot)$ from the blind unforgeability security game. Next, in the case above, we have that $1 \gets \Ver(\pk, s, m_0, \sigma_0)$ and $1 \gets \Ver(\pk, s, m_1, \sigma_1)$, which implies that 
    \begin{align*}
        1 &\gets \BUS.\Ver(\pk_\BUS, (s, m_0, r^0, r_1^0), \sigma_\BUS^0), \text{ and }\\
        1 &\gets \BUS.\Ver(\pk_\BUS, (s, m_1, r^1, r_1^1), \sigma_\BUS^1)
    \end{align*}
    Furthermore, in the case above, we have that $r_1^0 \neq \cO_1(s, m_0, r^0)$ or $r_1^1 \neq \cO_1(s, m_1, r^1)$, which implies that
    \begin{align*}
        (s, m_0, r^0, r_1^0) &\in B, \text{ or }\\
        (s, m_1, r^1, r_1^1) &\in B
    \end{align*}
    Therefore, if the case above occurs, then for some $i \in \bit$, $(s, m_i, r^i, r_1^i), \sigma_\BUS^i$ represent a solution to the blind unforgeability game. The probability that $\cA$ outputs such a solution is negligible, due to the blind-unforgeability of $\BUS$.
    \bhaskar{Add more detail.}
\end{proof}

\paragraph{Hybrid 7.} This hybrid is the same as hybrid 6 except that $\cO_\sigma$ is no longer blinded on set $B$.

On each input $(s, m, r, r_1)$, $\cO_\sigma$ always answers the query with $\sigma_\BUS = \BUS.\Sign(\sk_\BUS, (s, m, r, r_1); H_2(s, m, r, r_1))$.

\begin{claim}
    For any QPT adversary $\cA$ and any $\secp \in \bbN$, $\Pr[\text{Hybrid }6(\secp, \cA) \to 1] = \Pr[\text{Hybrid }7(\secp, \cA) \to 1]$.
\end{claim}
\begin{proof}
The proof is essentially the same as the proof of \cref{thm:OSS-security-proof-hybrids-4-5}.

The only difference between the hybrids is how $\cO_\sigma$ acts when its input satisfies $(s, m, r, r_1) \in B$. In hybrid 6, $\cO_\sigma$ outputs $\bot$ on such an input whereas in hybrid 7, $\cO_\sigma$ computes $\BUS.\Sign$.

In hybrids 6 and 7, it is not possible to query $\cO_\sigma$ on an input $(s, m, r, r_1) \in B$. $\cO_\sigma$ is only queried by $O^\SEQ_f$, which computes $r_1$ as $r_1 = \cO_1(s, m, r) = \PRF(k_1, (s, m, r))$. Therefore $\cO_\sigma$ is only queried on superpositions of inputs $\notin B$.

Therefore, the $O^\SEQ_f$ oracle in hybrids 6 and 7 are perfectly indistinguishable, so the two hybrids have the same probability of outputting $1$.
\end{proof}

\paragraph{Hybrid 8.} This hybrid is the same as hybrid 7 except that $\PRF(k_1, \cdot)$ is replaced with a (truly) random oracle $H_1(\cdot)$.

Let $H_1$ be a random oracle mapping $\cS_\secp \times \cM_\secp \times \bit^\secp \to \bit^\ell$. Now $\mathcal{O}_1$ is parametrized by $H_1$. On input $(s, m, r)$, $\cO_1$ outputs
\[r_1 = H_1(s, m ,r)\]

\begin{claim}
    For any QPT adversary $\cA$, there is a negligible function $\negl(\cdot)$ such that for any $\secp \in \bbN$, $\abs{\Pr[\text{Hybrid }7(\secp, \cA) \to 1] - \Pr[\text{Hybrid }8(\secp, \cA) \to 1]} \leq \negl(\secp)$.
\end{claim}
\begin{proof}
    This follows from the security of $\PRF$.
\end{proof}
    
\paragraph{Hybrid 9.} This hybrid is the same as hybrid 8, except that $H_1$ is implemented as a compressed oracle.

\begin{claim}
    For any QPT adversary $\cA$ and any $\secp \in \bbN$, $\Pr[\text{Hybrid }8(\secp, \cA) \to 1] = \Pr[\text{Hybrid }9(\secp, \cA) \to 1]$.
\end{claim}
\begin{proof}
    The random oracle is indistinguishable from the compressed oracle due to \cite{Zha18} lemma 4.
\end{proof}

Let us summarize the definition of $O_f^\SEQ$ in hybrid 9. $O_f^\SEQ$ has two compressed oracle databases $D, D_1$ ($D$ is the database for the oracle $H$, and $D_1$ is the database for the oracle $H_1$). When queried on a basis state $\ket{s,m,u,b}_\cQ \otimes \ket{D}_\cD \otimes \ket{D_1}_{\cD_1}$, the oracle responds to the query as follows:
    \begin{enumerate}
        \item If the database $D$ contains an entry of the form $(s', m', r')$ such that $s' = s$ but $m' \neq m$, then skip the following steps. Otherwise, continue.
        \item \label{step:hybrid-seq-oracle-with-serial-number-queries-CO}
        \begin{enumerate}
            \item Prepare the state $\ket{s, m, 0, 0, 0}$ in an ancillary register $\cQ'$. Query $H(s, m)$ by applying the compressed oracle unitary $\CO$ to $\cQ' \times \cD$. This maps $\cQ'$ to a superposition over states of the form $\ket{s, m, r, 0, 0}$. 
            \item Query $H_1(s, m, r)$ by applying the compressed oracle unitary $\CO$ to $\cQ' \times \cD_1$. This maps $\cQ'$ to a superposition over states of the form $\ket{s, m, r, r_1, 0}$.
            \item Query $\cO_\sigma$ to compute $\sigma_\BUS = \cO_\sigma(s, m, r, r_1)$, and write the result to register $\cQ'$:
            \[\ket{s, m, r, r_1, 0}_{\cQ'} \to \ket{s, m, r, r_1, \sigma_\BUS}_{\cQ'}\]
        \end{enumerate}
        \item Apply the following isometry to registers $\cQ \times \cQ'$:
        \begin{align*}
            \ket{s, m, u, b}_\cQ \otimes \ket{s, m, r, r_1, \sigma_\BUS}_{\cQ'} \to \ket{s, m, u \oplus (r, r_1, \sigma_\BUS), b \oplus 1}_\cQ \otimes \ket{s,m,r, r_1, \sigma_\BUS}_{\cQ'}
        \end{align*}
        \item Uncompute \cref{step:hybrid-seq-oracle-with-serial-number-queries-CO}.
    \end{enumerate}

    Note that $\cO_\sigma$ does not query $H$ or $H_1$.

\paragraph{Hybrid 10.} This hybrid is the same as hybrid 9, except that right after $\cA$ outputs $(m_0, m_1, \sigma_0, \sigma_1, s)$, the challenger checks that the compressed oracle databases $(D, D_1)$ satisfy the following property: for every entry $(s, m, r, r_1) \in D_1$, there is the entry $(s, m, r) \in D$. If this check fails, then the output of the hybrid is $0$.

\begin{claim}
    For any QPT adversary $\cA$, there is a negligible function $\negl(\cdot)$ such that for any $\secp \in \bbN$, $\abs{\Pr[\text{Hybrid }9(\secp, \cA) \to 1] - \Pr[\text{Hybrid }10(\secp, \cA) \to 1]} \leq \negl(\secp)$.
\end{claim}

\jiahui{expanded the original proof to make it more formal}

\begin{proof}
    This follows from the compressed oracle chaining lemma
    (\cref{lem:compressed-chaining}).

    First, for either hybrid 9 or 10, from the beginning of the hybrid
    until right after $\cA$ outputs $(m_0,m_1,\sigma_0,\sigma_1,s)$,
    the only times that $H$ or $H_1$ are queried is during an execution
    of $O_f^\SEQ$. Furthermore, the forward computation
    in $O_f^\SEQ$ queries $H,H_1$ in a chained fashion, where the
    inputs and outputs of $H$ are fed as input to $H_1$. Specifically,
    the forward computation is:
    \begin{enumerate}
        \item Compute $r=H(s,m)$.
        \item Compute $r_1=H_1(s,m,r)$.
    \end{enumerate}
    Note that $\cO_\sigma$ does not query $H$ or $H_1$.
    After computing the signature and copying the output,
    uncomputing step
    \ref{step:hybrid-seq-oracle-with-serial-number-queries-CO}
    first applies the inverse signing computation, then queries
    $H_1$ to uncompute $r_1$, and finally queries $H$ to uncompute
    $r$. Thus the inverse calls occur in the reverse order; they are
    not another forward chained computation.

    Second, we verify that this complete procedure obeys
    the same norm estimate as the query procedure in
    \cref{lem:compressed-chaining}. Let us rename our variables to
    match the notation of \cref{lem:compressed-chaining}:
    \begin{align*}
        H &\to G,\\
        H_1 &\to H,\\
        ((s,m),\epsilon,r)&\to(x_g,y_1,y_2).
    \end{align*}
    Here $\epsilon$ denotes the empty string, so
    $\cY_1=\{\epsilon\}$.
    Additionally, take the function $f$ in
    \cref{lem:compressed-chaining} to be
    \[
        f(x_g,y_1\concat y_2)=f((s,m),r)=(s,m)=x_h.
    \]
    Note that $y_2$ comes from a set of size $|\cY_2|=2^\secp$.
    Let $t=\mathsf{poly}(\secp)$ bound the total number
    of complete calls to $O_f^\SEQ$ before $\cA$ outputs, including
    any calls made by the simulator. Each complete call uses at
    most two queries to $G$, two queries to $H$, and two calls to
    $\cO_\sigma$, counting both computation and uncomputation.
    Thus $t$ counts complete SEQ calls, not individual calls to
    $G$ or $H$.

    With these new variable names, on the branch where
    the SEQ check permits the query, $O_f^\SEQ$ proceeds as follows:
    \begin{enumerate}
        \item Compute $y_1\concat y_2\coloneqq G(x_g)$.
        \item Compute $x_h\coloneqq f(x_g,y_1\concat y_2)$.
        \item Compute $z\coloneqq H(x_h,y_2)$ in a
        work register. Here $z$ is the value denoted $r_1$ in the
        original notation.
        \item Compute the signature using $\cO_\sigma$,
        copy the output $(r,r_1,\sigma_\BUS)$ to the output register,
        and flip $b$, as in the definition of $O_f^\SEQ$ in hybrid 9.
        \item Uncompute the signature, then $z$, then
        $x_h$, and finally $y_1\concat y_2$. The two oracle calls in
        this last step are to $H$ and then $G$.
    \end{enumerate}

    % ADDED: check the controls, the signing computation, and the
    % reverse-order calls against the proof of the chaining lemma.
    Let $E$ be the projector defined in the proof of
    \cref{lem:compressed-chaining}: it projects onto database pairs
    such that every $(x_h\concat y_2,z)\in D_H$ has a matching
    $(x_g,y_1\concat y_2)\in D_G$ with
    $f(x_g,y_1\concat y_2)=x_h$. Write $\overline E=I-E$.

    The SEQ check is the first step in the description of
    $O_f^\SEQ$ in hybrid 9. For a fixed query $(s,m)$, it tests only
    whether a different cell $(s,m')$, with $m'\neq m$, is occupied.
    Its coherent computation and uncomputation therefore commute
    with $E$ and with $\Decomp_G$ at $x_g=(s,m)$. The signing
    computation, its inverse, and the output-copying operation
    leave $D_G,D_H$ and the input registers $(s,m,r)$ unchanged.
    Thus they also commute with $E$ and with $\Decomp_G$.

    Expand each of the two $G$ queries as
    $\Decomp_G\circ\CO'_G\circ\Decomp_G$, as in
    \cref{sec:compressed-oracle}. The $\Decomp$ after the
    forward copy and the decompression before the inverse copy
    commute with all operations between them and cancel, since
    $\Decomp_G^2=I$. Hence exactly two $\Decomp_G$ operations remain
    in each complete permitted query: one at the beginning and one
    at the end.

    To check the inputs to both $H$ queries in this representation,
    recall that at query boundaries each cell of a compressed
    database is orthogonal to its uniform non-$\bot$ state. For
    $D_G$, this holds initially and is preserved by compressed
    queries. It is also preserved by the SEQ check, which tests
    only whether cells are $\bot$ and does not distinguish their
    non-$\bot$ values. Consequently, the first $\Decomp_G$ leaves
    the queried cell non-$\bot$, and $\CO'_G$ copies its value
    $y_1\concat y_2$ into the work register. Throughout both the
    forward and inverse $H$ queries, $D_G$ therefore contains
    $(x_g,y_1\concat y_2)$, and the input to $H$ is
    $(f(x_g,y_1\concat y_2),y_2)$.
    Each of these $H$ queries changes only an already-witnessed
    cell of $D_H$, so it preserves $E$ and $\overline E$ on these
    states. The inverse call has the same property as the forward
    call. This is the argument of
    \cref{claim:chaining-co-H}, applied to both calls, not just the
    forward one. All other operations between the two remaining
    decompressions leave $D_G$ and $D_H$ unchanged.

    Applying \cref{claim:chaining-decomp} to these two $\Decomp_G$
    therefore gives, for every normalized state $\ket{\psi}$ reached
    before a complete query,
    \[
        \norm{\overline E\,O_f^\SEQ\ket{\psi}}
        \leq \norm{\overline E\ket{\psi}}
             +2\sqrt{\frac{2}{|\cY_2|}}.
    \]
    The same estimate holds for a superposition of permitted and
    skipped queries: the SEQ control commutes with the operators
    just considered, and the skipped branch is the identity.
    % Equivalently, apply decompression estimate on each of the two orthogonal control subspaces, using its homogeneous form for subnormalized vectors.

    Third,
    \cref{lem:compressed-chaining} implies that right after $\cA$
    outputs $(m_0,m_1,\sigma_0,\sigma_1,s)$, the state of the system
    gives squared norm at least $1-8t^2/|\cY_2|$ to
    databases $(D_G,D_H)$ that satisfy the following condition:
    for all $(x_h\concat y_2,z)\in D_H$, there exists an entry
    $(x_g,y_1\concat y_2)\in D_G$ such that
    $f(x_g,y_1\concat y_2)=x_h$.
    Indeed, both databases are initially empty, and
    operations between complete SEQ calls do not act on the
    databases. If $\ket{\psi_t}$ is the normalized joint state at
    this point, purified to include all measurement records, then
    \[
        \norm{\overline E\ket{\psi_t}}
        \leq 2t\sqrt{\frac{2}{|\cY_2|}},
        \qquad
        \norm{\overline E\ket{\psi_t}}^2
        \leq\frac{8t^2}{|\cY_2|}.
    \]

    %Fourth, let us translate back to our original notation.
    The above bound implies that right after $\cA$
    outputs $(m_0,m_1,\sigma_0,\sigma_1,s)$, the state of the system
    gives squared norm at least $1-8t^2/2^\secp$ to
    databases $(D,D_1)$ that satisfy the following condition:
    for all $(s,m,r,r_1)\in D_1$, there exists an entry
    $(s',m',r)\in D$ such that $(s,m)=(s',m')$.

    Fifth, in hybrid 10, the challenger checks that for every entry
    $(s,m,r,r_1)\in D_1$, there is the entry $(s,m,r)\in D$.
    The argument above implies that this check passes with
    overwhelming probability.
    More precisely, the accepting component is
    $E\ket{\psi_t}$, and
    \[
        \norm{\ket{\psi_t}-E\ket{\psi_t}}
        =\norm{\overline E\ket{\psi_t}}
        \leq 2t\sqrt{\frac{2}{2^\secp}}.
    \]
    Inserting this check and outputting $0$ on failure changes the
    probability of any subsequent acceptance by at most twice this
    vector distance, which is negligible for polynomial $t$.
    Since this check is the only difference between hybrids 9 and 10,
    this shows that the output distributions of hybrids 9 and 10
    are statistically close.
\end{proof}

\jiahui{Original proof has minor issues. Fixed.}

\begin{claim}
    For any QPT adversary $\cA$, there is a negligible function $\negl(\cdot)$ such that for any $\secp \in \bbN$, $\Pr[\text{Hybrid }10(\secp, \cA) \to 1] \leq \negl(\secp)$.
\end{claim}

\begin{proof}
    First, parse $\sigma_i=(r^i,r_1^i,\sigma_\BUS^i)$ for $i\in\bit$,
    as in item \ref{unforgeability-game-step:new-checks}; malformed
    outputs are rejected. The output of hybrid 10 is $1$ only if
    both of the following events occur:
    \begin{itemize}
        \item Event $E_1$: The challenger finds that for every entry
        $(s,m,r,r_1)\in D_1$, there is the entry $(s,m,r)\in D$.
        \item Event $E_2$: The challenger finds that $m_0\neq m_1$,
        $r_1^0=H_1(s,m_0,r^0)$, and $r_1^1=H_1(s,m_1,r^1)$.
    \end{itemize}
    We will show that
    \[
        \Pr[E_1\land E_2]\leq \frac{2}{2^\ell},
    \]
    which suffices to prove that hybrid 10 outputs $1$ with negligible
    probability. Here $H_1$ has range $\bit^\ell$, as specified in
    hybrid 8. For this upper bound, we may omit the ordinary
    signature-verification checks in
    item \ref{unforgeability-game-step:old-checks}, but still 
    requiring $m_0\neq m_1$. Those signature checks do not act on
    the databases and commute with the checks defining $E_1$ and
    $E_2$.

    Second, right after the challenger checks $(D,D_1)$, every
    database basis state with nonzero amplitude has the property
    that no two entries of $D$ have the same value of $s$.
    We prove this inductively. Before any queries have been made to
    $H$, the database $D$ is empty, so the property holds.
    Consider a query $(s,m)$ to $O_f^\SEQ$ and assume that the
    property holds at the start of the query. By the SEQ check in
    item \ref{step:seq-oracle-SEQ-check} of
    \Cref{fig:seq-oracle-with-serial-numbers}, also used in the
    description of $O_f^\SEQ$ immediately before hybrid 10, the
    oracle skips the query if $D$ contains an entry $(s,m',*)$ with
    $m'\neq m$. Otherwise, $D$ either has no entry with serial
    number $s$ or has its unique such entry at $(s,m)$.
    The queries to $H$, including uncomputation, modify only the
    cell at $(s,m)$. They may add or remove this entry or change its
    value, but cannot introduce two entries with the same serial
    number. The queries to $H_1$ and $\cO_\sigma$ do not
    modify $D$. Thus the property holds after the complete query,
    also for a superposition of query inputs. The final check
    defining $E_1$ is diagonal in the database basis, so it preserves
    this property.

    Third, the component on which the check defining $E_1$ succeeds
    is supported on databases $(D,D_1)$ with the following properties:
    \begin{enumerate}
        \item No two entries of $D$ have the same $s$-value.
        \item For every entry $(s,m,r,r_1)\in D_1$, there is the
        entry $(s,m,r)\in D$.
    \end{enumerate}
    These properties imply that $D_1$ does not contain both an entry
    $(s,m,*,*)$ and an entry $(s,m',*,*)$ with $m\neq m'$.
    Otherwise, $D$ would contain $(s,m,*)$ and $(s,m',*)$, contrary
    to the first property. This statement holds on each database
    basis state in the component; it does not assert that the
    databases have been fully measured.

    Fix a classical output tuple $(m_0,m_1,\sigma_0,\sigma_1,s)$ with
    $m_0\neq m_1$. Let $\ket{\varphi}$ be the unnormalized component
    obtained by keeping the outcome $E_1$ of the challenger's
    database check, starting from the normalized state conditioned
    on this fixed output tuple. Include any registers needed to
    purify the state. Thus $\norm{\ket{\varphi}}^2$ is the
    probability that the $E_1$ check succeeds for this fixed tuple,
    and is at most $1$. Define the two fixed inputs
    \[
        v_i:=(s,m_i,r^i)\qquad(i\in\bit).
    \]
    Since $m_0\neq m_1$, we have $v_0\neq v_1$.
    The two properties above imply that every database basis state
    occurring in $\ket{\varphi}$ has at least one of
    $D_1(v_0),D_1(v_1)$ equal to $\bot$.

    Decompose $\ket{\varphi}$ as
    \[
        \ket{\varphi}=\ket{\varphi_0}+\ket{\varphi_1},
    \]
    where $\ket{\varphi_0}$ consists of its database basis
    components satisfying $D_1(v_0)=\bot$, and
    $\ket{\varphi_1}$ consists of those satisfying
    $D_1(v_0)\neq\bot$. On the latter components,
    $D_1(v_1)=\bot$ necessarily holds. This is an orthogonal
    decomposition, so
    \[
        \norm{\ket{\varphi_0}}^2+
        \norm{\ket{\varphi_1}}^2
        =\norm{\ket{\varphi}}^2.
    \]
    We use this decomposition only in the calculation; the
    challenger does not measure which component occurs.

    Fourth, set $N:=|\bit^\ell|=2^\ell$. For each $i\in\bit$, let
    $V_i$ be the linear map that appends a fresh answer register in
    $\ket{0^\ell}$, makes a compressed-oracle query to $H_1$ at the
    fixed input $v_i$, and keeps only the answer $r_1^i$.
    We suppress the answer register after projecting it onto this
    fixed value. Thus $V_i$ describes the accepting part of the
    corresponding equality check in
    item \ref{unforgeability-game-step:new-checks}, and
    $\norm{V_i\ket{\chi}}\leq\norm{\ket{\chi}}$ for every vector
    $\ket{\chi}$.

    We first compute its action when $D_1(v_i)=\bot$.
    By the definition $\CO=\Decomp\circ\CO'\circ\Decomp$ in
    \Cref{sec:compressed-oracle}, for a database $d$ with
    $d(v_i)=\bot$ the initial decompression gives
    \[
        \Decomp_{v_i}\ket{d}
        =\frac{1}{\sqrt N}\sum_{a\in\bit^\ell}
          \ket{d\cup\{(v_i,a)\}}.
    \]
    Here $\Decomp_{v_i}$ acts on the database of $H_1$.
    Copying the value into the answer register, keeping the answer
    $r_1^i$, and applying the final decompression therefore gives
    \[
        V_i\ket{d}
        =\frac{1}{\sqrt N}\Decomp_{v_i}
          \ket{d\cup\{(v_i,r_1^i)\}}.
    \]
    Different residual databases $d$ in this calculation give
    orthogonal vectors before the final decompression, which is
    unitary. Consequently, for every vector $\ket{\chi}$ supported
    on $D_1(v_i)=\bot$, including vectors entangled with other
    registers,
    \[
        \norm{V_i\ket{\chi}}
        =\frac{1}{\sqrt N}\norm{\ket{\chi}}.
    \]

    Because $v_0\neq v_1$, the maps $V_0$ and $V_1$ act on different
    database cells and commute. Applying the preceding equality to
    the fixed absent cell in each component gives
    \begin{align*}
        \norm{V_1V_0\ket{\varphi_0}}
        &\leq\norm{V_0\ket{\varphi_0}}
         =\frac{\norm{\ket{\varphi_0}}}{\sqrt N},\\
        \norm{V_1V_0\ket{\varphi_1}}
        &=\norm{V_0V_1\ket{\varphi_1}}
         \leq\norm{V_1\ket{\varphi_1}}
         =\frac{\norm{\ket{\varphi_1}}}{\sqrt N}.
    \end{align*}
    By linearity, the triangle inequality, and Cauchy--Schwarz,
    \begin{align*}
        \norm{V_1V_0\ket{\varphi}}
        &\leq
          \norm{V_1V_0\ket{\varphi_0}}
          +\norm{V_1V_0\ket{\varphi_1}}\\
        &\leq
          \frac{\norm{\ket{\varphi_0}}+
                \norm{\ket{\varphi_1}}}{\sqrt N}\\
        &\leq
          \sqrt{\frac{2}{N}}\,
          \sqrt{\norm{\ket{\varphi_0}}^2+
                \norm{\ket{\varphi_1}}^2}\\
        &=\sqrt{\frac{2}{N}}\norm{\ket{\varphi}}.
    \end{align*}
    The squared norm on the left is the joint probability of the
    $E_1$ check and both equality checks succeeding for the fixed
    output tuple, with the ordinary signature checks omitted.
    It is at most
    \[
        \frac{2}{N}\norm{\ket{\varphi}}^2\leq\frac{2}{N}.
    \]
    The bound is uniform over the classical output tuple. Tuples
    with $m_0=m_1$, or malformed signatures, cannot satisfy $E_2$.
    Averaging over the output tuple and reinstating any omitted
    signature checks therefore gives
    \[
        \Pr[\text{Hybrid }10(\secp,\cA)\to1]
        \leq\Pr[E_1\land E_2]
        \leq\frac{2}{2^\ell}.
    \]
    By the choice of $\ell$ immediately before
    \Cref{fig:OSS-construction}, $\ell$ is the bitlength of a tuple
    $(s,m,r)$ with $r\in\bit^\secp$. Hence $\ell\geq\secp$, and
    $2/2^\ell\leq2^{1-\secp}$ is negligible. This proves the claim.
\end{proof}

The previous claims imply that for any QPT adversary $\cA$, there is a negligible function $\negl(\cdot)$ such that for any $\secp \in \bbN$, $\Pr[\text{Hybrid }1(\secp, \cA) \to 1] \leq \negl(\secp)$. Since hybrid 1 is the unforgeability game, this implies that the OSS scheme satisfies unforgeability.
\end{proof}

\ifllncs

\else\fi
\end{document}